\documentclass[dutch,english]{book}

\usepackage{graphicx}
\usepackage{amsmath}
\usepackage{amscd}
\usepackage{amsfonts}
\usepackage{amsthm}
\usepackage{fancybox}
\usepackage{psfrag}
\usepackage{babel}
\usepackage[a4paper=true,breaklinks=true]{hyperref}

\def\IN{\mathbb{N}}
\def\IZ{\mathbb{Z}}
\def\IR{\mathbb{R}}
\def\slash{\llap /}
\def\slashed#1{#1\!\slash\,}

\newcommand{\DOUBLEFIGURE}[5][ht]{\@dblfloat{figure}[#1]\centerline{%
        \parbox{.45\textwidth}{\centerline{\epsfig{file=#2}}}~~~~
        \parbox{.45\textwidth}{\centerline{\epsfig{file=#3}}}}
        \centerline{\parbox[t]{.45\textwidth}{\caption{#4}}~~~~
        \parbox[t]{.45\textwidth}{\caption{#5}}}\end@dblfloat}

\newcommand{\hepth}[1]{\href{http://xxx.lanl.gov/abs/hep-th/#1}{\tt hep-th/#1}}
\newcommand{\Math}[2]{\href{http://xxx.lanl.gov/abs/math.#1/#2}{\tt math.#1/#2}}

\newcommand\spires[1]{\href{http://www-spires.slac.stanford.edu/spires/find/hep/www?j=#1}}
\newcommand\npb[3]{\spires{NUPHA\%2CB#1\%2C#3}
        {{\it Nucl.\ Phys.\ }{\bf B #1} (#2) #3}}

\newcommand\prsla[3]{{\it Proc.\ Roy.\ Soc.\ Lond.\ }{\bf A #1} (#2) #3}
\newcommand\jhep[3]{{\spires{JHEPA\%2C#1\%2C#3}
{{\it J. High Energy Phys.\ }{\bf #1} (#2) #3}}}
\newcommand\prl[3]   {\spires{PRLTA\%2C#1\%2C#3}
        {{\it Phys.\ Rev.\ Lett.\ }{\bf #1} (#2) #3}}
\newcommand\plb[3]   {\spires{PHLTA\%2CB#1\%2C#3}
        {{\it Phys.\ Lett.\ }{\bf B #1} (#2) #3}}
\newcommand\mpla[3]  {\spires{MPLAE\%2CA#1\%2C#3}
        {{\it Mod.\ Phys.\ Lett.\ }{\bf A #1} (#2) #3}}
\newcommand\cmp[3]   {\spires{CMPHA\%2C#1\%2C#3}
        {{\it Commun.\ Math.\ Phys.\ }{\bf #1} (#2) #3}}
\newcommand\prd[3]   {\spires{PHRVA\%2CD#1\%2C#3}
        {{\it Phys.\ Rev.\ }{\bf D #1} (#2) #3}}
\newcommand\cqg[3]   {\spires{CQGRD\%2C#1\%2C#3}
        {{\it Class.\ and Quant.\ Grav.\ }{\bf #1} (#2) #3}}
\newcommand\prep[3]  {\spires{PRPLC\%2C#1\%2C#3}
        {{\it Phys.\ Rept.\ }{\bf #1} (#2) #3}}
\newcommand\ptp[3]   {\spires{PTPKA\%2C#1\%2C#3}
        {{\it Prog.\ Theor.\ Phys.\ }{\bf #1} (#2) #3}}
\newcommand\npps[3]  {\spires{NUPHZ\%2C#1\%2C#3}
        {{\it Nucl.\ Phys.\ }{\bf #1} {\it(Proc.\ Suppl.)} (#2) #3}}
\newcommand\jmp[3]   {\spires{JMAPA\%2C#1\%2C#3}
        {{\it J.\ Math.\ Phys.\ }{\bf #1} (#2) #3}}
\newcommand\atmp[3] {\spires{00203\%2C#1\%2C#3}
        {{\it Adv.\ Theor.\ Math.\ Phys.\ }{\bf #1} (#2) #3}}

\newcommand{\tr}{\mbox{${\rm tr \;}$}}
\newcommand{\Tr}{\mbox{${\rm Tr \;}$}}
\newcommand{\arctanh}{\mbox{${\rm arctanh \;}$}}
\newcommand{\tail}{\mbox{$({\rm tail})$}}
\newcommand{\rs}[1]{\raisebox{1.5ex}[0pt][0pt]{#1}}
\newcommand{\dual}[1]{{}^*\!#1}

\catcode`@=11
\@addtoreset{equation}{section}
\def\theequation{\thesection.\arabic{equation}}

\def\lhead{\@ifnextchar[{\@xlhead}{\@ylhead}}
\def\@xlhead[#1]#2{\gdef\@elhead{#1}\gdef\@olhead{#2}}
\def\@ylhead#1{\gdef\@elhead{#1}\gdef\@olhead{#1}}

\def\chead{\@ifnextchar[{\@xchead}{\@ychead}}
\def\@xchead[#1]#2{\gdef\@echead{#1}\gdef\@ochead{#2}}
\def\@ychead#1{\gdef\@echead{#1}\gdef\@ochead{#1}}

\def\rhead{\@ifnextchar[{\@xrhead}{\@yrhead}}
\def\@xrhead[#1]#2{\gdef\@erhead{#1}\gdef\@orhead{#2}}
\def\@yrhead#1{\gdef\@erhead{#1}\gdef\@orhead{#1}}

\def\lfoot{\@ifnextchar[{\@xlfoot}{\@ylfoot}}
\def\@xlfoot[#1]#2{\gdef\@elfoot{#1}\gdef\@olfoot{#2}}
\def\@ylfoot#1{\gdef\@elfoot{#1}\gdef\@olfoot{#1}}

\def\cfoot{\@ifnextchar[{\@xcfoot}{\@ycfoot}}
\def\@xcfoot[#1]#2{\gdef\@ecfoot{#1}\gdef\@ocfoot{#2}}
\def\@ycfoot#1{\gdef\@ecfoot{#1}\gdef\@ocfoot{#1}}

\def\rfoot{\@ifnextchar[{\@xrfoot}{\@yrfoot}}
\def\@xrfoot[#1]#2{\gdef\@erfoot{#1}\gdef\@orfoot{#2}}
\def\@yrfoot#1{\gdef\@erfoot{#1}\gdef\@orfoot{#1}}

\newdimen\headrulewidth
\newdimen\footrulewidth
\newdimen\plainheadrulewidth
\newdimen\plainfootrulewidth
\newdimen\headwidth
\newif\if@fancyplain \@fancyplainfalse
\def\fancyplain#1#2{\if@fancyplain#1\else#2\fi}

\footrulewidth\z@
\plainheadrulewidth\z@
\plainfootrulewidth\z@

\def\@fancyhead#1#2#3#4#5{#1\hbox to\headwidth{\vbox{\hbox
{\rlap{\parbox[b]{\headwidth}{\raggedright#2\strut}}\hfill
\parbox[b]{\headwidth}{\centering#3\strut}\hfill
\llap{\parbox[b]{\headwidth}{\raggedleft#4\strut}}}\headrule}}#5}

\def\@fancyfoot#1#2#3#4#5{#1\hbox to\headwidth{\vbox{\footrule
\hbox{\rlap{\parbox[t]{\headwidth}{\raggedright#2\strut}}\hfill
\parbox[t]{\headwidth}{\centering#3\strut}\hfill
\llap{\parbox[t]{\headwidth}{\raggedleft#4\strut}}}}}#5}

\def\headrule{{\if@fancyplain\headrulewidth\plainheadrulewidth\fi
\hrule\@height\headrulewidth\@width\headwidth \vskip-\headrulewidth}}

\def\footrule{{\if@fancyplain\footrulewidth\plainfootrulewidth\fi
\vskip-0.3\normalbaselineskip\vskip-\footrulewidth
\hrule\@width\headwidth\@height\footrulewidth\vskip0.3\normalbaselineskip}}

\def\ps@fancy{
\let\@mkboth\markboth
\@ifundefined{chapter}{\def\sectionmark##1{\markboth
{\ifnum \c@secnumdepth>\z@
 \thesection\hskip 1em\relax \fi ##1}{}}
\def\subsectionmark##1{\markright {\ifnum \c@secnumdepth >\@ne
 \thesubsection\hskip 1em\relax \fi ##1}}}
{\def\chaptermark##1{\markboth {\ifnum \c@secnumdepth>\m@ne
 \@chapapp\ \thechapter. \ \fi ##1}{}}
\def\sectionmark##1{\markright{\ifnum \c@secnumdepth >\z@
 \thesection. \ \fi ##1}}}
\def\@oddhead{\@fancyhead\relax\@olhead\@ochead\@orhead\hss}
\def\@oddfoot{\@fancyfoot\relax\@olfoot\@ocfoot\@orfoot\hss}
\def\@evenhead{\@fancyhead\hss\@elhead\@echead\@erhead\relax}
\def\@evenfoot{\@fancyfoot\hss\@elfoot\@ecfoot\@erfoot\relax}

\headwidth\textwidth}
\def\ps@fancyplain{\ps@fancy \let\ps@plain\ps@plain@fancy}
\def\ps@plain@fancy{\@fancyplaintrue\ps@fancy}

\catcode`\@=12

\begin{document}

\baselineskip13pt

\pagestyle{fancy}

\frontmatter

\begin{titlepage}

\begin{center}
\large{Vrije Universiteit Brussel\\
Faculteit Wetenschappen\\
Theoretische Natuurkunde\\}
\vspace{5.5cm}
\Large{Abelian and Non-abelian D-brane Effective Actions}\\
\vspace{12pt}
\large{\textsc{Paul Koerber}}\\
\vspace{12pt}
\end{center}
\vspace{4cm}
Promotor: Prof. Dr. A. Sevrin
\vspace{-2.1em}
\begin{flushright}
Proefschrift ingediend\\
met het oog op het\\
behalen van de graad van\\
Doctor in de Wetenschappen\\
\end{flushright}
\vspace{1.8cm}
\begin{center}
2004
\end{center}
\end{titlepage}

\newpage
\thispagestyle{empty}
\mbox{}
\vspace{14cm}\\
The cover photo shows the statue of Pythagoras in Pythagoreion, Samos.

\newpage
\thispagestyle{empty}
\mbox{}

\vspace*{7cm}
\begin{flushright}
{\it There is geometry in the humming of the strings.} \\
{\scriptsize \sc Pythagoras (569BC-475BC)}
\end{flushright}

\newpage
\thispagestyle{empty}
\mbox{}
\newpage

\section*{Acknowledgements}

The first words of thanks I extend to my thesis advisor, Alexander
Sevrin, for five years of fruitful cooperation before and during
the work on my Ph.D.\ thesis, for encouragement when the flood of
string theory was overwhelming at times, for his contagious
enthusiasm, fail-safe support and for proofreading the manuscript.
Further thanks goes to the members of my jury: Professors J.~Van Craen,
R.~Roosen, A.~Sevrin, F.~Lambert, P.~Di Vecchia,
M.~Henneaux and A.~Van Proeyen.

I give thanks to the usual audience of the seminars in Leuven and
Brussels for creating a congenial scientific atmosphere.
This includes the people from (or formerly from)
Leuven: (alphabetically) Bert Cosemans, Yves Demasure, Jos
Gheerardyn, Lennaert Huiszoon, Bert Janssen, Tassilo Ott, Geert Smet, Walter Troost,
Joris Van Den Bergh and Antoine Van Proeyen; the people from (or formerly
from) ULB:  Riccardo Argurio, Glenn Barnich, Xavier Bekaert, Giulio Bonelli, Nicolas Boulanger,
Sandrine Cnockaert, Sophie de Buyl and Marc
Henneaux; and of course the people from (or formerly from) VUB: Jo
Bogaerts, Lies De Foss\'e, Arjan Keurentjes, Stijn Nevens, Alexander Sevrin,
Jan Troost and Alexander Wijns.

Special thanks to Martijn Eenink and Mees de Roo --- both from Groningen ---
for productive collaborations, to Arjan Keurentjes, Jan Troost and Walter Troost for
their patient and adept explanations, to Bert Cosemans for being
my roommate at several conferences and to Martijn Eenink,
Stijn Nevens, Geert Smet, Joris Van Den Bergh and Alexander Wijns
for enjoyable times at those same conferences and for becoming friends.

For inspiring discussions at the ``office'', I would like to thank
Arjan Keurentjes, Stijn Nevens and Alexander Wijns while for all
further practical matters Rudi Vereecken --- our secretary --- was
the man to turn to. For the wearing task of proofreading the
manuscript, next to Alexander Sevrin the following people deserve
my gratitude: Malcolm Fairbairn, Alexander Wijns and the members
of the jury.

As for the people not related to physics, but who still
contributed, indirectly, to the coming about of this thesis, I
would like to mention my parents and my girlfriend Isabel for
their unceasing support.

On the financial level, this work would not be possible without the
Research Assistantship Grant (Aspirant FWO) I received from the
Fund for Scientific Research in Flanders (F.W.O.-Vlaanderen, research project G.0034.02).
Furthermore, our group is supported by the Free University of Brussels (Vrije Universiteit
Brussel), by the Belgian Federal Science Policy Office
through the Interuniversity Attraction Pole P5/27 and
by the European Commission RTN programme
HPRN-CT-2000-00131,
{\em The Quantum Structure of Space-time and the Geometric Nature of
Fundamental Interactions}, in which our group is associated
to the University of Leuven.

\subsection*{Trademark Notice}

``Java and all Java-based marks'' are trademarks or
registered trademarks of Sun Microsystems, Inc. in the United
States and other countries.

\newpage
\thispagestyle{empty}

\parskip 6pt plus 1pt minus 1pt

\lhead[\fancyplain{}{}]{\fancyplain{}{}}

\rhead[\fancyplain{}{}]{\fancyplain{}{}}

\chead{\fancyplain{}{\bf Contents}}

\cfoot{\fancyplain{}{}}

\tableofcontents

\newpage
\thispagestyle{empty}

\mainmatter

\parskip 5pt plus 1pt minus 1pt

\lhead[\fancyplain{}{\bf\thepage}]{\fancyplain{}{}}

\chead[\fancyplain{}{\bf\leftmark}]{\fancyplain{}{\bf\rightmark}}

\rhead[\fancyplain{}{}]{\fancyplain{}{\bf\thepage}}

\cfoot{\fancyplain{\bf\thepage}{}}

\chapter{Introduction}
\label{introduction}

\section*{Is String Theory a Fundamental Theory of Nature?}

An optimist would claim so. He would point out that it
incorporates all the rules of quantum field theory while at the
same time it is a theory of gravity.  That the combination of
those two is indeed a unique property was learned by trial and
(much) error through previous attempts to reconcile the two most
important theories of established physics: general relativity
describing gravity and quantum field theory describing the other
forces. The problem of earlier theories of quantum gravity is
their non-renormalizability, whereas string theory
is in fact UV finite. If pressed a little bit, the optimist would
have to admit that there exist other attempts to pull off the
task of quantizing gravity, like loop quantum gravity, but he would gladly continue that
string theory not only incorporates gravity, but also in a very
natural way gauge theories and, under the condition of accepting
supersymmetry, matter i.e.\ fermions. Indeed, all the types of
particles that make up the Standard Model --- gauge bosons, Higgs
bosons and fermions --- have together with the graviton their place
in the oscillation spectrum of the string. Moreover, consistency is
so restrictive that there are only five different perturbative string theories,
which are --- together with 11-dimensional supergravity ---
related to each other by dualities.  So in a way string theory is
unique.

A pessimist would claim that string theory fails the above question
in two ways: firstly it is not even a (complete) theory and
secondly it does not describe nature.  It is true that
perturbative string theories are in fact merely a tool to construct
the Feynman diagrams of some deeper theory. From our experience with
quantum field theories we know that such expansions miss important
physics, one notorious example being confinement.
In the same way perturbative string theories only provide part of the whole picture;
they lack non-perturbative information.
However, in recent years a lot of progress has been made on this point.
For instance, dualities linking the different string theories
relate in some cases weak coupling to strong coupling so that
perturbative calculations in the first theory provide non-perturbative
information on the second theory. Furthermore, an important class of
non-perturbative solitons has been found: branes.
The fundamental degrees of freedom of
the underlying (non-perturbatively valid) theory, baptized M-theory,
may remain largely unknown, at least most people working
on string theory are now convinced that this theory must exist.
Rather than being a completed theory, M-theory remains very much
work in progress.

The pessimist would hit harder if he were to bring up the lack of
experimental evidence. Indeed, despite all of the promises string theory does not
make a single hard verifiable prediction. Neither do rival theories
of quantum gravity. It is possible to construct configurations in string
theory that resemble to a high extent the Standard Model,
for instance by using intersecting D-branes. However, as of yet there is no
way to single out these models as a preferred vacuum. This touches on another
weak point: through the web of dualities, in theory there may be a unique string theory,
in practice there is an overabundance of possible vacua.
Indeed, we can study string theory around a certain background, but we lack a kind of
global potential function assigning values to all possible backgrounds and allowing us
to select the ground state.
Under these circumstances it is not possible to fulfill an old promise
and make predictions about all or some of the free parameters of the Standard
Model.

Since string theory leans heavily on supersymmetry the discovery, hopefully
at LHC, of supersymmetric partners of particles in the Standard Model is
looked forward to. However, supersymmetry is shared with many other theories,
most notably the Supersymmetric Standard Model, so that the experimental
discovery of supersymmetry would hardly be full confirmation of string theory.
Because of the extremely high energies involved, perhaps the future of experimental
verification of string theory lies not in particle accelerators, but in
its astrophysical and cosmological predictions. At this point the optimist would remark that
string theory has already passed an important
test in (partially) solving a theoretical problem that arises when describing
a typical general relativistic object, a black hole, in a quantum mechanical
way: it succeeds in calculating the semi-classically predicted entropy of a
(nearly) supersymmetric black hole by counting its microstates.
Unfortunately, many tough nuts rest to crack in these domains like the
explanation of the observed small positive cosmological constant and
the construction of string theory in time-dependent backgrounds.
Most probably, it is just the fate of contemporary fundamental physics that
the time span between theory and experimental verification becomes
ever larger.

However, the discussion so far left out that string theory is sometimes
an incredibly powerful tool in other fields of physics and mathematics.
In this small space we can only give a few examples.
Most notably there is the connection with gauge theories.  It turns out that
many properties of gauge theory have a geometric interpretation in terms
of D-branes.  Some time ago 't Hooft noted that the large N limit
of gauge theories very much looked like a string theory.  A first
concrete realization of such a connection was the AdS/CFT correspondence,
which states that $N=4$ Super-Yang-Mills theory is dual to string
theory on $AdS_5 \times S^5$.
Other examples are the incorporation of Montonen-Olive duality of
gauge theory in the larger S-duality of string theory and the
recent advances in the non-perturbative calculation of the chiral sector of
$N=1$ Super-Yang-Mills.  Then there is the natural incorporation
in string theory of non-(anti)commutative field theories and, perhaps
more exotic, the connection between topological strings and knot theory.
Again, the pessimist would be quick to point out that many of these
connections are not proven in a strict mathematical sense. Indeed,
for instance the AdS/CFT correspondence and S-duality are in fact conjectures, but
in the meantime an impressive amount of indirect evidence has been found.


While the argument between the optimist and the pessimist could go on
for a while, I would like to note that string theory is in any case a
generating functional of a lot of beautiful problems, which is reason enough
for me to work on it. So let us get started!

\section*{Plan}

Chapter \ref{landscape} opens with a broad overview of string theory.
It glances over perturbative string theories first and then turns to
non-perturbative effects with an emphasis on D-branes.  The last part
of the chapter introduces the subject of this thesis, the low-energy effective
action of a single D-brane (abelian) and of multiple coinciding D-branes
(non-abelian).  It gives an overview of past attempts and successes in
constructing these effective actions. The method developed in this thesis
is based on an extension of the instanton equations.

Chapter \ref{equations} gradually builds up. It starts with the
familiar instanton equations in four-dimensional Yang-Mills theory and
first extends them to more than four dimensions. It turns out that
there are two important types: the {\em complex} and the {\em
octonionic} equations. In the complex case we find the holomorphicity
condition and a condition called after Donaldson, Uhlenbeck and Yau (DUY).
The original work in this chapter starts
when these equations are in turn extended from Yang-Mills to
Born-Infeld theory.

In chapter \ref{solutions} we review solutions to these equations and show that they have
an interpretation in D-brane physics. As such it is reasonable to require these configurations
to be solutions of the full D-brane effective action, which we aim to construct.
The configuration of intersecting D-branes will be important later on
when we devise a check on our action.  We comment on solutions without small field strength
limit and outline future work on the octonionic generalization of the BIon.

Chapter \ref{abelianBI} treats the approach of \cite{uniqueabelian} to construct the
abelian Born-Infeld action from the requirement that solutions of the extended complex equations
should solve the equations of motion. Starting from the action in \cite{wyllard}, we also derive the
all-order in $\alpha'$ result for the terms in the DUY condition with 4 derivatives.
Although the calculation is involved, the result can be written in a compact and elegant way.

Chapter \ref{nonabelianBI} summarizes papers \cite{alpha3} and \cite{alpha4}, where the non-abelian
D-brane effective action was constructed up to order $\alpha'{}^4$.  The computer program written to perform
the large calculations necessary to obtain these results can be found on the website \cite{manual},
together with a concise manual and a large sample of its output.

In chapter \ref{checks} we describe the spectrum check, performed in \cite{testalpha3} and \cite{testfermion},
on both the bosonic sector as well as the fermionic sector found by the Groningen group in \cite{groningen}.
The check is based on the spectrum of strings stretching between intersecting D-branes.

While all these chapters constitute a whole, chapter \ref{boundarysusy} lies somewhat outside the
main line of development. Still, the ultimate goal of the efforts in this chapter is again the
construction of the D-brane effective action, but this time using the old method of requiring Weyl invariance
of the non-linear $\sigma$-model.  We want to work
in $N=2$ superspace because we expect the calculations to be easier.  Since there were some problems in the
past constructing a superspace in the presence of
boundaries, we had to sort this out first, which is precisely the content of \cite{susyboundary} and this chapter.

Note that all papers that I co-authored are indicated by letters in contrast to all other referred
articles, which are labelled by numbers. \cite{corfutalk} is a contribution to the proceedings of a workshop
in Corfu, where I presented a talk, and summarizes articles \cite{uniqueabelian}, \cite{alpha3} and \cite{testalpha3}.

\newpage
\thispagestyle{empty}

\chapter{A Landscape Picture}
\label{landscape}

In this chapter we will give a short review of string
theory. Its purpose is to introduce to the non-expert
the broad context for the research in this thesis
and at the same time it is an excellent opportunity for the author to
go through the vast ``general knowledge'' of string theory again
some four years after his first exposure to it.
Regrettably, it would be virtually impossible to
provide the non-expert with all the necessary background to
understand the technical details in the following chapters.
This would require an introduction as thick as the
excellent reviews \cite{bookGSW}, \cite{bookpolchinski}
or \cite{cvj}, to which I humbly refer.

In the first section we
treat string theory in the most strict sense, i.e.\ as one of the
five original perturbative string theories. These are auxiliary
quantum field theories on the world-sheet of a one-dimensional
object --- a string --- that provide the technology
to calculate interaction amplitudes perturbatively in the
coupling constant $g_s$.  The bottom line is that they define
the theory essentially as a set of Feynman diagrams.

However, we know from quantum field theory that
a perturbative expansion is usually not the whole story.
At strong coupling, $g_s \gg 1$, non-perturbative
effects show up, some of the most famous of which are D-branes \cite{polchinski}.
These can be compared to solitons in quantum field theory.
Next to D-branes we will talk in the second section
about the equally famous dualities that exist between different
string theories and in some cases even relate strong coupling to weak
coupling so that the scope of the perturbative analysis can be extended to these
strong coupling regimes through calculations in the dual weak coupling regimes.  We also comment
on the status of the non-perturbatively defined underlying theory, {\em M-theory}. It will become
clear that this is still very much work in progress --- although very exciting.

The subject of this thesis will be the effective action on the world-volume of
D-branes.  In the last section of this chapter we will give a definition
and outline the possible ways of calculating it.

\section{Perturbative String Theory}

\subsection{The Polyakov Action and its Symmetries}

\begin{figure}[!t]
\centering
\setlength{\fboxsep}{10pt}%
\shadowbox{%
\begin{minipage}{.9\textwidth}
\begin{center}
\psfrag{t}{$\tau$}
\psfrag{s}{$\sigma$}
\psfrag{a}{(a)}
\psfrag{b}{(b)}
\psfrag{c}{(c)}
\psfrag{d}{(d)}
\includegraphics[scale=0.44]{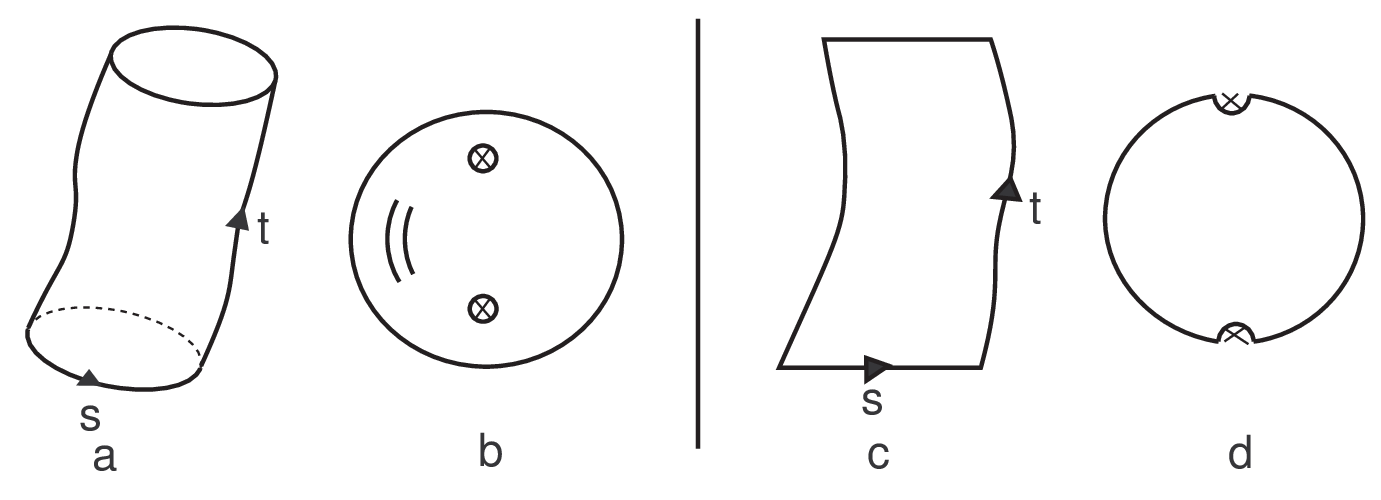}
\caption[String world-sheets embedded in space-time.]{String world-sheets embedded in space-time.  The world-sheet is parameterized
by the coordinates $\tau$ and $\sigma$. Each point $(\tau,\sigma)$ on the world-sheet
is at point $X^{\mu}(\tau,\sigma)$ in space-time. (a) World-sheet of a propagating closed string. (b)
By using the symmetries of the world-sheet (Weyl and reparameterization invariance) surface (a) can be mapped into
a sphere with standard metric. The incoming and outgoing string states are then mapped to a
tiny circle which shrinks to a point when we take the original states to infinity.
They can then be replaced by a point-like insertion, a {\em vertex operator}, indicated
by a cross. (c) World-sheet of an open string. (d) By using again the symmetries the world-sheet of (c) can
be mapped into a disk. The incoming and outgoing string states are replaced by
vertex operators.  Comparing (b) and (d) we note that open string world-sheets have boundaries, in contrast to closed string world-sheets.}
\label{worldsheet}
\end{center}
\end{minipage}
}
\end{figure}

Perturbative string theories are quantum field theories on the world-sheet of the string, which is
the two-dimensional surface swept out when the string moves in space-time as shown in
figure \ref{worldsheet}.
We parameterize it with the coordinates $\tau=\sigma^0$ (time) and $\sigma=\sigma^1$ (length). Thinking
about the analogue situation of a particle in relativity, the most natural Poin\-ca\-r\'e-invariant
action that springs to mind is to take the area of the world-sheet. This leads to the
{\em Nambu-Goto action} \cite{nambugoto}, which is cumbersome to quantize because
it contains a square-root. Fortunately, by introducing an auxiliary world-sheet metric $h_{ab}$,
it is possible to construct a quadratic action, the {\em Polyakov action}\footnote{This action
was found by Deser and Zumino \cite{dz} and by Brink, Di Vecchia and Howe \cite{bdh} in
generalizing to a theory with local world-sheet supersymmetry.  However, it is usually named after
Polyakov \cite{polyakov} because he used it in his path integral treatment of string theory.}.
Integrating out the auxiliary metric, one recovers the Nambu-Goto action.
The Polyakov action reads
\begin{equation}
S_{\text{P}}[X,h_{ab}] = S_X[X,h_{ab}] - \lambda \chi,
\end{equation}
with
\begin{equation}
\begin{split}
S_X[X,h_{ab}] & = - \frac{1}{4\pi\alpha'} \int d\tau d\sigma \sqrt{-h} h^{ab} \eta_{\mu\nu}\partial_a X^{\mu} \partial_b X^{\nu}, \\
\chi & = \frac{1}{4\pi} \int d\tau d\sigma \sqrt{-h} R,
\end{split}
\label{polyaction}
\end{equation}
where the coordinates $X^{\mu}(\tau,\sigma)$ parameterize the embedding of the world-sheet into target space-time.
$h$ is the determinant of the metric $h_{ab}$ while $R$ denotes the two-dimensional Ricci scalar. The
world-sheet metric $h_{ab}$ has Minkowski signature so that the eigenvalue associated to the timelike dimension
is negative. The parameter $\alpha'$ of dimension length-squared defines
the fundamental string length $\alpha'=l_s^2$. Sometimes the parameter $T=\frac{1}{2\pi\alpha'}$, which
can be interpreted as the {\string tension}, is used instead.
$\chi$ is a total derivative and thus a topological invariant.  As we will soon discuss, it plays
an important role in the string loop expansion.

Figure \ref{worldsheet} shows two types of strings. Indeed,
taking periodic boundary conditions $X^{\mu}(\tau,\sigma+2\pi)=X^{\mu}(\tau,\sigma)$ leads to
closed strings while Neumann boundary conditions at the endpoints $\partial_{\sigma} X^{\mu}|_{0,\pi}=0$ result
in open strings\footnote{For closed strings the length is usually normalized to $2\pi$, for open strings to $\pi$.}.
The latter boundary conditions prevent momentum from flowing off the ends of the string so that the ends are
free. We will see later on that Dirichlet boundary conditions are also possible and correspond to open strings
ending on D-branes.
\begin{figure}[t]
\centering
\setlength{\fboxsep}{10pt}%
\shadowbox{%
\begin{minipage}{.9\textwidth}
\begin{center}
\psfrag{a}{(a)}
\psfrag{b}{(b)}
\includegraphics[scale=0.67]{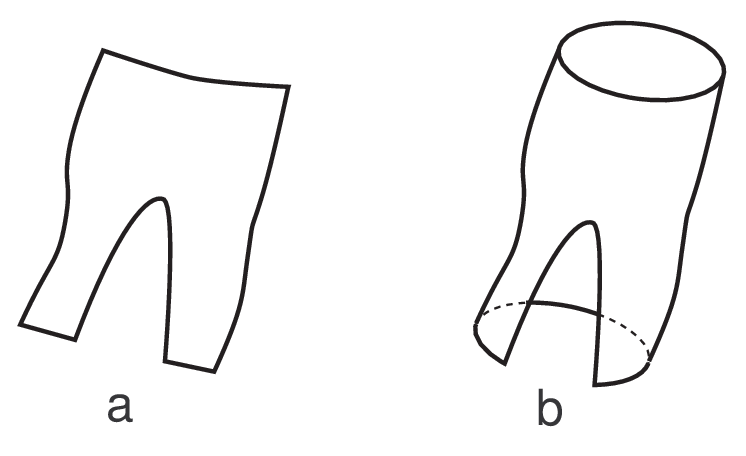}
\caption[Open strings imply closed strings]{Open strings imply closed strings.
Indeed, the process of joining
two open strings (a) looks locally, in the neighbourhood of the interaction, the same as the process of joining the ends of one open
string to form a closed string (b). Allowing the one without the other would require some
non-local constraint.}
\label{openclosed}
\end{center}
\end{minipage}
}
\end{figure}
As figure \ref{openclosed} shows, string theories with open strings will always contain closed strings as well.

Classically strings wave. For closed strings the left- and right-moving waves are independent
while for open strings the boundary conditions force the left- and right-moving modes to combine into
standing waves. Projecting onto states that are invariant under world-sheet parity introduces two more
types of strings: {\em unoriented} open and closed strings.

The Polyakov action is not only invariant under world-sheet general coordinate transformations, but also under a rescaling
of the world-sheet metric $h_{ab} \rightarrow \Lambda h_{ab}$, called a {\em Weyl transformation}.
The latter property is unique to two world-sheet
dimensions, i.e.\ it would not be true for higher-dimensional membranes.
These three local symmetries, reparameterizations of the two coordinates and Weyl symmetry, can, barring topological subtleties, be used to put $h_{ab}=\eta_{ab}$.
Even after choosing this {\em conformal gauge}, a combination of Weyl and reparameterization invariance
remains as a classical symmetry: the {\em conformal symmetry}, which is generated by the infinite-dimensional
{\em Virasoro algebra}. As such, perturbative string theory belongs to the class of conformally invariant
field theories in two dimensions, which are in some cases exactly solvable because of their high amount of
symmetry (see \cite{ginsparg} for a review on bosonic conformal field theory and \cite{fms} for supersymmetric
conformal field theory with applications to string theory).  Due to subtleties of normal ordering the quantum Virasoro algebra will in
general differ from its classical counterpart by the introduction of a {\em central charge} $c$.

\subsection{Spectrum}

A first interesting result of quantizing action \eqref{polyaction} is the spectrum of physical
states and their masses.  Indeed, the quantized oscillation modes of the string determine the
particle content of our theory!
As usual in theories with local symmetries the spectrum contains unphysical states, which
have to be separated from the physical ones.  This requires some technical machinery
(light-cone gauge, old covariant quantization or BRST quantization) that we will not delve into.

Table \ref{openstringspectrum} lists
the lowest mass levels for the open bosonic string. To each of the particles in the table
a field in the low-energy effective theory is associated.
\begin{table}
\begin{equation*}
\begin{array}{|c|c|l|c|}\hline\hline
\text{State} & \rule[-2mm]{0mm}{6mm} \text{Conditions} & (\text{Mass})^2 & \text{Field} \\ \hline
|0;k \rangle & \rule[-2mm]{0mm}{6mm} & M^2=-k^2=-\frac{1}{\alpha'} & T \\
|e;k \rangle & \quad \begin{array}{l}e_{\mu}k^{\mu}=0\\\rule[-2mm]{0mm}{2mm} e_{\mu} \cong e_{\mu} + k_{\mu}\end{array}\;
 & M^2=-k^2=0 & A_{\mu}\\ \hline\hline
\end{array}
\end{equation*}
\caption[The lowest mass states for the open bosonic string.]{The lowest mass states for the open bosonic string. We find a tachyon with low-energy field $T$
and a photon(gluon) with field $A_{\mu}$. $e_{\mu}$ is the polarization vector.}
\label{openstringspectrum}
\end{table}
In the first line one finds a scalar particle with negative mass-squared,
the {\em tachyon}. The presence of such a particle indicates, just as in the case
of the Higgs field, that one is perturbing around an unstable vacuum.  If there is no
other --- stable --- vacuum, the energy is unbounded from below and the theory is inconsistent.
For the open bosonic string theory it is conjectured that a true vacuum exists. Indeed, in modern language,
there is a space-filling D-brane present on which
open strings can end.  We will elaborate on D-branes in section \ref{dbranes}.
The tachyon indicates that this D-brane is not stable: it can decay such that a closed string vacuum results.
The difference in energy density between the two vacua is thus equal to the D-brane tension.
Evidence for this conjecture is found in bosonic string field theory \cite{wittenSFT}\cite{tachyon}.
In any case, in supersymmetric string theory the tachyon can consistently be removed
from the spectrum by the Gliozzi-Scherk-Olive(GSO)-projection \cite{GSO}.

Of more interest to us is the second particle, which has all the
properties of an abelian gauge boson $A_{\mu}$ i.e. it is massless
and gauge symmetry is implemented as in the Gupta-Bleuler treatment
of electrodynamics: $e_{\mu} k^{\mu}=0$ is the Lorentz gauge and $e_{\mu} \cong e_{\mu} + k_{\mu}$
is the residual symmetry. In fact, the simplest way to handle the local symmetries of the
Polyakov action and obtain these conditions for
physical states, old covariant quantization, is very similar to Gupta-Bleuler
quantization. By assigning $N$ different labels to the endpoints of
open strings, Chan-Patton factors \cite{CPfactors}\cite{neveuscherk}, it is
possible to
introduce a $U(N)$ gauge group\footnote{Unoriented string theories
will lead to the gauge groups $SO(N)$ or $USp(2N)$.}. Indeed,
labelling the endpoints of open strings in this way, one
introduces $N^2$ different types of strings filling out the
adjoint representation of $U(N)$.  In modern language this
amounts to introducing $N$ D-branes; see figure \ref{D-branes}.
The bulk part of this thesis will be about constructing an
effective action for this non-abelian gauge particle.

For an example of a spectrum calculation see section \ref{spectrumstringtheory}, where the spectrum is
constructed for open strings with endpoints
on two different intersecting D-branes in type II string theory --- which we will introduce in a moment ---
both in the bosonic and the fermionic sector.

For completeness we state the bosonic closed string spectrum in table \ref{closedstringspectrum}.
\begin{table}
\begin{equation*}
\begin{array}{|c|c|l|c|} \hline\hline
\text{State} & \rule[-2mm]{0mm}{6mm} \text{Conditions} & (\text{Mass})^2 & \text{Field(s)} \\ \hline
|0;k \rangle & & \rule[-2mm]{0mm}{6mm} M^2=-k^2=-\frac{4}{\alpha'} & T \\
|l;k \rangle &  \quad \begin{array}{l}l_{\mu\nu}k^{\mu}=l_{\mu\nu}k^{\nu}=0\\l_{\mu\nu} \cong l_{\mu\nu} + a_{\mu}k_{\nu}+k_{\mu}b_{\nu}
\\\rule[-2mm]{0mm}{2mm} a_{\mu}k^{\mu}=b_{\mu}k^{\mu}=0\end{array} & M^2=-k^2=0 & G_{\mu\nu},B_{\mu\nu},\Phi \\ \hline\hline
\end{array}
\end{equation*}
\caption[The lowest mass states for the closed bosonic string.]{The lowest mass states for the closed bosonic string.
We find a tachyon with field $T$ and a reducible tensor representation resulting
in the fields $G_{\mu\nu}$,$B_{\mu\nu}$ and $\Phi$. $l_{\mu\nu}$ is the polarization tensor.}
\label{closedstringspectrum}
\end{table}
The first particle is again a tachyon $T$, but this time there is no other stable vacuum known.
The second line comprises three different irreducible representations of the Little group --- which is in the
massless case $SO(D-2)$ --- namely a traceless symmetric representation (the graviton with field $G_{\mu\nu}$), an anti-symmetric
representation (with field $B_{\mu\nu}$) and a scalar representation (the dilaton $\Phi$).

Since the bosonic spectrum contains tachyons --- for which for the closed string no stable
vacuum is known --- and no fermions, one should in fact turn to world-sheet supersymmetric
string theories.
For the world-sheet fermions there are two types of boundary conditions,
Ramond(R) and Neveu-Schwarz(NS).  This RNS-model \cite{RNS}
comes with a world-sheet supersymmetric extension of the Polyakov action \cite{dz}\cite{bdh}.
\begin{table}[!t]
\begin{center}
\begin{tabular}{|c|c|c|c|}\hline\hline
Name & Type of Strings & Bosonic Spectrum & Susy\\ \hline
Type I & unoriented open + closed & \rule[-2mm]{0mm}{6.5mm} $G,\Phi,A^{SO(32)}_{[1]},C_{[2]}$ & $N=1$ \\
Type IIA & closed strings & $G,B_{[2]},\Phi,C_{[1]},C_{[3]}$ & $N=2$ \\
Type IIB & closed strings & $G,B_{[2]},\Phi,C_{[0]},C_{[2]},C^+_{[4]}$ & $N=2$ \\
Type HE & heterotic closed strings & $G,B_{[2]},\Phi,A^{E8\times E8}_{[1]}$ & $N=1$\\
Type HO & heterotic closed strings & \rule[-2.5mm]{0mm}{2.5mm} $G,B_{[2]},\Phi,A^{SO(32)}_{[1]}$ & $N=1$\\ \hline \hline
\end{tabular}
\caption[The five consistent supersymmetric perturbative string theories.]{The five consistent supersymmetric perturbative string theories.
They all live in 10 space-time dimensions.
Choosing NS boundary conditions
in both the right-moving and the left-moving sector (NS-NS) leads for type II string theories to the same spectrum
as the bosonic closed string: $G, B_{[2]},\Phi$.  Choosing R-R boundary conditions
leads to antisymmetric tensors of different dimensions: $C_{[n]}$, $n$ even in type IIB and odd
in type IIA. As an unoriented theory, Type I does not include $B_{[2]}$, but adds an antisymmetric
2-tensor $C_{[2]}$ from the R-R sector instead. It introduces the gauge boson $A_{[1]}$
with gauge group $SO(32)$ via Chan-Patton factors.  Heterotic theories \cite{heterotic} combine
the left-moving side from bosonic string theory with the right moving side from supersymmetric
string theory.  The central charge mismatch is filled in by introducing extra left-moving fields with
gauge group indices.\\Space-time fermions come from the mixed sectors, NS-R or R-NS, of the closed
string or from the $R$ sector of the open string. The heterotic theories
and type I contain one gravitino and one spinor, while the type II theories have two gravitinos,
two spinors and a double amount of supersymmetry.  However, for type IIA the spinors are of different
chirality while for type IIB they have the same chirality. The same applies to the
gravitinos. This implies that type IIB is chiral theory, i.e. it has definite handedness%
.
}
\label{susystringtheories}
\end{center}
\end{table}
It is remarkable that by introducing world-sheet fermions and world-sheet
supersymmetry, one obtains in the end space-time fermions in the spectrum and space-time
supersymmetry! Here the GSO-projection plays a non-trivial role, for instance in preserving
the spin-statistics theorem. Space-time supersymmetry thus comes in a rather roundabout way, a problem
on which we comment later on. See table~\ref{susystringtheories} for the particle spectrum of
the five perturbative superstring theories.

The introduction of many strings will back-react on
the background in which a ``probe'' string moves. Indeed, in the presence of a coherent state of
gravitons $G_{\mu\nu}(X)-\eta_{\mu\nu}$, antisymmetric tensor particles $B_{\mu\nu}(X)$ and dilatons $\Phi(X)$,
the Polyakov action reads
\begin{equation}
\begin{split}
& S_{\text{P,curv.}}^{\text{bulk}}[X,h_{ab}] = \\
& - \frac{1}{4\pi\alpha'} \int d\tau d\sigma \sqrt{-h} \left( \left(h^{ab} G_{\mu\nu}(X) + \epsilon^{ab} B_{\mu\nu}\right) \partial_a X^{\mu} \partial_b X^{\nu}
+ \alpha' R \Phi(X) \right).
\label{polyactioncurved1}
\end{split}
\end{equation}
Here $\epsilon^{ab}$ is a tensor normalized so that $\sqrt{-h}\epsilon^{01}=1$.
This action describes a set of scalar fields --- the $X^{\mu}$ --- of which the kinetic term
has a field dependent coefficient; it contains $G_{\mu\nu}(X)$. This kind of action is often called
a {\em non-linear $\sigma$-model}.
For open strings, the following boundary term is added:
\begin{equation}
S_{\text{P,curv.}}^{\text{boundary}}[X,h_{ab}]= - \int d \sigma^a A_{\mu} \partial_a X^{\mu}
= - \int d \tau \left(\left. A_{\mu} \partial_{\tau} X^{\mu}\right|_{\sigma=0} - \left. A_{\mu} \partial_{\tau} X^{\mu}\right|_{\sigma=\pi}\right),
\label{polyactioncurved2}
\end{equation}
where $A_{\mu}(X)$ corresponds to a coherent state of gauge bosons.
In chapter \ref{boundarysusy} we will study the supersymmetric version of this action in the presence
of a boundary.

\subsection{Amplitudes}

Let us now take a look at an $n$-point scattering amplitude, i.e.\ an amplitude with $n$ incoming and outgoing
string states at infinity.  As explained in figure \ref{worldsheet},
one can after an appropriate transformation localize these states at a point and replace them by the
insertion of a vertex operators ${\cal V}_{(j_i,k_i)}$ where $j_i$ and $k_i$ denote the
internal state and momentum of the $i$th external string.
The $n$-point amplitude is given schematically
by the following path integral over the coordinates $X(\tau,\sigma)$ and the world-sheet metric $h(\tau,\sigma)$
\footnote{In the Polyakov path-integral approach a Wick rotation is first made so that the signature
of the world-sheet becomes Euclidean.  The path integral over all metrics is then better defined.},
\begin{equation}
S_{j_1,\ldots,j_n}(k_1,\ldots,k_n) = \sum_{\text{topologies}} \int \frac{[dX dh]}{V_{\text{diff}\times\text{Weyl}}}
\exp\left(-S_X - \lambda \chi \right) \prod_{i=1}^n {\cal V}_{(j_i,k_i)} \, .
\label{amplitude}
\end{equation}

\begin{figure}[!t]
\centering
\setlength{\fboxsep}{10pt}%
\shadowbox{%
\begin{minipage}{.9\textwidth}
\begin{center}
\psfrag{1}[c]{$\chi=2$}
\psfrag{2}[c]{$\chi=0$}
\psfrag{3}[c]{$\chi=-2$}
\psfrag{4}[c]{$\chi=1$}
\psfrag{5}[c]{$\chi=0$}
\psfrag{6}[c]{$\chi=-1$}
\psfrag{7}[c]{$\chi=-1$}
\psfrag{a}{(a)}
\psfrag{b}{(b)}
\includegraphics[scale=0.43]{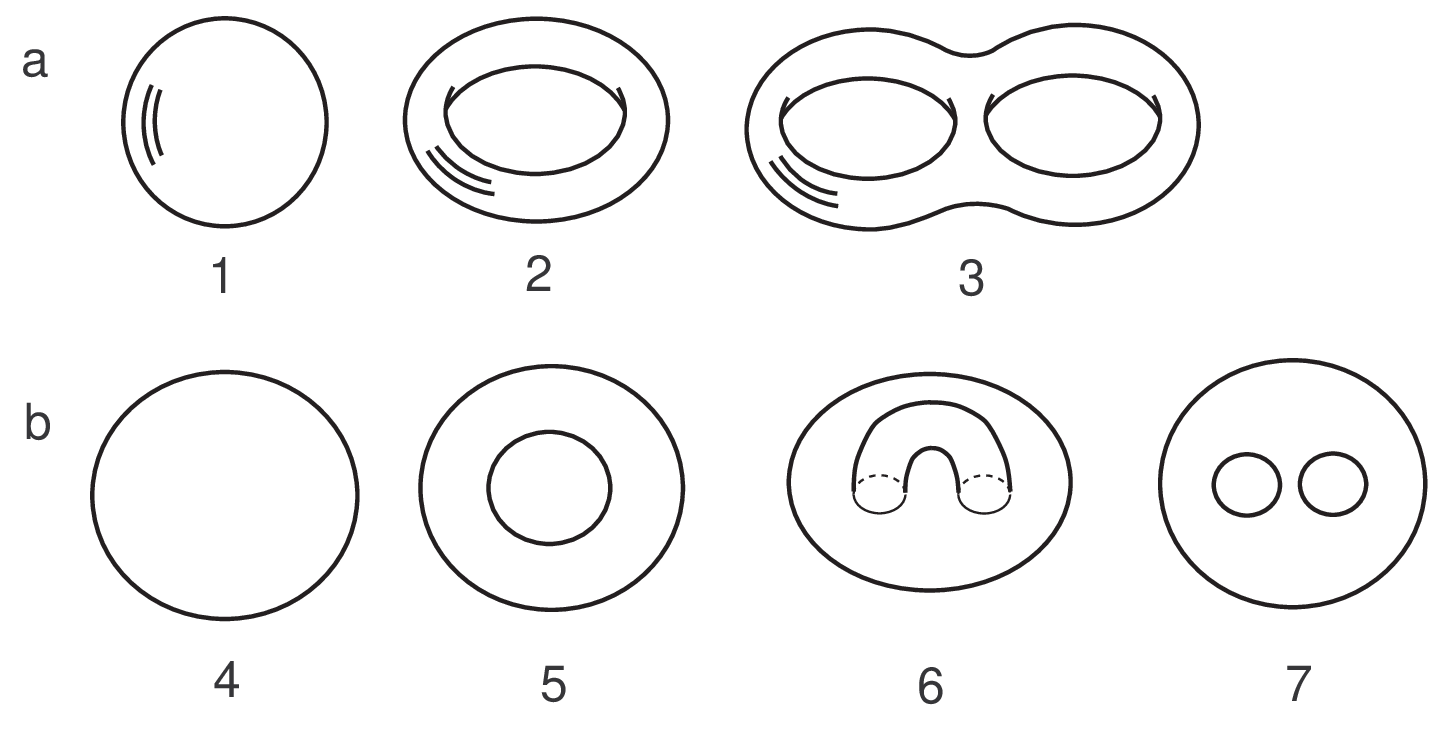}
\caption[Vacuum amplitudes]{Vacuum amplitudes (that means without vertex operators).
(a) Amplitudes without boundaries: zero (sphere), one (torus) and two loops.
These amplitudes involve only closed strings.
(b) Amplitudes with boundaries: zero loops (disc), one open loop (annulus), one closed
loop (disc+handle or torus+hole) and two open loops.
For unoriented string theories one should include
unoriented surfaces as well, which are obtained from an oriented surface by cutting a hole an sewing
in a M\"obius strip. This is called adding a {\em cross-cap} to the surface.}
\label{stringloops}
\end{center}
\end{minipage}
}
\end{figure}

As we mentioned already, $\chi$ depends only on the topology of the world-sheet;
it is the Euler number given by
\begin{equation}
\chi=2-2g-b \, ,
\end{equation}
where $g$ is the number of handles and $b$ the number of holes in the world-sheet.
Taking the string coupling constant to be $g_s=e^{\lambda}$, the different topologies
are weighted with $g_s^{2g+b-2}$. See figure \ref{stringloops} for examples.
In other words, each closed string loop --- a handle --- introduces a factor of $g_s^{2}$
and each open string loop --- a hole --- introduces a factor of $g_s$.
It follows that $g_c \sim g^2_o \sim g_s$ where $g_c$ is the closed string coupling
constant and $g_o$ the open string coupling constant.
From the last term in the action \eqref{polyactioncurved1} we see that the constant part of the dilaton, $\Phi_0$,
sets $g_s=\exp{\Phi_0}$. Therefore, the string coupling $g_s$ is not a fundamental
constant, but instead set by the vacuum expectation value of the dilaton.

In the path integral \eqref{amplitude} we have to divide out the local symmetries
to avoid overcounting equivalent configurations.  To do this properly one has to
use the Faddeev-Popov procedure 
and introduce ghost fields for every
local invariance; to be specific: a $bc$-ghost system for the Weyl/diffeomorphism
invariance and a $\beta\gamma$-ghost system for the local world-sheet supersymmetry of
the superstring.
But the Weyl symmetry is not guaranteed to be present in the quantum theory. Indeed,
in flat space one can show that the Weyl anomaly is equal to
\begin{equation}
\langle T^a{}_a \rangle = -\frac{c_{\text{tot}}}{12} R \, ,
\end{equation}
where $T_{ab}$ is the world-sheet energy-momentum tensor, $R$ the Ricci scalar
and $c_{\text{tot}}$ the total central charge
appearing in the quantum Virasoro algebra. Classically, the energy-momentum
tensor should be traceless in a Weyl invariant theory so that the central charge really measures the
departure from this.
The total central charge is the sum of the central charges
of the matter sector and the ghost sector.
As an example, for the bosonic string $c_{\text{ghost}}=-26$ and $c_{X}=D$ such
that the number of dimensions has to be $D=26$.

In curved space, i.e.\ using the non-linear $\sigma$-model \eqref{polyactioncurved1},
quantum Weyl invariance requires the renormalization group $\beta$-functions to vanish, which
leads to conditions on $G_{\mu\nu}$, $B_{\mu\nu}$ and $\Phi$.  For the open string one
must add the boundary term \eqref{polyactioncurved2} such that one obtains in the end an additional condition on $A_{\mu}$.
The solutions of all these conditions are the backgrounds in which strings can consistently propagate.
These equations can be considered as target space-time equations of motion
and one can wonder whether they could be derived from an action principle.
In fact, as we will see this is one of the ways to calculate the effective action.

This is a good point to point out to the reader that so far two coupling constants for two entirely
different quantum theories were in fact introduced:
\begin{itemize}
\item $\alpha'$: controls the {\em auxiliary} 2-dimensional quantum theory on the world-sheet.
In a flat background and some other special backgrounds this theory can be completely solved (to all orders in $\alpha'$).
In a generic curved background this is not possible anymore and one has to calculate for instance the
$\beta$-functions order per order in $\alpha'$. This corresponds to higher derivative terms
in the effective action.
\item $g_s$: controls the loop expansion in the underlying target space theory. Only recently
explicit calculations have been successfully completed up to two loops \cite{hoker}.
\end{itemize}

The attractiveness of string theory lies in the fact that
all the amplitudes are at each order UV finite \cite{mandelstam}.  The amplitudes will contain divergences but they can
be related to poles of intermediate particles going on-shell, which means that these particles propagate
over long distances. So these divergences are in fact IR effects. Only
UV divergences would signify a break-down of the theory while IR divergences can be dealt with
precisely as in field theory.

\subsection{Problems}
\label{problems}

Perturbative string theories have a number of well-known problems:
\begin{itemize}
\item Abundance of possible vacua. We have seen above that the Polyakov action
could be defined in a background curved space-time and that the $\beta$-functions
constrain the background fields.  Still, there is an enormous amount
of freedom and many vacua are possible.  To make contact with the real world, one would
look for vacua that look like a product of 4-dimensional Minkowski space and a 6-dimensional
compact and tiny internal space. The geometry of the 6-dimensional internal manifold
determines the physics in the 4 non-compact dimensions to a high degree.  The most promising of such compactifications
are the ones that preserve (part of) supersymmetry because firstly the strategy is to work towards
the Supersymmetric Standard Model to avoid the hierarchy problem. The remaining supersymmetry is then broken
at a lower energy level.
A second and more pragmatic reason is that a state of unbroken supersymmetry in many cases also solves
the equations of motion.  The geometry of the 6-dimensional manifold is then {\em Calabi-Yau}.

Many vacua resemble our world i.e.\
4-dimensional flat Minkowski, 6 dimensions compactified and more or less the Standard
Model spectrum of particles.  However, we have no way to single out these vacua
from first principles.
Moreover, most of these vacua have {\em moduli}
i.e. free parameters characterizing the internal manifold that do not seem to change the (effective) energy
of the vacuum: they correspond to {\em flat directions} of the effective potential.
If these moduli are 4-dimensionally space-time dependent, they give rise to extra massless
scalar fields that are not observed. A possible solution is to fix these moduli at certain values
by considering compactifications with fluxes \cite{fluxes}.  This means that expectation
values for some of the antisymmetric
tensors given in table \ref{susystringtheories} are turned on.

Alternative views on the vacua problem are based on statistical physics \cite{douglas}
or even, partially, on the anthropic principle \cite{anthropic}.

\item Curved backgrounds. The conformal field theory on the world-sheet can only be exactly worked out in
a few special backgrounds among which are flat space, pp-waves
and group manifolds. Worse yet, for the definition of the world-sheet action
there were in the supersymmetric case two candidates, the RNS (Ramond-Neveu-Schwarz)
and the Green-Schwarz action \cite{greenschwartz}\footnote{The RNS action is treated in all
the reviews \cite{bookGSW}\cite{bookpolchinski}\cite{cvj}, while for the GS action
you have to look in \cite{bookGSW}.}. They have complementary features.  The RNS action can be
quantized covariantly and calculating scattering amplitudes is easier since the powerful machinery
of conformally invariant field theories can be used. As a drawback, space-time
supersymmetry is somewhat hidden and, even more seriously, R-R backgrounds cannot be incorporated
at all. The Green-Schwarz action on the other hand has manifest space-time supersymmetry
and it is possible to incorporate R-R backgrounds.  The most severe drawback is that
it can only be quantized in the light-cone gauge. In this light-cone gauge both actions can be shown to
be equivalent.
Fortunately, rather recently a manifestly supersymmetric action and a way to covariantly quantize it was discovered
by Berkovits \cite{berkovitsaction} and elaborated upon by Grassi, Policastro, Porrati
and van Nieuwenhuizen \cite{vannieuwenhuizenaction}.
\item Only perturbative definition. The most severe objection in fact is that for a
quantum field theory a perturbative definition is hardly enough. Indeed, for string theories
it can be shown that the perturbative expansion looks like \cite{grossperiwal}
\begin{equation}
\sum_{l=0}^{\infty} g_s^l {\cal O}(l!) \, ,
\end{equation}
and diverges. This means that the sum
cannot be unambiguously evaluated without non-perturbative information.  One
expects non-perturbative effects of the order
\begin{equation}
\exp\left(-{\cal O}\left(1/g_s\right)\right) \, .
\end{equation}
Obviously, the perturbative expansion is not a good description anymore when
$g_s \rightarrow \infty$. This kind of behaviour is in fact very common in quantum field theory where
effects of the order
$\exp\left(-{\cal O}\left(1/g^2\right)\right)$ are typically encountered.
There the perturbative approach can be saved by introducing
classical instanton backgrounds around which to perturb and then summing over all possible instantons.
\end{itemize}
The first two problems could be summarized by saying that we lack a {\em background-independent formulation}.
Together with the last problem this strongly suggests that perturbative string theory is only a tool that
can probe aspects of a more fundamental theory.

\section[More Between Heaven and Earth than Strings]{More Between Heaven and Earth than\\Strings}

\subsection{T-duality}

Surprisingly, it is possible that two string theories of different perturbative type and/or in
different backgrounds are completely equivalent.  These transformations between equivalent
theories generally define a discrete group and are called {\em dualities}.  A first
example is T-duality \cite{tdualityreview}, which relates space-time geometries possessing a compact isometry
group.  T-duality is a perturbative duality in the sense that it is valid order per order
in the loop expansion in $g_s$.

We consider the simplest case, bosonic theory in flat space with one dimension, say the $l$th, compactified on a circle
with radius $R$,
\begin{equation}
X^l \cong X^l + 2 \pi R \, .
\end{equation}
The compact isometry is of course translation along $X^l$. The mass-shell condition for an
open string now reads
\begin{equation}
M^2=\Big(\frac{n}{R}\Big)^2+\Big(\frac{wR}{\alpha'}\Big)^2+ \text{(oscillator contr.)} \, .
\label{massformula}
\end{equation}
As before --- when we calculated the spectrum --- the oscillation contribution
depends on which quantum state of the string is excited i.e.\ which particle it represents.
As for the particle, there is furthermore a contribution from the (center of
mass) momentum, which is discretized because of
the periodic boundary conditions.
Indeed, the operator $\exp \left( 2 \pi i R p_l\right)$ translates the string once around the
periodic dimension and must
leave states invariant so that
\begin{equation}
p_l = \frac{n}{R} \, .
\label{momentquant}
\end{equation}
The states labelled by $n$ are
called {\em Kaluza-Klein} states. Since strings have a tension,
there is an additional contribution proportional to the length of
the string. Closed strings can wrap around the compactified
dimension. The states labelled by $w$, the {\em winding number},
correspond to strings winding $w$ times around the compact
dimension. The winding number is conserved as closed strings cannot unwind
without breaking. We see that the mass does not change
if we send $R \rightarrow \frac{\alpha'}{R}$ and interchange
Kaluza-Klein states with winding states: $n \leftrightarrow w$.
This still holds if we consider the oscillation modes as well
so that we found a duality of the full world-sheet conformal
field theory i.e.\ the theory on $R$ and the theory on
$\hat{R}=\frac{\alpha'}{R}$ are completely equivalent. This means
that closed strings cannot probe distances smaller than
$\sqrt{\alpha'}$, a typical phenomenon that will appear in more
complex geometries as well.

Consider now open strings. In this case the momentum in the
compactified dimension is still quantized. On the other hand, because of
their tension open strings will contract as much as possible and unlike
for wrapped closed strings, there is nothing to prevent that.
So there are Kaluza-Klein states, but there is only zero winding.
Applying T-duality as for the closed string, interchanging winding and momentum,
one finds winding states with zero momentum.
Since open strings differ only at the endpoints from closed strings, we want
to explain this using only the endpoints: fixing the endpoints at a certain position
in the direction $X^l$, the string cannot have (center of mass) momentum in this direction
anymore. On the other hand, fixing the endpoints prevents the open string from
contracting and winding is possible. In this interpretation the endpoints of the open
string are in the T-dual picture constrained to a hyperplane $X^l=x^l$.
Put in another way, T-duality interchanges Neumann
boundary conditions, allowing the ends to move freely, with Dirichlet boundary conditions, constraining the ends.
The hyperplanes on which strings end were given the name D(irichlet)-branes in \cite{firstDbranes}.

Let us discuss in more detail the case of open strings in a constant diagonal $N \times N$ background\footnote{Contrary to
the rest of the thesis
the gauge field is hermitian in this discussion.}:
\begin{equation}
\begin{split}
A_l = \text{diag}_A \left[- \frac{\theta_A}{2 \pi R} \right] = -i \Lambda^{-1}\frac{\partial \Lambda}{\partial X^l}, \\
\Lambda(X^l) = \text{diag}_A \left[ \exp \left(-\frac{i \theta_A}{2\pi R}\right) \right],
 \end{split}
\end{equation}
where the $\theta_A$ are constants and $\text{diag}_A$ indicates a diagonal matrix of which the element at position $AA$
is given in square brackets and $A$ runs from $1$ to $N$.
Locally this is pure gauge, but not globally since the gauge parameter picks up a phase
\begin{equation}
\text{diag}_A\left[- i \theta_A \right]
\end{equation}
under $X^l \rightarrow X^l + 2 \pi R$.
The canonical momentum $p_l$ conjugate to the center of mass position of the string is given by
\begin{equation}
p_{l} = \int_0^\pi d \sigma \frac{\partial {\cal L}}{\partial \dot{X}^{l}} = \frac{1}{2\pi\alpha'} \int_0^\pi d \sigma \; \partial_{\tau} X_{l} + A_{l}|_{\pi} - A_{l}|_{0} \, ,
\end{equation}
where we have used the Polyakov action \eqref{polyaction} with the boundary term \eqref{polyactioncurved2}.
The momentum is quantized as in eq.~\eqref{momentquant}.
Consider in the T-dual picture a string stretching from D-brane $A$ to D-brane $B$.
We define the ``momentum'' $v_l$ appearing in the mode expansion of the string as
\begin{equation}
v_{l} = \frac{1}{2\pi\alpha'} \int_0^\pi d \sigma \; \partial_{\tau} X_{l},
\end{equation}
so that
\begin{equation}
v_{l} = \frac{n}{R} + A_{AA} - A_{BB} = \frac{1}{2\pi\alpha'} (2 \pi n + \theta_B - \theta_A)\hat{R} \, .
\end{equation}
The distance in the $l$th direction between the endpoints is
\begin{equation}
\hat{X}(\pi) - \hat{X}(0) = \int_0^{\pi} d \sigma \; \partial_{\sigma} \hat{X}_{l} = \int_0^{\pi} d \sigma \; \partial_{\tau} X_{l} = 2 \pi \alpha' v_l = (2 \pi n + \theta_B - \theta_A)\hat{R} \, ,
\end{equation}
where the hat denotes quantities in the T-dual picture and we have used that T-dualizing interchanges Neumann and Dirichlet boundary
conditions:
\begin{equation}
\partial_{\sigma} \hat{X}_{l} = \partial_{\tau} X_{l} \, .
\end{equation}
The distance between the two endpoints $A$ and $B$ is invariable;
they are indeed fixed at hyperplanes at positions $\theta_A$ and $\theta_B$
respectively. In the T-dual picture $n$ becomes the winding number. The open string analogue of eq.\ \eqref{massformula}
reads
\begin{equation}
M^2=v_l^2+ \text{(oscillator contr.)} = \frac{\left(2 \pi n + \theta_B - \theta_A\right)^2 \hat{R}^2}{(2\pi\alpha')^2} + \text{(oscillator contr.)} \, .
\label{massformula2}
\end{equation}
If the oscillators do not contribute, i.e. for the lowest lying modes, the mass is proportional to
the string length. The minimal mass is attained when $n=0$ and is then proportional to the distance between the branes.

\subsection{D-branes}
\label{dbranes}

Let us now consider open strings in the presence of a $p$-dimensional D-brane, which
can be obtained by applying T-duality in $9-p$ dimensions.
Consequently, one has next to Neumann boundary conditions in $p+1$ dimensions,
Dirichlet boundary conditions in the remaining $9-p$ dimensions:
\begin{equation}
\partial_{\sigma} X^{\mu} = 0, \,\,\, \mu=0,\ldots,p \qquad
X^{i} = x^i , \,\,\, i=p+1,\ldots,9 \, .
\end{equation}
Since T-duality interchanges Neumann and Dirichlet boundary conditions, it is now obvious
that T-duality in a direction perpendicular to a D$p$-brane results in a D$(p+1)$-brane,
while T-duality in a longitudinal direction results in a D$(p-1)$-brane.
The presence of a D$p$-brane will break translation invariance in the $9-p$ transversal
directions and it will break Lorentz invariance $SO(9,1)$ to $SO(p,1)\times SO(9-p)$. As a consequence
the spectrum of massless states from table \ref{openstringspectrum} changes into:
\begin{equation}
A_{\mu}, \, \mu=0,\ldots 9 \longrightarrow \left\{
\begin{array}{l}A_{a}, \, a=0,\ldots,p\\\Phi_i,\, i=p+1,\ldots,9\end{array}\right. \, .
\label{Adimred}
\end{equation}
$A_{a}$ describes a $p$-dimensional gauge theory living on the brane while the $\Phi_i$ are
the Goldstone bosons associated with the breaking of translational symmetry.  They are collective
coordinates describing the position of the branes consistent with the fact that D-branes are
in fact dynamical objects. Indeed, we have already seen that a background value for $A_{\mu}$
has in the T-dual picture the interpretation of the position of the D-brane.

Although D-branes were already discovered from the above T-duality argument in \cite{firstDbranes},
it was not until \cite{polchinski} that it was realized that D-branes were in fact a missing link
as they can act as sources for the R-R fields $C_{[n]}$, which fundamental strings cannot.
Indeed, the world-volume of a p-brane couples in a natural way to a (p+1)-form
potential as follows
\begin{equation}
\int_{V_{p+1}} C_{[p+1]} \, ,
\label{branecoupling}
\end{equation}
where the integral is over the D-brane world-volume $V_{p+1}$.
In fact, this is an {\em electric} coupling, but the same potential $C_{[p+1]}$ can couple
{\em magnetically} to a D$(6-p)$-brane as follows:  the Hodge dual of the field strength
$F_{[p+2]}=d C_{[p+1]}$ in 10 dimensions is a (8-p)-form $(\dual{F})_{[8-p]}$, which
has a (7-p)-form $C'_{[7-p]}$ as potential; the latter (7-p)-form is suitable
for coupling to a D$(6-p)$ brane.

Considering the possible R-R forms in table
\ref{susystringtheories}, we see that IIB theories should contain D$(-1)$- (D-instantons),
D$1$-, D$3$-, D$5$- and D$7$-branes, while IIA theories contain D$0$-, D$2$-, D$4$- and D$6$-branes.
Since T-duality sends a D$p$-brane to a D$(p+1)$- or a D$(p-1)$-brane,
we must conclude that it also interchanges type IIA and type IIB theories, which is indeed
the case.  We learn in addition that there must be a D$9$-brane in IIB, which we will discuss at
the end of the section, and a D$8$-brane in IIA. These are not associated to propagating states,
so they do not show up in the spectrum.

\begin{figure}[tp]
\centering
\setlength{\fboxsep}{10pt}%
\shadowbox{%
\begin{minipage}{.9\textwidth}
\begin{center}
\psfrag{s=t}{$\sigma=\tau'$}
\psfrag{t=s}{$\tau=\sigma'$}
\psfrag{i}[c]{$\mu_p,T_p$}
\psfrag{p}[c]{$\frac{(\kappa_{10})^2}{G_{9-p}(y)}$}
\psfrag{C}[c]{\text{Closed Channel}}
\psfrag{O}[c]{\text{Open Channel}}
\psfrag{a}{(a)}
\psfrag{b}{(b)}
\psfrag{c}{(c)}
\psfrag{1}{1}
\psfrag{2}{2}
\includegraphics[scale=0.54]{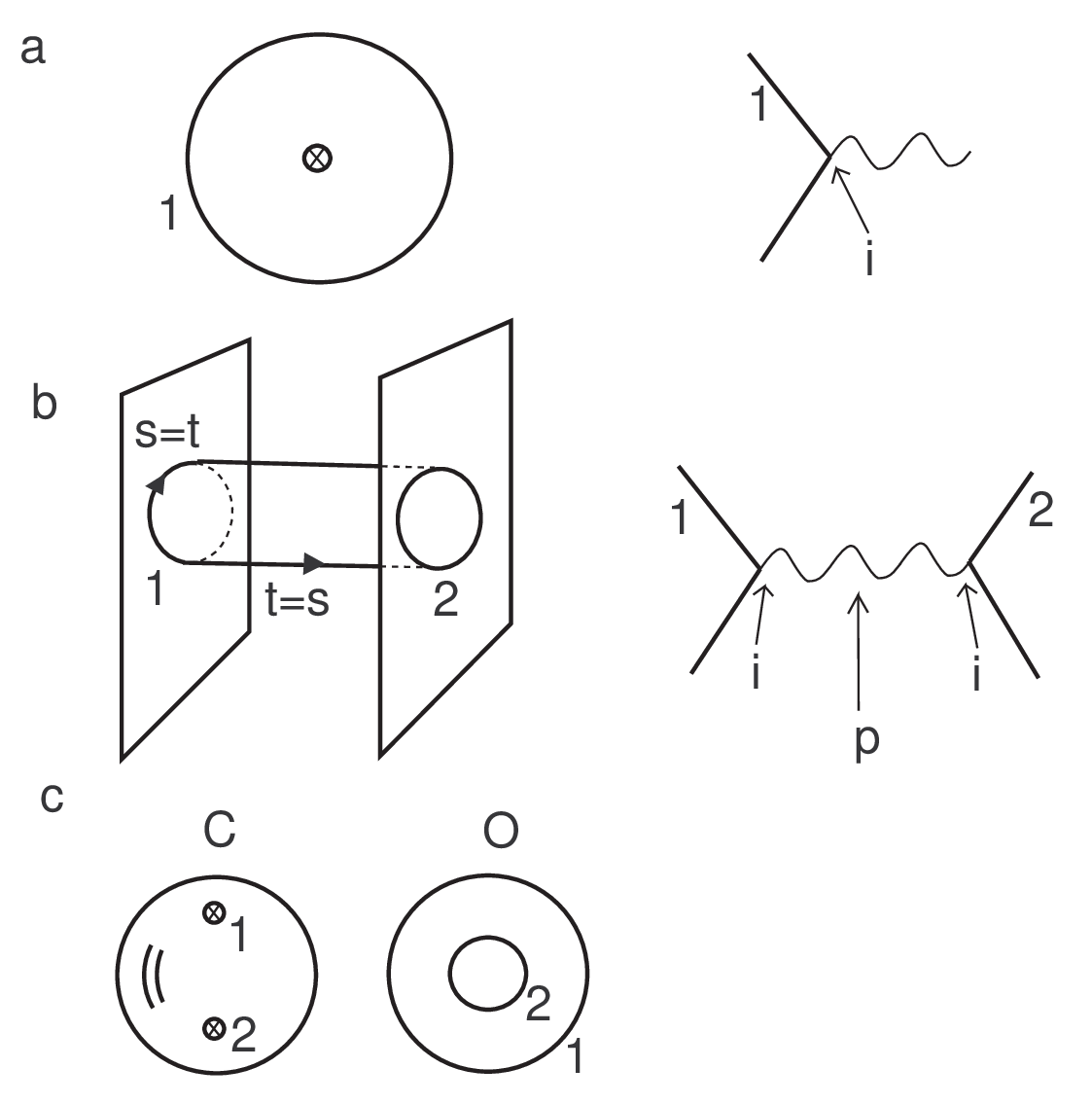}
\caption[D-brane couplings]{D-brane couplings. The numbers 1 and 2 label the two D$p$-branes. (a) The {\em tadpole}: coupling of a D-brane to a closed string state (NS-NS or R-R).
Left: string theory diagram. Right: the corresponding low-energy field theory diagram.
In the field theory the coupling to an NS-NS particle (graviton or dilaton) contributes a factor of $T_p$,
the coupling to an R-R particle a factor of $\mu_p$.
(b) Interaction between two D$p$-branes via a closed string state. Left: string theory diagram. Right: field
theory diagram.  The propagator in the field theory contributes $\frac{(\kappa_{10})^2}{G_{9-p}(y)}$.
(c) The string theory diagram of (b) can be interpreted in two ways. In the closed string
channel as a tree level diagram (a sphere with an incoming and an outgoing D-brane state) and in the open string channel as a one loop diagram
(an annulus).}
\label{dbranecoupling}
\end{center}
\end{minipage}
}
\end{figure}

In \cite{polchinski} the coupling of a D-brane to NS-NS and R-R states of closed strings
was calculated. By comparison to the corresponding low-energy field theory amplitudes,
its tension $\tau_p$ and its charge $\mu_p$ can be determined.
This could be done directly from diagram \ref{dbranecoupling}a, but because
of normalization issues of both open and closed string vertex operators, it is easier
to calculate diagram \ref{dbranecoupling}b.  The diagram shows two D$p$-branes feeling each
other's force by exchanging a closed string. In this {\em closed
string channel} we label the world-sheet coordinates as $(\tau,\sigma)$. But, by world-sheet duality
the diagram can also be looked at in terms of an open string with endpoints on the two D$p$-branes, moving
around in a loop.  World-sheet time and length are interchanged in the {\em open string channel}
$(\tau',\sigma')=(\sigma,\tau)$.
In this channel the diagram has no vertex operators and can be most easily calculated
with the aid of the Coleman-Weinberg formula
\begin{equation}
{\cal A}= \int_0^{\infty} \frac{dt}{2t} \Tr \left(\frac{1+\exp(\pi i F)}{2} e^{-2\pi\alpha' t L_0}\right)
\end{equation}
Here $t$ is a modulus characterizing the cylinder: it measures the periodicity of $\tau'$,
$\tau' \cong \tau'+2 \pi t$.  This modulus represents the only degree of freedom that is left over from
the integral over all possible metrics on a cylinder $[dh]$, when reparameterization and Weyl invariance
are taken into account.  Tr denotes the trace over all possible open string states, both from the
NS sector and from the R sector. The GSO projection $\frac{1}{2}\left(1+\exp\pi i F\right)$, where $F$ is the
world-sheet fermion number, projects out unwanted states like the tachyon.  We see that each open string state is propagated
by the world-sheet Hamiltonian $\alpha' L_0$ over a distance $2\pi t$ and is thus running around
the cycle of the cylinder.

Taking the limit $t \rightarrow 0$ amounts to an infinitesimally small loop in the open string channel,
clearly an UV effect.  However, through a Weyl transformation we can blow up the radius of the loop, which since
the Weyl transformation acts isotropically results at the same time in an infinitely long cylinder.
So in the closed string channel this is in fact an IR limit as the closed string states propagate over
a long distance. The amplitude will then
be dominated by the exchange of the lowest mass closed string states,
\begin{equation}
\begin{split}
{\cal A} & = {\cal A}_{\text{NS-NS}} + {\cal A}_{\text{R-R}} = 0 \\
         & \stackrel{t\rightarrow 0}{\approx} (1-1)i V_{p+1} 2\pi (4\pi^2\alpha')^{3-p}G_{9-p}(y) \, ,
\end{split}
\end{equation}
with $G_{d}(y)=\frac{1}{4\pi^{d/2}}\Gamma(d/2-1)y^{2-d}$ the scalar Green's function in $d$ dimensions.
Working out the formula would reveal that the attractive force from the NS-NS sector (graviton and dilaton) is exactly
cancelled by the repulsive force of the R-R sector: ${\cal A}_{\text{NS-NS}}=-{\cal A}_{R-R}$.
This is typical for supersymmetric states ---  D-branes preserve half of the supersymmetry --- as the contribution of the fermions
exactly cancels the contribution of the bosons. In calculating tension and charge we will
be interested in the two contributions separately.

Consider now the low-energy action
\begin{equation}
{\cal S} = {\cal S}_{\text{NS-NS}}+{\cal S}_{\text{R-R}}+{\cal S_{\text{DBI}}} \, ,
\end{equation}
with
\begin{equation}
\begin{split}
{\cal S}_{\text{NS-NS}} & = \frac{e^{-2\Phi_0}}{2(\kappa_{10})^2} \int d^{10} x (-\tilde{G})^{1/2} \left(R-\frac{1}{2}\partial_{\mu}\Phi \partial^{\mu}\Phi\right) + \cdots \, ,\\
{\cal S}_{\text{R-R}} & = -\frac{1}{4\kappa_{10}^2} \int d^{10} x (-G)^{1/2} \left|F_{[p+2]}\right|^2 + \mu_p \int_{V_{p+1}} C_{p+1} \, .
\end{split}
\end{equation}
The action ${\cal S}_{\text{NS-NS}}$ consists of the Einstein-Hilbert action and a kinetic term for the dilaton while
${\cal S}_{\text{R-R}}$ consists of a kinetic term for the R-R field and the coupling eq.~\eqref{branecoupling}.
${\cal S}_{\text{DBI}}$ is the Dirac-Born-Infeld action \eqref{DBI}\footnote{${\cal S}_{\text{NS-NS}}$
contains the metric $\tilde{G}$ in the {\em Einstein frame}, which is rescaled with respect to the metric
as it appears in string theory with a dilaton dependent factor. In this way one obtains the standard form of the Einstein-Hilbert action.
In ${\cal S}_{\text{R-R}}$ and $S_{\text{DBI}}$ (see eq.~\eqref{DBI}) on the other hand the metric appears in the string frame.}.
We read off that the gravitational coupling constant is $\kappa=\kappa_{10}e^{\Phi_0}=\kappa_{10} g_s $. The R-R p-form charge
is $\mu_p$ and the tension appearing in front of the D-brane effective action is $\tau_p$.
Computing the corresponding diagrams in the field theory and comparing to the string theory
result, one finds
\begin{equation}
\begin{split}
\tau_p & =\frac{\pi^{1/2}}{\kappa}\left(4\pi^2\alpha'\right)^{\frac{3-p}{2}} = \frac{\pi^{1/2}}{\kappa_{10}g_s}\left(4\pi^2\alpha'\right)^{\frac{3-p}{2}} = \frac{T_p}{g_s} \\
\mu_p & = T_p = \frac{\pi^{1/2}}{\kappa_{10}} \left(4\pi^2\alpha'\right)^{\frac{3-p}{2}} \, .
\end{split}
\label{tensioncharge}
\end{equation}
Next to the physical tension we have introduced $T_p=\tau_p g_s$, where the $g_s^{-1}$ dependence is removed.

The D-brane charges must obey a Dirac quantization condition.
Indeed, consider a D$p$-brane, which is an electric source for $C_{[p+1]}$.
By Gauss' law the charge of this brane $\mu_p$
must be proportional to the flux $Q$ through a $8-p$ sphere surrounding the brane:
\begin{equation}
\mu_p = \frac{1}{2 \kappa_{10}^2} Q = \frac{1}{2 \kappa_{10}^2}\int_{S^{8-p}} \dual{F}_{[p+2]} = \frac{1}{2 \kappa_{10}^2} \int_{S^{7-p}} C_{[7-p]},
\end{equation}
since $\dual{F}_{[p+2]}=F_{[8-p]}=d C_{[7-p]}$ everywhere, except on a Dirac string.  $S^{7-p}$ is a sphere
surrounding the Dirac string. For the Dirac string to be invisible for a ``magnetic'' D$(6-p)$-brane
travelling around the Dirac string, we need that $\exp(i\mu_{6-p}Q)=1$ so that
\begin{equation}
2\kappa_{10}^2 \mu_p \mu_{6-p} = 2\kappa_{10}^2 \mu_e \mu_m  =  2 \pi n \, .
\end{equation}
For the D-brane charge, eq.~\eqref{tensioncharge}, this is indeed satisfied with $n=1$.
The 3-point graviton amplitude learns that $\kappa=2\pi g_c \sim g_s$.
Commonly one chooses the additive normalization of the dilaton field $\Phi_0$ such that
\begin{equation}
\kappa^2 = \frac{1}{2}\left(2\pi\right)^7 g_s^2 \alpha'^4 \, ,
\end{equation}
from which follows
\begin{equation}
\tau_p = g_s^{-1} \mu_p = \frac{1}{g_s (2\pi)^p \alpha'^{(p+1)/2}} \, .
\end{equation}
The ratio of the tension of the fundamental string $T=\frac{1}{2\pi\alpha'}$ to the D1-brane
(D-string) is then precisely $g_s$. As a further application note that the lowest-order approximation of $S_{\text{DBI}}$ is
Yang-Mills theory with coupling constant
\begin{equation}
\frac{1}{g_{\text{YM}}^2}=\tau_p(2\pi\alpha')^2 = \frac{1}{g_s(2\pi)^{p-2}\alpha'{}^{\frac{p-3}{2}}} \, .
\label{ymcoupling}
\end{equation}

The tension $\tau_p$ of a D-brane is of the order $e^{-\Phi_0}=g_s^{-1}$
as could be expected since its coupling to the graviton is described by the disc diagram of figure
\ref{dbranecoupling}a.  As a consequence plugging in a D-brane background in the path integral
will lead to the non-perturbative effects of the order $\exp(-1/g_s)$ that
were anticipated in section~\ref{problems}. To have a non-zero contribution these D-branes have
to be localized in both time and space: D(-1)-branes (D-instantons) \cite{dinstantons} or wrapped D$p$-branes in
Euclidean space-time \cite{deuclidean}.

As mentioned already, a D-brane breaks half of the supersymmetry. The fact that it does not break
all supersymmetry makes it a BPS(Bogomolny-Prasad-Sommerfield)-state.  Central charges appear
in the supersymmetry algebra \cite{wittenolive}. For the anticommutator of a left-moving and
a right-moving supersymmetry charge, these are precisely the R-R charges.
In the BPS-mechanism, invariance of the D-brane under a supersymmetry will force
the mass to be equal to the charge, which explains
$\mu_p = T_p$ in eq.~\eqref{tensioncharge}.  We will later take another approach to BPS-states where
we will show that they are at an global minimum of the action \cite{bogo}.

\begin{figure}[!t]
\centering
\setlength{\fboxsep}{10pt}%
\shadowbox{%
\begin{minipage}{.9\textwidth}
\begin{center}
\psfrag{a}{(a)}
\psfrag{b}{(b)}
\includegraphics[scale=0.49]{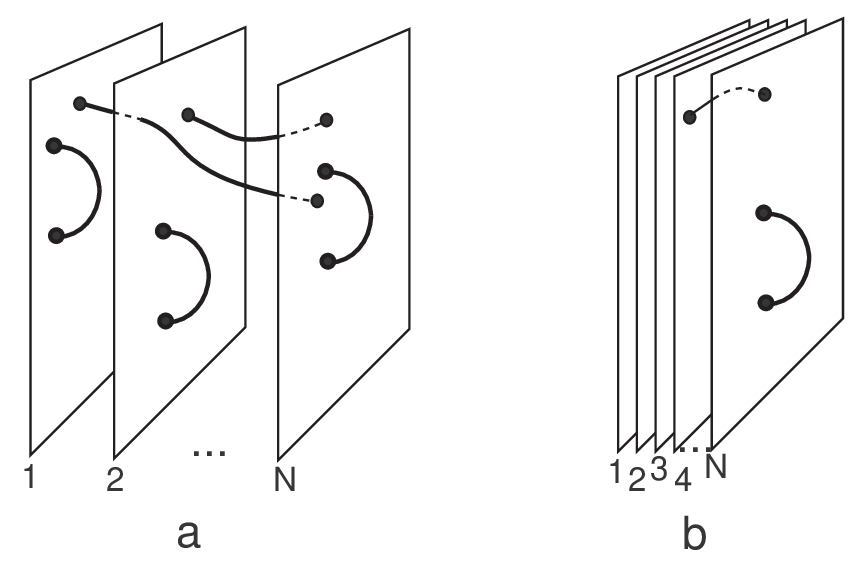}
\caption[D-branes]{$N$ D-branes. (a) Non-coinciding D-branes: the only massless states
arise from strings beginning and ending on the same brane since strings beginning and ending
on different branes have a mass proportional to the distance between the branes.  The gauge group is $\prod_N U(1)$.
(b) Coinciding D-branes: next to strings
beginning and ending on the same brane, the lowest-level states of strings beginning and ending on different branes
become massless as well, adding up to a total of $N^2$ different states.  These fill out
the adjoint representation of $U(N)$, the new gauge group. Gauge groups of the type $SO(N)$ can be introduced by
adding orientifold planes.}
\label{D-branes}
\end{center}
\end{minipage}
}
\end{figure}

Labelling the endpoints of open strings by introducing $N$ Chan-Patton labels translates
in the new language into introducing $N$ D-branes. For each brane, we have a copy of the gauge sector
originating in strings beginning and ending on that brane. In this way we end up with the gauge group
$U(1)\times U(1)\times \cdots \times U(1)$.  As we saw in eq.~\eqref{massformula2} the low lying modes of a string
have masses proportional to the length of the string so that if the D-branes coincide, there are extra massless states coming from
strings beginning and ending on different coinciding D-branes.  Keeping track of the orientation of strings, we count
a total of $N^2$ massless states making up the adjoint representation of $U(N)$, which will be the new gauge group.
All this is pictured in figure~\ref{D-branes}.
Giving a vacuum expectation value to some (or all) of the diagonal elements of $\Phi_i$
lets the branes move apart and breaks the $U(N)$ gauge group.
This provides a geometrical picture of the Higgs effect.

Pushing the coinciding D-brane picture a little bit further we arrive at an interesting point. Indeed, considering
the way that the collective coordinates $\Phi_i$ of the branes originate under T-duality
from dimensional reduction of the gauge potential $A_{\mu}$ as in eq.~\eqref{Adimred}, we must
conclude that they also take values in the adjoint of $U(N)$ i.e.\ the space-time coordinates denoting the positions of
the D-branes are enlarged to matrices \cite{wittenmatrixval}.  The interpretation of the diagonal elements is easy:
they are the positions of each of the $N$ D-branes, but the off-diagonal components remain mysterious.
The resulting {\em non-commutative geometry} leads to very interesting effects such as dielectric branes
and fuzzy funnels \cite{myersreview}.
Difficult questions that further arise are what happens when the background fields become dependent
on these matrix-valued coordinates \cite{dgeometry}\cite{raamsdonk} and how one can define matrix general coordinate
invariance \cite{deboer}.  There seem to be still a lot of unresolved issues.

From the previous discussion it becomes clear that introducing open strings with
free endpoints into the theory is in fact the same as inserting a D$9$-brane. Obviously,
when there are $N$ space-filling D$9$-branes, they are always coinciding. In type I theory there are
16 D$9$-branes and one orientifold O$9$-plane.  The latter produces another 16 image
D$9$-branes, makes the theory unoriented and changes the gauge group to $SO(32)$.
It can be shown that the orientifold plane has a negative RR charge of $-2^4=-16$
so that its charge exactly cancels the charge of the $16$ original D$9$-branes.  This is necessary for the
consistency of the theory because if there were a net charge, the flux lines would have nowhere
to go. This is a modern interpretation of the one-loop infinity cancellation in \cite{GSanomaly}.

Apart from the microscopic description of D-branes as hyperplanes on which strings can end, they
also have a description as solutions of the equations of motion of the low-energy effective
theory: IIA or IIB supergravity, where they look like higher-dimensional
extensions of charged black holes, black $p$-branes \cite{pbranes}.  In fact, they were proposed first in this picture
as the sought after sources of the R-R fields.
The connection between the two pictures can be made, for instance by extracting the long-range
background field surrounding a set of D$p$-branes from amplitude calculations and
comparing to the black $p$-brane solution \cite{pbranescat}.

This is a good point to stress the power
of string theory and D-branes as a tool to study gauge theories. We have already seen how coinciding
D-branes generate a non-abelian gauge group and how D-branes moving apart provide a geometrical picture
of the Higgs effect. There are many other examples but let us list only few. We will see in chapter \ref{solutions} how
a D3-D1 intersection provides a realization of the Dirac monopole \cite{bion1}\cite{bion2}. More or less
in the same vain D$1$-branes stretching between separated D$3$-branes provide a geometric picture of the mathematical ADHM classification
of instanton solutions \cite{diaconescu}. In chapter \ref{checks} we will see how the quantization
of magnetic flux on a torus has a natural explanation in terms of the T-dual picture of intersecting
branes. The most famous example is of course the Maldacena conjecture \cite{maldacenaconj}, which provided
the first concrete realization of an observation made by 't Hooft that the large $N$ limit of gauge theories
looks like a string theory \cite{thooftlargeN}. The argument relies on the equivalence between D$3$-branes and black $3$-branes.
Indeed, the first are in the low-energy limit described by $N=4$ Super-Yang-Mills theory while the latter
are in the low-energy limit described by string theory on the near-horizon region of the $3$-brane,
which is $AdS_5 \times S^5$. Interestingly, the range of validity of perturbative Yang-Mills
theory, $g^2_{\text{YM}} N \ll 1$, is completely opposite to the range of validity where we can use
classical supergravity (instead of the whole string theory) in the dual picture, $g^2_{\text{YM}} N \gg 1$.
This means we can study the strongly coupled gauge theory by means of classical supergravity which
is ample motivation to look for the gravity dual of other gauge theories. On the other hand,
it makes the conjecture hard to prove.
Another more recent example is the work of Dijkgraaf and Vafa, where string theory provided
the motivation for an algorithm to construct the {\em exact} effective superpotential of the chiral
glueball superfield in  $N=1$ gauge theories via perturbative calculations in an auxiliary matrix model \cite{dijkgraaf}.
Only later on were these results also proven in field theory \cite{dijkgraafproof}. Another important development is
the natural incorporation of non-commutative gauge theories in string theory as a low-energy limit
when the NS-NS background field $B_{[2]}$ is turned on \cite{seibergwitten}. More recently, it was discovered
that playing the same trick with the R-R background field $C_{[2]}$ switched on leads to non-(anti)commutative supersymmetric gauge theories \cite{nonanti}.

Although the focus is clearly on D-branes, we will briefly review some other non-perturbative results.

\subsection{Non-perturbative Dualities}

We have seen that T-duality relates type IIA and type IIB string theory.  In fact, T-duality also
relates the two heterotic string theories.  Still these are perturbative dualities.
However, there are also dualities relating a strongly coupled theory to a weakly coupled
theory \cite{wittenduality}\cite{hulltownsend}.
One example is electromagnetic duality in Yang-Mills theory, which interchanges light
electric charges with heavy magnetic monopoles \cite{montonenolive}. In fact, this only works out well
if there is enough supersymmetry, namely in $N=4$ Super-Yang-Mills \cite{osborn}.
In that case it turns out to be part of a discrete $SL(2,\IZ)$ group of dualities, $S$-duality.

Type IIB string theory is self-dual under a similar $SL(2,\IZ)$ duality group.
Under this duality fundamental strings and D$1$-branes are interchanged.
In fact, since type IIB string theory contains $N=4$ Super-Yang-Mills theory on a D$3$-brane, it incorporates the
S-duality of Yang-Mills \cite{emduality}\cite{tseytlinselfdual}!  Type I $SO(32)$ and heterotic $SO(32)$ also turn out to be related by S-duality.

Special cases are the strong coupling limit of type IIA and heterotic $E_8 \times E_8$ theory. Focusing on type IIA
we take $g_s \rightarrow \infty$ and find that the lowest mass states are in fact the D0-branes.
The mass of a system of $n$ (marginally) bound D$0$-branes is
\begin{equation}
n \tau_0 = \frac{n}{g_s \alpha'^{1/2}} \, .
\end{equation}
This evenly spaced spectrum looks like a Kaluza-Klein spectrum with periodic dimension
\begin{equation}
R_{10} = g_s \alpha'^{1/2} \, .
\end{equation}
So one could make the bold assumption that as $g_s \rightarrow \infty$ this dimension
is decompactified  and one ends up with an 11-dimensional theory. The low-energy
effective theory would then be nothing but 11-dimensional supergravity. Eleven dimensions is the
maximum number of dimensions allowing a (locally) supersymmetric field theory and this theory
is unique. The assumption can be justified by arguing on low-energy effective
actions and by comparing the BPS spectra of the theories.  We have seen that string theory
in 11 dimensions is not consistent, but in fact the strong coupling limit of type IIA is not
a string theory anymore! In fact little is know about this theory and its fundamental degrees
of freedom beyond the low-energy limit and its ties to type IIA.  It has been given the name {\em M-theory}.

Let us briefly mention the M-theory origin of the various branes of type IIA string theory.
The bosonic sector of 11-dimensional supergravity consists of the metric $G$ and a 3-form.
The black brane that is electrically charged under the 3-form is called the M2-brane while the magnetically
charged object is the M5-brane. D0-branes are in the strong coupling limit, as we have just discussed,
Kaluza-Klein states. The fundamental type IIA strings are M2-branes wrapped on the 11th dimension while
the D2-branes originate in transversal M2-branes. The D4-branes are wrapped M5-branes while the NS5-branes
(the magnetic duals of the fundamental strings) are transversal M5-branes. D6-branes, as the magnetic dual
of D0-branes, have as strong coupling limit Kaluza-Klein monopoles. The M-theory origin of D8-branes is
still unclear.

In \cite{matrixtheory} a proposal was made was for the microscopic description of M-theory based on
D0-branes: Matrix-theory.
Usually, the name M-theory is not only used for the strong coupling limit of type IIA, but more generally
for the (unknown) underlying theory incorporating the five perturbative string theories, and M-theory in the
narrow sense.

\subsection{Problems Revisited}

After all we have learned about non-perturbative string theory the question arises whether we
have made any progress in solving the problems listed in section \ref{problems}.
As for the last problem we know now the nature of some of the non-perturbative instanton effects: these
are the D-instantons or wrapped Euclidean D-branes.  Furthermore, we have become more convinced that their indeed exists an underlying description,
M-theory although we do not know the details. The background problem still stands.  We know now that
some backgrounds are related by perturbative or non-perturbative dualities to other backgrounds, but the multitude
of different backgrounds remains.  As for the inability to make calculations in other than a few backgrounds, we
find a new tool in dualities, but on the other hand the problem becomes worse as we realize that there
are corners of the moduli space of backgrounds where strings are not even the right degrees of freedom.

\section{D-brane Effective Actions}

\subsection{Definition}

String theory can be thought of as an infinite-component field theory.
Indeed, every particle in the spectrum, massless or massive, corresponds to a field. This approach
is taken seriously in string field theory \cite{wittenSFT}.
However, as massive particles can only be detected at very high energies, characterized
by the string scale $\alpha'{}^{-1/2}$, for phenomenology we will be exclusively interested in
the massless\footnote{The way in which
these particles acquire their measured masses is still unknown.} particles, of which fortunately there are only a finite number.
For the open superstring ending on a D$p$-brane these will be the ``gluon'' $A_{a}$, transverse
scalars $\Phi_i$ and their fermionic superpartners; see table \ref{openstringspectrum} and eq.~\eqref{Adimred}.

The {\em low-energy effective action} is defined as the result of ``integrating out'' all the massive and massless modes
circulating in loops.  Moreover, we only allow the massless modes as external states. This means that the effective
action should generate at tree-level an S-matrix, which reproduces the string-theory
S-matrix for massless external states.  We can expect the effective action to be non-local, i.e.\ it will contain
an infinite number of derivatives, and we can expect it to be highly complicated.  Remarkably, for the abelian
open superstring there exists a relatively simple closed expression valid to all orders
in $\alpha'$: the Dirac-Born-Infeld action. The catch is that this expression is only valid
in the {\em slowly varying field strength limit}. You could compare this success with the gravitational
sector, deriving from the closed string, where only the first few orders in $\alpha'$ are known.
The results are less spectacular in the non-abelian case however where achievements are only
perturbative in $\alpha'$.

An inconvenient property of the effective action is its ambiguity.
Indeed, in field theory the following equivalence theorem exists: different Lagrangians lead to the
same on-shell S-matrix, in our case equal to the string theory S-matrix, if there exists a
{\em field redefinition} $\phi=\phi' + T(\phi')$ transforming these Lagrangians into each other.
Therefore we have to be careful when comparing results in the literature that look at first sight
very different.

\subsection{Dirac-Born-Infeld and Wess-Zumino}
\label{DBIWZ}

Let us consider the abelian case and take the limit of constant gauge field strengths.
Under these conditions the bosonic part of the
effective action for the fields coupling to a D$p$-brane can be obtained to all orders in $\alpha'$
and consists of two contributions: the Dirac-Born-Infeld term and the Wess-Zumino term.
The Dirac-Born-Infeld \cite{BornInfeld}\cite{dirac} term reads
\begin{equation}
S_{\text{DBI}} = - T_p \int d^{p+1} x \, e^{-\Phi} \left[ -\det\left(P[G+B]_{ab}+2\pi\alpha' F_{ab}\right)\right]^{1/2} \, ,
\label{DBI}
\end{equation}
with $T_p$ the already calculated D-brane tension, $G$ the metric, $B$ the NS-NS 2-form, $\Phi$ the
dilaton (see table~\ref{closedstringspectrum}) and $F$ the field strength of $A$ (see table~\ref{openstringspectrum}), $F=dA$.
Furthermore,
\begin{equation}
P[T]_{a_1\ldots a_n} = \frac{\partial X^{\mu_1}}{\partial x^{a_1}} \cdots \frac{\partial X^{\mu_n}}{\partial x^{a_n}} T_{\mu_1\cdots\mu_n}
\end{equation}
denotes the pullback of the tensor $T$. We have seen that the D-brane also couples to the R-R fields $C_{[n]}$.
These interactions are described by the Wess-Zumino term\footnote{We leave out
terms arising in a massive IIA background \cite{groningenTdual}\cite{GHTWZ}.}:
\begin{equation}
S_{\text{WZ}} = \mu_p \sum_n \int_{V_{p+1}} P\left[C_{[n]} e^{B}\right] e^{2\pi\alpha' F},
\label{WZ}
\end{equation}
where $\mu_p$ is the already calculated charge. Here all multiplications are in fact wedges of forms.
The formula should be interpreted as follows: take an allowed R-R field $C_{[n]}$, i.e.\ even for type IIB and odd
for type IIA, then select from the expansions of the exponentials a form with the appropriate dimension
$p+1-n$ so that we can integrate over the world-volume of the D-brane. In this way, one finds
the coupling considered in eq.~\eqref{branecoupling} as the leading term.
The Wess-Zumino term was first introduced in \cite{douglasWZ}
based on observations of the D-brane boundary state in the presence of a constant gauge field in \cite{li}.
An easy way to see that we must add other terms to the term in eq.~\eqref{branecoupling} is
through an argument based on invariance under T-duality \cite{groningenTdual}\cite{alvarezTdual}.
Moreover, in \cite{greenWZ} it was shown that the Wess-Zumino term is necessary to cancel the anomaly of chiral
fermions on intersections of branes or on branes wrapping cycles on curved manifolds.

This action is invariant under a number of local symmetries. First of all, there is the
freedom to reparameterize world-volume as well as space-time coordinates,
then there is not only the gauge symmetry of $A_{a}$
but also the gauge symmetries of $B$ and $C_{[n]}$. The latter are realized as follows.
The bulk field $B$ and the boundary field $A$ appear in the combination ${\cal F}=P[B]+2\pi\alpha' F$
in the world-sheet action.
We see from the string world-sheet action eqs.~\eqref{polyactioncurved1} and \eqref{polyactioncurved2}
that the tensor gauge symmetry associated to $B$,
\begin{equation}
\delta B= d \xi \, ,
\end{equation}
where $\xi$ is a 1-form, must because of the boundary be completed with
\begin{equation}
\delta A= -\frac{P[\xi]}{2\pi\alpha'} \, .
\end{equation}
Indeed, the combination ${\cal F}$ is then invariant under
the tensor gauge symmetry and it is this combination that must appear in the effective action as well.
In addition, the Wess-Zumino action is invariant under the following collective gauge transformation of the R-R
fields \cite{GHTWZ}:
\begin{equation}
\delta C=d \mu + H \wedge \mu + \lambda e^B
\end{equation}
where $C=\sum_n C_{[n]}$, $H=dB$ and the forms $\mu=\sum_n \mu_{[n-1]}$ and the
scalar $\lambda$ generate the gauge transformation.
It is understood that forms of the appropriate dimensions are selected to match the dimensions
on both sides.

The main theme of this thesis will be to loosen the two assumptions and try to calculate the effective
D-brane action in the abelian case with derivative corrections or in the non-abelian case.
It turns out that in the non-abelian case there is no appropriate analogue of the slowly varying field strength limit
so that derivative corrections must always be included.

As the dependence on $B$ can be found from the dependence on $F$, we will put $B=0$.
Furthermore, we will work in a flat background $G_{\mu\nu}=\eta_{\mu\nu}$ and we will
not study corrections involving derivatives of pullbacks of the bulk fields $G_{\mu\nu}, B_{\mu\nu}, \Phi, C_{[p+1]}$
\cite{curvaturecorr}.
Next, we first employ space-time diffeomorphisms to align the fiducial world-volume of the D-brane
along $X^i=0$ with $i=p+1,\ldots,9$ and then world-volume diffeomorphisms to match the world-volume
coordinates with the remaining space-time coordinates, $X^a=x^a$.
This gauge is called the {\em static gauge}.  As a consequence the induced metric reads $P[G]_{ab}=\eta_{ab}+\partial_a \Phi^i \partial_b \Phi_i$,
where the $\Phi^i=X^i$ are the scalars that describe the transverse position of the D-brane.
Furthermore, we will usually consider D$9$-branes or D$p$-branes where we put $\Phi^i=0$.
The pull-backs become trivial and the Dirac-Born-Infeld action reduces to its Born-Infeld form.
The expression for lower-dimensional D-branes including the $\Phi^i$ can be derived from dimensionally reducing the expression for
the D$9$-brane.

\subsection{Ancient History of Born-Infeld}

Before moving on it is interesting to have a look at the reason why the
Born-Infeld action was introduced in the first place \cite{BornInfeld}, namely
as an improvement of the Maxwell action which would allow the energy of the field configuration
of an electric point source to be finite. At that time people were worried that using the Maxwell equations it
is impossible to describe an electron as a point source without having not only a singularity
for the electromagnetic field at the origin, but also infinite energy.  Only much later on
the principles of renormalization were discovered, solving the issue in a completely different way.

The Born-Infeld action in $3+1$ flat dimensions can be written as
\begin{equation}
S_{\text{BI}} = - T_4 \int d^4 x \sqrt{1 + \frac{1}{2} b^2 F_{\mu\nu} F^{\mu\nu} - b^4 \left( \frac{1}{4} F_{\mu\nu} \dual{F}^{\mu\nu} \right)^2 } \, ,
\end{equation}
where $T_4=1/b^2$ is the tension, $\dual{F}^{\mu\nu}=\frac{1}{2}\epsilon_{\mu\nu\lambda\kappa} F^{\lambda\kappa}$ the
dual field strength and we replaced the dimension length-squared constant $2\pi\alpha'$ by $b$.
The Bianchi identities are the same as in Maxwell theory, $dF=0$, while the equations of motion become
$d^*G=0$ with
\begin{equation}
G^{\mu\nu} = \frac{2}{T_4 b^2}\frac{\partial {\cal L}_{\text{BI}}}{\partial(F_{\mu\nu})} =
\frac{-F^{\mu\nu}+\dual{F}^{\mu\nu}\; b^2\left(\frac{1}{4} F_{\lambda\kappa}\dual{F}^{\lambda\kappa}\right)}{\sqrt{1 + \frac{1}{2} b^2 F_{\rho\xi} F^{\rho\xi} - b^4 \left( \frac{1}{4} F_{\rho\xi} \dual{F}^{\rho\xi} \right)^2 }} \, .
\label{consti}
\end{equation}
Defining the electromagnetic fields $E_i$, $B_i$, $D_i$ and $H_i$ for $i=1,\ldots,3$ as follows:
\begin{equation}
\begin{split}
E_i & = F_{0i}, \quad B_i = \frac{1}{2} \epsilon_{ijk} F^{jk}, \\
D^i & = G^{0i}, \quad H^i = - \frac{1}{2} \epsilon^{ijk} G_{jk},
\end{split}
\end{equation}
the set of field equations is exactly the same as for Maxwell theory in a substance with
dielectric constant and susceptibility being certain functions of the fields.
Indeed, the constitutive equations, which can be derived from eq.~\eqref{consti}, read
\begin{equation}
\begin{split}
D_i & = \frac{E_i + B_i \left(\vec{E}\cdot\vec{B}\right)}{\sqrt{1+b^2\left(\vec{B}^2-\vec{E}^2\right)-b^4 \left(\vec{E}\cdot \vec{B}\right)^2}} \; , \\
H_i & = \frac{B_i - E_i \left(\vec{E}\cdot\vec{B}\right)}{\sqrt{1+b^2\left(\vec{B}^2-\vec{E}^2\right)-b^4 \left(\vec{E}\cdot \vec{B}\right)^2}} \; .
\label{consti2}
\end{split}
\end{equation}

Putting the magnetic field $B$ to zero we find for the Lagrangian
\begin{equation}
{\cal L}_{\text{BI}} = - T_4 \sqrt{1- b^2 \vec{E}^2}
\end{equation}
which sets a maximal value $|\vec{E}|\leq E_{\text{max}}$ with $E_{\text{max}}=1/b$ just as there is an upper limit
for the velocity in special relativity. In fact, if $\vec{E}$ is constant, after T-duality along
the direction of $\vec{E}$ the speed of the brane is precisely $b\vec{E}$ so that
the upper limit on the electric field follows from the upper limit on the velocity.
Inserting a point source at the origin $\rho(\vec{r})=q \delta(\vec{r})$  we find for the radial component of $D$
\begin{equation}
D_{r} = \frac{q}{4 \pi r^2} \, .
\end{equation}
Eq.~\eqref{consti2} becomes
\begin{equation}
D_r = \frac{E_r}{\sqrt{1- b^2 E_r^2}}
\end{equation}
such that
\begin{equation}
E_r = \frac{q}{4 \pi \sqrt{r^4 + \left(\frac{bq}{4\pi}\right)^2}} \, .
\end{equation}
Note that $E_r$ attains the maximal value $E_{\text{max}}$ at the origin.
We can also calculate the energy:
\begin{equation}
H = \int d^3 x \left( T_4 b^2 E_i D^i - {\cal L}_{\text{BI}} - T_4 \right) = \frac{q^{\frac{3}{2}}b^{-\frac{1}{2}} \Gamma(\frac{1}{4})^2}{12 \pi} \approx 0.349 \; q^{\frac{3}{2}}b^{-\frac{1}{2}} \, .
\end{equation}
Here we subtracted the zeroth order contribution. The result is then indeed finite!

Another interesting point is that the usual electromagnetic duality can be generalized to
Born-Infeld theory in the following way \cite{emduality}:
\begin{equation}
\begin{split}
F_{\mu\nu} \longrightarrow \cos \alpha \; F_{\mu\nu} + \sin \alpha \; \dual{G}_{\mu\nu} \, , \\
G_{\mu\nu} \longrightarrow \cos \alpha \; G_{\mu\nu} + \sin \alpha \; \dual{F}_{\mu\nu} \, .
\end{split}
\label{emduality}
\end{equation}
The equations of motion are invariant under this transformation.
As a collorary, for self-dual field configurations $F_{\mu\nu} = \pm \dual{F}_{\mu\nu}$
(in Euclidean Born-Infeld theory) this transformation implies that the equations of motion
(and thus the Lagrangian) reduce to
the equations of motion and the Lagrangian of Maxwell theory.
For instance the non-abelian
Born-Infeld theory with the symmetric trace prescription has electromagnetic duality.
As a consequence, instanton
equations in four dimension do not receive $\alpha'$ corrections. If electromagnetic duality,
eq.~\eqref{emduality}, is to be kept, this is expected to hold even when derivative corrections
are taken into account. In \cite{thorlacius} it is shown via the $\beta$-function method that
a related set of solutions does not acquire corrections if the full D-brane effective action is
taken into account. As we will see it is a totally different matter when there are more than four
dimensions.

\subsection{Different Roads to Rome}
\label{roads}

The Born-Infeld action for open strings in $10$ dimensions --- in modern language the D$9$-brane action --- was first
obtained in \cite{fradkintseytlin1} and \cite{ACNY} and extended to lower-dimensional branes in \cite{leigh}.
In \cite{polchinski} it was realized that D-branes take a prominent place as non-perturbative solitons, which
led to a renewed interest in D-brane effective actions. The supersymmetric version
\cite{susyBIcederwall}\cite{susyBIaganagic}\cite{susyBIbergshoeff} comes about as follows.
First a manifestly $N=2$ supersymmetric action with an additional local fermionic symmetry,
{\em $\kappa$-symmetry}, is written down.  The cancellation of terms under the $\kappa$-transformation involves a subtle
interplay between the Born-Infeld and the Wess-Zumino part.  Fixing $\kappa$-symmetry with a gauge choice leads to
a supersymmetric version of the Born-Infeld action. The supersymmetry transformation takes on a complicated form since
it contains a compensating $\kappa$-transformation to stay in the $\kappa$-gauge.

However, as shown in \cite{bilal1} physically it is hard to make sense of the limit of
large but slowly varying fields. Indeed, small derivatives imply that the fields stay large
over an extended region. An estimate of the total energy indicates that gravitational
effects, present in the closed sector of string theory, can no longer be neglected and the system
is at risk of collapsing to a black hole. So derivative corrections should be taken
into account. Let us introduce the notation ${\cal L}_{(r,q)}$ for the
contributions to the Lagrangian from derivative terms of the form
\begin{equation}
g_{\text{YM}}^{-2} \, \alpha'{}^{q+r} \partial^{2q} F^{r+2} \, .
\label{termform}
\end{equation}
Derivative corrections were first studied in \cite{abelian4derivative}. There it was shown that
for the superstring ${\cal L}_{(r,1)}=0, \, \forall r$ so that the first correction
has four derivatives.  The contribution ${\cal L}_{(2,2)}$ to the Born-Infeld part
was calculated in the same article.
Only much later all terms with four derivatives, $\sum_{r=2}^{\infty} {\cal L}_{(r,2)}$, were found in \cite{wyllard}
and a conjecture has been made for terms with more derivatives \cite{wyllard2}. The conjecture gives a prescription
to calculate derivative corrections to the Born-Infeld part from derivative corrections to the Wess-Zumino part.
However, the conjecture is not entirely well-defined (there are ordering ambiguities) and since the results for
the Wess-Zumino term with more than four derivatives --- calculated in \cite{wyllard} --- are incomplete, the results
following from the conjecture for the Born-Infeld term are also expected to be only partial.
All terms in the Born-Infeld part of $\sum_{q=2}^{\infty} {\cal L}_{(2,q)}$ were found in \cite{groningen3} and the procedure therein
was extended to the non-abelian
case in \cite{brasil2}, but there it leads only to partial information. Despite the limitations of the conjecture
in \cite{wyllard2}, the terms predicted by it with six derivatives and four field strengths agree with \cite{groningen3}.

As we have seen, when several branes coincide the result is a non-abelian gauge theory where the
zeroth order in $\alpha'$ is the Yang-Mills action \cite{wittenmatrixval}.  However, compared
to the abelian case much less progress has been made; there is no all-order in $\alpha'$ result.
Firstly, it is not known how to order the --- now non-commuting --- field strengths and covariant derivatives
and secondly new identities of the form
\begin{equation}
\left[D_{\mu},D_{\nu}\right] F_{\rho\lambda} = \left[F_{\mu\nu},F_{\rho\lambda}\right] \, ,
\end{equation}
appear, relating commutators of covariant derivatives to commutators of field strengths. It turns out that there is no unambiguous way
to take the slowly varying field strength limit without taking at the same time the abelian limit!
Up to order $\alpha'^2$ the non-abelian effective action can be extracted
from the abelian one by symmetrizing over the gauge indices \cite{grosswitten}\cite{tseytlinvectorfield}.

In \cite{tseytlinSTr} a truncation of the non-abelian effective action was proposed
where only the terms are kept that come from symmetrizing over the gauge indices (Tseytlin's {\em symmetrized
trace} proposal). Soon, it turned out that this proposal could not capture essential physics; it made
the wrong prediction for the spectrum of strings stretching between intersecting branes \cite{HTspectrum}\cite{DST}.
The discrepancy starts at $\alpha'{}^4$.
Actually, already at order $\alpha'{}^3$ corrections had been found in \cite{kitazawa},
contradicting the symmetrized trace prescription that only produces terms at even order in $\alpha'$
because of the antisymmetry of $F$.  However, the terms in \cite{kitazawa} turned out to be wrong
in the sector sensitive to the 5-point amplitude, but not to the 4-point amplitude.
Indeed, in string theory there are only contributions to the
spectrum of intersecting branes from even orders in $\alpha'$ and these terms did make
a non-zero contribution. So the search was on for the correct order $\alpha'{}^3$
contribution.

We computed in \cite{alpha3} the correct bosonic terms at order $\alpha'{}^3$ and in \cite{testalpha3} checked that our terms
did not contribute to the spectrum.  Furthermore, our result was confirmed
in \cite{groningen},\cite{sym1},\cite{sym2} and \cite{brasil}. We also computed the bosonic terms at
order $\alpha'{}^4$ in \cite{alpha4}.  These terms did pass the spectrum test \cite{testalpha4},
which is rather interesting because this is the first order where the symmetrized trace prescription fails
this test.

After this brief overview of the developments, let us now list the different ways --- used by the different groups
above --- of obtaining the action, eqs.~\eqref{DBI} and \eqref{WZ}, and their (derivative and/or non-abelian) corrections.
We make a basic distinction between direct methods, which calculate
the action straightforwardly from string theory and indirect methods, which use a symmetry or other property
the action should have.  The disadvantage of the indirect methods is that in most cases the action
is not completely fixed by requiring the desired property and typically unknown coefficients remain.

\subsubsection*{Direct Methods}

\begin{itemize}
\item String S-matrix method \cite{neveuscherk}\cite{Smatrix}.
The first approach follows immediately from the definition of the effective action.
The idea is to calculate $n$-point scattering amplitudes in perturbative string theory.
Then the most general gauge-invariant Lagrangian at the appropriate order is constructed
or an appropriate ansatz is made. Next its unknown coefficients
are fixed by comparing the on-shell scattering amplitudes with the results of string theory.
In fact, we have used this method at a rudimentary level to determine $\tau_p$ and $\mu_p$ in section
\ref{dbranes}. Since an $n$-point amplitude can only probe terms in the effective action containing up to $n$
gauge potentials $A_{\mu}$ and thus $n$ field strengths $F_{\mu\nu}$, the method is
perturbative in the number of field strengths although with a good ansatz it is possible
to construct an infinite series of derivative corrections.  With an ansatz based on supersymmetry,
in \cite{groningen3} the maximum information was extracted from the 4-point amplitude when
the Born-Infeld part of $\sum_{q=2}^{\infty} {\cal L}_{(q,2)}$ was calculated.

Because of its perturbative nature in $F$, the method is not powerful enough even to produce the abelian Born-Infeld term.
In the non-abelian case it has been more successful mostly because in that case the other available methods are
not as powerful either.  In \cite{grosswitten} and \cite{tseytlinvectorfield} the bosonic sector up
to order $\alpha'^2$ was calculated from the 4-point amplitude.  For the fermionic
sector --- in the meantime already known from methods based on supersymmetry \cite{bergshoeffnoether} ---
this method was used in \cite{metsaevsusy} and \cite{groningensevrin}.
Partial information on order $\alpha'{}^3$ and $\alpha'{}^4$ was obtained from the same 4-point
function in \cite{bilal1}.  But the full calculation of the bosonic sector at $\alpha'{}^3$ via the
S-matrix requires a 5-point amplitude.  This was successfully completed in \cite{brasil},
correcting an older result in \cite{kitazawa}.
\item Partition function approach. This method was developed in \cite{fradkintseytlin0} where it was
realized that the Polyakov path-integral with background fields produced --- barring renormalization
subtleties described in more detail in \cite{tseytlinsubtle}\cite{abelian4derivative} --- the effective action.  This was based
on the observation that the background fields in the Polyakov action eqs.~\eqref{polyactioncurved1} and \eqref{polyactioncurved2}
act as sources for the corresponding particles and functional derivatives with respect to these fields pull
down the appropriate vertex operators. In fact, the Born-Infeld term was first
calculated with this method in \cite{fradkintseytlin1} and it was shown that there are no corrections
with two derivatives for the superstring in \cite{abelian4derivative}.  The method has been extended to lower-dimensional branes in \cite{tseytlinselfdual}.
\item Boundary state formalism.  Here the boundary state representing the D-brane is constructed \cite{boundarystate} and
then the coupling to the NS-NS or R-R bulk fields is calculated.  This method was used in
\cite{wyllard} to construct all terms with four derivatives in both the Born-Infeld
and the Wess-Zumino part in the abelian case.  When the path-integral representation of the boundary state is used, it becomes clear
that the method is related to the partition function approach.
\item Requiring Weyl invariance of the non-linear $\sigma$-model.  Here one looks at the action for an
open string in a curved background, eqs.~\eqref{polyactioncurved1} and \eqref{polyactioncurved2}.
Then one requires the Weyl anomaly of the $\sigma$-model to vanish, which amounts to putting
the $\beta$-functions to zero.  The resulting equations are equivalent to equations of motion deriving from
an effective action. In this way the abelian Born-Infeld
action was obtained for the D$9$-brane with a $\sigma$-model 1-loop calculation in \cite{ACNY} and for lower-dimensional branes in \cite{leigh}.
A $\sigma$-model 2-loop calculation for the bosonic string was performed in \cite{tseytlin2loop}, yielding corrections with two
derivatives.  The $\beta$-functions are insensitive to the topology of the world-sheet so that they contain
no information about higher string loops. Nevertheless, a proposal to generalize the Weyl invariance method beyond the string tree-level was made in
\cite{weylloop} based on \cite{FS}. The equivalence of the effective action derived in the Weyl invariance approach to the S-matrix effective action is to
our knowledge not yet (rigorously) proven.
\end{itemize}

\subsubsection*{Indirect Methods}

These are based on requiring

\begin{itemize}
\item Supersymmetry. Since world-sheet supersymmetric string theories eventually lead to space-time supersymmetry, the effective
action has to be supersymmetric.  One should realize however that as the effective action receives
corrections, the supersymmetry transformations take on corrections as well and can be very complicated.

If there is more than one supersymmetric invariant, they will all get arbitrary coefficients
in this method and the effective action will only be fixed modulo these coefficients. In \cite{bergshoeffnoether}
an iterative method based on supersymmetry, the {\em Noether method}, was used to construct
the non-abelian terms at order $\alpha'{}^2$ including quadratic fermionic terms. This method was
continued up to order $\alpha'{}^3$ in \cite{groningen} and in the abelian case up to order $\alpha'{}^4$
in \cite{groningen2}. An alternative method, based on superspace Bianchi identities and perturbing conventional
constraints in superspace, was used to construct all terms at order $\alpha'{}^2$
including quartic terms in the fermions \cite{goteborg}.

In yet another approach the effective action for $N=4$ Super-Yang-Mills is calculated.  If there is a unique deformation
of Super-Yang-Mills theory, this deformation should be as well the D$3$-brane
effective action as the effective action for $N=4$ Super-Yang-Mills. Corrections which agree with
the results in \cite{alpha3} and \cite{groningen} up to order $\alpha'{}^3$
were derived in this way in \cite{sym1} and the corrected version of \cite{sym2}.
\cite{sym1} contains results up to order $\alpha'{}^4$, but they are in a form
that cannot be easily compared to our results in \cite{alpha4}. Results at order $\alpha'{}^3$ and partial results
at order $\alpha'{}^4$ were also derived in \cite{sym3}.

Requiring the existence of $\kappa$-symmetry was attempted in the non-abelian case in \cite{nonabeliankappa}, but this
seemed to be problematic.
\item The existence of a certain class of BPS solutions. This is the method used in this thesis in
chapters \ref{abelianBI} and \ref{nonabelianBI}.
We will come back to it in much more detail later. It demands the existence of a certain type of BPS solutions and seems to be
related to the supersymmetry method.  Using this method we succeeded in being the first to calculate
the bosonic sector of the non-abelian D-brane effective action at order $\alpha'{}^3$ \cite{alpha3} and at
order $\alpha'{}^4$ \cite{alpha4}.

\item T-duality, duality, Seiberg-Witten map.  The requirement of T-duality combined with
Lorentz covariance is enough to construct the Born-Infeld part.  The requirement of Montonen-Olive
duality of the D3-brane also puts strong constraints on the action \cite{emduality}\cite{tseytlinselfdual}.
But these requirements do not seem to be strong enough to determine derivative corrections or non-abelian
extensions. Constraints based on the Seiberg-Witten map have been used in \cite{cornalba}.
In \cite{wyllard2} Wyllard has used the Seiberg-Witten map to check his derivative corrections \cite{wyllard}.
S-duality between $SO(32)$ type I en heterotic string theory on the other hand was used
to investigate the terms with six field strengths in \cite{stieberger}.
\end{itemize}

\newpage
\thispagestyle{empty}


\chapter{Soliton Equations in Yang-Mills and Born-Infeld Theory}
\label{equations}
\chaptermark{Soliton Equations in YM and BI Theory}

In this chapter we will generalize the familiar instanton
equations in 4-dimensional Euclidean Yang-Mills theory, first to
more than four dimensions and then to Born-Infeld
theory. We will use the fact that these generalized instanton
equations should put the equations of motion to zero as a constraint to
construct the D-brane effective action in chapters \ref{abelianBI}
and \ref{nonabelianBI}.
Our interest lies thus ultimately in D-brane physics.  The
action for a D$p$-brane can be derived from the supersymmetric
action of a D$9$-brane by dimensional reduction. The actions in
this chapter should be regarded as truncations of D$p$-brane
effective actions where we put all transverse scalars to
zero\footnote{A little nuance is in order. The {\em form} of the equations
is the same for ``instanton'' as for ``monopole'' type
configurations if we take $F_{ij}=D_i \Phi_j$ where $j$ is a
reduced dimension. So we do not have to put the scalars
$\Phi^j=0$, but we can instead hide them in the notation $F_{ij}$.}.
Keeping this 10-dimensional origin in mind we can rely on
supersymmetry for certain lines of reasoning. Furthermore, we will
only consider time-independent magnetic field strengths and thus
work with the $p$-dimensional spatial part of the Lagrangians.
Basically, we are only interested in the {\em form} of the
equations here and we will not study solutions until the next
chapter.

See appendix \ref{conventions} for the conventions about
field strengths, complex coordinates and $\Gamma$-matrices.\footnote{Especially,
note that we work with anti-hermitian gauge field
strengths, which introduces an extra minus sign for the action compared to the
hermitian conventions commonly used in the physics literature. Furthermore, the fact that we work
with the spatial part of a Minkowski action instead of an Euclidean
action puts in another minus sign.}

\section{4d}

We start with the familiar instanton equations in $4$
dimensions. However, since these 4 dimensions should in our case,
as explained, be interpreted as the spatial part of a
higher-dimensional action with Minkowski signature it is more appropriate to
refer to them as soliton equations.

The usual Yang-Mills action,
\begin{equation}
- {\cal S}_{\text{YM}} = H = -\frac{1}{4g_{\text{YM}}^2} \int d^4 x \, \Tr F_{ij} F^{ij} \, ,
\end{equation}
has equations of motion
\begin{equation}
D^{i}F_{ij} = 0 \, .
\label{eomYM}
\end{equation}
Configurations satisfying the (anti-)self-duality condition,
\begin{equation}
F_{ij}=\pm \dual{F}_{ij}=\pm \frac{1}{2}\epsilon_{ijkl}F^{kl} \, ,
\label{selfdual}
\end{equation}
automatically solve the equations of motion \eqref{eomYM} by
means of the Bianchi identities
\begin{equation}
D^i F_{ij} = \pm \frac{1}{2} \epsilon_{ijkl} D^i F^{kl}=0 \, .
\end{equation}

In generalizing, we will focus on two aspects of this solution.
First of all, these solutions satisfy a Bogomolny bound \cite{bogo}:
\begin{equation}
\begin{split}
H & = -\frac{1}{4g_{\text{YM}}^2} \int d^4 x \, \Tr F_{ij} F^{ij} \\
  & = -\frac{1}{8g_{\text{YM}}^2} \int d^4 x \, \Tr (F \mp \dual{F})_{ij} (F \mp \dual{F})^{ij}
      \mp \frac{1}{8g_{\text{YM}}^2} \int d^4 x \, \epsilon^{ijkl} \Tr F_{ij} F_{kl} \\
  & \ge \mp \frac{1}{2g_{\text{YM}}^2} \int \Tr F \wedge F = \frac{8\pi^2}{g_{\text{YM}}^2} |k_{\text{inst}}| \, ,
\end{split}
\label{4dbogobound}
\end{equation}
where we know from the theory of characteristic classes that $k_{\text{inst}}=-\frac{1}{16\pi^2} \int \Tr F \wedge F$ is an
integer labelling the second Chern class of the bundle. It depends only on topological properties
of the bundle. The inequality follows because the
first term in the second line is as a sum of squares always positive. Hence the configurations
satisfying eq.~\eqref{selfdual} are not only a {\em local} minimum of the energy but a {\em global}
minimum for all field configurations belonging to the same topological class.
The well-known finite action, single instanton ($k_{\text{inst}}=1$) solutions to the (anti-)self-duality
condition \eqref{selfdual} in $SU(2)$ Yang-Mills were found in \cite{BPSTmonopole}. Static solutions --- solitons
--- in $3+1$ dimensions in Yang-Mills theory with a Higgs field were found by Prasad and Sommerfield \cite{prasad}
expounding on the monopole ansatz of 't Hooft \cite{hooftmono} and Polyakov \cite{polmono}.  One can consider the
Higgs fields as arising from dimensionally reducing an extra dimension $F_{i4}=D_i \Phi$ in which case
the field strength is again (anti)-self-dual and the bound is saturated.

Secondly, for a configuration not to break all supersymmetry the supersymmetry
transformation of the gaugino\footnote{The transformation of the gauge field contains the gaugino.
Since we consider bosonic backgrounds, it vanishes without further constraints.} has to vanish for some spinors $\epsilon$ and $\epsilon_{NL}$:
\begin{equation}
\delta \chi = \epsilon_{\text{NL}} + \frac{1}{2} F_{ij} \Gamma^{ij} \epsilon = 0 \, .
\end{equation}
The $\epsilon_{\text{NL}}$ generates the so called {\em non-linear supersymmetry} and can always be adjusted
so that the $u(1)$ part of $F$ does not break supersymmetry. Indeed, a single D-brane keeps
half of the supersymmetry, no matter what vacuum expectation value the field strength takes.
Henceforth we take $F$ in $su(N)$ and proceed with the linear supersymmetry
\begin{equation}
\delta \chi = \frac{1}{2} F_{ij} \Gamma^{ij} \epsilon = 0 \, .
\end{equation}
Using the (anti-)self-duality eq.~\eqref{selfdual} and the following
property of $\Gamma$ matrices
\begin{equation}
\Gamma^{i_1\ldots i_n}\Gamma_{1\ldots p}= \frac{(-1)^{\frac{n(n-1)}{2}}}{(p-n)!} \epsilon^{i_1\ldots i_p}\Gamma_{i_{n+1}\ldots i_p} \, ,
\end{equation}
where $p$ is the number of dimensions so that in this case $p=4$,
it is easy to show that for unbroken supersymmetry the generator $\epsilon$ must obey
\begin{equation}
\delta \chi = \frac{1}{2}F_{ij}\Gamma^{ij} \, \frac{1 \mp \Gamma_{1234}}{2} \epsilon = 0 \, ,
\label{gauginovar2}
\end{equation}
for the self-dual (upper sign) and anti-self-dual (lower sign) solution respectively. As a consequence half of the
supersymmetry is left unbroken. Indeed, in the (anti-) self-dual case the generators with
positive (negative) chirality put variation \eqref{gauginovar2} to zero. Let us stress that if
we talk henceforth about a certain fraction of the total supersymmetry, what we regard as ``total'' is the
amount of supersymmetry left unbroken by a single D-brane i.e.\ only the linear supersymmetry, which
is half of the bulk supersymmetry.

Solutions saturating a Bogomolny bound and leaving unbroken a non-trivial subalgebra
of the full supersymmetry algebra are called {\em Bo\-go\-molny-Pra\-sad-Som\-mer\-field(BPS) states}
and the conditions for the bound {\em BPS equations}.
We see from eq.~\eqref{4dbogobound} that the Hamiltonian is then equated to a term which
only depends on topological properties of the field configuration i.e.\ transformation functions
of the bundle and boundary conditions.  This term is called a {\em topological charge}.  In the case of 4-dimensional
Yang-Mills, this charge is given by the second Chern class.

\section{More than Four!}
\label{instantonhigher}

Let us generalize the two properties of the previous section
to dimensions $p>4$ and start with the second one:
the vanishing of the linear supersymmetry transformation of the gaugino,
\begin{equation}
\delta \chi = \frac{1}{2} F_{ij} \Gamma^{ij} \epsilon = 0 \, .
\label{gauginovar3}
\end{equation}
Given a certain $\epsilon$ this should be regarded as an equation for $F$.
While the original symmetry of the action is $SO(p)$, eq.~\eqref{gauginovar3}
has a smaller symmetry group $G$, the group that leaves $\epsilon$ invariant\footnote{When acting
on spinors one should more properly speak about $Spin(p)$ since $SO(p)$ has no spinor representations.
However, we will not be concerned about global properties. So we will be sloppy about this matter and
write $SO(p)$ everywhere.}.
We recognize in $F_{ij}\Gamma^{ij}$ the generator of an infinitesimal
rotation in the spinor representation. Consequently the solutions for $F$
of eq.~\eqref{gauginovar3} are specified precisely by the generators of the
algebra of $G$.
For the classification of these solutions we therefore arrive naturally at a (for instance in
special holonomy) very well studied mathematical question: which subgroups of
$SO(p)$ have invariant spinors.
Later on, we will study two cases
in more detail: $SU(k) \subset SO(2k)$ (the complex case) and $SO(7)_{\pm}\subset SO(8)$ (the
octonionic case).
For our purposes these cases are the most interesting because they leave the least
supersymmetry and put the most severe constraints on the actions we will try to construct.
Actually, even the equations of the complex case can --- in dimensions lower than $8$ --- be obtained from the equations
of the octonionic case by putting some field strength components to zero so that
the octonionic case imposes the most severe constraints.  The complex equations on the
other hand are easier to work with and can be used in all even dimensions, even in 10 and more dimensions
although the supersymmetry argument is strictly speaking not valid anymore.
The complex equations are the ones we will put into use to construct
the (non-abelian) D-brane effective action.
For now we try to keep our derivation as general as possible.

We indicate (one of) the $G$-invariant spinor(s) with $|\eta_0\rangle$, which we take to be commuting
and orthonormal, $\langle \eta_0|\eta_0\rangle=1$, and move on to the demonstration
of the Bogomolny bound.  Using the following property of $\Gamma$ matrices
\begin{equation}
\Gamma^{ij}\Gamma^{kl}=\eta^{jk}\eta^{il}-\eta^{jl}\eta^{ik} + \eta^{jk}\Gamma^{il}+\eta^{il}\Gamma^{jk}
-\eta^{jl}\Gamma^{ik}-\eta^{ik}\Gamma^{jl}+ \Gamma^{ijkl} \, ,
\label{gammaprop}
\end{equation}
one can ``square'' the gaugino variation \eqref{gauginovar3}
\begin{equation}
F_{ij}\Gamma^{ij}F_{kl}\Gamma^{kl} = -2 F_{ij}F^{ij} + F_{ij}F_{kl}\Gamma^{ijkl} \, .
\label{varsquare}
\end{equation}
The trick is to ``sandwich'' this expression between $\langle \eta_0|$ and $|\eta_0\rangle$,
\begin{equation}
\begin{split}
H = & - \frac{1}{4g_{\text{YM}}^2} \int d^p x \, \Tr F_{ij} F^{ij} \\
= & \, \frac{1}{8g_{\text{YM}}^2} \int d^p x \left[ \sum_A \left|F_{kl}^A \Gamma^{kl}|\eta_0\rangle\right|^2
-\Tr F_{ij}F_{kl} \langle \eta_0 | \Gamma^{ijkl}| \eta_0 \rangle \right] \\
\ge & - \frac{1}{2g_{\text{YM}}^2} \int \Tr F \wedge F \wedge \dual{T} \, ,
\label{BPSboundhigher}
\end{split}
\end{equation}
so that we find a Bogomolny bound. In the above $A$ runs over the gauge indices.
Furthermore, we introduced the tensor $T_{ijkl}=\langle \eta_0| \Gamma_{ijkl}|\eta_0\rangle$,
which is obviously invariant under $G$, and
used $(F_{ij} \Gamma^{ij}|\eta_0\rangle)^\dagger=-\langle \eta_0|F_{ij}\Gamma^{ij}$.
We will show below that the last term in eq.~\eqref{BPSboundhigher} is a topological charge\footnote{Although
we did not study this in detail, we expect that the topological
charge also appears in the supersymmetry algebra as a central charge \cite{wittenolive}. Indeed, in
the supersymmetry algebra of 6-dimensional \cite{olivealgebra} and 10-dimensional Super-Yang-Mills theory appears
a topological current which looks suspiciously like the topological charge above.},
depending only on the topological
class of $F$ {\em and} $T$ so that configurations satisfying
\begin{equation}
F_{ij} \Gamma^{ij} |\eta_0\rangle = 0 \, ,
\label{gauginovar4}
\end{equation}
are, just as in the 4-dimensional case, at a global minimum of the energy inside their topological class.

In the following we will show that this equation is in fact equivalent to the one appearing in the pioneering
work of Corrigan et al.~\cite{shm}.  There the following first-order
differential equation is proposed as a generalization of eq.~\eqref{selfdual}:
\begin{equation}
F_{ij} = \frac{1}{2\lambda} U_{ijkl}F^{kl} \, ,
\label{corrigan}
\end{equation}
where $U$ is a totally antisymmetric tensor and $\lambda$ a constant. A relationship like this again implies
by way of the Bianchi identities that the Yang-Mills equations are satisfied.
Applying eq.~\eqref{corrigan} twice and demanding consistency, one
can show that only certain values for $\lambda$ are allowed.
Using eq.~\eqref{corrigan} in the action,
\begin{equation}
H(F^{\text{sol}}) = - \frac{1}{4g_{\text{YM}}^2} \int d^p x \, \Tr F^{\text{sol}}_{ij}F^{\text{sol},ij} =
- \frac{1}{2g_{\text{YM}}^2\lambda} \int \Tr F^{\text{sol}} \wedge F^{\text{sol}} \wedge \dual{U} \, ,
\end{equation}
we see that the solutions of eq.~\eqref{corrigan} saturate the Bogomolny bound of eq.~\eqref{BPSboundhigher} for
$U=T$ and $\lambda=1$, which implies immediately eq.~\eqref{gauginovar4}.
The converse can be proved by multiplying eq.~\eqref{gauginovar4}
with $\langle \eta_0|\Gamma^{kl}$ and using eq.~\eqref{gammaprop}.
A problem in this derivation is the presence of terms containing two
$\Gamma$-matrices, $\langle \eta_0|\Gamma_{ij}|\eta_0 \rangle$.  However, in
\cite{bak} it is shown from the reality of the field strengths $F^A_{ij}$ that if $|\eta_0\rangle$
is an invariant spinor then so must $C|\eta_0\rangle^*$,
\begin{equation}
F_{ij} \Gamma^{ij} |\eta_0\rangle = 0 \Rightarrow F_{ij} \Gamma^{ij} C|\eta_0\rangle^* = 0 \, ,
\label{conjspinor}
\end{equation}
where $C$ is the charge conjugation matrix:
\begin{equation}
C=\left(\Gamma_{1\ldots p}\right)^k \Gamma_2 \ldots \Gamma_{2k} \, ,
\end{equation}
so that $\left(\Gamma_i\right)^T=\left(\Gamma_i\right)^*=C^{\dagger}\Gamma_i C$.
Replacing in the above derivation $|\eta_0\rangle$ with $C|\eta_0\rangle^*$ changes the
sign of terms with two $\Gamma$ matrices with respect to the other terms so that these terms
must vanish separately.

We note that the other values
of $\lambda\neq 1$ that appear in \cite{shm} do not correspond to an equation like
\eqref{gauginovar4} and that it is possible to establish the Bogomolny bound directly from
eq.~\eqref{corrigan} \cite{acharyaBPS}\cite{bak}.

Finally, we will proof that a term of the form
\begin{equation}
W[F] = \int P(F) \wedge \dual{T}
\label{topocharge}
\end{equation}
only depends on
topological information (transformation functions of the bundle, boundary conditions) and
is invariant under small deformations of the connection $A$.
Here $P(F)$ is an {\em invariant polynomial} i.e.\ a polynomial in the
components of $F$ such that $P(g^{-1}Fg)=P(F)$ where $g \in U(N)$, the gauge group.
An example is indeed $P(F)=\Tr F \wedge F$, but also the expressions that show
up in section~\ref{BPSBI}. One can show that
\begin{align}
d P(F) & = 0 \, , \label{invpolprop1} \\
P(F')-P(F) & = d Q \, , \label{invpolprop2}
\end{align}
where $F$ and $F'$ are the curvatures of two connections $A$ and $A'$ on the
same bundle.
The proof is standard, see e.g.~\cite{egh}.
From eq.~\eqref{invpolprop2} follows
\begin{equation}
W[F']-W[F] = \int d\left(Q \wedge \dual{T}\right) \, ,
\label{topodif}
\end{equation}
if $d^*T=0$. We will see that
in the complex case $\dual{T}=K \wedge \cdots \wedge K$ and in the octonionic
case $\dual{T}=\pm T$ so that this condition becomes $dK=0$ or $dT=0$ respectively.
If the manifold has no boundaries or $Q=0$ on the boundary, which is the
case if $F=F'$ on the boundary, expression~\eqref{topodif} vanishes. This proves
that eq.~\eqref{topocharge} is invariant under deformation of $A$.
On the other hand, let $K'=K+dV$ or $T'=T+dV$, then eq.~\eqref{invpolprop1}
shows that $W[F]$ does not change if there are no boundaries or
$V=0$ on the boundary. Therefore  $W[F]$ only depends on
the cohomology class of $K$ and $T$ respectively.

\subsection{The Complex Case}

We consider the subgroup $SU(k) \subset SO(2k)$. In this case there exists
an $SU(k)$ invariant 2-form $K$ satisfying $K_{ij}K^{jk}=-\delta_i^k$. If integrable it allows
us to introduce complex coordinates. There are two invariant spinors,
namely the ``empty'' state $|0\rangle$ satisfying
\begin{equation}
\Gamma_{\alpha}|0\rangle = 0, \quad \forall \alpha\in\{1,\ldots,k\}\, ,
\end{equation}
and the completely filled state $|\!\uparrow \rangle =2^{-\frac{k}{2}} \Gamma_{\bar{1}}\cdots\Gamma_{\bar{k}}|0\rangle$.
Since the dimension of the spinor representation\footnote{Without using a Weyl condition, so
we count here both negative and positive chirality spinors.} is $2^k$ the fraction of unbroken supersymmetry
is $1/2^{k-1}$.
From $C=\left(\Gamma_{1\ldots p}\right)^k \Gamma_2 \ldots \Gamma_{2k}$
follows $C|0\rangle^*= (-i)^k |\!\uparrow \rangle$ such that according to eq.~\eqref{conjspinor}
if eq.~\eqref{gauginovar4} holds for $|0\rangle$, it must also hold for $|\!\uparrow\rangle$.
In complex coordinates the equations for the invariant spinors become
\begin{equation}
\left(2 \, F_{\alpha\bar{\beta}} \Gamma^{\alpha\bar{\beta}} + F_{\alpha\beta} \Gamma^{\alpha\beta} + F_{\bar{\alpha}\bar{\beta}} \Gamma^{\bar{\alpha}\bar{\beta}}\right) |0,\uparrow \rangle =0 \, ,
\end{equation}
where $|0,\uparrow\rangle$ means that the equation is valid for $|0\rangle$ and $|\!\uparrow\rangle$.
From this follows that, since each term has to vanish separately,
\begin{equation}
F_{\alpha\beta}=F_{\bar\alpha\bar\beta}=0, \label{hol}
\end{equation}
and
\begin{equation}
\sum_{\alpha} F_{\alpha\bar{\alpha}} \equiv F_{\alpha\bar{\alpha}} = 0. \label{duy0}
\end{equation}
Eq.~\eqref{hol} is the condition for a holomorphic vector bundle while eq.~\eqref{duy0}
is called the {\em Donaldson-Uhlenbeck-Yau(DUY)} or {\em stability condition} \cite{duy}.
It can be easily shown that these conditions together solve the equations of motion: see eq.~\eqref{YMcomplexeom}.

Furthermore, we note that
\begin{equation}
\begin{split}
K_{ij} & =-i \langle 0 |\Gamma_{ij}|0\rangle = i\langle \uparrow\!|\Gamma_{ij}|\!\uparrow \rangle, \\
T_{ijkl} & = -\frac{1}{2}(K \wedge K)_{ijkl}= \langle 0,\uparrow\! | \Gamma_{ijkl} | 0,\uparrow \rangle ,
\end{split}
\label{complexgamma}
\end{equation}
which can easily be calculated in complex coordinates. The more general expressions
\begin{equation}
\begin{split}
\langle 0|\Gamma_{i_1\ldots i_{2n}}|0\rangle & = \frac{i^n}{n!}\left(K \wedge \cdots \wedge K\right)_{i_1\ldots i_{2n}},\\
\langle \uparrow\!|\Gamma_{i_1\ldots i_{2n}}|\!\uparrow \rangle & = \frac{(-i)^n}{n!}\left(K \wedge \cdots \wedge K\right)_{i_1\ldots i_{2n}},\\
{\Big.}^{\textstyle *}\!\left(\frac{i^l}{l!} \prod_{1}^l K \right) & =i^k(-1)^l \left( \frac{i^{k-l}}{(k-l)!} \prod_1^{k-l} K \right)
\label{complexgammageneral}
\end{split}
\end{equation}
will be useful later on. From this follows that when the Bogomolny bound is saturated the
following term remains
\begin{equation}
H = \frac{(-1)^k}{2 g^2_{\text{YM}}(k-2)!} \int \Tr F \wedge F \wedge \prod_{1}^{k-2} K \, .
\end{equation}
This term is topological if $dK=0$, which
means that $K$ should be a K\"ahler form.

\subsection{The Octonionic Case}
Next we consider a subgroup $SO(7)_{\pm} \subset SO(8)$.
Some explanation is in order since $SO(7)$ can be embedded
into $SO(8)$ in three essentially different ways. $SO(8)$ has
three 8-dimensional representations transforming
into each other under triality:
a positive- and a negative-chirality spinor representation, and the vector representation.
Each of the ways of embedding $SO(7)$ reduces only one
of the three $8$-dimensional representations to $1+7$
while the other two remain irreducible. We are interested
in the two embeddings $SO(7)_{\pm}$ leaving a positive- respectively
negative-chirality spinor $|\eta_{\pm}\rangle$ invariant.
Since there is only one invariant spinor, the fraction of unbroken supersymmetry is $1/16$.

Let us first explain how this case got its name. Single out one direction, say the 8th.
Eq.~\eqref{corrigan} becomes:
\begin{equation}
F_{8i} = \frac{1}{2} T_{8ijk} F^{jk},
\end{equation}
where $T_{ijkl}$ is 4-form invariant under $SO(7)_{\pm}$.
Now $f_{ijk}=T_{8ijk}$ is invariant under the subgroup of $SO(8)$ that leaves invariant
a spinor {\em and} a vector: this is precisely $G_2$, the automorphism group of the octonions.
Define $o = o^8 + \sum_{i=1}^7 o^i e_i$ with $o^i \in \IR$ for $i \in \{1,\ldots,8\}$
and where the $e_i$ satisfy
\begin{equation}
e_i e_j = - \delta_{ij} + f_{ijk} e_k \, .
\end{equation}
It can be shown that since $f_{ijk}$ is invariant under precisely $G_2$,
this defines the octonionic algebra \cite{octonions}. The $e_i$ are then the unit imaginary
octonions.

In the remainder of this section we focus on the positive-chirality spinor.
For a concrete representation of $|\eta_+\rangle$ and
the corresponding antisymmetric tensor $T$ we introduce
complex coordinates. This means choosing an $SU(4)$ within
$SO(7)_+$. As a consequence the concrete representation
of $T$ in complex coordinates will only be invariant under this $SU(4)$
and not under the whole of $SO(7)_+$. The most general
form of $|\eta_+\rangle$ consistent with $SU(4)$ invariance,
the reality condition $C|\eta_+\rangle^*=|\eta_+\rangle$
and the normalization $\langle \eta_+|\eta_+ \rangle=1$ is
\begin{equation}
|\eta_+\rangle =\frac{1}{\sqrt{2}} \left( e^{i\alpha}|0\rangle + e^{-i\alpha}|\!\uparrow \rangle \right) \, .
\end{equation}
By choosing adapted complex coordinates we can always take $\alpha=0$.

In complex coordinates $T_{ijkl}=\langle \eta_+|\Gamma_{ijkl}|\eta_+\rangle$ reads
\begin{equation}
\begin{split}
T_{\alpha_1\bar{\alpha}_2\alpha_3\bar{\alpha}_4} & =
\delta_{\alpha_1\bar{\alpha}_2}\delta_{\alpha_3\bar{\alpha}_4}
-\delta_{\alpha_1\bar{\alpha}_4}\delta_{\alpha_3\bar{\alpha}_2} \, , \\
T_{\alpha_1\alpha_2\alpha_3\alpha_4} & = 2 \epsilon_{\alpha_1\alpha_2\alpha_3\alpha_4} \, , \\
T_{\bar{\alpha}_1\bar{\alpha}_2\bar{\alpha}_3\bar{\alpha}_4} & =
2 \epsilon_{\bar{\alpha}_1\bar{\alpha}_2\bar{\alpha}_3\bar{\alpha}_4} \, , \\
\end{split}
\end{equation}
and antisymmetric combinations.
Using this concrete representation one easily finds for the square of $T$
\begin{equation}
T_{ijkl}T^{klmn} = -4 \, T_{ij}{}^{mn}+12 \, \delta_{[i}^m\delta_{j]}^n \, ,
\end{equation}
which can be used to derive the Bogomolny bound directly
from eq.~\eqref{corrigan} \cite{acharyaBPS}.
In complex coordinates eq.~\eqref{corrigan} takes the form
\begin{align}
\sum_{\alpha} F_{\alpha\bar{\alpha}} & = 0 \, , \label{duyocto} \\
F_{\alpha_1\alpha_2} & = \frac{1}{2} \epsilon_{\alpha_1\alpha_2\alpha_3\alpha_4} F^{\alpha_3\alpha_4} \qquad \text{and (cc)} \label{holoocto}\, .
\end{align}
The DUY condition is the same as in the complex case while the holomorphicity condition is deformed.
In the next section we will lift both the complex equations and the octonionic equations to Born-Infeld theory.

\section{Lifting the Soliton Equations to Born-Infeld}
\label{BPSBI}

As the Born-Infeld action is only valid for the abelian case, for the remainder
of this chapter we take either as gauge group $U(1)$ or either we only allow the field strengths
to take values in the Cartan subalgebra.  When we evaluate the group trace $\Tr$ in the latter case, the
action will just be a sum of $N$ single brane actions. We will not explicitly denote this summation.
Evaluating the gauge trace as in eq.~\eqref{generatortrace} introduces an extra minus sign.

From \cite{susyBIcederwall}\cite{susyBIaganagic}\cite{susyBIbergshoeff} we borrow the follow identity, which we
adapt to our conventions\footnote{For both the complex and the
octonionic case we can take the dimension of the D-brane to be even $p=2k$.}:
\begin{equation}
\left(\rho^{(2k)}_E + \rho^{(2k)}_O\right)\left(\rho^{(2k)}_E - \rho^{(2k)}_O\right)
= \det \left(\delta + 2\pi\alpha'F\right) \, ,
\label{gammasquared}
\end{equation}
with
\begin{equation}
\rho^{(2k)}_{E} \pm \rho^{(2k)}_{O} = \sum_{n=0}^k \frac{(\pm 1)^{(k-n)}(2\pi\alpha')^n}{2^n n! (2(k-n))!}
F_{i_1 i_2}\ldots F_{i_{2n-1}i_{2n}} \Gamma_{i_{2n+1}\ldots i_{2k}} \epsilon^{i_1\ldots i_{2k}} \, ,
\end{equation}
and
\begin{equation}
\left[\rho^{(2k)}_E,\rho^{(2k)}_O\right] = 0 \, .
\label{EOcomm}
\end{equation}
The subscripts $E/O$ indicate whether the expressions contain $4n$ or $(4n+2)$
$\Gamma$-matrices respectively. Later on it will be more important whether the expressions
are even or odd in the number of field strengths $F$.  We will indicate the latter distinction with
the subscripts $E_F/O_F$.

In the above mentioned papers eq.~\eqref{gammasquared} was used in the context of manifestly $N=2$ supersymmetric
actions to construct a $\kappa$-invariant action for D-branes. The matrix $\rho$ plays a prominent
role in the definition of $\kappa$-symmetry.
The Born-Infeld action follows when
gauge fixing $\kappa$-symmetry and putting all fermions and scalars to zero.
However,
since eq.~\eqref{gammasquared} is more generally valid, the Bogomolny bound below could be
used in more general cases.

For the proof we base ourselves on \cite{susyBIaganagic} and use the rotation invariance of both sides of eq.~\eqref{gammasquared}
to rotate to a special coordinate system where $F_{2i-1,2i}=-F_{2i,2i-1}=f_i$
and all other components zero so that
\begin{equation}
\begin{split}
\rho^{(2k)}_{E} \pm \rho^{(2k)}_{O} & =
\sum_{n=0}^k \sum_{\stackrel{i_{1}  < \cdots <i_{n}}
                                {i_{n+1}< \cdots <i_{k}}}
(\pm 1)^{k-n} (2\pi\alpha')^n f_{i_1} \ldots f_{i_n} \Gamma_{2i_{n+1}-1,2i_{n+1}} \ldots \Gamma_{2i_k-1,2i_k} \\
& = \prod_{i=1}^k \left(2\pi\alpha' f_i \pm \Gamma_{2i-1,2i} \right) \, ,
\end{split}
\end{equation}
where $(i_1,\ldots,i_k)$ is a permutation of the numbers $(1,\ldots,k)$.
From this follows indeed
\begin{equation}
\begin{split}
\left(\rho^{(2k)}_E + \rho^{(2k)}_O\right)\left(\rho^{(2k)}_E - \rho^{(2k)}_O\right)
& = \prod_{i=1}^k \left( 1 + (2\pi\alpha'f_i)^2 \right) \\
& = \det (\delta + 2\pi\alpha' F) \, .
\end{split}
\end{equation}
Swapping the two factors on the left hand side does not change anything in the derivation
so that eq.~\eqref{EOcomm} follows.

Sandwiching $\det\left(\delta+2\pi\alpha'F\right)$ between $\langle \eta_0|$ and $|\eta_0\rangle$ with
$\langle \eta_0|\eta_0\rangle=1$ and where $|\eta_0 \rangle$ is again a spinor
invariant under a certain subgroup of $SO(2k)$, we can construct two Bogomolny bounds:
\begin{equation}
\begin{split}
H_{\text{BI}} & = \tau_{2k} \int d^p x \, \sqrt{\det{\left(\delta+2\pi\alpha'F\right)}} \\
              & = \tau_{2k} \int d^p x \, \sqrt{\langle\eta_0|
          \left(\rho^{(2k)}_E + \rho^{(2k)}_O\right)\left(\rho^{(2k)}_E - \rho^{(2k)}_O\right)
          |\eta_0\rangle}\\
          & = \tau_{2k} \int d^p x \, \Bigg(|\langle\eta_0|\rho^{(2k)}_E|\eta_0\rangle|^2
          +|\langle\eta_0|\rho^{(2k)}_O|\eta_0\rangle|^2 \\
          & \qquad \qquad \qquad +\sum_{\eta\neq\eta_0} |\langle\eta|\rho^{(2k)}_E|\eta_0\rangle|^2
          +\sum_{\eta\neq\eta_0} |\langle\eta|\rho^{(2k)}_O|\eta_0\rangle|^2\Bigg)^{\textstyle\frac{1}{2}}, \\
\end{split}
\label{bogoboundBI}
\end{equation}
where we inserted
a complete set $\sum_{\eta} |\eta \rangle \langle \eta|=1$ and
used the hermiticity properties
$\left(\rho^{(2k)}_E\right)^{\dagger}=\rho^{(2k)}_E$ and
$\left(\rho^{(2k)}_O\right)^{\dagger}=-\rho^{(2k)}_O$.
$\tau_{2k}$ is the D-brane tension introduced in eq.~\eqref{tensioncharge}.
From this follow the bounds
\begin{equation}
\begin{split}
H_{\text{BI}} & \ge \tau_{2k} \int d^p x \, |\langle\eta_0|\rho^{(2k)}_E|\eta_0\rangle| \\
          & \ge \tau_{2k} \int d^p x \, |\langle\eta_0|\rho^{(2k)}_O|\eta_0\rangle|.
\end{split}
\end{equation}
Writing out\footnote{From this point on we use the subscripts $E_F/O_F$ explained below
eq.~\eqref{EOcomm}.} $\int d^p x \; |\langle \eta_0 | \rho^{(2k)}_{E_F}|\eta_0 \rangle|$ and
$\int d^p x |\langle \eta_0 | \rho^{(2k)}_{O_F}|\eta_0 \rangle|$, the $\Gamma$-matrices sandwiched between $\langle \eta_0|$ and
$|\eta_0\rangle$ will become antisymmetric tensors. Both terms are of the form eq.~\eqref{topocharge}
and are thus topological charges.
So there are in fact two possible bounds and two sets of equations that
saturate these bounds:
\begin{subequations}
\begin{align}
\label{solodd}
\langle \eta_0 |\rho^{(2k)}_{O_F} | \eta_0 \rangle & =
\langle \eta |\rho^{(2k)}_{O_F} | \eta_0 \rangle  =
\langle \eta |\rho^{(2k)}_{E_F} | \eta_0 \rangle = 0, \quad \forall |\eta\rangle \neq |\eta_0 \rangle \, ,  \\
\text{or} \qquad
\label{soleven}
\langle \eta_0 |\rho^{(2k)}_{E_F} | \eta_0 \rangle & =
\langle \eta |\rho^{(2k)}_{O_F} | \eta_0 \rangle =
\langle \eta |\rho^{(2k)}_{E_F} | \eta_0 \rangle = 0, \quad  \forall |\eta\rangle \neq |\eta_0 \rangle \, .
\end{align}
\end{subequations}
The two sets differ only by their first equation.
Note that only the first set, where the first equation is {\em odd} in the field strength $2\pi\alpha'F$,
has a limit for small field strengths. Indeed, in the second set the first equation starts with (schematically)
$1+(2\pi\alpha'F)^2$ so that solutions must have $2\pi\alpha'F = {\cal O}(1)$. We will
study solutions of both types in chapter \ref{solutions}.

After gauge fixing $\kappa$-symmetry, the supersymmetry variation of the gaugino
is given by \cite{susyBIaganagic}\cite{angles2}
\begin{equation}
\delta \chi = \epsilon_- + \Gamma_0 \left(\Gamma^{(2k)}_{E_F} + \Gamma^{(2k)}_{O_F}\right) \epsilon_+ \, ,
\end{equation}
where we introduced
\begin{equation}
\Gamma^{(2k)}_{E_F/O_F} = \frac{1}{\sqrt{\det(\delta+2\pi\alpha'F)}} \rho^{(2k)}_{E_F/O_F} \, .
\label{kappagamma}
\end{equation}
The part of the supersymmetry transformation not depending on $F$ is the higher
order generalization of the {non-linear supersymmetry}, which we encountered in
Yang-Mills theory. For solutions of eqs.~\eqref{solodd} and \eqref{soleven} the
part of the supersymmetry variation dependent on $F$ vanishes if we take $\epsilon_+=|\eta_0\rangle$.
In the next chapter we will see for concrete examples
that this is indeed the requirement for having a supersymmetric configuration of D-branes.

\subsection{The Complex Case}

We introduce again the empty state $|0\rangle$ and
the completely filled state $|\!\uparrow\rangle$, which are both invariant
under $SU(k) \subset SO(2k)$ and take $|\eta_0\rangle = |0 \rangle$ or
$|\eta_0\rangle = |\!\uparrow \rangle$. If the field strengths are holomorphic,
eq.~\eqref{hol}, then $\langle \eta_0 |\rho^{(2k)}_{E_F/O_F} | \eta \rangle=0$
for all $\eta \neq \eta_0$. So the equations common to both cases \eqref{solodd} and \eqref{soleven}
are already satisfied. To examine the remaining equation let us calculate
the $F$-dependent part of the supersymmetry
transformation:
\begin{multline}
\label{susyDUY}
(\Gamma^{(2k)}_{E_F} + \Gamma^{(2k)}_{O_F}) |0,\uparrow \rangle =
(-i)^k \prod_{\alpha=1}^k \frac{2\pi\alpha' f_{\alpha}^c + \Gamma_{\alpha\bar{\alpha}}}
{\sqrt{1-(2\pi\alpha'f_{\alpha}^c)^2}} | 0,\uparrow \rangle \\
= (\mp i)^k \prod_{\alpha=1}^k \sqrt{\frac{1\pm f_{\alpha}^c}{1\mp f_{\alpha}^c}} | 0,\uparrow \rangle
= (\mp i)^k \exp \left( \pm \tr \arctanh 2 \pi \alpha' F^c \right) | 0,\uparrow\rangle \, ,
\end{multline}
where the upper and lower signs are for the cases $|0\rangle$ and $|\!\uparrow \rangle$ respectively and the matrix $F^c$ is the complexified version of $F$ defined in eq.~\eqref{complexification}.
For the intermediate steps we have used the $SU(k)$ invariance to diagonalize $F$: $F^c_{\alpha\alpha}=F_{\alpha\bar{\alpha}}$
and all others components zero.
For definiteness we focus on the case $|\eta_0\rangle=|0\rangle$. Using eq.~\eqref{bogoboundBI} it is easy to show that:
\begin{subequations}
\begin{align}
\langle 0|\Gamma^{(2k)}_{E_F} + \Gamma^{(2k)}_{O_F}) |0 \rangle = \pm i^k
& \Longleftrightarrow \langle 0 | \rho^{(2k)}_{O_F} | 0 \rangle = 0 \, ,\\
\langle 0|\Gamma^{(2k)}_{E_F} + \Gamma^{(2k)}_{O_F}) |0 \rangle = \pm i^{k+1}
& \Longleftrightarrow \langle 0 | \rho^{(2k)}_{E_F} | 0 \rangle = 0 \, .
\end{align}
\end{subequations}
Note that in both equations the left hand side is a pure phase.
If $k$ is even, the piece with $\Gamma^{(2k)}_{E_F}$ provides the real part and the piece
with $\Gamma^{(2k)}_{O_F}$ the imaginary part; if $k$ is odd, it is the other way round.
Finally we find for the case of eq.~\eqref{solodd} and eq.~\eqref{soleven} respectively:
\begin{subequations}
\begin{align}
\label{duyBI}
\exp \left( \tr \arctanh 2 \pi \alpha' F^c \right) = \pm 1 & \Longleftrightarrow \tr \arctanh 2 \pi \alpha' F^c = \pi i n, \quad n \in \IZ \, , \\
\label{duyBI2}
\exp \left( \tr \arctanh 2 \pi \alpha' F^c \right) = \pm i & \Longleftrightarrow \tr \arctanh 2 \pi \alpha' F^c = \frac{(\pm 1 + 4n)i \pi}{2}, \quad n \in \IZ \, ,
\end{align}
\end{subequations}
Only the first equation has a small field strength limit. We must then put $n=0$ and find that
\begin{subequations}
\begin{align}
\label{holBI}
& F_{\alpha\beta} = F_{\bar{\alpha}\bar{\beta}} =0 \, , \\
\label{duyBI3}
& \left[\arctanh 2 \pi \alpha' F\right]_{\alpha\bar{\alpha}}  =0 \, .
\end{align}
\end{subequations}
The latter equation is the higher-order generalization of the DUY condition of eq.~\eqref{duy0}
while eq.~\eqref{hol} does not acquire any corrections.

From eq.~\eqref{complexgammageneral} we
find for the two topological charge terms corresponding to the cases of eqs.~\eqref{solodd}
and \eqref{soleven} respectively:
\begin{subequations}
\begin{align}
\label{topoeven}
\int \left| \langle 0 | \rho^{(2k)}_{E_F} | 0 \rangle \right| & =
\pm \int \sum_{t=0}^{\lfloor k/2 \rfloor} \frac{(-1)^t}{(2t)!} \left( \bigwedge_1^{2t} F \right) \wedge \exp K \nonumber \\
& = \pm \int \cos F \wedge \exp K \, \\
\label{topoodd}
\int \left| \langle 0 | \rho^{(2k)}_{O_F} | 0 \rangle \right| & =
\pm \int \sum_{t=0}^{\lfloor (k-1)/2 \rfloor} \frac{(-1)^t}{(2t+1)!} \left( \bigwedge_1^{2t+1} F \right) \wedge \exp K \nonumber \\
& = \pm \int \sin F \wedge \exp K \, ,
\end{align}
\end{subequations}
where the plus or minus sign should be chosen as to make the expression positive and
just as many $K$s are taken from the expansion of the exponential as needed to obtain the right dimension for the integral.

\subsection{The Octonionic Case}
\label{octonionicBI}

We focus on the case with small field strength limit, i.e.\ eq.~\eqref{solodd} with $n=0$ and take $\eta_0=\eta_+$.
For the octonionic case $k=4$ so that
\begin{subequations}
\begin{align}
\rho^{(8)}_{O_F} & = \pi \alpha' \left(F_{ij} \Gamma_{1\ldots 8} - \frac{(2\pi\alpha')^2}{2^3 3!} \epsilon_{iji_1\ldots i_6} F^{i_1 i_2} F^{i_3 i_4} F^{i_5 i_6} \right) \Gamma^{ij}, \\
\label{octoBIeven}
\rho^{(8)}_{E_F} & = \Gamma_{1\ldots 8} + \frac{(2\pi\alpha')^2}{2^2 2! 4!}F_{i_1i_2}F_{i_3i_4} \Gamma_{i_5\ldots i_8} \epsilon^{i_1\ldots i_8} + \frac{(2\pi\alpha')^4}{2^4 4!}F_{i_1i_2}F_{i_3i_4}F_{i_5i_6}F_{i_7i_8} \epsilon^{i_1\ldots i_8}.
\end{align}
\end{subequations}
If we define
\begin{equation}
\hat{F}_{ij}=F_{ij} - \frac{(2\pi\alpha')^2}{2^3 3!}\epsilon_{iji_1\ldots i_6} F^{i_1 i_2} F^{i_3 i_4} F^{i_5 i_6} \, ,
\end{equation}
we see that the first part of eq.~\eqref{solodd}, $\rho^{(8)}_{O_F}|\eta_+ \rangle=0$, is in fact exactly the same
as the lowest order eq.~\eqref{gauginovar4}, but with $F$ replaced by $\hat{F}$.
Thus we find eq.~\eqref{corrigan} --- in complex coordinates eqs.~\eqref{duyocto} and \eqref{holoocto} --- where $F$ is replaced by $\hat{F}$.
However, there is a further equation $\langle \eta | \rho^{(8)}_{E_F} | \eta_+ \rangle=0$ for all $\eta\neq \eta_+$. Using eq.~\eqref{varsquare}
it simplifies to
\begin{equation}
\langle \eta | F_{i_1i_2} F_{i_3i_4} \Gamma^{i_1i_2i_3i_4} | \eta_+ \rangle =0 \Leftrightarrow
\langle \eta | \left(F_{ij} \Gamma^{ij}\right)^2 | \eta_+ \rangle =0 \, .
\end{equation}
Since $\langle \eta_+|\Gamma^{kl}$ spans all $\eta \neq \eta_+$ we find that it is also equivalent to
\begin{equation}
[F,F^{\text{dual}}]^{kl}=0 \, ,
\end{equation}
with
\begin{equation}
F^{\text{dual}}_{ij}=\frac{1}{2} T_{ijkl} F^{kl} \, .
\end{equation}
We see immediately that it is a sort of weaker version of eq.~\eqref{corrigan} for $F$ i.e.\
it is implied by this equation, but the converse is not true. 
Summarizing we get the equations
\begin{equation}
\begin{split}
\hat{F} & =\hat{F}^{\text{dual}} \, , \\
[F,F^{\text{dual}}] & = 0 \, .
\end{split}
\label{octoeq}
\end{equation}
From eq.~\eqref{octoBIeven} we find for the
topological charge term
\begin{equation}
\int \left| \langle 0 | \rho^{(8)}_{E_F} | 0 \rangle \right| =
\pm \int \left( d x^1 \ldots d x^8 + \frac{(2 \pi \alpha')^2}{2} F \wedge F \wedge T + \frac{(2 \pi \alpha')^4}{4!} F \wedge F \wedge F \wedge F\right).
\end{equation}

We will use the complex equations to construct the D-brane effective action in chapters \ref{abelianBI} and
\ref{nonabelianBI}. In the next chapter we will look for solutions of the complex equations both with and without
small field strength limit. We will also describe the search for an octonionic BIon, which should be a solution
of the eqs.~\eqref{octoeq}.

\newpage
\thispagestyle{empty}

\chapter{Brany Realizations}
\label{solutions}

In this chapter we will have a look at concrete solutions of the BPS equations
and their interpretation in string theory.  In the first section configurations
with constant field strengths will be studied. In the T-dual picture these look like
intersecting D-branes.  These examples are taken from \cite{angles1}\cite{jabbari}\cite{moduliangles}\cite{angles2}\footnote{For
intersecting M-brane configurations see \cite{Mbrane}.}
and are reviewed here because they will be used in chapter \ref{checks} for the spectrum check.
We will show that these configurations
satisfy the BPS equations which were found from the Born-Infeld action in the previous chapter.
It is interesting to see that solutions without small field strength limit \cite{wittenBPS} can also be easily
constructed.

In the second section we comment on the octonionic BIon solution, which is still work in progress.

\section[Constant Field Strengths and Intersecting Branes]{Constant Field Strengths and Intersecting\\Branes}

\subsection{Coinciding D-branes with Constant Field Strengths}

Simple examples of BPS states are easy to construct in D-brane physics.
Let us start with one D-brane at rest without field strengths. We discussed before
that a single D-brane already breaks half of the $N=2$ supersymmetry of type II string theory.
Indeed, for a linear combination $\epsilon_+ Q^+ + \epsilon_- Q^-$ of the supercharges
originating from the left- and right-movers on the string world-sheet to be unbroken,
the following condition must be satisfied\footnote{Both $\epsilon_+$ as $\epsilon_-$
are $d=10$ Majorana-Weyl spinors. We take
$\Gamma_{11} \epsilon_+ = \epsilon_+$ so we can get rid of the $\Gamma_{11}$s when convenient.
The $\pm$ is related to the choice of orientation of the brane.},
e.g.\ \cite[chapter 13]{bookpolchinski},
\begin{equation}
\epsilon_- = \pm \prod_{j=p+1}^9 \left(\Gamma_j\Gamma_{11}\right) \epsilon_+ = \pm \Gamma_D \epsilon_+ ,
\label{Dunbrokensusy}
\end{equation}
where the brane is at rest and aligned along the spatial directions $1,\ldots,p$ and
$\Gamma_j \Gamma_{11}$ acts as the parity operation $x^j \rightarrow - x^j$ on
spinors.
Following \cite{boundaryfield}\cite{moduliangles}\cite{wittenBPS} we interpret the matrix $\Gamma_D$
as the spinor representation of the orthogonal transformation that world-sheet modes
undergo when they are reflected from the end of the string. In fact, in chapter 8
this orthogonal transformation will be calculated in the more general case of a D-brane with constant field strength $B$.
According to eq.~\eqref{bcfermions} on vectors it acts as
\begin{equation}
\left(P_+\frac{1-B}{1+B} - P_-\right)^i{}_j \, ,
\label{bmatrix}
\end{equation}
where $P_+$ and $P_-$ are projectors on the directions parallel and transverse to the brane respectively.
Putting $B=0$ we see that all Dirichlet directions,
i.e.\ the directions spanned by $P_-$, take on a minus sign.  The product of these reflections in the
Dirichlet directions is on spinors indeed represented by $\Gamma_D$.

Let us restrict to even-dimensional branes $p=2k$ and
consider two coinciding D$(2k)$-branes with constant magnetic field
\begin{equation}
F_{ij}= f_{ij} \left(i \sigma_3\right)=
i \left(
   \begin{array}{cc}
     f_{ij} & 0 \\
     0 & -f_{ij}
   \end{array}
  \right).
  \label{fdef}
\end{equation}
In the following we derive the conditions under which some supersymmetry is left unbroken.
The interpretation of the matrix orthogonally transforming the right-moving modes with respect to the left-moving
modes remains valid in the presence of constant field strengths \cite{boundaryfield}.
In eq.~\eqref{bmatrix} we put $B_{ij}=\pm 2\pi\alpha' f_{ij}$ for the first and second brane
respectively and find the following conditions for supersymmetry on both branes
\begin{equation}
\epsilon_- = \Gamma_D \rho(L) \epsilon_+, \qquad
\epsilon_- = \Gamma_D \rho(L^{-1}) \epsilon_+,
\end{equation}
where we indicate the spinor representation of the rotation $L=\frac{1-2 \pi \alpha'  f}{1+ 2 \pi \alpha' f}$
by $\rho(L)$.
Combining the two conditions
we find that $\epsilon_+$ generates an unbroken supersymmetry if
\begin{equation}
\rho(L)^2 \epsilon_+ = \epsilon_+ \, .
\label{BPScond}
\end{equation}
To proceed it is convenient to define the {\em angle matrix} $\phi$ as follows
\begin{equation}
L=\frac{1-2\pi f}{1+2\pi f}=\exp(-2 \, \arctanh 2 \pi f) = \exp -\phi \, .
\end{equation}
This enables us to write down explicitly the spinor representation of $L$ as
\begin{equation}
\rho(L)= \pm \exp \left(\frac{1}{4}\phi_{ij}\Gamma^{ij}\right) \, .
\label{spinorrep}
\end{equation}
In the above all functions of matrices are defined by their series expansion.
We can always rotate to a canonical coordinate frame where the $f$ and $\phi$ matrices take
block diagonal forms
\begin{equation}
\begin{split}
f_{2i-1,2i} & =-f_{2i,2i-1} = f_i \, , \\
\phi_{2i-1,2i} & = -\phi_{2i,2i-1}=\phi_i = 2 \arctan(2\pi\alpha' f_i) \, ,
\label{angles}
\end{split}
\end{equation}
and all other components zero.
From eq.~\eqref{spinorrep} we see that the eigenvalues of $\rho(L)^2$ are
\begin{equation}
\exp (2i \sum_{i=1}^k s_i \phi_i) \, ,
\end{equation}
where $s_i\in \{-\frac{1}{2},+\frac{1}{2}\}$. The number of unbroken supersymmetries for
a certain configuration of angles is given by the number
of $+1$ eigenvalues. See table \ref{BPSangles} for an overview.
\begin{table}
\begin{center}
\begin{tabular}{|c|l|c|c|}\hline\hline
$k$ &BPS Condition & Invariance Group & Susys\\ \hline\hline
2 &$\phi_1 \pm \phi_2=2\pi n$ & $SU(2)$ & 8\\ \hline
3 &$\phi_1 \pm \phi_2 \pm \phi_3=2\pi n$ & $SU(3)$ & 4\\ \cline{2-3}
  &$\phi_1 \pm \phi_2=2\pi n$, \, $\phi_3=0$ & $SU(2)$ & 8 \\ \hline
4 &$\phi_1 \pm \phi_2 \pm \phi_3 \pm \phi_4=2\pi n$ & $SU(4)$  & 2\\ \cline{2-3}
  &$\phi_1 \pm \phi_2=2\pi n$, $\phi_3 \pm \phi_4=2\pi m$ & $SU(2)\times SU(2)$ & 4 \\ \cline{2-3}
  &$\phi_1 \pm \phi_2 \pm \phi_3 =2\pi n$, $\phi_4=0$ & $SU(3)$ & 4 \\ \cline{2-3}
  &$\phi_1 \pm \phi_2=2\pi n$, \, $\phi_3=\phi_4=0$ & $SU(2)$ & 8
\\ \hline\hline
\end{tabular}
\caption{BPS conditions and the number of unbroken supersymmetries}
\end{center}
\label{BPSangles}
\end{table}
In short the generic BPS condition leaving the least supersymmetry reads
\begin{equation}
\sum_{i=1}^k \phi_i = 2 \sum_{i=1}^k \arctan 2\pi\alpha' f_i = 2 \pi n, \quad n \in \IZ,
\label{BPSgeneric}
\end{equation}
where we chose all plus signs by choosing an appropriate coordinate frame i.e.\
we can flip the sign in front of each $\phi_i$ by swapping $x^{2i-1} \leftrightarrow x^{2i}$.
It leaves a fraction $1/2^{k-1}$ of the supersymmetry of a single D-brane unbroken.

Alternatively, we can follow a more mathematical approach.  The presence of D-branes
along $2k$ dimensions breaks the Lorentz group to
\begin{equation}
SO(9,1) \rightarrow SO(2k) \times SO(9-2k,1) \, ,
\end{equation}
where $SO(9-2k)$ is the Lorentz group in the time and Neumann directions and $SO(2k)$
the rotation group in the Dirichlet directions. $SO(2k)$ is the part were the action is.
The condition for unbroken supersymmetry
\eqref{BPScond} says that $L$ must belong to a subgroup of $SO(2k)$ that has
invariant spinors $\epsilon_+$. This is the case if $L \in SU(k)$ for {\em some} complex structure\footnote{In fact,
in the case of constant field strengths $L\in U(k)$ gives no condition at all because
an arbitrary rotation is always in some $U(k)$ subgroup. We can find this $U(k)$ by first going to
the canonical frame where $\phi$ is block diagonal and choosing the usual complex coordinates.
It is a totally different matter when the field strength is not constant because at each point
the rotation $L$ must be in the same $U(k)$ corresponding to the same complex structure. This requirement
will lead to integrability conditions.
Moreover, in a non-flat metric background the BPS condition on the supergravity side will restrict
to covariantly constant spinors. In e.g.\ a Calabi-Yau manifold this will already
single out an $SU(k)$ to which our rotation $L$ must belong. Similar consideration apply
to other special holonomies with covariantly constant spinors such as $SO(7)_{\pm} \subset SO(8)$, which was
studied in the previous chapter.}.
In complex coordinates with respect to this complex structure follows
\begin{equation}
f_{\alpha\beta}=f_{\bar{\alpha}\bar{\beta}}=0 \, ,
\label{holobrane}
\end{equation}
and
\begin{equation}
\begin{split}
\det \left( \frac{1-2 \pi \alpha' f^c}{1+ 2 \pi \alpha' f^c} \right) = 1,
\end{split}
\end{equation}
where the matrix $f^c$ is the complexification of $f$ defined in eq.~\eqref{complexification}.
From the latter condition follows
\begin{equation}
\exp \tr \ln \frac{1- 2\pi f^c}{1+ 2 \pi f^c} = \exp \left(- 2\, \tr \arctanh 2 \pi f^c \right)=1\, ,
\end{equation}
from which we find in the end the condition:
\begin{equation}
2\, \tr \arctanh 2 \pi f^c = 2 \pi i n \, .
\label{DUYbrane}
\end{equation}
In the case of block diagonal $f_c$, i.e.\ $f^c_{\alpha\alpha}=if_{\alpha}$ and
all other components zero, this is equivalent to eq.~\eqref{BPSgeneric}.
Eqs.~\eqref{holobrane} and \eqref{DUYbrane} are exactly the same as eqs.~\eqref{hol} and \eqref{duyBI}
obtained in the previous chapter in Born-Infeld theory. Moreover, in \cite{angles2} it was proven that the matrix
$\rho(L)^2$ in eq.~\eqref{BPScond}, found in string theory, is modulo a constant prefactor equal to the
matrix $\Gamma_{E_F}+\Gamma_{O_F}$ in eq.~\eqref{kappagamma}, found in Born-Infeld theory.

As for the counting of unbroken supersymmetries observe that under
\begin{equation}
SO(9,1) \rightarrow SO(2k) \times SO(9-2k,1) \rightarrow SU(k) \times SO(9-2k,1) \, ,
\end{equation}
a positive-chirality Weyl spinor of $SO(9,1)$ transforms as follows
\begin{equation}
16 \rightarrow (2^{k-1},2^{4-k})+(2^{k-1}{}',2^{4-k}{}') \rightarrow
\left\{ \begin{array}{cc} 2(1,2^{4-k})+\cdots & \text{for $k$ even} \\
                          (1,2^{4-k})+(1',2^{4-k}{}')+ \cdots & \text{for $k$ odd}\end{array} \right. ,
\end{equation}
where the accents denote negative-chirality spinors. We find again that the fraction of
unbroken supersymmetries is $2 \cdot 2^{4-k}/16=1/2^{k-1}$.

\subsection{T-dual Picture: Intersecting Branes}
\label{intersectingbranes}

It is possible to make a T-duality transformation so that the constant field strengths
$f_{ij}$ completely disappear. Instead, the resulting configuration can be interpreted in terms of
D-branes intersecting at angles.  First make a coordinate rotation such that $f_{ij}$
takes the canonical block diagonal form of eq.~\eqref{angles}.
Then make a gauge choice such that the potentials have the form:
\begin{equation}
A_{2i-1}=0,\quad A_{2i}=F_{2i-1\,2i}x^{2i-1}.\label{pot}
\end{equation}
T-dualizing along the directions $2,4,\ldots,2k$ we end up with two D$k$-branes
along the hyperplanes
\begin{equation}
X^{2i}=- 2\pi\alpha' i A_{2i} = \pm 2 \pi \alpha' f_i x^{2i-1}\; , \label{trc}
\end{equation}
where the plus and minus sign is for the first and second D-brane respectively.
From eq.~\eqref{trc} we find that the
two D$k$-branes are at angles
\begin{equation}
\begin{split}
\phi_i & = \arctan(2\pi\alpha' f_i )-\arctan(-2\pi\alpha' f_i) \\
& = 2 \arctan(2\pi\alpha' f_i) \, ,
\label{relfangle}
\end{split}
\end{equation}
These are the angles introduced in eq.~\eqref{angles} for which we have just found a physical
interpretation.
If needed we can rotate back to the original frame where the
relation \eqref{relfangle} reads in matrix notation
\begin{equation}
2 \pi \alpha' f = \tanh \frac{\phi}{2} \, ,
\label{matrixrelangles}
\end{equation}
where $\phi$ is again the angle matrix.

Deriving the BPS condition in this T-dual picture will lead to the same results,
e.g.\ \cite[chapter 13]{bookpolchinski}.

\subsection{Energy}

The calculation of the energy provides
a nice illustration of the formulae of this and the previous chapter.
If we take $f$ in eq.~\eqref{fdef} to be block-diagonal, the energy
is given by
\begin{equation}
H_{\text{BI}} = 2 \tau_{2k} \int d^{2k} x \; \sqrt{\det\left(1 + 2 \pi \alpha' F \right)} = 2 \tau_{2k} V_{2k} \prod_{i=1}^k \sqrt{1 + (2\pi\alpha' f_i)^2} \, ,
\label{simpleenergy}
\end{equation}
where $\tau_{2k}$ is the D-brane tension defined in eq.~\eqref{tensioncharge}.
The factor $2$ arises from summing over the $2$ branes and $V_{2k}$ is the $2k$-dimensional volume. In flat
space this volume is infinite, but if we wrap the branes on $k$-tori, as we will in section \ref{setup}, the volume
becomes
\begin{equation}
V_{2k} = (2\pi)^{2k} \prod_{i=1}^{k} L_{2i-1} L_{2i} \, ,
\end{equation}
where $L_{j}$ is the radius of the $j$th cycle of the torus.

In the T-dual picture of intersecting branes the energy is proportional to the area of the branes,
\begin{equation}
H^{\text{T}}_{\text{BI}} = 2 \tau_k V_k \prod_{i=1}^k \sqrt{1 + \tan^2 \frac{\phi_i}{2}} \, ,
\end{equation}
where $V_k$ is the volume in the dimensions $1,3,\ldots,2k-1$ of the dual torus. From eq.~\eqref{relfangle}
and the transformation property of the D-brane tension $\tau_p=T_p e^{-\Phi}$, e.g.~\cite[section 8.7]{bookpolchinski},
\begin{equation}
\tau_{2k}=\tau_k \prod_{i=1}^{k} (2\pi L_{2i})^{-1} \, ,
\end{equation}
we see that the energy in the T-dual picture indeed equals the energy in the original picture.

With a vacuum state $|0\rangle$ corresponding to the standard complex coordinates eq.~\eqref{cc}, we find for this simple configuration:
\begin{subequations}
\begin{align}
\label{rhoodd}
\left| \langle 0 | \rho^{(2k)}_{O_F} | 0 \rangle \right| & = \sum_{i_1} f_{i_1} - \sum_{i_1<i_2<i_3} f_{i_1}f_{i_2}f_{i_3} + \cdots \, , \\
\label{rhoeven}
\left| \langle 0 | \rho^{(2k)}_{E_F} | 0 \rangle \right| & = 1 - \sum_{i_1 < i_2} f_{i_1}f_{i_2} + \sum_{i_1<i_2<i_3<i_4} f_{i_1}f_{i_2}f_{i_3}f_{i_4} - \cdots \, .
\end{align}
\end{subequations}
With eqs.~\eqref{simpleenergy},\eqref{rhoodd} and \eqref{rhoeven} you can check eq.~\eqref{bogoboundBI}
for this simple configuration. If the DUY expression eq.~\eqref{rhoodd} is zero, we
find for the energy the integral of eq.~\eqref{rhoeven}. For instance if $k=4$ this reads
\begin{equation}
H_{\text{BI}} = \tau_8 \int \left( 1 - f_1 f_2 - f_1 f_3 - f_1 f_4 - f_2 f_3 - f_2 f_4 - f_3 f_4 + f_1 f_2 f_3 f_4 \right) \, .
\end{equation}

\subsubsection*{Derrick's Theorem}

Derrick's theorem \cite{derrick} states that there do not exist solitons with finite
energy in pure Yang-Mills theory except for non-abelian configurations
in $p=4$, where $p$ is the number of {\em spatial} dimensions.
The argument rests on the invariance of the energy $H= \int d^p x F_{ij}F^{ij}$ under a scale
transformation. If $A_i(x)$ is a static solution of the equations of motion, the energy
should be stationary under taking
\begin{equation}
A_i(x) \rightarrow a^{\lambda} A_i(ax) \, ,
\label{derricktransf}
\end{equation}
with $a$ close to $1$. The field strength transforms as
\begin{equation}
F_{ij}(x) \rightarrow a^{1+\lambda} \left( \partial_{(a x^i)} A_j(ax) - \partial_{(a x^j)} A_i{(ax)}\right) + a^{2 \lambda} [A_i(ax),A_j(ax)] \, .
\label{derricktransf2}
\end{equation}
If we want the field strength to transform uniformly, we have to take $\lambda=1$.\footnote{In the abelian case
we can take any value for $\lambda$ so that in the end we can proof there are no solitons in any dimension.}
With this choice follows
\begin{equation}
F_{ij}(x) \rightarrow a^2 F_{ij}(ax),
\end{equation}
so that
the energy functional becomes
\begin{equation}
H(a)= \int d^p(ax) \; a^{-p+4} F_{ij}(ax)F^{ij}(ax)= a^{-p+4} H \, .
\end{equation}
This is stationary, $\left.\frac{\partial H}{\partial a}\right|_{a=1}=0$, precisely if $p=4$.

However, Derrick's theorem does not apply to the solutions in this chapter because they have infinite energy.
One could wrap these solutions on a torus, but then a transformation like eq.~\eqref{derricktransf} would
violate flux quantization (for more on flux quantization see section \ref{setup}).

Since moreover Derrick's theorem seems not to apply to Born-Infeld theory, a more interesting question is whether there
are genuine finite energy solitons of the abelian Born-Infeld action or of the non-abelian
generalization in dimensions other than $p=4$.  If they exist, these solutions cannot have
small field strengths, $\alpha'F \ll 1$, because in that case one could make an expansion in $\alpha'$ and apply
Derrick's theorem to the lowest order, which is Yang-Mills theory.

\subsection{Solutions Without Small Field Strength Limit}

In \cite{wittenBPS} configurations were studied that do not have a small field strength limit.
One interesting configuration is the D0-D6 state in the presence of a non-zero field strength.
The condition for preserving supersymmetry reads on the D0 and D6 brane respectively:
\begin{equation}
\begin{split}
\text{D}0: \qquad \epsilon_- & = \prod_{j=1}^9 \left(\Gamma_j\Gamma_{11}\right) \epsilon_+ \, , \\
\text{D}6: \qquad \epsilon_- & = \prod_{j=7}^9 \left(\Gamma_j\Gamma_{11}\right) \rho(L) \epsilon_+ \, ,
\end{split}
\end{equation}
which leads, using $\Gamma_{11}\epsilon_+=\epsilon_+$, to
\begin{equation}
\epsilon_+ = \Gamma_1 \ldots \Gamma_6 \rho(L) \epsilon_+ \, .
\end{equation}
In fact, $\Gamma_1 \ldots \Gamma_6$ is the spinor representation of a reflection or equivalently
of a rotation over $\pi$ in the planes $(1,2),(3,4)$ and $(5,6)$:
\begin{equation}
\Gamma_1 \ldots \Gamma_6 = \exp \left( \frac{\pi}{2}\left(\Gamma_{12}+\Gamma_{34}+\Gamma_{56}\right)\right) \, .
\end{equation}
Taking the field strengths block diagonal and introducing the angles $\phi_i$ as in eq.~\eqref{angles},
we find for the condition for supersymmetry
\begin{equation}
\phi_1 + \phi_2 + \phi_3 = \pi \, ,
\label{exoticsusy}
\end{equation}
where by choosing the right orientation of the $(1,2),(3,4)$ and $(5,6)$ planes we arranged for all
signs to be positive.
This correspond to eq.~\eqref{duyBI2} with $n=0$, which puts $\langle 0 | \rho^{(2k)}_{E_F}| 0\rangle$ to
zero instead of $\langle 0 | \rho^{(2k)}_{O_F}| 0\rangle$ and has no small field strength limit.
For $k=2$ and $k=4$ one can similarly construct D2-D4, D2-D8 and D6-D8 configurations.

The above D0-D6 configuration can be T-dualized along the 1,3,5 directions to obtain
two D3-branes at angles $\phi'_i$ with
\begin{equation}
\tan \phi_i' = -\frac{1}{f_i} \Longrightarrow \phi'_i = \frac{\pi-\phi_i}{2} \, .
\end{equation}
Condition \eqref{exoticsusy} becomes condition \eqref{BPSgeneric} with $n=1$ so that these
configurations are T-dual to the previous ones.

\section{Further Research}

In this section we briefly sketch ongoing research on another type of solutions, which are generalizations
of the {\em magnetic BIon} configuration \cite{bion1}\cite{bion2}. This configuration consists of a single D3-brane with
a stack of coinciding D1-branes perpendicular to it. The BIon is a solution of the abelian gauge theory living
on the D3-brane world-volume. It satisfies
\begin{equation}
\partial_i \Phi = - \frac{1}{2} \epsilon_{ijk} F^{jk}, \qquad i,j,k\in \{1,2,3\} \, ,
\label{monopole}
\end{equation}
where $i,j,k$ label the spatial coordinates on the D$3$-brane world-volume and $\Phi$ is the scalar
describing the position of the D3-brane in the fourth dimension. This equation follows from
dimensionally reducing eq.~\eqref{selfdual}.
Requiring spherical symmetry we find the solution
\begin{equation}
\begin{split}
\Phi & =\frac{c}{r} \, , \\
F_{ij} & = c \; \epsilon_{ijk} \frac{x^k}{r^3} \, .
\label{monopolesol}
\end{split}
\end{equation}
This is nothing but the Dirac monopole! Dirac quantization requires:
\begin{equation}
c=n/2, \qquad n \in \IZ \, .
\end{equation}
From the expression for $\Phi$ we see that the D3-brane has an infinite spike interpreted as the
D1-branes sticking out.  The calculation of the D1-brane charge as in \cite{hashtaylor}
supports the claim that there are indeed $n$ D1-branes.

Eq.~\eqref{monopole}, with 4-dimensional origin, does not acquire Born-Infeld corrections. In fact,
in \cite{thorlacius} it was proven that the solution given by eq.~\eqref{monopolesol} does not deform, even for
the full abelian effective action with derivative corrections. It would be interesting to
study the octonionic generalization
\begin{equation}
\partial_i \Phi = - \frac{1}{2} f_{ijk} F^{jk}, \qquad i,j,k\in \{1,\ldots,7\} \, ,
\label{octomonopole}
\end{equation}
which as we saw in section \ref{octonionicBI} does get corrections. $f_{ijk}$ are the octonionic
structure constants, totally antisymmetric, and in our conventions:
\begin{equation}
f_{127}=f_{163}=f_{154}=f_{253}=f_{246}=f_{347}=f_{567}=1 \, ,
\end{equation}
and all others zero.
A D7-D5 intersection is described by:
\begin{equation}
\begin{split}
\Phi & =\frac{c}{R} \, , \\
F_{i_1i_2} & = c \; P_-^{j_1}{}_{i_1} P_-^{j_2}{}_{i_2} P_-^{j_3}{}_{i_3} f_{j_1j_2j_3} \frac{x^{i_3}}{R^3} \, ,
\end{split}
\end{equation}
with $P_-^i{}_j$ the projection on the space transverse to the D5-brane and $R$ the transverse distance
$R=P_-^i{}_j x^j P_-^k{}_i x_k$. If we take for instance the D5 lying along $(34568)$ such that
$P_-^1{}_1=P_-^2{}_2=P_-^7{}_7=1$ and all other components zero, we obtain a solution
of eq.~\eqref{octomonopole} which is a straightforward generalization of eq.~\eqref{monopolesol}.
To get more general octonionic solutions, one takes a linear combination of multiple D5-branes which
are rotated under $G_2$ angles with respect to the first one.

The interesting region is of course where the D5-branes intersect. In that region there will be higher-order
corrections, which are presently under study.

In summary, in this chapter we have seen that there are supersymmetry preserving D-brane configurations
that correspond to the BPS equations of chapter \ref{equations}. As such it seems legitimate to take
as a constraint on the D-brane effective action that these BPS equations should automatically solve
its equations of motion.

\newpage
\thispagestyle{empty}

\chapter{Abelian Born-Infeld and Beyond}
\label{abelianBI}

In chapter \ref{equations} we studied solitons in Yang-Mills
theory saturating a Bogomolny bound and preserving a
certain fraction of the supersymmetry while in chapter \ref{solutions} we saw examples of these BPS states.
In this chapter we focus on what we called the
{\em complex case} where the BPS equations consist of the holomorphicity condition \eqref{hol} and the DUY condition \eqref{duy0}.
We saw, also in chapter \ref{equations}, that the DUY condition
takes on $\alpha'$ corrections when turning to the abelian
Born-Infeld action; it is perturbed to eq.~\eqref{duyBI3}.

In the previous chapters we took the Born-Infeld action as a starting point.
Here we turn the argument around. We will look for a higher order action that
allows as a solution these BPS states.
In the first section we start with the Yang-Mills action and work order by order in $\alpha'$,
constructing at the same time the action {\em as well} as the deformation of the DUY condition.
In the abelian case and neglecting derivative terms, the unique answer is the Born-Infeld action.
The abelian case should be seen as a toy model where the principles
can be explained and all calculations are easily done by hand.
The non-abelian case, treated in the next chapter, is a lot more complex
although the basic principle is the same.

In contrast to the octonionic case which is specific to eight dimensions, the BPS equations of the complex
case can be put into use in all even (spatial) dimensions; even in more than eight dimensions
where there is strictly speaking no supersymmetry anymore. In this and the following chapter
we make the implicit assumption that the BPS equations should imply the equations of motion in all
these dimensions (so even in more than eight), which means in practice that we will not use any
identities between terms that are specific to certain dimensions.
This is compatible with T-duality which requires the lower-dimensional actions to follow
from the higher-dimensional ones by dimensional reduction so that, putting the scalars to zero, they
should have the same form.

Since these BPS states are connected to the existence of supersymmetry, we believe that our
constructive approach is in fact a shortcut to the Noether method, briefly discussed in section \ref{roads}.
This method tries to construct the action iteratively by requiring supersymmetry.  Although more involved the
Noether method has one advantage over our algorithm in that the fermionic terms are also determined.

In the second section we study derivative corrections. First we give a precise definition of the slowly
varying field strength limit. Next we base ourselves on the result in \cite{wyllard} to calculate 4-derivative
corrections to the DUY condition to {\em all} orders in $\alpha'$. Although the calculation itself is highly
involved, the result is very elegant.

\section{Abelian Born-Infeld}

\subsection{First Steps}

Starting with the Yang-Mills action,
\begin{equation}
S_{\text{YM}} = \frac{1}{4g_{\text{YM}}^2} \int d^{p+1} x \; \Tr F_{\mu\nu} F^{\mu\nu} \, ,
\end{equation}
we find the equations of motion
\begin{equation}
D^{\mu} F_{\mu\nu}=0 \, .
\end{equation}
We only turn on magnetic fields in $2k$ dimensions and
introduce complex coordinates eq.~\eqref{cc}. The equations of motion become
\begin{equation}
\begin{split}
0 & = D_{\bar{\alpha}} F_{\alpha\bar{\beta}} + D_{\alpha} F_{\bar{\alpha}\bar{\beta}} \\
  & = D_{\bar\beta}F_{\alpha \bar\alpha }+2D_{\alpha }F_{\bar\alpha\bar\beta}, \qquad (\text{and cc}),
\label{YMcomplexeom}
\end{split}
\end{equation}
where we used the Bianchi identities.  The equations of motion are indeed satisfied
if we plug in the holomorphicity condition~\eqref{hol} and the DUY condition~\eqref{duy0}.

A natural question that arises is whether we can deform the Yang-Mills
action in such a way that solutions to these BPS equations remain solutions of the equations of motion.
Though the discussion so far
holds for both the abelian (Maxwell) as well as the non-abelian (Yang-Mills) case, we focus in
the remainder of this chapter on the abelian case. This means, as before, that we consider a $U(1)$
gauge theory or, alternatively, require the magnetic fields to
take values in the Cartan subalgebra. When we evaluate the group trace in the latter case the
action will just be a sum of $N$ single brane actions. We will not explicitly denote this summation.
According to eq.~\eqref{generatortrace}, evaluating the group trace introduces an extra minus sign.

In addition, we assume that the field strengths vary slowly. In other
words, we add terms polynomial in the field strength to the action and
ignore terms containing derivatives of the field strength. We will loosen this restriction
in the next section.

Under these assumptions we arrive at the following most general
Lagrangian term through order $\alpha'{}^2$:
\begin{equation}
\frac{1}{g_{\text{YM}}^2}\left(\frac{1}{4} \tr F^2 + \lambda_{(0,1)} \tr F^4 + \lambda_{(2)} \left(\tr F^2\right)^2 + {\cal O}\left(\alpha'{}^4 F^6\right)\right) \, ,
\label{lagrangian2}
\end{equation}
from which follow the equations of motion
\begin{equation}
\partial^{\mu_1} F_{\mu_1 \nu}+ 8 \lambda_{(0,1)} \, \partial^{\mu_1} (F^3_{\mu_1\nu})+
 8 \lambda_{(2)} \,\partial^{\mu_1} (F_{\mu_1 \nu}\tr F^2)+{\cal O}(\alpha'{}^4 \partial F^5)=0 .
\label{eom2}
\end{equation}
Here we introduced the notation
\begin{equation}
\begin{split}
\tr F^{2l} & = F_{\mu_1\mu_2}F^{\mu_2\mu_3} \cdots F_{\mu_{2l-1}\mu_{2l}}F^{\mu_{2l}\mu_1} \, , \\
F^{2l-1}_{\mu\nu} & = F_{\mu\mu_2}F^{\mu_2\mu_3} \cdots F_{\mu_{2l-1}\nu} \, ,
\label{tracenot}
\end{split}
\end{equation}
and $\lambda_{(0,1)}$ and $\lambda_{(2)}$ are arbitrary real coefficients, the notation
of which will become clear later on. Due to the antisymmetry of $F$ all contributions at odd orders
in $\alpha'$ vanish.

Passing to complex coordinates we find while implementing the holomorphicity condition~\eqref{hol}
\begin{equation}
\partial_{\bar{\alpha}}F_{\alpha\bar{\beta}} + 8 \lambda_{(0,1)} \partial_{\bar{\alpha}} F^3_{\alpha\bar{\beta}}
+ 16 \lambda_{(2)} \partial_{\bar{\alpha}} \left(F^2_{\gamma\bar{\gamma}}F_{\alpha\bar{\beta}}\right) + {\cal O}(\alpha'{}^4 \partial F^5) =0 \, ,
\end{equation}
where we used the notation
\begin{equation}
F^{l}_{\alpha\bar{\gamma}} = F_{\alpha\bar{\alpha}_2}F_{\alpha_2\bar{\alpha}_3} \cdots F_{\alpha_{l-1}\bar{\alpha}_l}F_{\alpha_{l}\bar{\gamma}} \, .
\end{equation}
We will sometimes call $F^l_{\alpha\bar{\alpha}}$, with summation over $\alpha$, an l-loop.
Finally, using the Bianchi identities we obtain
\begin{align}
& \partial_{\bar{\beta}} \left( F_{\alpha\bar{\alpha}} + \frac{8 \lambda_{(0,1)}}{3} F^3_{\alpha\bar{\alpha}}\right) && \text{DUY} \nonumber \\
& + \left( 4 \lambda_{(0,1)} + 16 \lambda_{(2)} \right) \partial_{\bar{\gamma}} F^2_{\alpha\bar{\alpha}} F_{\gamma\bar{\beta}} && \rightarrow 0 \nonumber \\
& + 16 \lambda_{(2)} F^2_{\gamma\bar{\gamma}} \partial_{\bar{\beta}} F_{\alpha\bar{\alpha}} && \text{1-loop I} \nonumber \\
& + 8 \lambda_{(0,1)} F^2_{\gamma\bar{\beta}} \partial_{\bar{\gamma}} F_{\alpha\bar{\alpha}} + {\cal O}(\alpha'{}^4 \partial F^5) = 0 && \text{1-loop II} \, .
\label{eom2complex}
\end{align}
The first line can be made to vanish if we deform the DUY condition \eqref{duy0} to
\begin{equation}
F_{\alpha \bar\alpha}+\frac{8\lambda_{(0,1)}}{3} F^3_{\alpha\bar\alpha} + {\cal O}(\alpha'{}^4F^5)=0 \, ,
\label{duy1}
\end{equation}
while the second line vanishes if
\begin{equation}
\lambda_{(2)} = -\frac{1}{4} \lambda_{(0,1)} \, .
\label{cond2}
\end{equation}
Using eq.~\eqref{duy1} we see that the third and fourth line vanish at {\em this} order,
but will make a contribution at order $\alpha'{}^4 \partial F^5$ containing the coefficient $\lambda_{(0,1)}$.
The conditions at order $\alpha'{}^4 \partial F^5$ will relate $\lambda_{(0,1)}$ to the arbitrary coefficients at
that order. The terms in the third and fourth line are called 1-loop terms because they contain
$F_{\alpha\bar{\alpha}}$. We learn from the example that these kind of terms, which contain the beginning of a DUY condition,
are responsible for conditions between coefficients at different orders.

We see that at this point there are two undetermined coefficients left:
an overall multiplicative constant ($\frac{1}{g_{\text{YM}}^2}$)
and $\lambda_{(0,1)}$. The latter can be absorbed in a rescaling of the field $F$.
Although the result does not look spectacular at this order --- only one
coefficient is fixed, namely $\lambda_{(2)}$ by eq.~\eqref{cond2} --- we will see that at higher
orders no new undetermined coefficients are introduced so that the resulting
action is essentially unique.
Rescaling $F$ appropriately we can put
\begin{equation}
\lambda_{(0,1)}=\frac{(2\pi\alpha')^2}{8} \, ,
\end{equation}
and reproduce the Born-Infeld Lagrangian through this order
\begin{equation}
\frac{1}{g_{\text{YM}}^2}\left(\frac{1}{4} \tr F^2 + \frac{(2\pi\alpha')^2}{8} \tr F^4 - \frac{(2\pi\alpha')^2}{32} \left(\tr F^2\right)^2 + {\cal O}\left(\alpha'{}^4 F^6\right)\right) \, .
\label{lagrangian2b}
\end{equation}

In a similar way, one can push this calculation an order further by adding
the most general Lagrangian terms through order six in $F$ and again requiring that
the (deformed) BPS solutions solve the equations of motion.
In this calculation one needs that the pieces $F_{\alpha\bar{\alpha}}$
in the two last terms of eq.~\eqref{eom2complex} get completed to eq.~\eqref{duy1}.
In the end one finds that the coefficients are completely fixed
and indeed lead to the Born-Infeld action
through sixth order in $F$. Furthermore the DUY
condition acquires an order $\alpha'{}^4 F^5$ correction:
\begin{equation}
F_{\alpha \bar\alpha}+\frac{(2\pi\alpha')^2}{3} F^3_{\alpha
\bar\alpha} + \frac{(2\pi\alpha')^4}{5} F^5_{\alpha\bar\alpha}
+{\cal O}(\alpha'{}^6F^7)=0.
\end{equation}

These results raise the suspicion that the Born-Infeld action is, in the slowly varying field strength limit, the {\em
only} deformation of Maxwell that allows
for BPS solutions of the form
eqs.~\eqref{hol} and \eqref{duy0}. Furthermore one expects that
the holomorphicity condition \eqref{hol} remains unchanged
while the DUY condition~\eqref{duy0} receives
$\alpha '$ corrections. The proof of this will be the subject of the next
subsection.

\subsection{Proof for All Orders}
\label{proofallorders}

In this section we will construct the unique deformation of the Maxwell action
that allows for BPS solutions that are in leading order given by eqs.~\eqref{hol}
and \eqref{duy0}.
Consider a general term in the higher order Lagrangian
\begin{equation}
\frac{1}{g_{\text{YM}}^2} \, \lambda_{(p_1,p_2,\ldots,p_n)} \, (\tr F^2)^{p_1} (\tr F^4)^{p_2}
\ldots (\tr F^{2n})^{p_n} \, ,\quad p_i\in\IN,\ \forall i
\in\{1,\cdots,n\},
\end{equation}
where $\lambda_{(p_1,p_2,\ldots,p_n)}\in\IR$.  Hence the notation $\lambda_{(0,1)}$
and $\lambda_{(2)}$ in eq.~\eqref{lagrangian2}.
Dropping the prefactor $\frac{1}{g_{\text{YM}}^2}\,\lambda_{(p_1,\ldots,p_n)}$ for a while, this term contributes
to the equations of motion by
\begin{equation}
\label{gentermeom}
\sum_{j=1}^{n} 4 j \, p_j \partial^{\mu} \left( \left(F^{2j-1}\right)_{\mu\nu}
        (\tr F^2)^{p_1} (\tr F^4)^{p_2} \ldots (\tr F^{2j})^{p_j - 1}
        \ldots (\tr F^{2n})^{p_n} \right) \, .
\end{equation}
Passing to complex coordinates we obtain
\begin{equation}
\label{gentermcomp}
\sum_{j=1}^{n} 4 j \, p_j \, 2^{p_1+\cdots+p_n-1} \partial_{\bar{\alpha}}
\left( \left(F^{2j-1}\right)_{\alpha\bar{\beta}}
      (F^2)^{p_1}_{\gamma_1\bar{\gamma}_1} \ldots (F^{2j})^{p_j-1}_{\gamma_j\bar{\gamma}_j} \ldots (F^{2n})^{p_n}_{\gamma_n\bar{\gamma}_n} \right) \, .
\end{equation}
Using the Bianchi identities and the holomorphicity condition, eq.~(\ref{hol}), we find for the action of the
derivative operator $\partial_{\bar{\alpha}}$ on $(F^{2j-1})_{\alpha\bar{\beta}}$:
\begin{equation}
\partial_{\bar{\alpha}} (F^{2j-1})_{\alpha\bar{\beta}}
 = \sum_{h=1}^{2j-2} \frac{1}{h} \left( \partial_{\bar{\alpha}}
F^{h}_{\gamma\bar{\gamma}} \right) (F^{2j-1-h})_{\alpha\bar{\beta}}+
\frac{1}{2j-1}\partial_{\bar\beta}(F^{2j-1}_{\gamma\bar{\gamma}})\, .
\end{equation}
Implementing this result in eq.~\eqref{gentermcomp} yields
\begin{equation}
\label{gentermexp}
\begin{split}
& \begin{split}
 \sum_{j=1}^{n} 4 j \, p_j \, 2^{p_1+\cdots+p_n-1}
 \Biggl( & \biggr(\sum_{h=1}^{2j-2} \frac{1}{h}
\left( \partial_{\bar{\alpha}} F^{h}_{\gamma\bar{\gamma}} \right)(F^{2j-1-h})_{\alpha\bar{\beta}}
 + \frac{1}{2j-1}
 \partial_{\bar\beta}(F^{2j-1}_{\gamma\bar{\gamma}})\biggr) \\
& \times (F^2)^{p_1}_{\gamma_1\bar{\gamma}_1}  \ldots (F^{2j})^{p_j-1}_{\gamma_j\bar{\gamma}_j} \ldots (F^{2n})^{p_n}_{\gamma_n\bar{\gamma}_n}
\end{split} \\
& + \sum_{\stackrel{\scriptstyle g=1}{g \neq j}}^n p_g \left(\partial_{\bar{\alpha}} F^{2g}_{\gamma\bar{\gamma}}\right)
\left(F^{2j-1}\right)_{\alpha\bar{\beta}} (F^2)^{p_1}_{\gamma_1\bar{\gamma}_1}
\ldots (F^{2j})^{p_j-1}_{\gamma_j\bar{\gamma}_j} \ldots
(F^{2g})^{p_g-1}_{\gamma_g\bar{\gamma}_g} \ldots (F^{2n})^{p_n}_{\gamma_n\bar{\gamma}_n} \\
& + (p_j -1) \theta(p_j-1) \left(\partial_{\bar{\alpha}} F^{2j}_{\gamma\bar{\gamma}}\right)
\left(F^{2j-1}\right)_{\alpha\bar{\beta}} (F^2)^{p_1}_{\gamma_1\bar{\gamma}_1} \ldots (F^{2j})^{p_j-2}_{\gamma_j\bar{\gamma}_j}
\ldots (F^{2n})^{p_n}_{\gamma_n\bar{\gamma}_n}  \Biggr)\, .
\end{split}
\end{equation}
Let us study the different types of terms in the equations of motion in order to determine
the coefficients such that the (deformed) BPS configurations are solutions.

\begin{enumerate}
\item Terms of the form $\partial_{\bar{\beta}} F^{2r-1}_{\alpha\bar{\alpha}}$. There is one of
these terms at each order and they add up to
\begin{equation}
4 \partial_{\bar{\beta}} \left( \lambda_{(1)} F_{\alpha \bar\alpha } +
\frac{2}{3}\lambda_{(0,1)} (F^3)_{\alpha \bar\alpha } +
\frac{3}{5} \lambda_{(0,0,1)}
(F^5 )_{\alpha \bar\alpha }
+ \cdots \right) \, .
\end{equation}
The leading order vanishes because of eq.~(\ref{duy0}).
It is clear that the all-order expression should vanish by itself
thereby giving the deformed DUY condition:
\begin{equation}
\lambda_{(1)} F_{\alpha \bar\alpha } +
\frac{ 2}{3}\lambda_{(0,1)} (F^3)_{\alpha \bar\alpha } +
\frac{ 3}{5} \lambda_{(0,0,1)}
(F^5 )_{\alpha \bar\alpha }
+ \cdots =0\, .  \label{duyg}
\end{equation}
\item Terms of the form $\left( \partial_{\bar{\alpha}} F^{2r}_{\gamma\bar{\gamma}} \right)
(F^{2l-1})_{\alpha\bar{\beta}} \tail$ where
\begin{equation}
\tail= (F^2)^{p_1} (F^4)^{p_2}\ldots (F^{2n})^{p_n} \, .
\end{equation}
As these terms involve traces over even powers of the field strength, they
can never be cancelled by a condition like eq.~\eqref{duyg}, so they should
cancel order by order among themselves.
If we look at the first versus the two last terms of
eq.~\eqref{gentermexp}, we see immediately that a term of this form originates
from two different terms in the action, namely $(\tr F^{2l+2r}) \tail$ and
$(\tr F^{2l}) (\tr F^{2r})\tail$.  Suppose first that $l \neq r$.
Requiring such a term to vanish results according to the first two terms
in eq.~\eqref{gentermexp} in the following condition:
\begin{equation}
\begin{split}
\label{coeffcond1}
&(l+r)(p_{l+r}+1) \lambda_{(\ldots,p_l,\ldots,p_r,\ldots,p_{l+r}+1,
\ldots)} + \\
&4 l r(p_l+1)(p_r+1) \lambda_{(\ldots,p_l+1,\ldots,p_r+1,
\ldots,p_{l+r},\ldots)} =0 \, .
\end{split}
\end{equation}
When $l = r$, we find analogously, but now using the first and the third term in eq.~(\ref{gentermexp}):
\begin{equation}
\label{coeffcond2}
(p_{2l}+1) \lambda_{(\ldots,p_l,\ldots,p_{2l}+1,\ldots)}+
2 l (p_l+2)(p_l+1) \lambda_{(\ldots,p_l+2,\ldots,p_{2l},\ldots)} =0 \, .
\end{equation}
These two conditions are enough to determine all coefficients {\em within} a
certain order if one is known.
We give an example of the chain of relations at order $F^{8}$:
\setlength{\unitlength}{1mm}
\begin{center}
\begin{picture}(120,25)
\put(10,20){\makebox(0,0){$(4,0,0,0)$}}
\put(30,20){\vector(-1,0){10}}
\put(25,22){\makebox(0,0)[b]{(\ref{coeffcond2})}}
\put(40,20){\makebox(0,0){$(2,1,0,0)$}}
\put(60,20){\vector(-1,0){10}}
\put(55,22){\makebox(0,0)[b]{(\ref{coeffcond2})}}
\put(70,20){\makebox(0,0){$(0,2,0,0)$}}
\put(90,20){\vector(-1,0){10}}
\put(85,22){\makebox(0,0)[b]{(\ref{coeffcond2})}}
\put(100,20){\makebox(0,0){$(0,0,0,1)$}}
\put(40,2){\makebox(0,0){$(1,0,1,0)$}}
\put(40,6){\vector(0,1){10}}
\put(42,11){\makebox(0,0)[l]{(\ref{coeffcond1})}}
\put(92,16){\vector(-3,-1){42}}
\put(71,11){\makebox(0,0)[r]{(\ref{coeffcond1})}}
\end{picture}
\end{center}

The conditions eqs.~\eqref{coeffcond1} and \eqref{coeffcond2} lead
to the following solution for the coefficients
\begin{equation}
\label{factorsx}
\lambda_{(p_1,p_2,\ldots,p_n)}=\frac{(-1)^{\left(\sum_j p_j\right)+1}}{4^{\sum_j p_j}} \frac{1}{p_1 ! \ldots p_n !} \,
                               \frac{1}{1^{p_1} \ldots n^{p_n}} \, C_{\sum_j j p_j},
\end{equation}
where $C_{\sum_j j p_j}\in\IR$ are unknown constants and all summations over $j$ run from $1$ to $n$. To fix these unknowns we need conditions
between coefficients of {\em different} orders.
\item Terms of the form $\left( \partial_{\bar{\alpha}} F^{2r-1}_{\gamma\bar{\gamma}} \right)
\left(F^{2s}\right)_{\alpha\bar{\beta}}  \tail$.
They relate different
orders in $F$.  \sloppy{These terms each originate from only
one term in the Lagrangian, namely $\left(\tr F^{2(r+s)}\right) \tail$.}
The only way to cancel these terms is by virtue of eq.~\eqref{duyg}.
Using eqs.~\eqref{gentermexp} and \eqref{factorsx} we find that such a term appears
in the equations of motion as
\begin{equation}
\frac{(-1)^{\sum_l p_l}}{2^{\sum_l p_l}
\prod_l (p_l!)\prod_l l^{p_l}}\frac{C_{\left(\sum_l l p_l\right)+(r+s)}}{2r-1}
\left( \partial_{\bar{\alpha}} F^{2r-1}_{\gamma\bar{\gamma}} \right)
\left(F^{2s}\right)_{\alpha\bar{\beta}}  \tail ,
\end{equation}
where all summations and products over $l$ run from $1$ to $n$ and
the $p_l$ are the (tail) values. For a
given (tail) and a given $s$, the sum over $r$ of such terms has to vanish through the use
of eq.~(\ref{duyg}).  This leads to the conditions
\begin{equation}
\frac{C_{r+s}}{C_s}=\frac{C_{r+1}}{C_1}, \quad  \forall r,s \in \IN,
\end{equation}
such that the $C_r$ form a geometric series.
This determines all unknowns $C_r$ in terms of two:
\begin{equation}
C_r=C_2\left(\frac{C_2}{C_1}\right)^{r-2}, \quad r\geq 3.\label{rels}
\end{equation}
We still have the freedom to rescale the field strength in the equations of
motion by an arbitrary factor. Furthermore, the equations of
motion are only determined modulo an arbitrary multiplicative factor which reflects the
fact that we can only determine the action modulo an overall
multiplicative factor. This freedom can be used to put $C_1=1$ and $C_2=(2 \pi \alpha')^2$.
Combining this with eq.~(\ref{rels}), we find
\begin{equation}
C_r=(2 \pi \alpha')^{2r-2}, \quad\forall r \geq 1. \label{ctfix}
\end{equation}
\end{enumerate}
The action is now completely fixed as well as the DUY condition.
The latter becomes
\begin{equation}
\begin{split}
\label{pertinst}
0 & = F_{\alpha \bar\alpha } + \frac{(2\pi\alpha')^2}{3} (F^3)_{\alpha \bar\alpha } +
\frac{(2\pi\alpha')^4}{5} (F^5)_{\alpha \bar\alpha } + \cdots \\
  & = \mbox{tr}\, {\rm arctanh} \, 2\pi\alpha' F^c,
\end{split}  
\end{equation}
where $F^c$ is the complexification of $F$ defined in eq.~\eqref{complexification} and the trace is taken over the complex indices. This expression exactly matches eq.~\eqref{duyBI3}.

To complete the proof we still have to show that the coefficients
eq.~\eqref{factorsx} with eq.~\eqref{ctfix} are indeed generated by the expansion
of the Born-Infeld Lagrangian. For this we rewrite the Born-Infeld Lagrangian as
\begin{equation}
\begin{split}
{\cal L}_{\text{BI}} =&-\frac{1}{(2\pi\alpha' g_{\text{YM}})^2}\sqrt{\det(\delta^\mu {}_\nu+2 \pi \alpha' F^\mu {}_\nu)} \\
=&\frac{1}{g_{\text{YM}}^2}\sum_{r=0}^{\infty} \frac{(-1)^{r+1}}{4^r r!} (\tr F^2 +
\frac{(2\pi\alpha')^2}{2}\tr F^4 + \cdots +
           \frac{(2\pi\alpha')^{2p-2}}{p} \tr F^{2p} + \cdots )^r \, .
\label{BIabelian}
\end{split}
\end{equation}
A general term in the abelian Born-Infeld Lagrangian
\begin{equation}
\frac{1}{g_{\text{YM}}^2}\lambda_{(p_1,p_2,\ldots,p_n)}^{BI} \, (\tr F^2)^{p_1} (\tr F^4)^{p_2}
\ldots (\tr F^{2n})^{p_n} \, ,
\end{equation}
originates from the $r$th term in the Taylor expansion,
with $r$ given by
\begin{equation}
r=p_1 + p_2 + \cdots + p_n \, .
\end{equation}
Hence the coefficients become
\begin{equation}
\label{BIfactors}
\lambda_{(p_1,p_2,\ldots,p_n)}^{BI}=\frac{(-1)^{r+1}(2\pi\alpha')^m}{4^r}
\frac{1}{p_1 ! \ldots p_n !} \, \frac{1}{1^{p_1} \ldots n^{p_n}} \, ,
\end{equation}
where $m=2\left(\sum_j j p_j\right)-2$ is the order in $\alpha'$.

\subsection{With Hindsight}

The above subsections provide an algorithm to {\em construct}
the D-brane effective action order by order in $\alpha'$.
In the next section we will see that the procedure
is still valid when we abandon the slowly varying field strength limit and
in the next chapter we tackle the non-abelian case.  There
the answer was {\em not} known previously.

However, in the simple case of Born-Infeld we do not need
the constructive algorithm to show that the BPS states are solutions.
We already saw that they satisfy a Bogomolny bound in chapter \ref{equations}, but now we will show it
immediately by means of the equations of motion.
We consider the Born-Infeld Lagrangian given by eq.~\eqref{BIabelian}
and define $h$ as the determinant $h=\det \left(\eta_{\mu\nu}+2 \pi \alpha' F_{\mu\nu} \right)$
and $h^{\mu\nu}$ as the inverse matrix of $\eta_{\mu\nu}+2 \pi \alpha' F_{\mu\nu}$.
We can easily derive the following properties
\begin{align}
\partial_{\rho} h^{\mu\nu} & = - 2 \pi \alpha' h^{\mu\tau}\partial_{\rho}F_{\tau\sigma}h^{\sigma\nu} \\
\partial_{\rho} h^{\mu\nu}_A & = - 2 \pi \alpha' h^{\mu\tau}_S\partial_{\rho}F_{\tau\sigma}h^{\sigma\nu}_S
- 2\pi\alpha' h^{\mu\tau}_A\partial_{\rho}F_{\tau\sigma}h^{\sigma\nu}_A \\
\partial_{\rho} h^{\mu\nu}_S & = - 2 \pi \alpha' h^{\mu\tau}_S\partial_{\rho}F_{\tau\sigma}h^{\sigma\nu}_A
- 2\pi\alpha' h^{\mu\tau}_A\partial_{\rho}F_{\tau\sigma}h^{\sigma\nu}_S \, ,
\end{align}
where $h^{\mu\nu}_A$ and $h^{\mu\nu}_S$ indicate the antisymmetric and symmetric parts
respectively.
The equations of motion of ${\cal L}_{\text{BI}}= - \tau_p \sqrt{-h}$,
where $\tau_p$ is the tension defined in eq.~\eqref{tensioncharge}, are
\begin{multline}
\partial_{\mu} \left(\sqrt{-h} h_A^{\mu\nu}\right) =
2\pi\alpha' \frac{\sqrt{-h}}{2} h^{\sigma\delta}_A\partial_{\mu}F_{\delta\sigma}h_A^{\mu\nu} \\
- 2\pi\alpha' \sqrt{-h} h^{\mu\delta}_A\partial_{\mu}F_{\delta\sigma}h_A^{\sigma\nu}
- 2\pi\alpha' \sqrt{-h} h^{\mu\delta}_S\partial_{\mu}F_{\delta\sigma}h_S^{\sigma\nu} = 0 \, .
\end{multline}
The first two terms cancel because of the Bianchi identities while the latter reads
in complex coordinates and after using condition \eqref{hol} and the Bianchi identities
\begin{equation}
\begin{split}
& - h^c \; \partial_{\bar{\alpha}} \left(2\pi\alpha' F + \frac{(2\pi\alpha')^3}{3}F^3+\frac{(2\pi\alpha')^5}{5}F^5\right)_{\gamma\bar{\gamma}} h_{S,\alpha\bar{\beta}} = \\
& - h^c \; \partial_{\bar{\alpha}} \left( \tr {\rm arctanh} \; \left(2 \pi \alpha' F^c\right)\right) h_{S,\alpha\bar{\beta}} = 0 \, ,
\end{split}
\end{equation}
from which we immediately read off the DUY condition \eqref{pertinst}. Here $h^c$ is the determinant of the complexification,
defined in eq.~\eqref{complexification}, of the matrix $\eta_{\mu\nu} + F_{\mu\nu}$.

\section{Derivative Corrections}
\label{abeliander}

In the previous section we neglected derivative terms. The following argument
shows that this is indeed a consistent truncation.
The abelian effective Lagrangian consists of terms of the form
$l_{(r,q)}=g_{\text{YM}}^{-2} \, \alpha'{}^{r+q} \partial^{2q} F^{r+2}$.
Now consider the rescaling \eqref{derricktransf} where
we put $\lambda=-1$:
\begin{equation}
A_i(x) \rightarrow a^{-1} A_i(ax) \, .
\label{slowlyvarying}
\end{equation}
It follows from eq.~\eqref{derricktransf2} that the field strength $F_{\mu\nu}$ stays invariant.
Every derivative on field strengths on the other hand introduces a factor of $a$ so that
the general term transforms as
\begin{equation}
l_{(r,q)} \rightarrow a^{2q} \, l_{(r,q)}.
\label{slowlyvarying2}
\end{equation}
Taking $a \rightarrow 0$, we can make an expansion in the number of derivatives.
This is called the {\em slowly varying field strength limit}. The zeroth order result is the Born-Infeld
Lagrangian.

We will show in the next chapter that our algorithm can be used to construct derivative
terms as well. There we will tackle at once the non-abelian case with derivative corrections.
However, in the abelian case we do not need to start from scratch since Wyllard has already
calculated the 4-derivative corrections to all orders in $\alpha'$ in \cite{wyllard}. This is the first correction
to the Born-Infeld term since it is known that the 2-derivative corrections vanish \cite{abelian4derivative}.
His action reads
\begin{equation}
\begin{split}
S_{\text{Wyl}} = - \tau_{p} \int d^{p+1} x \; \sqrt{-h} \bigg[ 1 +
\frac{1}{96} \big( & -h^{\mu_4\mu_1}h^{\mu_2\mu_3}h^{\rho_4\rho_1}h^{\rho_2\rho_3}S_{\rho_1\rho_2\mu_1\mu_2}S_{\rho_3\rho_4\mu_3\mu_4} \\
&  + \frac{1}{2}
h^{\rho_4\rho_1}h^{\rho_2\rho_3}S_{\rho_1\rho_2}S_{\rho_3\rho_4}\big)+{\cal O}(\partial^6)\bigg] \,,
\end{split}
\end{equation}
where
\begin{equation}
\begin{split}
S_{\rho_1\rho_2 \mu_1\mu_2} & =
\left(2 \pi \alpha'\right)^2 \partial_{\rho_1}\partial_{\rho_2}F_{\mu_1\mu_2} + 2\;(2\pi\alpha')^3 h^{\nu_1 \nu_2}\partial_{\rho_1}
F_{[\mu_1|\nu_1} \partial_{\rho_2|}F_{\mu_2]\nu_2}\, , \\
S_{\rho_1\rho_2} & = h^{\mu_1\mu_2}S_{\rho_1\rho_2\mu_1\mu_2} \, .
\end{split}
\end{equation}
As Wyllard himself already noted in \cite{wyllard2}, the action can be written as
\begin{equation}
S_{\text{Wyl}} = - \tau_p \int d^{p+1} x \; (1+P) \; \sqrt{-h} \, ,
\end{equation}
where we defined
\begin{equation}
\begin{split}
& P = \frac{1}{48} \; S^{\rho_1}{}_{\rho_2 \mu_1\mu_2} S^{\rho_2}{}_{\rho_1 \mu_3\mu_4}
\frac{\delta}{\delta (2\pi\alpha' F_{\mu_1\mu_2})} \frac{\delta}{\delta (2\pi\alpha' F_{\mu_3\mu_4})}, \\
& S^{\rho_1}{}_{\rho_2 \mu_1\mu_2} = h^{\rho_1\rho_3} S_{\rho_3\rho_2 \mu_1\mu_2}.
\end{split}
\end{equation}

If we apply eq.~\eqref{hol},
the equations motions of this action can after a very tedious --- and computerized --- calculation be written
as
\begin{equation}
- h^c \; \partial_{\bar{\alpha}} \left({\cal D}^{(q=0)}+{\cal D}^{(q=2)}\right) h_{S,\alpha\bar{\beta}}
- h^c \; \partial_{\bar{\alpha}} {\cal D}^{(q=0)} (\text{tail})_{\alpha\bar{\beta}} = 0 \, ,
\label{eomder}
\end{equation}
where ${\cal D}^{(q=0)}$ and ${\cal D}^{(q=2)}$ are the DUY contributions with zero and four derivatives respectively
\begin{subequations}
\begin{align}
{\cal D}^{(q=0)} & = \left[{\rm arctanh} \; \left(2 \pi \alpha' F\right)\right]_{\gamma\bar{\gamma}} \, ,\\
{\cal D}^{(q=2)} & = -\frac{1}{48} \, S_{\rho_1\rho_2\alpha_1\bar{\alpha}_2} S_{\rho_3\rho_4\alpha_3\bar{\alpha}_4} h^{\rho_2\rho_3}h^{\rho_4\rho_1}\left( h^A_{\alpha_2\bar{\alpha}_3} h^S_{\alpha_4\bar{\alpha}_1} + h^S_{\alpha_2\bar{\alpha}_3} h^A_{\alpha_4\bar{\alpha}_1}\right) \, .
\end{align}
\end{subequations}
Note that there is some freedom in ${\cal D}^{(q=2)}$ since one could add to it terms containing ${\cal D}^{(q=0)}$ and
compensate these by adding appropriate terms to $(\text{tail})$. We took ${\cal D}^{(q=2)}$ to be as simple as possible.
Remarkably ${\cal D}^{(q=2)}$ can be rewritten using the operator $P$:
\begin{equation}
{\cal D}^{(q=2)} = P \; {\cal D}^{(q=0)} \, .
\end{equation}
From eq.~\eqref{eomder} follows that
\begin{subequations}
\begin{align}
F_{\alpha\beta} = F_{\bar{\alpha}\bar{\beta}} = 0 \, , \\
 (1+P) \left[\arctanh 2 \pi \alpha' F\right]_{\alpha\bar{\alpha}} = 0
\end{align}
\end{subequations}
imply the equations of motions to this order in the derivatives. The latter equation is the 4-derivative generalization of eq.~\eqref{duyBI3}.
If one assumes that eq.~\eqref{susyDUY}, which we derived in the context of the Born-Infeld action without derivative corrections, can be still taken seriously
when there are 4-derivative corrections, one finds to this order:
\begin{equation}
\langle 0 | \Gamma^{(2k)}_{E_F} + \Gamma_{O_F}^{(2k)}| 0 \rangle = (-i)^k \exp \left( {\cal D}^{(q=0)}\right)\left(1+{\cal D}^{(q=2)}\right) \, .
\end{equation}
$\Gamma^{(2k)}_{E_F}$ and $\Gamma^{(2k)}_{O_F}$ can be found by taking the real and imaginary parts of the right hand side i.e.\
again the pieces containing even and odd numbers of $F$.
We know from \cite{susyBIcederwall}\cite{susyBIaganagic} and \cite{susyBIbergshoeff} that this $\Gamma$-matrix
plays a central role in the construction of $\kappa$-symmetry. So this expression could help to construct a $\kappa$-symmetric
and supersymmetric effective action with derivative corrections.

Furthermore, Wyllard proposed the following conjecture for the terms with more derivatives:
\begin{equation}
S_{\text{Wyl-conj}} = -\tau_p \int d^{p+1} x \exp \left(\sum_{n\ge 2}\frac{\zeta(n)}{n(2\pi)^n} \tr \left( S_{\mu_2\mu_1} \frac{\delta}{\delta (2 \pi \alpha' F_{\mu_1\mu_2})}\right)^n\right) \sqrt{- h} \, ,
\end{equation}
where the $\left[S_{\mu_1\mu_2}\right]^{\rho_1}{}_{\rho_2}=S^{\rho_1}{}_{\rho_2 \mu_1\mu_2}$ are considered as matrices over the $\rho$ indices and the trace is also over the $\rho$ indices.
The conjecture is not completely well-defined in the sense that
there is some ambiguity in the ordering of the derivatives
to $F_{\mu_1\mu_2}$.
It is not difficult to guess the corresponding conjecture for the DUY condition:
\begin{equation}
\exp \left(\sum_{n\ge 2}\frac{\zeta(n)}{n(2\pi)^n} \tr \left( S_{\mu_2\mu_1} \frac{\delta}{\delta (2 \pi \alpha' F_{\mu_1\mu_2})}\right)^n\right) \left[\arctanh 2 \pi \alpha' F\right]_{\alpha\bar{\alpha}} = 0\, .
\end{equation}
Finally, we hope that we can extend the result with four derivatives to the non-abelian
case using techniques which will be introduced in the next chapter.

This is all currently under investigation!

\newpage
\thispagestyle{empty}

\chapter{Non-abelian D-brane Effective Action}
\label{nonabelianBI}

Motivated by the successes in the abelian case we make the
basic assumption that the D-brane effective action should also in the non-abelian case
allow these BPS states as a solution of the classical equations of motion.
The algorithm to construct the non-abelian effective action is a generalization of
the method applied in the previous section to the abelian case.

In the first section we discuss why the complexity
in the non-abelian case is far greater.  As a result computations made by hand became unfeasible,
forcing us to write a computer program tailored to the task at hand.

In the second section we describe the algorithm in detail.
Roughly speaking, the program constructs the most general Lagrangian and
deformation of the DUY condition at each order in $\alpha'$.
Subsequently it imposes that field strength configurations satisfying eq.~\eqref{hol}
and the generalized DUY condition solve the equations of motion.
To impose eq.~\eqref{hol}, it is again very convenient to work in complex coordinates
both for the DUY condition and the equations of motion.
Putting the equations of motion to zero modulo the DUY condition generates a set of equations since the coefficient of
each {\em independent} term has to be zero.  From these conditions we can fix the coefficients
of the Lagrangian {\em as well} as the coefficients of the DUY
deformation.  Afterwards, we study which Lagrangian coefficients are not affected by field
redefinitions.
\begin{figure}[!t]
\centering
\setlength{\fboxsep}{10pt}%
\shadowbox{%
\begin{minipage}{.9\textwidth}
\begin{center}
\psfrag{m}[]{\shortstack{Most general\\Lagrangian\\ $l$ coeff.}}
\psfrag{e}[cl]{Euler-Lagrange}
\psfrag{E}[cl]{Eom}
\psfrag{1}[]{{\footnotesize 1-loop terms}}
\psfrag{d}[]{\shortstack{Most general\\DUY\\ $d$ coeff.}}
\psfrag{t}[cl]{{\footnotesize Subtract ``derived''}}
\psfrag{s}[cl]{{\footnotesize Subtract ``principal''}}
\psfrag{0}[tc]{\footnotesize \shortstack{Must be 0\\DUY $\rightarrow$ solution\\fix $l$ and $d$}}
\psfrag{f}[]{\shortstack{Most general\\field redefinition\\ $f$ coeff.}}
\psfrag{R}[]{\shortstack{Resulting\\Lagrangian}}
\psfrag{a}[cr]{{\footnotesize Apply}}
\psfrag{l}[]{\shortstack{Comparison\\other results\\literature}}
\includegraphics[scale=0.6]{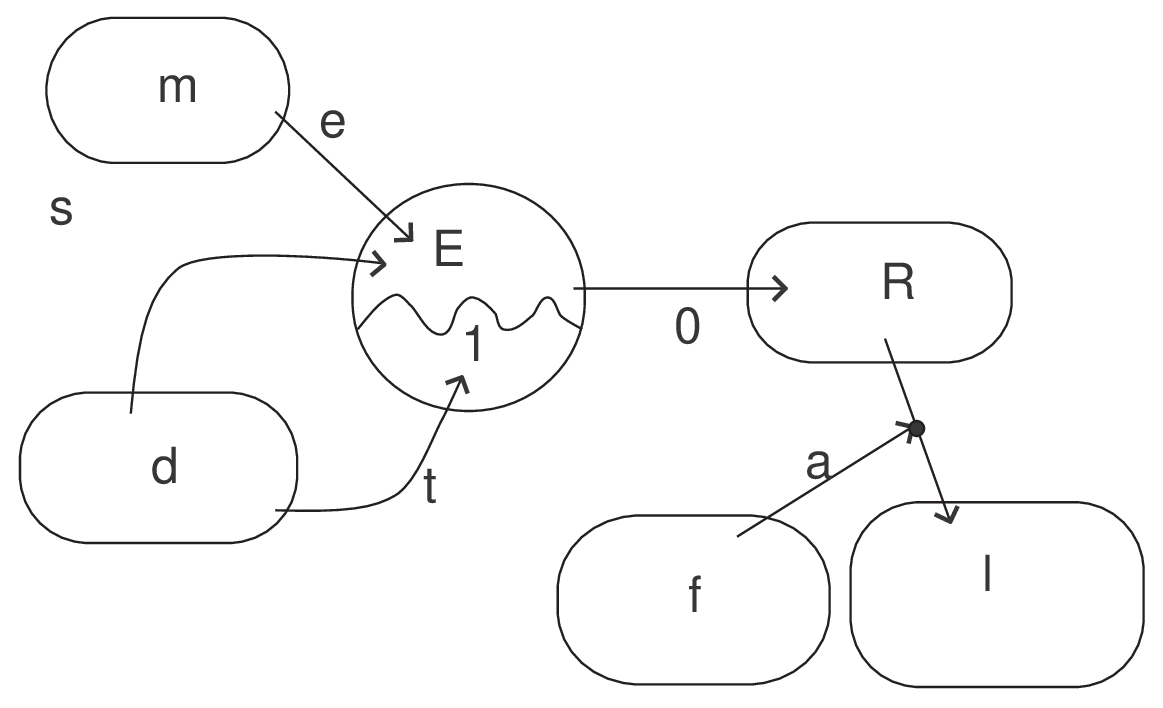}
\caption{A rough flowchart of the calculations in the non-abelian case.}
\label{flowchart}
\end{center}
\end{minipage}
}
\end{figure}
At the end of the section we introduce much-needed optimizations.

In the last section we study the results and compare to other results in the literature. We will see that the existence of these BPS states
is again very restrictive.  Nevertheless, there is a free parameter left at order $\alpha'{}^3$,
which can be fixed from string theory amplitude calculations.  We explain why we expect more of these
parameters popping up at higher orders.

\section{With a Little Help from the Computer}
\label{nacomplexity}

In the abelian case we could make a slowly varying
field approximation and expand in the number of derivatives. To zeroth order we could
neglect the derivatives altogether. This turns out to be no longer possible.
Reexamining the slowly varying field strength limit, we consider
the transformation eq.~\eqref{derricktransf} and find for the covariant
derivative and the field strength
\begin{equation}
\begin{split}
D_{\mu} & = \partial_{\mu} + [A_{\mu}, \cdot] \rightarrow a (\partial_{(a x^\mu)}+ a^{\lambda-1} [A_{\mu},\cdot]), \\
F_{\mu\nu} & = \partial_{\mu} A_{\nu} - \partial_{\nu} A_{\mu} + [A_{\mu},A_{\nu}] \rightarrow
a^{\lambda+1} (\partial_{(a x^\mu)} A_{\nu} - \partial_{(a x^\nu)} A_{\mu}) + a^{2\lambda} [A_{\mu},A_{\nu}] \, .
\end{split}
\end{equation}
In the limit $a\rightarrow 0$ we find $l_{(r,q)}\rightarrow a^{2q+(\lambda+1)(r+2)} l_{(r,q)}$
if $\lambda \ge 1$ and $l_{(r,q)}\rightarrow a^{2(r+q+2)\lambda} l_{(r,q)}$ if $\lambda \le 1$.
To make an expansion in $q$ but not in $r$, both cases are useless.
Intuitively speaking, identities of the form
\begin{equation}
[D_{\mu},D_{\nu}]\cdot = [F_{\mu\nu}, \cdot]
\end{equation}
prevent a limit that would allow an unambiguous expansion in the number of derivatives.
Indeed, making the covariant derivatives very small would at the same time make
the theory nearly abelian.

Including covariant derivatives greatly adds to the complexity of the problem.
Firstly, there will be many more possibilities for the
most general terms in the Lagrangian, DUY condition and equations of motion.
Secondly, not all these terms will be independent due to partial integration,
Bianchi and $[D,D]\cdot = [F,\cdot]$ identities. Thirdly, field redefinitions
make physically equivalent Lagrangians look different. Suppose we perform a field redefinition
\begin{equation}
A_{\mu} \rightarrow A_{\mu} + {\cal F}_{\mu} = A_{\mu} +  2 \pi \alpha' f_{(0,0,0)}  D^{\nu} F_{\nu\mu} + \cdots \, ,
\label{generalFR}
\end{equation}
where ${\cal F}_{\mu}$ is the most general gauge invariant expression transforming as a Lorentz
covariant vector.  Such a field redefinition leads to a different Lagrangian, nevertheless corresponding
to the same S-matrix. As long as we neglected derivatives on field strengths, this was not an issue since, as we
can see from eq.~\eqref{generalFR}, a field redefinition will introduce derivative terms.

Table \ref{complexity} shows how the complexity grows with the order.
\begin{table}
\begin{center}
\begin{tabular}{|c|ccccc|} \hline\hline
Order & \rule[-2mm]{0mm}{6mm}\makebox[1cm]{$\alpha'^0$} & \makebox[1cm]{$\alpha'^1$} & \makebox[1cm]{$\alpha'^2$} &
\makebox[1cm]{$\alpha'^3$} & \makebox[1cm]{$\alpha'^4$} \\ \hline
Lagrangian & 1$\mid$1 & 7$\mid$2 & 73$\mid$7 & 980$\mid$36 & 15890$\mid$300 \\
Unaffected FR & 1 & 1 & 4 & 13 & 96 \\
FR & 1$\mid$1 & 9$\mid$5 & 93$\mid$36 & 1179$\mid$329 & not calc. \\
DUY & 1$\mid$1 & 4$\mid$3 & 29$\mid$15 & 184$\mid$27 & 4022$\mid$246 \\
Eom & 1 & 7 & 57 & 156 & 1818 \\
Equations & 1 & 7 & 57 & 156 & 1816 \\ \hline\hline
\end{tabular}
\caption[Number of terms in the Lagrangian, equations of motion, DUY conditions and field redefinitions]{
Number of terms in the Lagrangian (both the total number and the number of terms unaffected by
field redefinitions), equations of motion, DUY condition and field redefinitions.
The numbers are formatted as follows: \#(total terms)$\mid$\#(independent terms).
Using various optimizations only the independent terms in the equations of motion
were constructed. The last row shows the number of final equations. Each independent term in
the  equations of motion is expected to lead to a final equation, but in the $\alpha'^4$ case
two terms contained a trivial equation ($0=0$).}
\label{complexity}
\end{center}
\end{table}
Nevertheless our computer program is able to calculate, fully automatically, the effective
Lagrangian up to order $\alpha'^4$.
This seems to be the limit both due to computer time and, perhaps even more restrictive,
computer memory\footnote{At the moment only partial results at order $\alpha'{}^5$ are possible.}.  Already for this result, many optimizations
were needed.  We will discuss for instance the use of the Sym ordering prescription and
antisymmetrized Bianchi identities.

The computer language of choice is Java, which as a modern object-oriented programming language proved to
us more user friendly than the in physics more commonly used C or Fortran.  The program runs on the
command line and produces as output \LaTeX~files, ready to copy and paste from. We refer to the separate
user manual for a more extended description \cite{manual}.

Having said this, we can delve into the rather technical details.

\section{More Details than You Asked For}

First we describe the different kinds of terms and how the program separates the
dependent ones from the independent ones. Then we explain the algorithm in detail and
finally we discuss an ordering principle for the terms and other optimizations.

\subsection{Types of Terms}
The program distinguishes 4 kinds of terms, of which the properties are listed in table \ref{typeproperties}.
\begin{table}[!t]
\begin{center}
\begin{minipage}{\textwidth}
\begin{tabular}{|c|c|c|c|c|}
\hline\hline
& & Field & DUY & Equations \\
\rs{Properties} & \rs{Lagrangian} & redefinitions & deformation & of motion \\
\hline
Group trace & & &  & \\
in front\footnote{So there is cyclic symmetry.} & \rs{yes} & \rs{no} & \rs{no} &
\rs{no} \\ \hline
Covariance or & Lorentz & Lorentz & $U(k)$ & $U(k)$\\
invariance & invariant & vector & invariant & vector
\\ \hline
Free index & no & yes & no & yes \\
\hline
Complex & & & & \\
coordinates & \rs{no} & \rs{no} &
\rs{yes} & \rs{yes} \\
\hline
\# field strengths\footnote{The order $\alpha'{}^m$ satisfies $m=r+q$.}
& r+2 & r+1 & r+1 & r+1 \\
\# derivatives
& 2q & 2q+1 & 2q & 2q+1 \\ \hline
Type of& Bianchi & Bianchi & Bianchi & Bianchi \\
identities           & PI\footnote{PI: partial integration identities.},
DDF\footnote{DDF: identities of the type $[D,D]\cdot=[F,\cdot]$.}
 & DDF$^d$ & DDF$^d$ & DDF$^d$ \\
\hline
Used & ${\cal L}$ & ${\cal F}$ & ${\cal D}$ & ${\cal E}$ \\
Symbols\footnote{The curly type is used for denoting all terms at a certain
order, the small type
in the labelling of the individual terms or coefficients.} & $l$ & $f$ & $d$ & $e$ \\
\hline\hline
\end{tabular}
\end{minipage}
\caption{Properties of the different types of
terms.}
\label{typeproperties}
\end{center}
\end{table}
At each order $\alpha'{}^m$ and for each type the program will:
\begin{enumerate}
\item Construct all possible terms with the appropriate Lorentz or $U(k)$ invariance or covariance, build
out of field strength tensors $F$ and covariant derivatives.  These terms are classified by the numbers $r$ and $q$, where the
order in $\alpha'$ is given by $m=r+q$, $r$ determines the number of field strengths and $q$ the number of derivatives as indicated
in table~\ref{typeproperties}.
For the Lagrangian we assume that
there is a single group trace in front, in agreement with tree level string perturbation theory.
\item Construct all possible identities between those terms. These are the partial integration identities,
Bianchi identities and identities that follow from $[D,D]\cdot = [F,\cdot]$.
The latter read in their most general form:
\begin{equation}
\begin{split}
[D_{\mu_1}D_{\mu_2}\ldots D_{\mu_{n-2}}F_{\mu_{n-1}\mu_n},D_{\mu_{n+1}}\ldots D_{\mu_{n+l-2}}F_{\mu_{n+l-1}\mu_{n+l}}] & = \\
[D_{\mu_1},[D_{\mu_2},\ldots [D_{\mu_{n-2}},[D_{\mu_{n-1}},D_{\mu_n}]]\ldots]]D_{\mu_{n+1}}\ldots D_{\mu_{n+l-2}}F_{\mu_{n+l-1}\mu_{n+l}}. &
\end{split}
\end{equation}
\item Solve those (linear) identities and thus separate the linear dependent terms from the
independent ones, forming a basis.  The basis terms of the Lagrangian and the DUY deformations
multiply each an arbitrary coefficient and these form the unknowns of our problem.  The coefficients of the terms
in the equations of motion on the other hand are obtained
by varying the Lagrangian and subtracting DUY conditions. They will be function of the unknowns. During the elimination process the program
has to be careful to keep track of them in the following way.  Suppose $T^1, T^2, \ldots, T^n$
are terms in the equation of motions, which read:
\begin{equation}
\sum_{j=1}^n c_j T^j = 0 \, .
\end{equation}
i.e.\ term $T^j$ carries the coefficient $c_j$.  Now, if $T^i$ is
eliminated by means of the, say, $[D,D]\cdot = [F,\cdot]$ identity
\begin{equation}
T^i = \sum_{j \neq i} d_j T^j \, ,
\end{equation}
it has to transfer its coefficient $c_i$ to the other terms so that the
equations of motion transform into
\begin{equation}
\sum_{j \neq i} (c_j + d_j c_i) T^j =0 \, .
\end{equation}

In a similar manner the program keeps track of the effect of field redefinitions on the terms in the
Lagrangian.
\end{enumerate}
With all this technology our program allows to reexpress the different results in the literature in the same basis, making the
problem of comparing easy.

\subsection{The Algorithm}

So now we know how the program separates the independent terms from the total set of terms --- and
this with care for the coefficients that these terms carry --- we describe the complete
algorithm:
\begin{enumerate}
\item Construct all possible {\em independent} terms in the Lagrangian and assign to each of these terms
an arbitrary coefficient $l_{(r,q,x)}$ where $r$ and $q$ indicate the number of field strengths and derivatives while
$x$ labels the individual terms\footnote{Actually, the Java program has a slightly more complicated scheme
for labelling the different terms which is explained in \cite{manual}. However, we do not want to overload the
discussion here.}.

As an extra complication we must keep track of which coefficients
can be changed by a field redefinition
\begin{equation}
A_{\mu} \rightarrow A_{\mu} + {\cal F}_{\mu},
\end{equation}
where ${\cal F}_{\mu}$ is a linear combination of all possible {\em independent} field redefinition terms at
the appropriate order.  To each term in the Lagrangian we assign a coefficient, indicating how the term changes
under the most general field redefinition. During the elimination process these coefficients are handled in the same way as the
coefficients in the equations of motion.  In the end we find all independent terms in the
Lagrangian and how they change under field redefinitions.  We call the terms that are affected by
a field redefinition {\em FR changeable} while the terms with zero coefficient change are {\em FR invariant}.
We want all the coefficient changes under field redefinitions to be independent so that the resulting FR changeable terms
do not hide a FR invariant combination.  In this way we maximize the number of FR invariant
basis terms.  To obtain this goal we use the following rule of thumb during the elimination process:
always eliminate terms with zero coefficient changes first. Indeed, eliminating a term with a non-zero coefficient change
would spread this coefficient change over all terms in the identity.  Of course, this rule of thumb
does not guarantee that all coefficient changes are independent, which still has to be checked.
\label{lagstep}
\item Construct all possible {\em independent} field redefinition terms for use in step
 \ref{lagstep} at higher orders. Assign to each of these terms the arbitrary coefficient $f_{(r,q,x)}$.
\item Construct all possible {\em independent} terms in the DUY condition and assign to each of these
terms an arbitrary coefficient $d_{(r,q,x)}$.
\item Construct all possible terms in the equations of motion.
\item The coefficients of those terms in the equations of motions will have
three contributions:
\begin{enumerate}
\item The coefficients obtained from varying the terms in the Lagrangian.
These contain the arbitrary Lagrangian coefficients $l_{(r,q,x)}$.  Note that in the
non-abelian case there is also a contribution from varying the covariant derivatives.
\item \label{DUYprincipal} Subtraction of the deformed DUY condition:
\begin{equation}
D_{\bar{\beta}} ( F_{\alpha_1\bar{\alpha}_1} + {\cal D}_{(2)} \cdots + {\cal D}_{(m-1)} +
{\cal D}_{(m)}) = 0 \, .
\end{equation}
The first term cancels the contribution of the Lagrangian to the equations of
motion at order $\alpha'{}^0$.  The coefficients of ${\cal D}_{(2)}$ to ${\cal D}_{(m-1)}$
were determined at lower order while ${\cal D}_{(m)}$ contains the unknown coefficients
$d_{(r,q,x)}$. Only this last piece --- which is the non-abelian equivalent of the first type of terms in the
enumeration of section~\ref{proofallorders} --- adds to the equations of motion at the present order $\alpha'{}^m$.
The calculation at order $\alpha'{}^3$ reveals that ${\cal D}_{(1)}$ is zero.
\item \label{DUYtimesso} The equations of motion must be zero when applying the DUY condition so that they must,
in general, look like the DUY condition multiplied by some piece on the left and on the right. We will denote these
pieces by (head) and (tail). Note that we treated already in point \ref{DUYprincipal} the case where both (head) and (tail)
vanish. We add the following term:
\begin{equation}
o_{(r,q,x)} {\rm (head)} \,(F_{\alpha\bar{\alpha}} + {\cal D}_{(2)} + \cdots) {\rm (tail)} = 0 \, ,
\end{equation}
and demand that there exist values for $o_{(r,q,x)}$ so that the equations of motion disappear.
At the lowest order these terms will
contribute to so called {\em 1-loop terms}, i.e.\ terms containing $F_{\alpha\bar{\alpha}}$, an index loop consisting
of only one field strength.  At this lowest order each 1-loop term fixes an $o_{(r,q,x)}$ coefficient.
Since these terms obviously contribute at higher orders too, they are the source of relations between
coefficients at different orders. They are the non-abelian equivalent of the third type of terms in the enumeration of
section~\ref{proofallorders}.

Since the highest order we will ever calculate is $\alpha'{}^4$ and
there is no ${\cal D}_{(1)}$ contribution to the DUY condition, in practice, we can disregard
1-loop terms at order $\alpha'{}^3$ and $\alpha'{}^4$ altogether. These terms are therefore not counted
in table~\ref{complexity}. Likewise, we can ignore terms with more than one 1-loop at lower orders.
\end{enumerate}
\item Calculate the independent terms in the equations of motion.  Of course, the coefficients obtained
in step 5 must be kept track of appropriately.
\item Put the coefficients of the independent terms in the equation of motion to zero. From this set
of equations, which we call the {\em final set of equations}, we can solve for the unknowns $l_{(r,q,x)}$, $d_{(r,q,x)}$ and $o_{(r,q,x)}$.
As already mentioned, at orders $\alpha'{}^3$ and $\alpha'{}^4$ we do not consider 1-loop terms and thus we do not
introduce $o$ coefficients.

The number of equations (approximately the number of independent terms in the equations of motion)
is far greater than the number of unknowns so that in the generic case
the set would only admit the zero solution.  E.g. at order $\alpha'{}^4$ you
can infer from table \ref{complexity} that there are 1816 equations in 546 unknowns.
You could say that a non-trivial solution has to be protected
by some underlying principle.  This actually provides a check on our calculation\footnote{Earlier versions of the computer program which contained bugs did produce a zero
result!} and our initial assumptions.

The coefficients of these final equations all originate from operations that can only result in rational
numbers.  However, unknowns that are not fixed at a certain order, undetermined coefficients in the result,
can appear with a power or multiplied with another undetermined coefficient in the final equations at higher order.
This happens via the mechanism explained in step \ref{DUYtimesso}, when an undetermined coefficient is present
both in the lower order equations of motion as in the DUY deformation.
When solving the equations of motion we will try to treat them as independent parameters, not as unknowns.  At the low orders we
considered, this was not an issue since there was essentially only the one free parameter
at order $\alpha'{}^3$.  At higher orders it could be possible that these undetermined coefficients are fixed.
As we will see in section \ref{undetermined} we
can easily predict which undetermined coefficients will appear at higher orders and we will speculate about possible
relations among them.
\item Finally, we should only consider {\em FR invariant} terms when comparing to other results in the
literature.
\end{enumerate}

\subsection{Ordering Principle and Optimizations}
\label{organization}

One of the hardest parts is solving the set of identities to find the independent terms of each type.
One thing we can do is try to reduce the set of terms and identities right from the start.  It is actually
possible to get rid of the $[D,D]\cdot = [F,\cdot]$ identities using the following
scheme.

Consider any linear combination of terms.
Now, start with the terms without derivatives and fully symmetrize in the field strengths.
In this process we use the $[D,D]\cdot=[F,\cdot]$ identities to convert the introduced commutators
into derivatives\footnote{This means that we push the $[D,D]\cdot=[F,\cdot]$ identity
to the left. An attentive reader could wonder why we do not push the identity to the right
and use symmetrized derivatives.  This leads indeed to fewer terms and fewer identities
because in this case there are no antisymmetrized Bianchi identities (see eq.~\eqref{antisymbianchi}).
The resulting Lagrangian, however, will be in a strange basis and difficult to interpret.
Nevertheless, there is no reason not to use this scheme for the equations of motion. In fact, the only
remaining identities for the equations of motion, the Bianchi identities, contain in complex
coordinates only two terms so that they really only put the one equal to the other.  In this case it is
possible to directly construct the independent terms.  That is why we gave for the equations of motion
only the number of independent terms in table \ref{complexity}.}.
Next, we turn to the terms with two derivatives an again fully symmetrize
them, whereby a term of the form $DF$ or $D^2F$ is considered as a single entity.
Again, terms with more derivatives are added as compensation.
We proceed in this fashion order by order in the number of derivatives.
Since the resulting terms are symmetrized in the field strengths, all ``non-abelianality'' sits in the
covariant derivatives.  The operation of symmetrizing in the field strengths is called Sym, in
combination with the group trace we will use STr.
All terms, whether they be part of the Lagrangian, DUY condition, field
redefinitions or equations of motion, will be symmetrized in this way.

A major advantage is that terms can now be unambiguously classified by the number
of derivatives since the only identities connecting
terms with a different number of derivatives were the $[D,D]\cdot = [F,\cdot]$ identities.
In addition, the non-abelian calculation now follows the abelian one more closely since
symmetrized terms in the Lagrangian lead to symmetrized terms in the equations of motion.
Except for the obvious fact that the derivatives are non-commuting there are however some
other differences:
\begin{enumerate}
\item There are extra identities because in symmetrizing we only used up part of the
$[D,D] \cdot = [F,\cdot]$ identities.  An example will clarify this.
Only one of the two identities
\begin{equation}
\begin{split}
[F_{\mu_1\mu_2}, F_{\mu_3\mu_4}] & = [D_{\mu_1}, D_{\mu_2}] F_{\mu_3\mu_4}, \\
[F_{\mu_3\mu_4}, F_{\mu_1\mu_2}] & = [D_{\mu_3}, D_{\mu_4}] F_{\mu_1\mu_2}
\end{split}
\end{equation}
is used to commute $F_{\mu_1\mu_2}$ and $F_{\mu_3\mu_4}$.  The other one remains in the form
\begin{equation}
[D_{\mu_1}, D_{\mu_2}] F_{\mu_3\mu_4} + [D_{\mu_3}, D_{\mu_4}] F_{\mu_1\mu_2} = 0 \, .
\end{equation}
This kind of identities is related to antisymmetry, as in the case above, or to Jacobi identities of $F$-commutators.
In general they read
\begin{equation}
\begin{split}
& [D_{\mu_1}, [ \ldots [D_{\mu_{l-2}}, [D_{\mu_{l-1}},D_{\mu_l}]]\ldots]] D_{\mu_{l+1}} \ldots D_{\mu_{l+n-2}} F_{\mu_{l+n-1}\mu_{l+n}} + \\
& [D_{\mu_{l+1}}, [ \ldots [D_{\mu_{l+n-2}}, [D_{\mu_{l+n-1}},D_{\mu_{l+n}}]]\ldots ]] D_{\mu_1} \ldots D_{\mu_{l-2}} F_{\mu_{l-1}\mu_l} = 0 \, ,
\end{split}
\label{antisymbianchi}
\end{equation}
and we call them {\em antisymmetrized Bianchi identities}.
\item \label{nonabelianmech} Consider the deformed DUY condition
\begin{equation}
F_{\alpha\bar{\alpha}} + \text{Sym} \, {\cal D}_{(2)} + \text{Sym} \, {\cal D}_{(3)} + \cdots = 0 \, ,
\end{equation}
where Sym denotes the above defined symmetrization.
Somewhere in the equations of motions we will find $\text{Sym} \left\{F_{\alpha\bar{\alpha}} T \right\}$
where $T$ contains, say, $n$ field strengths.
For this piece of the equations of motion to be zero when using the stability condition, the term
$\text{Sym} \left\{\text{Sym}\left({\cal D}_{(2)}\right) T\right\}$ must also be present in the equations of motion at higher order.
When applying the outer symmetrization $\text{Sym}\left({\cal D}_{(2)}\right)$ is considered as one block
which means that the factors therein stay together. Therefore we must
further symmetrize this term by mixing those factors among the factors of $T$.  Unlike in the
abelian case terms with more derivatives are thus introduced.  Interestingly, the number of additional derivatives is always a multiple
of four. Indeed, both $\text{Sym} \left\{\text{Sym}\left({\cal D}_{(2)}\right) T\right\}$ and $\text{Sym} \left\{ {\cal D}_{(2)} T \right\}$
are symmetric under the reversion of all factors so that their difference must contain an {\em even} number of commutators
of field strengths.

The mechanism of {\em distributing the DUY condition} presented here thus introduces extra derivatives and prevents the symmetrized trace Born-Infeld to have
BPS solutions\footnote{We mean here in a generic even dimension. Indeed, the symmetrized
trace Born-Infeld action has, together with many other possibilities, BPS solutions in 4 dimensions.
In that case the BPS equation does not
acquire corrections and is just the (anti)-self-duality equation.}.
\end{enumerate}

After we found the result, we used a second organizing principle to simplify even further.
We tried to use basis terms with as much ``groups'' of nested covariant derivative commutators
as possible.  Each group corresponds, using a $[D,D]\cdot=[F,\cdot]$ identity, to a commutator of $F$s or equivalently to
an algebra structure constant, which can be put in front.
To clarify this correspondence, we write \eqref{lagalpha3} in different ways using $[D,D]\cdot=[F,\cdot]$ identities:
\begin{equation}
\begin{split}
 [D_{\mu_3},D_{\mu_2}] D_{\mu_4} F_{\mu_5\mu_1} D^{\mu_5} [D^{\mu_4},D^{\mu_3}] F^{\mu_1\mu_2}  & = \\
 [F_{\mu_3\mu_2},D_{\mu_4} F_{\mu_5\mu_1}]D^{\mu_5} [F^{\mu_4\mu_3},F^{\mu_1\mu_2}] & = \\
 [D_{\mu_4},[D_{\mu_5},D_{\mu_1}]]F_{\mu_3\mu_2} D^{\mu_5} [D^{\mu_1},D^{\mu_2}]F^{\mu_4\mu_3} & \, .
\end{split}
\end{equation}
Both the first and the third term contain two groups of commutators. Indeed, the nested
commutators $[D_{\mu_4},[D_{\mu_5},D_{\mu_1}]]$ in the third term count as one group since
they originate from one commutator of field strengths as is clear from the second term.
Although we pushed the $[D,D]\cdot=[F,\cdot]$ identities to the right initially,
in this way we can easily see which terms can be pushed how far in the other direction if necessary.
This comes in handy when comparing to results from string amplitude calculations where the
number of field strengths is important. Indeed, if a term contains $r$ field strengths,
it cannot contribute to $r'$-point amplitudes for $r' < r$.

\section{Results and Limitations of the Algorithm}
\label{nonabelianresults}

\subsection{Results}

The above described algorithm and ordering prescription was pursued up to order
$\alpha'{}^4$.
The effective action, modulo field redefinitions, is given by
\begin{equation}
{\cal L}=\frac{1}{g_{\text{YM}}^2}\left({\cal L}_{(0)}+{\cal L}_{(2)}+{\cal L}_{(3)}+{\cal L}_{(4)}\right),
\end{equation}
where the leading term is simply
\begin{equation}
\label{lagalpha0}
{\cal L}_{(0)} = - \Tr\, \left\{\frac{1}{4}\tr F^2\right\},
\end{equation}
the Yang-Mills Lagrangian with coupling $g_{\text{YM}}$.
As usual $\Tr$ indicates the trace over the gauge indices and $\tr$ over the Lorentz indices
as in eq.~\eqref{tracenot}.
Then we have
\begin{equation}
\label{lagalpha2}
{\cal L}_{(2)}=(2 \pi \alpha')^2 \; \mbox{STr} \left\{\frac{1}{8} \tr F^4
- \frac{1}{32} \left(\tr F^2\right)^2
\right\},
\end{equation}
where STr denotes the symmetrized trace prescription.
Note that at this point, we fixed the overall multiplicative
factor in front of the action as well as the scale of the
gauge fields using the parameters $g_{\text{YM}}$ and $\alpha'$.
The next term is
\begin{equation}
\label{lagalpha3}
{\cal L}_{(3)}= (2 \pi \alpha')^3 K \; \Tr\left\{[D_{\mu_3},D_{\mu_2}] D_{\mu_4} F_{\mu_5\mu_1} \, D^{\mu_5} [D^{\mu_4},D^{\mu_3}] F^{\mu_1\mu_2} \right\},
\end{equation}
where $K$ is an undetermined coefficient.
Finally, the fourth order term is completely
determined and it is given by
\begin{equation}
{\cal L}_{(4)} = (2 \pi \alpha')^4 \left({\cal L}_{(4,0)} + {\cal L}_{(2,2)} + {\cal L}_{(0,4)} \right) \, ,
\label{lagalpha4}
\end{equation}
with
\begin{equation}
\begin{split}
{\cal L}_{(4,0)} & =  -\mbox{STr} \left( \frac{1}{12} \tr F^6
                 - \frac{1}{32} \tr F^4 \; \tr F^2
                  + \frac{1}{384} \left(\tr F^2\right)^3 \right) , \\
\end{split}
\label{lagalpha4e1}
\end{equation}
\begin{equation}
\begin{split}
{\cal L}_{(2,2)} & = -\frac{1}{48} \; \mbox{STr} \Big( -2 \, F_{\mu_1\mu_2}D^{\mu_1}D_{\mu_6}D_{\mu_5}F^{\mu_2\mu_3}D^{\mu_6}F_{\mu_3\mu_4}F^{\mu_4\mu_5} \\
               & - F_{\mu_1\mu_2}D_{\mu_5}D_{\mu_6}F^{\mu_2\mu_3}D^{\mu_6}D^{\mu_1}F_{\mu_3\mu_4}F^{\mu_4\mu_5} \\
               & + 2 \, F_{\mu_1\mu_2}\left[D^{\mu_6},D^{\mu_1}\right] D^{\mu_5}F^{\mu_2\mu_3}F_{\mu_3\mu_4}D^{\mu_4}F_{\mu_5\mu_6} \\
               & + 3 \, D^{\mu_4}D^{\mu_5}F_{\mu_1\mu_2}F^{\mu_2\mu_3}\left[D^{\mu_6},D^{\mu_1}\right]F_{\mu_3\mu_4}F_{\mu_5\mu_6} \\
               & + 2 \, D^{\mu_6}\left[D^{\mu_4},D^{\mu_5}\right]F_{\mu_1\mu_2}F^{\mu_2\mu_3}D^{\mu_1}F_{\mu_3\mu_4}F_{\mu_5\mu_6} \\
               & + 2 \, D_{\mu_6}D^{\mu_5}F^{\mu_1\mu_2}\left[D^{\mu_6},D_{\mu_1}\right]F_{\mu_2\mu_3}F^{\mu_3\mu_4}F_{\mu_4\mu_5} \\
               & + 2 \, \left[D_{\mu_6},D^{\mu_1}\right]D_{\mu_3}D^{\mu_4}F_{\mu_1\mu_2}F^{\mu_2\mu_3}F_{\mu_4\mu_5}F^{\mu_5\mu_6} \\
               & + \left[D_{\mu_6},D^{\mu_4}\right]F_{\mu_1\mu_2}F^{\mu_2\mu_3}\left[D_{\mu_3},D^{\mu_1}\right]F_{\mu_4\mu_5}F^{\mu_5\mu_6} \Big) , \\
\end{split}
\label{lagalpha4e2}
\end{equation}
\begin{equation}
\begin{split}
{\cal L}_{(0,4)} & = -\frac{1}{1440} \; \mbox{STr} \Big( D_{\mu_6} [D^{\mu_4},D^{\mu_2}]D_{\mu_5} D^{\mu_5} [D^{\mu_1},D^{\mu_3}] D^{\mu_6} F_{\mu_1\mu_2} F_{\mu_3\mu_4} \\
                 & + 4 \, D^{\mu_2} D_{\mu_6} [D^{\mu_4},D^{\mu_1}][D_{\mu_5},[D^{\mu_6},D^{\mu_3}]] D^{\mu_5} F_{\mu_1\mu_2} F_{\mu_3\mu_4}  \\
                 & + 2 \, D^{\mu_2} [D^{\mu_6},D^{\mu_4}][D_{\mu_6},D^{\mu_1}] D_{\mu_5} [D^{\mu_5},D^{\mu_3}] F_{\mu_1\mu_2} F_{\mu_3\mu_4} \\
                 & + 6 \, D^{\mu_2} [D_{\mu_6},D^{\mu_4}]D_{\mu_5}[D^{\mu_6},D^{\mu_1}][D^{\mu_5},D^{\mu_3}] F_{\mu_1\mu_2} F_{\mu_3\mu_4} \\
                 & + 4 \, D_{\mu_6} D_{\mu_5} [D^{\mu_6},D^{\mu_4}][D^{\mu_5},D^{\mu_1}][D_{\mu_4},D_{\mu_3}] F_{\mu_1\mu_2} F^{\mu_2\mu_3} \\
                 & + 4 \, D_{\mu_6} D_{\mu_5} [D^{\mu_4},D^{\mu_2}][D^{\mu_6},D^{\mu_1}][D^{\mu_5},D^{\mu_3}] F_{\mu_1\mu_2} F_{\mu_3\mu_4} \\
                 & + 4 \, D_{\mu_6} [D_{\mu_5},D^{\mu_4}][D^{\mu_3},D^{\mu_2}][D^{\mu_5},[D^{\mu_6},D^{\mu_1}]] F_{\mu_1\mu_2} F_{\mu_3\mu_4} \\
                 & + 2 \, [D_{\mu_6},D^{\mu_1}][D^{\mu_2},D^{\mu_6}][D^{\mu_5},D^{\mu_4}][D_{\mu_5},D^{\mu_3}]F_{\mu_1\mu_2}F_{\mu_3\mu_4} \Big) \, .
\end{split}
\label{lagalpha4e3}
\end{equation}
\begin{figure}[tp]
\centering
\setlength{\fboxsep}{10pt}%
\shadowbox{%
\begin{minipage}{.93\textwidth}
\centerline{\parbox{.45\textwidth}{\centerline{\includegraphics[scale=0.27]{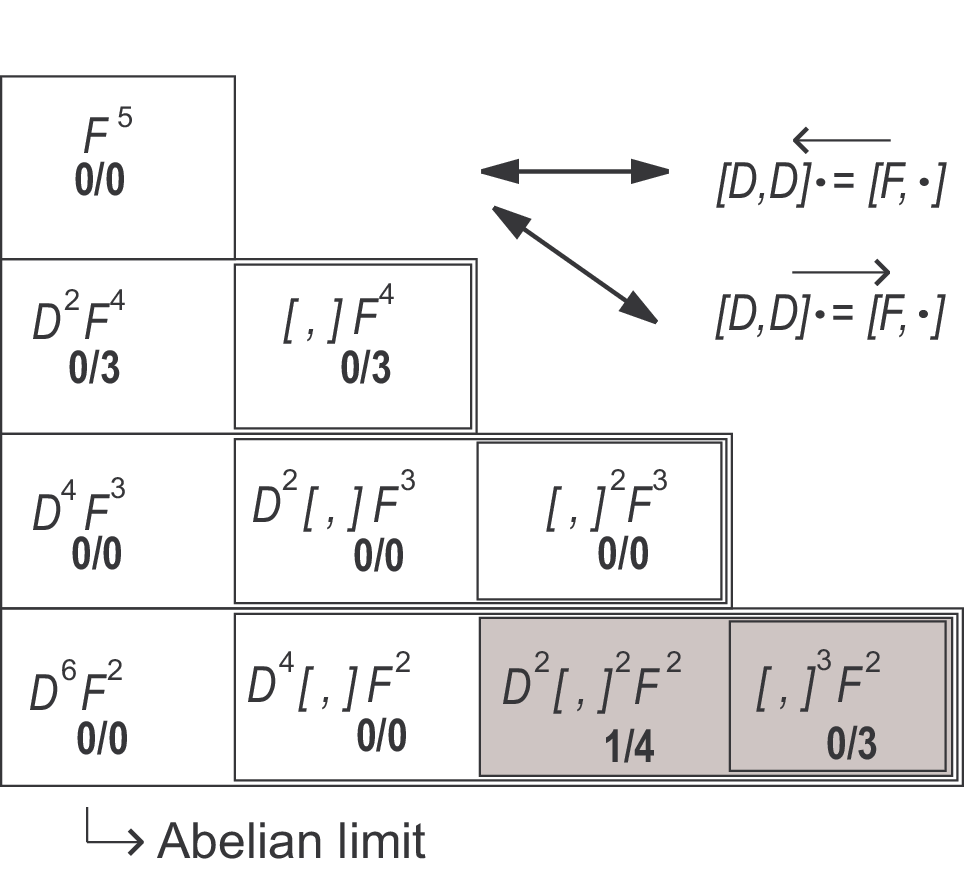}}}~~~~
\parbox{.45\textwidth}{\centerline{\includegraphics[scale=0.27]{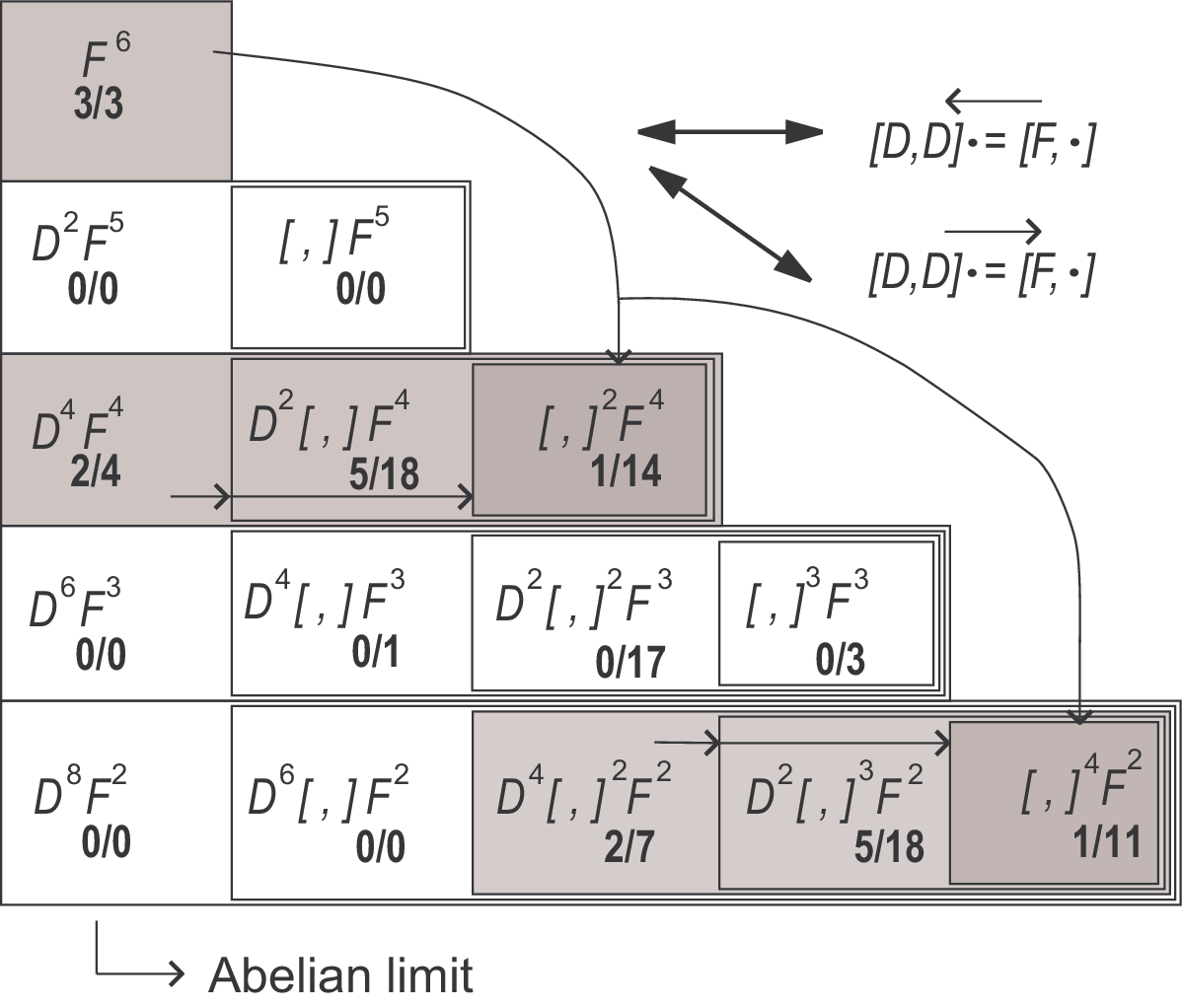}}}}
\centerline{\parbox[t]{.45\textwidth}{\caption[Basis terms at order $\alpha'{}^3$.]{\label{schemealpha3}Basis terms at order $\alpha'{}^3$.
The horizontal classification is by the number of derivatives if the $[D,D]\cdot=[F,\cdot]$ are
pushed to the left as in the text. The vertical classification is by the number of
commutator groups of $D$s.  Note that when pushing the $[D,D]\cdot=[F,\cdot]$ identities to the right,
you can find terms with the same number of derivatives on the diagonals as indicated.
The numbers indicate (\#non-zero coefficients in the result)/(\#all FR invariant basis terms). Boxes
with a non-zero contribution to the result are grey.}}~~~~
\parbox[t]{.45\textwidth}{\caption[Basis terms at order $\alpha'{}^4$.]{\label{schemealpha4}Basis terms at order $\alpha'{}^4$. See also
the caption of figure \ref{schemealpha3}. The arrows indicate how
the abelian result transforms in the non-abelian result. Partial
integration and Bianchi identities mix terms with different
numbers of commutator groups.  Hence the horizontal arrows.  For
the diagonal arrows see the mechanism of ``distributing the DUY condition'' explained
in point 2 of section
\ref{organization}. Using our method in the abelian case would lead
to two independent groups of terms: the Born-Infeld part and the
$\partial^4 F^4$ part.  In the non-abelian case these two groups meet
in the dark grey coloured boxes. As a result the undetermined coefficient
is fixed.}}}
\end{minipage}}
\end{figure}

We notice that the answer is surprisingly unique although there
are a number of undetermined coefficients creeping in.
The two parameters $g_{\text{YM}}$ and $2 \pi \alpha'$,
related to an overall rescaling of the action and a rescaling of the gauge field, are
inevitable.  By comparing to the calculation of closed string exchange between
two D-branes, we found in eq.~\eqref{ymcoupling} of section \ref{dbranes}:
\begin{equation}
g^2_{\text{YM}} = \tau_p^{-1} (2 \pi \alpha')^{-2} = g_s (2 \pi)^{p-2} \alpha'{}^{(p-3)/2} \, ,
\end{equation}
where $g_s$ is the string coupling and $\alpha'$ is the string length squared.
This leaves the parameter $K$ at order $\alpha'{}^3$, which can be fixed by comparing it for instance
to the partial result in \cite{bilal1}, which was
obtained from a string 4-point scattering amplitude calculation,
\begin{equation}
(2\pi\alpha')^3 K = 4 \alpha'{}^3 \zeta(3) \, .
\end{equation}
It has been shown by Ap\'ery \cite{zeta3} that $\zeta(3)$ is not a rational number,
which means that it could never be
produced by our algorithm at this order.  So it is most fortunate that it leaves an undetermined
coefficient here because it is the only way to accommodate for this irrational number!
We will discuss this undetermined coefficient and the ones appearing at higher orders more systematically in the next
subsection.

Furthermore, note that the result eq.~\eqref{lagalpha3} contains 2 structure constants so that it obviously disappears in the
abelian limit. The result at order $\alpha'{}^3$ was confirmed by a calculation using as a constraint supersymmetry \cite{groningen},
by a direct string 5-point amplitude calculation \cite{brasil} and by
$N=4$ Super-Yang-Mills effective action calculations \cite{sym1}\cite{sym2}\cite{sym3}.

Turning to the $\alpha'{}^4$ result we see that the terms with zero derivatives, ${\cal L}_{(4,0)}$,
form the symmetrized trace Born-Infeld at this order while the abelian limit of the terms with 4 derivatives, the
first two terms of ${\cal L}_{(2,2)}$, agree with the results in \cite{abelian4derivative}, \cite{wyllard} and
\cite{groningen2}.  If we use our method in the abelian case at order $\alpha'{}^4$, these
terms have an undetermined coefficient since terms without derivatives
have no way of communicating with terms with derivatives!  Indeed, in the abelian case the whole
contribution of terms with four derivatives, of which ${\cal L}_{(2,2)}$ is the first term, has an undetermined
overall factor; see section \ref{abeliander}. However, in the non-abelian case, terms of the
form $D^4 F^4$ and $F^6$ do communicate by means of the mechanism of ``distributing the DUY condition'', which
is explained in section \ref{organization}, point \ref{nonabelianmech} such that
the coefficient is fixed at precisely the right value.  We will come back to this in the next subsection.

Note that there are no terms with two derivatives nor with six derivatives.  If there were such terms, they would have had an
undetermined overall coefficient since the symmetrized trace cannot communicate by means
of ``distributing the DUY condition'' with terms in which the number of derivatives is not a multiple of four.
In the next section we will argue from the symmetries of the string $r$-point amplitude that these terms should indeed be absent.

In table \ref{complexity} we see that obtaining this result required solving 1816 equations in
546 unknowns.  The fact that there is a non-zero solution of this highly overdetermined set of equations
provides in itself a strong check on the result.  Moreover, another check based on the spectrum of strings
ending on intersecting branes will be the subject of the next chapter.

Figures \ref{schemealpha3} and \ref{schemealpha4} summarize the classification of terms at orders
$\alpha'{}^3$ and $\alpha'{}^4$.
Obviously, there is still basis freedom left so that some terms in the result can be moved to the right.
E.g.\ result \eqref{lagalpha3} can also be written as a sum of terms with $2$ and $3$ structure constants (as in \cite{groningen}), however
not with only terms with $3$ structure constants. Hence the grey area in figure \ref{schemealpha3}.

\subsection{The Dance of the Undetermined Coefficients}
\label{undetermined}

\begin{figure}[t!]
\centering
\setlength{\fboxsep}{10pt}%
\shadowbox{%
\begin{minipage}{.9\textwidth}
\begin{center}
\psfrag{0}[][]{$\alpha'{}^0$}
\psfrag{1}[][]{$\alpha'{}^1$}
\psfrag{2}[][]{$\alpha'{}^2$}
\psfrag{3}[][]{$\alpha'{}^3$}
\psfrag{4}[][]{$\alpha'{}^4$}
\psfrag{5}[][]{$\alpha'{}^5$}
\psfrag{6}[][]{$\alpha'{}^6$}
\psfrag{p0}[cr][cr]{$q=0$}
\psfrag{p1}[cr][cr]{$q=1$}
\psfrag{p2}[cr][cr]{$q=2$}
\psfrag{p3}[cr][cr]{$q=3$}
\psfrag{p4}[cr][cr]{$q=4$}
\psfrag{p5}[cr][cr]{$q=5$}
\psfrag{p6}[cr][cr]{$q=6$}
\psfrag{r0}{$r=0$}
\psfrag{r1}{$r=1$}
\psfrag{r2}{$r=2$}
\psfrag{r3}{$r=3$}
\psfrag{r4}{$r=4$}
\psfrag{r5}{$r=5$}
\psfrag{r6}{$r=6$}
\psfrag{P0}[][]{$1$}
\psfrag{P2}[][]{$\pi^2$}
\psfrag{P4}[][]{$\pi^4$}
\psfrag{P6}[][]{$\pi^6$}
\psfrag{N}[][]{$0$}
\psfrag{Z3}[][]{$\zeta(3)$}
\psfrag{Z5}[][]{$\zeta(5)$}
\psfrag{ZP3}[][]{$\pi^2\zeta(3)$}
\psfrag{Z32}[][]{$\zeta(3)^2$}
\includegraphics[scale=0.5]{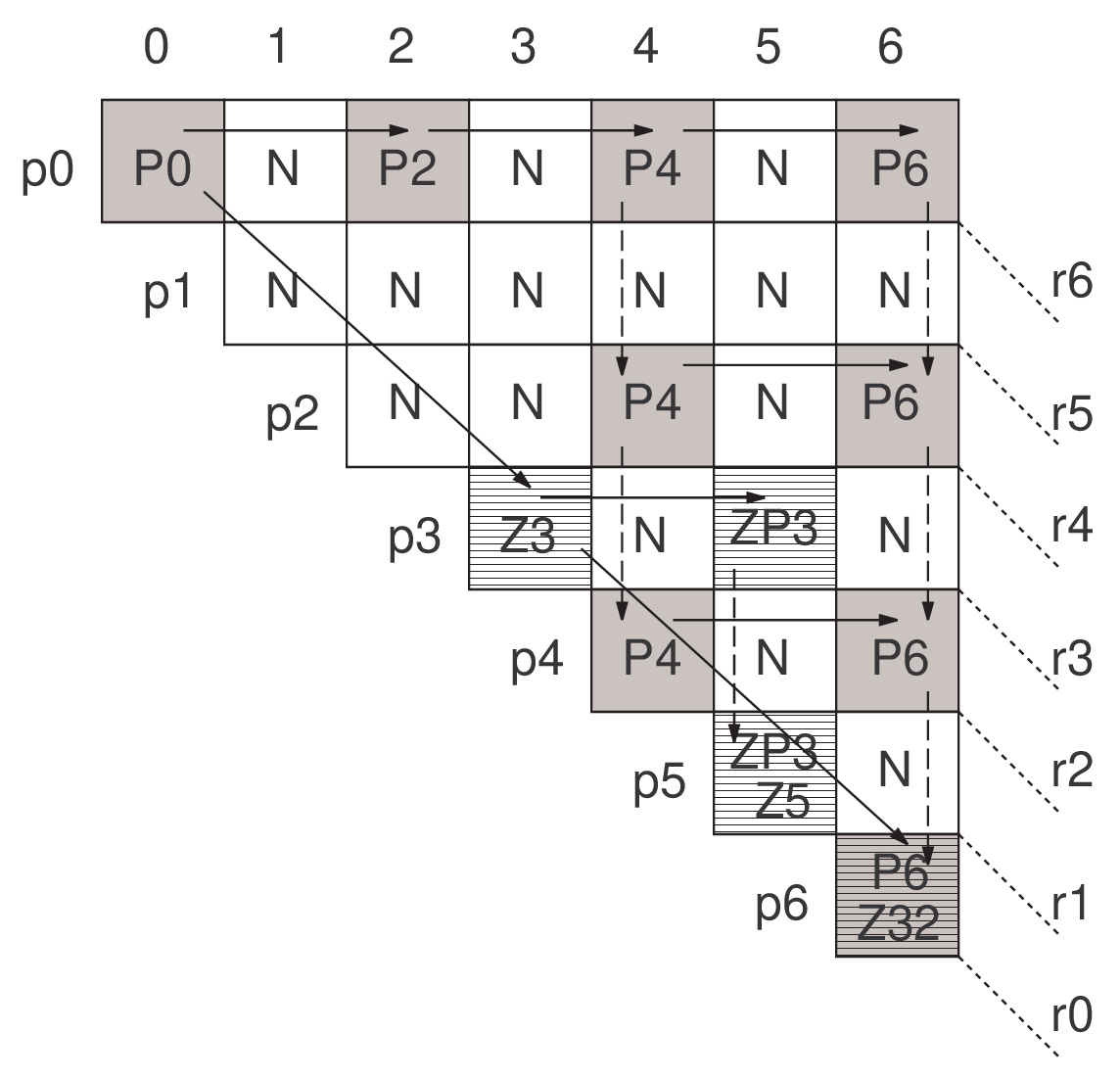}
\caption[The dance of the undetermined coefficients in the non-abelian case.]{The dance of the undetermined coefficients in the non-abelian case.
The terms are taken to be symmetrized in the field strengths as explained in
section \ref{organization}. The boxes are then labelled by $(r,q)$ where
$2q$ is the number of derivatives and $r+2$, along the diagonals, the number
of field strengths. The order in $\alpha'^m$ satisfies $m=r+q$. We have filled in
the values --- modulo a {\em rational} factor --- of the undetermined
coefficients arising from the string non-abelian $4$-point amplitude. The values in each column should be
multiplied by the indicated power of $\alpha'$.}
\label{nonabelianhigh}
\end{center}
\end{minipage}
}
\end{figure}
Let us have a detailed look at the undetermined coefficients. In the abelian case they were already extensively discussed
in \cite{groningen2} in the context of the Noether method.  The Noether method tries to construct the effective action by
requiring supersymmetry. It constructs iteratively the Lagrangian as well as the supersymmetry transformation.
Our method is very much related to the Noether method and presumably
it will leave exactly the same coefficients unfixed.  The role of the supersymmetry transformation is in our method
played by the DUY condition.

In the following we will take the Noether method point of view where an undetermined coefficient
means that there is an independent supersymmetric invariant and
extend the discussion of \cite{groningen2} to the non-abelian case.
We extracted the information for figure \ref{nonabelianhigh} from our result in the previous section and from the treatment of the string
non-abelian 4-point amplitude in \cite{brasil2}.  Since in the following discussion we are only interested in which
coefficients could be produced or be related to each other by our method, all values are modulo rational factors.
First of all there is the overall coefficient, which we have put to $1$ in the figure.
Every box that contains a non-zero contribution ${\cal L}_{(r,q)}$ to the Lagrangian has a contribution ${\cal D}_{(r,q)}$ to
the DUY condition which in principle contains the same undetermined coefficients.
The DUY condition will relate coefficients in different boxes
as follows: if there is an undetermined coefficient $\lambda_{(r,q)}$ in the Lagrangian ${\cal L}_{(r,q)}$
and an undetermined coefficient $\lambda_{(r',q')}$ in ${\cal L}_{(r',q')}$ and thus in ${\cal D}_{(r',q')}$ the coefficient
$\lambda_{(r,q)}\lambda_{(r',q')}$ will appear in ${\cal L}_{(r+r',q+q')}$.
The coefficient $\lambda_{(2,0)}$ of ${\cal D}_{(2,0)}$ --- indicated by
the horizontal arrows --- remains undetermined by our method because it can be changed by rescaling the field strength. In string
theory it is proportional to $(\pi\alpha')^2$. The action of ${\cal D}_{(2,0)}$ --- e.g. applying ${\cal D}_{(2,0)}$ twice must give the
same coefficient as applying ${\cal D}_{(4,0)}$ once --- fixes all coefficients in the first row: $\lambda_{(2r,0)}=(\alpha'\pi)^{2r}$ and $\lambda_{(2r+1,0)}=0$.
The first row is the symmetrized trace Born-Infeld.

In the abelian case the overall coefficient of ${\cal L}_{(2,2)}$ is undetermined because
it can always be rescaled by applying eq.~\eqref{slowlyvarying} as we have seen when we discussed the slowly varying field strength limit.  However,
in the non-abelian case terms with a different number of derivatives can communicate by means of the mechanism of
``distributing the DUY condition'', explained in point \ref{nonabelianmech} of section \ref{organization};
we indicate this with the vertical dashed arrows.  This mechanism fixes the coefficient $\lambda_{(2,2)}$ to $(\alpha'\pi)^4$, precisely the value predicted by
the string 4-point amplitude.  The fact that this coefficient is not free and in particular cannot be set to zero means
that, assuming that our method is equivalent to the Noether method, the non-abelian symmetrized trace Born-Infeld is {\em not supersymmetric}
without adding derivative terms. By the action
of ${\cal D}_{(2,0)}$ and the vertical arrows the coefficients in the grey boxes are fixed.

We saw that at $\alpha'{}^3$ there is an undetermined coefficient which is fixed by string theory to $\alpha'{}^3 \zeta(3)$. This irrational
number is called {\em Ap\'ery's constant} \cite{zeta3}.  By the action of the vertical and horizontal arrows $\alpha'{^5} \pi^2 \zeta(3)$
appears in box $(0,5)$, but there the string 4-point amplitude predicts also a new coefficient: $\alpha'{}^5 \zeta(5)$.  Now
we arrive at something interesting. There are two possibilities: either terms in that box
with a different coefficient do not ``see'' each other in which case there are two supersymmetric invariants, either the terms need each other
to form one supersymmetric invariant, in which case we would find that $\frac{\zeta(5)}{\pi^2\zeta(3)}$ is a rational number.  Since
very little is know about the Riemann zeta function of odd integers, this would
be really spectacular.  Probably we should be more conservative and assume the hypothesis that there are two
supersymmetric invariants corresponding to these two coefficients. The same goes through for the box $(0,6)$ where $\zeta(3)^2$ and $\pi^6$ come together:
either the terms with $\zeta(3)$ form a supersymmetric invariant independent from the Born-Infeld supersymmetric invariant or either $\frac{\zeta(3)}{\pi^3}$
is rational. When looking at the full 4-point string scattering
amplitude, at each odd order $\alpha'{}^{2r+1}$ we see a factor $\alpha'{}^{2r+1}\zeta(2r+1)$.
Assuming that there are no relations with rational coefficients between all these factors, there must start at least a
new supersymmetric invariant at each odd order.

In the abelian case, all contributions ${\cal L}_{(r,1)}$ are zero as well as all contributions that would come from odd-point
amplitudes.  The results up to order $\alpha'{}^4$ indicate that both assertions hold for the non-abelian case as well.
The second assertion needs a little clarification.
I do not claim that the odd-point amplitudes in the non-abelian case are zero
(take for instance the 5-point amplitude \cite{brasil}), only that if you use the symmetrization in the field strengths,
the boxes that in the abelian case would correspond to odd-point functions are zero. In other words all boxes with odd $r$
are zero.
This second assertion we can prove by considering the general form of an $n$-point tree scattering amplitude:
\begin{equation}
S_{A_1\ldots A_n}\left(j_{1},\ldots,j_{n}\right)= \sum_{P} \Tr\left( T_{A_{P(1)}}, \ldots ,T_{A_{P(n)}} \right) A\left( j_{P(1)},\ldots,j_{P(n)} \right) \, ,
\end{equation}
where $T_A$ are the gauge generators, the sum runs over all $(n-1)!$ cyclically inequivalent permutations of the external states and the
primitive amplitude $A$, i.e.\ the $n$-point amplitude for open strings that do not carry gauge quantum numbers, has
cyclic symmetry. The external states are massless gauge bosons carrying quantum numbers $\left(A_i,j_i\right)$,
where $A_i$ is the gauge index and $j_i=(e_i,k_i)$ describes the polarization and momentum.
Consider now world sheet parity $\Omega: \sigma \rightarrow \pi - \sigma$ under which the vector state transforms
as $\Omega |e,k \rangle=-|e,k \rangle$ and the order of the external states is reversed. This is a symmetry of the primitive
amplitude such that
\begin{equation}
A\left(j_{1},\ldots,j_{n}\right)=\Omega A\left(j_{1},\ldots,j_{n}\right) = (-1)^n A\left(j_{n},\ldots,j_{1}\right) \, .
\end{equation}
Using this expression in the complete $n$-point amplitude we find
\begin{equation}
\begin{split}
S_{A_1\ldots A_n}\left(j_{1},\ldots,j_{n}\right) & = S_{A_n\ldots A_1}\left(j_{n},\ldots,j_{1}\right) \\
& = \sum_{P} \Tr\left( T_{A_{P(n)}}, \ldots ,T_{A_{P(1)}} \right) (-1)^n A\left( j_{P(1)},\ldots,j_{P(n)} \right) \\
& = (-1)^n S_{A_n\ldots A_1}\left( j_{1},\ldots,j_{n}\right)
\end{split}
\end{equation}
From this follows that $r=n-2-\text{\#comm}$ must be even, where \#comm is the number of commutators in the gauge
indices. Since these commutators are converted into commutators of derivatives by the symmetrizing operation,
$r+2$ is precisely the number of field strengths in the part of the effective action that first contributes
to the $n$-point amplitude under consideration.

Concluding, although our algorithm is very restrictive, it will, just as the Noether method, not be able to {\em completely}
fix the non-abelian D-brane effective action because there are many independent supersymmetric invariants, at least a new one at
every odd order. Indeed, let us take the extreme case and assume that there are only two supersymmetric invariants, the Yang-Mills action and the non-abelian
D-brane effective action. This would lead to the bold and rather unlikely statement that $\frac{\zeta(2r+1)}{\pi^{2r+1}}$ is rational for all $r$.

\newpage
\thispagestyle{empty}

\chapter{Checks and Balances: the Spectrum}
\label{checks}

In order to check the effective action obtained in section \ref{nonabelianresults} we
can put it into use to calculate the spectrum of off-diagonal fluctuations of the gauge field and
compare this to the spectrum of strings stretching between two intersecting D-branes (wrapped on tori).
This kind of check was first devised in \cite{HTspectrum} and further elaborated on in \cite{DST}.
In these articles an exact match was found between the diagonal fluctuations in the abelian Born-Infeld
theory and the spectrum of strings beginning and ending on the same brane. Since this case is settled we
do not discuss it here. However, a discrepancy was signalled between the spectrum of off-diagonal
fluctuations for the non-abelian Born-Infeld action with the symmetrized trace prescription \cite{tseytlinSTr} and
the spectrum of strings beginning
and ending on different branes.  The disagreement started at order $\alpha'{}^4$.  Our action does better.
It does not contribute to the spectrum at order $\alpha'{}^3$ (as expected from string theory) \cite{testalpha3} and there is
perfect agreement at order $\alpha'{}^4$ \cite{testalpha4}. Next to the bosonic fluctuations we also
consider the fermionic fluctuations up to order $\alpha'{}^3$ \cite{testfermion}.

\section{Set-up}
\label{setup}

The calculation of the spectrum only probes $U(2)$ sub-sectors of
the full $U(N)$ theory. This is clear in string theory where strings beginning
and ending on the same brane see only a $U(1)$ sub-sector
while strings stretching between two branes see the $SU(2)$ sector.
But it also holds in the effective action approach, see for instance \cite{spectrumtroost}
for an explanation.
So without loss of generality we take a $U(2)$ gauge field, which lives
on the world-sheet of {\em two} coinciding D$p$-branes. Next, we wrap $2k$ dimensions
of those branes on a $2k$-torus with radii $R_i=\frac{L_i}{2\pi}$, where $i=1,\ldots,2k$.
All the ``action'' will be in the compact directions where we switch on a
constant diagonal magnetic background field $F^{(0)}_{ij}=f_{ij}(i\sigma_3)$.
Flux quantization requires
\begin{equation}
f_{ij} = \frac{2\pi n_{ij}}{L_iL_j} , \qquad n_{ij}=-n_{ji}, \; n_{ij} \in \IZ .
\end{equation}
By an appropriate rotation we can diagonalize the field strength
\begin{equation}
f_{2i-1,2i} = -f_{2i,2i-1} = f_i = \frac{2\pi n_i}{L_{2i-1}L_{2i}} \, ,
\label{diagbackground}
\end{equation}
and all other components zero. Here we take all the $f_i$ (and thus $n_i$) to be positive.

\begin{figure}[!t]
\centering
\setlength{\fboxsep}{10pt}%
\shadowbox{%
\begin{minipage}{.9\textwidth}
\begin{center}
\psfrag{L2i}{$L_{2i}$}
\psfrag{b}{$L'_{2i}$}
\psfrag{Li}{$L_{2i-1}$}
\psfrag{-n}{$-n_i$}
\psfrag{n}{$n_i$}
\psfrag{f}[c]{flux quanta}
\psfrag{3}{$n_i=3$}
\psfrag{T}{T-dual}
\includegraphics[scale=0.5]{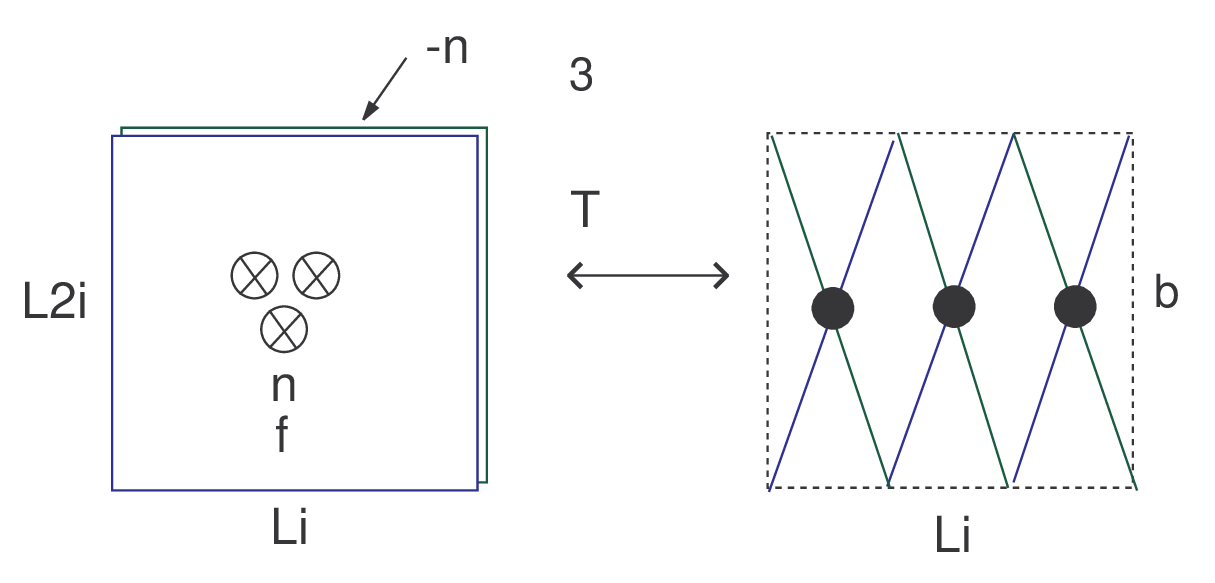}
\caption[Branes with flux quanta and intersecting branes.]{Branes with flux quanta on the left and
intersecting branes on the right. Both are represented in the covering space of the torus.
You can see the projection in the $(2i-1,2i)$ plane where the two D-branes on the left carry $n_i$ and $-n_i$ flux
quanta respectively. Since the field strength is proportional to the Pauli matrix $\sigma_3$, the
two branes carry opposite flux.
In the T-dual picture on the right the D-branes intersect $n_i$ times.}
\label{ibranes}
\end{center}
\end{minipage}
}
\end{figure}

In the T-dual picture we find, as in section \ref{intersectingbranes},
for the equations of the two branes
\begin{equation}
X^{2i} = - 2 \pi \alpha' i A^{(0)}_{2i} = \pm \frac{4\pi^2 \alpha'n_i}{L_{2i-1} L_{2i}}x^{2i-1} = \pm \frac{n_i L'_{2i}}{L_{2i-1}} x^{2i-1} \, ,
\end{equation}
where $L'_{2i} = 2\pi R'_{2i} = \frac{2\pi\alpha'}{R_{2i}}=\frac{4\pi^2 \alpha'}{L_{2i}}$
is the T-dual length of the torus in the $2i$-direction. The plus and minus signs are for the first and the
second brane respectively.
Flux quantization assures that the branes wrap an {\em integer} times the torus
in the $2i$-direction while they wrap once the $2i-1$ direction.  These branes are called
$(1,\pm n_i)$ branes.  The number of intersections between the two branes is given by
$Pf(n_{ij})=\prod_{i=1}^k n_i$. See figure~\ref{ibranes} for a picture of the set-up.

In complex coordinates the field strength satisfies
$F^{(0)}_{\alpha\beta}= F^{(0)}_{\bar\alpha\bar\beta}=0$, $F^{(0)}_{\alpha\bar\beta}=0$ for $\alpha\neq\beta$
and\footnote{We do
not sum over repeated indices corresponding to {\em complex}
coordinates in this chapter unless indicated otherwise.}
\begin{equation}
F^{(0)}_{\alpha\bar\alpha}= i\left(
   \begin{array}{cc}
     f^c_{\alpha} & 0 \\
     0 & -f^c_{\alpha}
   \end{array}
  \right),
\label{diagcomplexbackground}
\end{equation}
where the $f^c_\alpha$, $\alpha\in\{1,\cdots,k\}$ are imaginary
constants such that $-if^c_\alpha=f_\alpha \ge 0$.

\section{Spectrum from String Theory}
\label{spectrumstringtheory}

We perform the analysis of the spectrum
in the T-dual picture of ``intersecting branes'' where we can follow \cite{angles1}.
For our purposes we extend the analysis therein to the Ramond sector.
We calculate the spectrum of the superstring action, which reads in
superconformal gauge:
\begin{equation}
S_{\text{P}} = \frac{1}{2\pi} \int d \tau d \sigma \left( \frac{1}{2\alpha'} \left(\partial_{\tau} X^{\mu} \partial_{\tau} X_{\mu} -
\partial_{\sigma} X^{\mu} \partial_{\sigma} X_{\mu} \right) + 2 i \psi_- \partial_{\neq} \psi_- +2i \psi_+ \partial_{=} \psi_+ \right),
\label{susypolyakov}
\end{equation}
where the $X^{\mu}(\tau,\sigma)$ define as before the embedding of the string world-sheet into space-time
and the real world-sheet spinors $\psi_-$ and $\psi_+$ describe the right- and left-moving fermionic modes respectively.
We have defined $\partial_{\neq} = \frac{1}{2}(\partial_{\tau}+\partial_{\sigma})$ and $\partial_{=} = \frac{1}{2}(\partial_{\tau}-\partial_{\sigma})$.
The first brane is positioned along $1,3,\ldots,2k-1$ and the
second brane is rotated by angles $\phi_i \ge 0$ in the $(2i-1,2i)$ planes.
We introduce
\begin{equation}
\begin{split}
&\hat X{}^{2i-1}=\cos \phi_i \; X^{2i-1}+\sin \phi_i \; X^{2i},\quad
\hat X{}^{2i}=-\sin \phi_i \; X^{2i-1}+\cos \phi_i \; X^{2i}, \\
& \hat \psi{}_\pm^{2i-1}=\cos \phi_i \; \psi_\pm^{2i-1}+\sin \phi_i \; \psi_\pm^{2i},\quad
\hat \psi{}_\pm^{2i}=-\sin \phi_i \; \psi_\pm^{2i-1}+\cos \phi_i \; \psi_\pm^{2i},
\end{split}
\end{equation}
such that the $\hat{X}^{2i-1}$ are positioned along the second brane.
As already argued, we are only interested in strings stretching between the two branes.
Unlike the ``diagonal'' strings, those strings are forced by their tensions to be localized
at one of the intersections so that they will not see the torus geometry.
Therefore, we do not expect the dimensions $L_i$ to appear in the answer.
We impose the appropriate Neumann and Dirichlet boundary conditions
\begin{equation}
\begin{split}
\mbox{at }\sigma=0: \qquad & \partial_\sigma X^{2i-1}=0,\quad \partial_\tau X^{2i}=0, \\
& \psi_+^{2i-1}=\psi_-^{2i-1},\quad \psi_+^{2i}=-\psi_-^{2i}, \\
\mbox{at }\sigma=\pi: \qquad & \partial_\sigma \hat X{}^{2i-1}=0,\quad \partial_\tau \hat X{}^{2i}=0, \\
& \hat\psi{}_+^{2i-1}=\eta\hat\psi{}_-^{2i-1},\quad \hat\psi{}_+^{2i}=-\eta\hat\psi{}_-^{2i},
\end{split}
\end{equation}
where $\eta=+1$ or $\eta=-1$ in the Ramond and the Neveu-Schwarz sector respectively.
Upon solving the equations of motion and implementing the boundary conditions we get the following
mode expansion for the bosons:
\begin{equation}
\begin{split}
X^{2i-1} & = i \sqrt{\alpha'} \sum_{n\in\IZ} \left(
\frac{\alpha^i_{n_{+i}}}{n_{+i}}e^{-in_{+i}\tau}\cos{n_{+i}\sigma}+
\frac{\alpha^i_{n_{-i}}}{n_{-i}}e^{-in_{-i}\tau}\cos{n_{-i}\sigma}
\right), \\
X^{2i} & = i \sqrt{\alpha'} \sum_{n\in\IZ} \left(
\frac{\alpha^i_{n_{+i}}}{n_{+i}}e^{-in_{+i}\tau}\sin{n_{+i}\sigma}-
\frac{\alpha^i_{n_{-i}}}{n_{-i}}e^{-in_{-i}\tau}\sin{n_{-i}\sigma}
\right) ,
\end{split}
\end{equation}
where we introduced
\begin{equation}
\varepsilon_i = \frac{\phi_i}{\pi},\qquad n_{\pm i} = n\pm \varepsilon_i\mbox{ with }n\in\IZ.
\end{equation}
For the fermions we get in the Neveu-Schwarz sector
\begin{equation}
\begin{split}
\psi_\pm^{2i-1}&=\frac 1 2 \sum_{n\in\IZ+\frac{1}{2}}\left(
d^i_{n_{+i}}e^{-in_{+i}(\tau\pm\sigma)}+d^i_{n_{-i}}e^{-in_{-i}(\tau\pm\sigma)}\right), \\
\psi_\pm^{2i}&=\pm \frac i 2 \sum_{n\in\IZ+\frac{1}{2}}\left(
d^i_{n_{+i}}e^{-in_{+i}(\tau\pm\sigma)}-d^i_{n_{-i}}e^{-in_{-i}(\tau\pm\sigma)}
\right),
\end{split}
\end{equation}
and in the Ramond sector
\begin{equation}
\begin{split}
\psi_\pm^{2i-1} & = \frac 1 2 \sum_{n\in\IZ}\left(
d^i_{n_{+i}}e^{-in_{+i}(\tau\pm\sigma)}+d^i_{n_{-i}}e^{-in_{-i}(\tau\pm\sigma)}\right), \\
\psi_\pm^{2i} & = \pm \frac i 2 \sum_{n\in\IZ}\left(
d^i_{n_{+i}}e^{-in_{+i}(\tau\pm\sigma)}-d^i_{n_{-i}}e^{-in_{-i}(\tau\pm\sigma)}
\right).
\end{split}
\end{equation}
In quantizing eq.~\eqref{susypolyakov} we find the equal time (anti-)commutation relations
\begin{equation}
\begin{split}
\left[\partial_{\tau}X^i(\sigma),X^{j}(\sigma')\right] & =-2\pi i \alpha' \delta(\sigma-\sigma') \delta^{ij} \, , \\
\left\{\psi_{+}^{i}(\sigma),\psi_{+}^j(\sigma') \right\} = \left\{\psi_{-}^{i}(\sigma),\psi_{-}^j(\sigma') \right\} & = \pi \delta(\sigma-\sigma') \delta^{ij} \, ,
\end{split}
\end{equation}
and the rest vanishing.
The non-vanishing (anti-)commutation relations for the modes are then
\begin{equation}
\begin{split}
&[\alpha^i_{m_{+i}},\alpha^j_{n_{-j}}]=m_{+i}\delta_{m+n}\delta^{ij}, \\
& \{d^i_{m_{+i}},d^j_{n_{-j}} \}=\delta_{m+n}\delta^{ij}.
\end{split}
\end{equation}
We define the ground states in the Neveu-Schwarz and Ramond sector as follows:
\begin{subequations}
\begin{align}
& \alpha^i_{m+\epsilon_i} |0\rangle_{\text{NS}} = \alpha^i_{m+1-\epsilon_i} |0\rangle_{\text{NS}}=d^i_{r\pm \epsilon_i} |0\rangle_{\text{NS}}=0, \quad m \ge 0, \, r \ge 1/2, \\
& \alpha^i_{m+\epsilon_i} |0\rangle_{\text{R}} = \alpha^i_{m+1-\epsilon_i} |0\rangle_{\text{R}}=d^i_{m + \epsilon_i} |0\rangle_{\text{R}}= d^i_{m+1-\epsilon_i} |0\rangle_{\text{R}}= 0, \quad m \ge 0,
\end{align}
\end{subequations}
and the standard choice of annihilators in the transverse dimensions, which means in particular that
$|0\rangle_{\text{R}}$ carries a spinor representation of $SO(9-2k,1)$.
On physical states the positive frequency modes of the energy-momentum tensor must vanish.
These are the Virasoro generators $L_n$ for $n \ge 0$. To calculate the mass of a state, defined as
$M^2=-p^{\mu}p_{\mu}$, we
are interested in $L_0$ which is given by
\begin{equation}
\begin{split}
L_0 = & \, \alpha' l_0 + \alpha' p^{\mu}p_{\mu} + \frac{1}{2} \sum_{n \neq 0} : \alpha^{\mu}_n \alpha_{\mu,-n} : + \frac{1}{2} \sum_{r} r \, : d^{\mu}_{-r} d_{\mu,r} : \\
      & \, + \sum_{i=1}^k \sum_n : \alpha^i_{n+\varepsilon_i} \alpha^i_{-n-\varepsilon_i} : + \sum_{i=1}^k \sum_r \left(r-\varepsilon_i \right) : d^i_{-r+\varepsilon_i} d^i_{r-\varepsilon_i}: \, ,
\end{split}
\end{equation}
where the first line contains the vacuum energy $l_0$ and the (standard) contribution from the time dimension and the dimensions transverse to both branes, $\mu \in \{0,2k+1,\ldots,9-2k\}$,
while the second line is the contribution from the dimensions in which the branes are at angles. $n$ is integer while
$r$ is integer in the Ramond sector and half-integer in the Neveu-Schwarz sector.  The colons indicate normal ordering, that
is creation operators are moved to the left and annihilation operators to the right. The vacuum energy $l_0$ arises in normal
ordering and can be calculated by using the following regularized expressions:
for a boson with moding shifted
by $\varepsilon_i$:
\begin{equation}
\frac{1}{2\alpha'} \sum_{n=1}^{\infty} (n-\varepsilon_i) = \frac{1}{2\alpha'} \zeta(-1,\varepsilon_i)=
\frac{1}{\alpha'} \left(-\frac{1}{24}+\frac{1}{4} \varepsilon_i (1-\varepsilon_i)\right) \, ,
\end{equation}
and for a Ramond and Neveu-Schwarz fermion respectively:
\begin{equation}
\begin{split}
-\frac{1}{2\alpha'} \sum_{n=1}^{\infty} (n-\varepsilon_i) & = -\frac{1}{2\alpha'} \zeta(-1,\varepsilon_i)=
\frac{1}{\alpha'} \left(\frac{1}{24}-\frac{1}{4} \varepsilon_i (1-\varepsilon_i)\right) \, , \\
-\frac{1}{2\alpha'} \sum_{n=1}^{\infty} (n-\frac{1}{2}-\varepsilon_i) & = -\frac{1}{2\alpha'} \zeta(-1,\frac{1}{2}+\varepsilon_i) =
\frac{1}{\alpha'} \left(-\frac{1}{48}+\frac{1}{4} \varepsilon_i^2 \right) \, .
\end{split}
\end{equation}
Here we introduced the Hurwitz zeta function which is defined by
\begin{equation}
\zeta(s,\epsilon) = \sum_{n=0}^\infty \frac{1}{(n+\epsilon)^s} \, ,
\end{equation}
and regularized the infinite sums by taking the well-defined analytic
continuation to $s=-1$.
Just as in the case without magnetic fields, in the Ramond sector
the vacuum energy of the bosons is cancelled exactly by that of the fermions.
In the Neveu-Schwarz sector however, the vacuum energy is given by
\begin{multline}
l_0 = \frac{1}{\alpha'} \left( (8-2k) \left( - \frac{1}{24} - \frac{1}{48}\right) +
2 \sum_{i=1}^k \left( -\frac{1}{24}+\frac{1}{4} \varepsilon_i (1-\varepsilon_i)
-\frac{1}{48}+\frac{1}{4} \varepsilon_i^2 \right)\right) \\
= \frac{1}{2\alpha'} (-1 +\sum_{i=1}^k \varepsilon_i) \, .
\end{multline}
We have used that fact that as usual the ghost contributions cancel the effect of the time dimension and
the dimension longitudinal to the string.

In the Neveu-Schwarz sector the GSO projection removes the vacuum $|0\rangle_{\text{NS}}$ as well as its bosonic excitations.
The surviving lowest-lying states are then:
\begin{equation}
\begin{split}
& \prod_{i=1}^k (\alpha^i_{-\varepsilon_i})^{m_i}(d^j_{-\frac{1}{2}+\varepsilon_j})|0 \rangle_{\text{NS}}, \quad m_i \in\IN, \\
& \prod_{i=1}^k (\alpha^i_{-\varepsilon_i})^{m_i}(d^j_{-\frac{1}{2}-\varepsilon_j})|0 \rangle_{\text{NS}}, \quad m_i \in\IN,
\end{split}
\end{equation}
with mass given by
\begin{equation}
\label{massboson}
M^2 = \frac{1}{2\pi\alpha'}\left(\sum_i (2 m_i+1)\phi_i \mp 2 \phi_j \right) \, .
\end{equation}
With the minus sign and $m_i=0$ for all $i$ we find tachyonic modes unless
\begin{equation}
\sum_{i\neq j} \phi_i \ge \phi_j, \quad  \forall j .
\end{equation}
For $k=2$ this is equivalent to the BPS condition $\phi_1=\phi_2$, but for $k>2$
the BPS condition is stronger than
the condition for stability. In other words, there exist stable configurations, that is configurations
without tachyons, that
are not supersymmetric.

In the light-cone gauge the physical states in the Ramond sector that are not projected out by the GSO
projection are:
\begin{equation}
\label{statefermion}
\prod_{i=1}^k(\alpha^i_{-\varepsilon_i})^{m_i}(d^i_{-\varepsilon_i})^{l_i}|0\rangle_{\text{R}} , \quad m_i \in\IN, \,  l_i \in \{0,1\} ,
\end{equation}
where $|0\rangle_{\text{R}}$ carries a chiral spinor representation of $SO(8-2k)$.
The masses of these states, which in the absence of magnetic fields reduce to the gauginos,
are given by
\begin{equation}
\label{massfermion}
M^2=\frac{1}{2\pi \alpha'}\sum_{i=1}^k 2\left(m_i+l_i\right)\phi_i.
\end{equation}

\section{Spectrum from the Effective Action}

The mass formulae given in eqs.~(\ref{massboson}) and \eqref{massfermion} should be
reproduced by the spectrum of off-diagonal fluctuations in the effective action approach. Taking the effective action
given in eqs.~(\ref{lagalpha0}--\ref{lagalpha4e3}), one turns on the constant magnetic
background given by eq.~\eqref{diagbackground}.
Expanding the angles $\phi_i$ in powers of $\alpha'$ using eq.~\eqref{relfangle} gives
\begin{equation}
\frac{\phi_i}{2\pi\alpha'} = 2 \left( f_i - \frac{(2 \pi \alpha')^2 f_i^3}{3}+ \frac{(2\pi\alpha')^4 f_i^5}{5}+ {\cal O}\left((\alpha')^6 \right)\right) \, .
\label{anglealphaexp}
\end{equation}
From the orders in $\alpha'$ we read off that the terms linear in $f_i$ have to be
reproduced by ${\cal L}_{(0)}$, those cubic in $f_i$ by $ {\cal L}_{(2)}$,
while ${\cal L}_{(3)}$ should {\em not} contribute to the spectrum {\em at all}.
${\cal L}_4$ is responsible for the terms quintic in $f_i$.

The author collaborated in the spectrum check up to order $\alpha'{}^3$ both
for the bosonic and the fermionic terms while other members of the group pursued it up to
$\alpha'{}^4$ \cite{testalpha4}.  The latter is important because $\alpha'{}^4$ is the first order
where discrepancies appear between the string spectrum and the spectrum of the earlier proposed
symmetrized trace prescription, which consists of only the terms ${\cal L}_{(4,0)}$ \cite{HTspectrum}\cite{DST}.

We will describe the calculation of the Yang-Mills bosonic and fermionic spectrum in
detail and just state the results for the higher orders as the calculations become very involved.
We will also expand on how to deal with field redefinitions.

\subsection{The Yang-Mills Spectrum}

It will be convenient to introduce complex coordinates in which the constant background
field looks like eq.~\eqref{diagcomplexbackground}.
We split first the gauge fields as follows: $A= A^{(0)}+ \delta A$, where $A^{(0)}$
is the background and $\delta A$ the fluctuations. Here and in the next formulae
we suppress the complex index of the gauge field, which could be barred as well as unbarred.
We are only interested in the off-diagonal components of the gauge fields,
\begin{equation}
\delta A=
i\left(
   \begin{array}{cc}
     0 & \delta  A^+ \\
     \delta  A^- & 0
   \end{array}
  \right),
\end{equation}
as the diagonal fluctuations probe the abelian part of the action.
The background covariant derivative $D^{(0)}_{\alpha}$ acts on
the fluctuations as
\begin{equation}
D^{(0)}_{\alpha} \delta A =
i\left(
   \begin{array}{cc}
     0 & (\partial_{\alpha} + 2i A^{(0)}_{\alpha})\delta  A^+  \\
     (\partial_{\alpha} - 2i A^{(0)}_{\alpha}) \delta  A^- & 0
   \end{array}
  \right),
\end{equation}
and similarly for the covariant derivative with barred index.
This motivates us to introduce the following useful component background covariant derivatives:
\begin{equation}
{\cal D}_\alpha  \delta A^{\pm}=\left( \partial_\alpha \pm 2i A^{(0)}_\alpha \right) \delta  A^{\pm},\quad
{\cal D}_{\bar\alpha}  \delta A^{\pm}=\left( \partial_{\bar\alpha} \pm 2i A^{(0)}_{\bar\alpha} \right) \delta  A^{\pm}.
\label{covder}
\end{equation}
These satisfy the commutation relations
\begin{equation}
{[} {\cal D}_ \alpha , {\cal D}_ {\bar\beta} {]}=\pm 2i \delta _{ \alpha \beta }f^c_ \alpha ,\label{Dcomm}
\end{equation}
where the sign depends on whether the derivatives act on $\delta A^+$ or $\delta A^-$.
The spectrum of $ \delta A^+$ is equal to that of $ \delta  A^-$,
which reflects the two orientations of the strings stretching
between the two branes. We will investigate here
the spectrum of $ \delta A^+$.

Linearizing the equations of motion that follow from eq.~(\ref{lagalpha0}) we get
\begin{equation}
\begin{split}
0 & = \left({\cal D}^2+4i
f^c_\alpha \right) \delta A^+_\alpha
-\sum_{\beta=1}^k {\cal D}_\alpha ({\cal
D}_\beta \delta A^+_{\bar{\beta}}+{\cal D}_{\bar{\beta}}\delta
A^+_\beta)\;, \\
0 & =\left({\cal D}^2-4i
f^c_\alpha \right) \delta A^+_{\bar\alpha}
-\sum_{\beta=1}^k {\cal D}_{\bar\alpha }({\cal
D}_\beta \delta A^+_{\bar{\beta}}+{\cal D}_{\bar{\beta}}\delta
A^+_\beta)\;,
\label{Leading1}
\end{split}
\end{equation}
where
\begin{equation}
{\cal D}^2 \delta A^+= \left(\Box_{NC} + \sum_{\beta=1}^k{\cal D}_\beta {\cal
D}_{\bar{\beta}} +\sum_{\beta=1}^k{\cal D}_{\bar\beta} {\cal
D}_{{\beta}}\right),\label{Dkwad}
\end{equation}
and $\Box_{NC}$ denotes the d'Alambertian in the non-compact
directions.
Choosing the gauge
\begin{equation}
\sum_{\beta=1}^k ({\cal
D}_\beta \delta A^+_{\bar{\beta}}+{\cal D}_{\bar{\beta}}\delta
A^+_\beta)=0,
\label{gaugechoice}
\end{equation}
we can rewrite eq.~(\ref{Leading1}) as
\begin{equation}
\begin{split}
0 & = \left(\Box_{NC}+2\sum_{\beta=1}^k \left({\cal D}_{\bar{\beta}} {\cal D}_{\beta}+i f^c_\beta\right) +4i
f^c_\alpha \right) \delta A^+_\alpha ,\\
0 & = \left(\Box_{NC}+2\sum_{\beta=1}^k \left({\cal D}_{\bar{\beta}} {\cal D}_{\beta}+i f^c_\beta\right) -4i
f^c_\alpha \right) \delta A^+_{\bar\alpha}.
\label{leading2}
\end{split}
\end{equation}
In order to construct eigenfunctions of the operators following $\Box_{NC}$ in both equations we introduce creation and annihilation operators,
\begin{equation}
a_{\alpha} = i {\cal D}_{\alpha}, \qquad
a^{\dagger}_{\bar{\alpha}}= i {\cal D}_{\bar{\alpha}} \, ,
\end{equation}
which satisfy
\begin{equation}
[a_{\alpha},a^{\dagger}_{\bar{\beta}}]= - 2i f^c_{\alpha} \delta_{\alpha\bar{\beta}} = 2 f_{\alpha} \delta_{\alpha\bar{\beta}}.
\label{acomm}
\end{equation}
With these operators we can construct a complete set of
functions on the torus:
\begin{equation}
\phi _{\{m_1,m_2,\ldots,m_k\}}(z,\bar z) =
\left(a^\dagger_{\bar{z}^1}\right)^{m_1}\left(a^{\dagger}_{\bar{z}^2}\right)^{m_2}\ldots\left(a^{\dagger}_{\bar{z}^k}\right)^{m_k}
\phi _{\{0,0,\ldots ,0\}}(z,\bar z),
\end{equation}
where $\phi _{\{0,0,\ldots ,0\}}$ is defined through
\begin{equation}
a_{\alpha}\,\phi _{\{0,0,\ldots ,0\}}(z,\bar z)=0,\qquad \forall\, \alpha \in\{1,2,\ldots,k\}.
\label{grdst}
\end{equation}
The function $\phi _{\{0,0,\ldots ,0\}}(z,\bar z)$ is, modulo a discrete degeneracy, fully determined by eq.~(\ref{grdst}) and the requirement that it satisfies proper boundary conditions.
It was explicitly constructed in \cite{spectrumtroost} elaborating on the work of \cite{spectrumbaal}.
It is shown there that the degeneracy of the ground state is given by $Pf(n_{ij})=\prod_{i=1}^k n_i$, which is exactly the number
of intersection points of the two branes wrapping the torus. These are indeed the different places where the ``off-diagonal''
strings can be found.
Denoting the non-compact coordinates collectively by $y$ we make the Kaluza-Klein
expansion
\begin{equation}
\delta A^+(y,z,\bar z)=\sum_{(m_1,\ldots,m_k)\in\IN{}^{{}^k}} \delta A^{+\{m_1,\ldots,m_k\}}(y)
\,\phi _{\{m_1,\ldots,m_k\}}(z,\bar z).
\label{eigenfunctionexpboson}
\end{equation}
The masses of the modes $\delta A^{+\{m_1,\ldots,m_k\}}(y)$
can easily be determined.
Indeed, using eq.~\eqref{acomm} and $-if^c_{\alpha}=f_{\alpha}\ge 0$ one immediately gets
\begin{equation}
\left(\Box_{NC}-M^2\right) \delta A^{+\{m_1,\ldots,m_k\}}_{\alpha}(y)=0,
\end{equation}
with
\begin{equation}
M^2=2 \sum_{j=1}^k (2m_j +1)f_j + 4 f_i ,
\end{equation}
and
\begin{equation}
\left(\Box_{NC}-M^2\right) \delta A^{+\{m_1,\ldots,m_k\}}_{\bar{\alpha}}(y)=0,
\end{equation}
with
\begin{equation}
M^2=2 \sum_{j=1}^k(2m_j +1)f_j -4f_i ,
\end{equation}
which agrees with the leading term in
eq.~\eqref{massboson} using the expansion \eqref{anglealphaexp}.

Next, we turn our attention to the fermionic fluctuations:
\begin{equation}
\chi=
i\left(
   \begin{array}{cc}
     0 & \chi^+ \\
     \chi^- & 0
   \end{array}
  \right) .
\end{equation}
From the Dirac Lagrangian
\begin{equation}
{\cal L}^{F}_{(0)} = \frac{1}{2} \Tr \bar{\chi} \slashed{D} \chi\, ,
\end{equation}
where $\bar{\chi}=\chi^\dagger i \Gamma^0$,
follow readily the equations of motion,
\begin{equation}
\label{LeOreom}
\left(\partial\!\!\!/_{NC}+ {\cal D}\!\!\!\!/\right)\chi^+=0,
\end{equation}
where ${\cal D}$ is defined as in eq.~\eqref{covder}. The
equation for $\chi^-$ is similar.
Squaring the kinetic operator in eq.~(\ref{LeOreom}) we get
\begin{equation}
\label{eomsq}
\left(
\Box_{NC}+2\sum_{\beta=1}^{k}\left\{
{\cal D}_{\bar{\beta}} {\cal D}_{\beta}+if^c_\beta-if^c_\beta\Gamma_{\beta\bar\beta}
\right\}\right)\chi^+=0.
\end{equation}
Now we introduce the complete set of functions
$\psi_{\{(m_1,l_1),(m_2,l_2),\ldots,(m_k,l_k)\}}$ with
$m_i\in\IN$ and $l_i\in\{0,1\}$ for $i=1\ldots,k$ by
\begin{equation}
\psi_{\{(m_1,l_1),(m_2,l_2),\ldots,(m_k,l_k)\}} \equiv
\phi_{\{m_1,\ldots,m_k\}}\,\left(\Gamma_{\bar 1}\right)^{l_1}\left(\Gamma_{\bar 2}\right)^{l_2}
\cdots\left(\Gamma_{\bar k}\right)^{l_k}|0\rangle,
\end{equation}
where $|0\rangle$ is the spinor ``vacuum'' satisfying $\Gamma_{\alpha}|0\rangle=0,\,\forall \alpha$.
We expand the fermion field
\begin{equation}
\chi^+(y,z,\bar z)=\sum_{\{(m,l)\}}\chi^{+\{(m,l)\}}(y)\psi_{\{(m,l)\}}(z,\bar{z}),
\end{equation}
where $\{(m,l)\}\equiv\{(m_1,l_1),\,(m_2,l_2),\,\ldots(m_k,l_k)\}$ and
$y$ again collectively denotes the non-compact coordinates.
Using this Kaluza-Klein expansion one finds from eq.~(\ref{eomsq}) that the mass of
$\chi^{+\{(m,l)\}}(y)$
is given by
\begin{equation}
\label{masslofermion}
M^2=2\sum_{i=1}^k 2\left(
m_i+l_i
\right)f_i.
\end{equation}
This reproduces the mass of the fermionic modes found in eq.~\eqref{massfermion} through lowest
order in $\alpha'$.

\subsection{Field Redefinitions}

As an example, let us study the influence of the simplest possible field redefinition:
\begin{equation}
A_{\nu}  \rightarrow A_{\nu} + 2 \pi \alpha' \xi D^{\mu}F_{\mu\nu} \, .
\end{equation}
As a result of this redefinition the effective Lagrangian gains an $\alpha'$ correction
\begin{equation}
{\cal L}'_{(1)} = 2 \pi \alpha' \xi \Tr D^{\rho} F_{\rho\nu}D_{\mu} F^{\mu\nu} \, .
\end{equation}
Performing the spectrum calculation as above one finds the same result for the spectrum, i.e.\ one
recovers eq.~\eqref{leading2}, provided one redefines the fluctuations as
\begin{equation}
\delta A_{\alpha} \rightarrow \delta A_{\alpha} - 4 \pi \alpha' \xi
\left( {\cal D}^2 \delta A_{\alpha} + 2[F^{(0)}_{\alpha\bar{\beta}},\delta A_{\beta}]\right) \, ,
\end{equation}
and analogously for the barred indices.
Such correcting field redefinitions turn out to be unambiguously defined when demanding that the dispersion relation
should be of the form $(\Box_{NC}-M^2)\delta A =0$.  Crucial in finding the appropriate
field redefinition is the form of the derivatives with respect to the non-compact coordinates.

\subsection{Higher Orders}

At higher orders not only one has to apply an appropriate field redefinition,
but also the gauge condition will get corrections.
After a very long and tedious calculation \cite{testalpha4} one finds the following linearized equations of motion
for the fluctuations of the gauge field
\begin{multline}
\left[ \Box_{NC} + 2 \sum_{\beta=1}^k \left(1+ \frac{(2 \pi \alpha' f^c_\beta)^2}{3} +
\frac{(2\pi\alpha' f^c_\beta)^4}{5}\right)({\cal D}_{\bar{\beta}} {\cal D}_{\beta}
+if^c_\beta)+ \right. \\
\left. 4i\left(f^c_\alpha + \frac{(2\pi\alpha')^2 (f^c_\alpha)^3}{3} +
\frac{(2\pi\alpha')^4 (f^c_\alpha)^5}{5}\right) \right] \delta A^+_\alpha = 0,
\label{eom4}
\end{multline}
which, when expanding into the set of eigenfunctions $\phi_{\{m_1,\ldots,m_k\}}$ as in
eq.~\eqref{eigenfunctionexpboson}, reproduces exactly the correct spectrum.
This means in particular that although there is a non trivial contribution
to the effective action at order $\alpha'{}^3$, that contribution has no effect whatsoever on
the spectrum \cite{testalpha3} as predicted by string theory.

Non-abelian higher order corrections to the fermionic part of the effective action
can be found in \cite{metsaevsusy}\cite{bergshoeffnoether}\cite{groningensevrin}\cite{goteborg} (for order $\alpha'{}^2$) and \cite{groningen} (for order
$\alpha'{}^3$). The spectral test can only see the terms quadratic in the fermions.
Working out the calculation
as above one gets again an exact match
with the string theory spectrum in the Ramond sector. As the calculation is a straightforward generalization
of the lowest order case we do not repeat it here \cite{testfermion}.

In these calculations we
learned that the spectral test is weaker for the terms containing fermions than for
the purely bosonic terms.  Indeed, in \cite{STT} it was shown that demanding that the spectrum of the gauge field is correctly reproduced,
combined with the requirement that the abelian limit should agree with the known result, completely fixes
the bosonic part of the effective action through order $\alpha'{}^2$, which is not the case for the
fermionic terms.  From order $\alpha'{}^3$ and up the spectral test does not suffice to fix
the action, neither in the bosonic case nor in the fermionic case.  In \cite{STT} was studied how much of the
bosonic $\alpha'{}^4$ could be fixed by the spectrum calculation. However, since in the non-abelian case there
does not seem to be an equivalent of the slowly varying field limit, the assumption in that paper
that there would be no derivative corrections is manifestly too strong.

\section{Applications}

Intersecting brane configurations are very popular in D-brane phenomenology as a way to obtain
a particle spectrum as close to the Standard Model as possible, see \cite{uranga} for a review and \cite{blumenhagen}
for an extensive list of references.  Consider for instance two stacks of D6-branes intersecting over a
4-dimensional space-time and at angles $\phi_1$, $\phi_2$ and $\phi_3$ in the remaining 6 dimensions. In this case
$k=3$. A fortunate feature of this model, which was already noted in \cite{angles1}, is that it contains
a massless chiral fermion at the 4-dimensional intersection; see also eq.~\eqref{statefermion}.  If the two stacks
consist of $N_a$ and $N_b$ branes respectively, strings stretching between stack $a$ and stack $b$ lead to
a chiral fermion in the $(N_a,\bar{N}_b)$ representation of the gauge group $U(N_a)\times U(N_b)$ while strings
with the other orientation lead to a $(\bar{N}_a,N_b)$ chiral anti-fermion. Another nice feature is that the
different intersection points provide the different generations.

The necessity of introducing D6-branes instead of other-dimensional branes lies in the fact that any
continuous motion of the branes (preserving the same gauge symmetry) maintains the existence of the
intersection; in this way the chiral particles cannot suddenly become massive which would be
prohibited by an index theorem.

To address various issues like the hierarchy problem and obtaining stability, one would like to construct
supersymmetric configurations, which is much more difficult and requires, together with RR tadpole cancellation,
the introduction of O6-planes.

Furthermore, intersecting branes provide an interesting setting to study the dynamics of tachyon
condensation; for a review see \cite{epplelust}. Indeed, at small angles the Yang-Mills (with or without
higher order corrections) fluctuation analysis of this chapter can be used to study brane recombination \cite{hana1}\cite{naga2}.
This fluctuation analysis was carried through up to order $\alpha'{}^4$ in \cite{naga1} which provided another
(related) check on our D-brane effective action.

\newpage
\thispagestyle{empty}

\chapter{A Different Road: Supersymmetric $\sigma$-models with Boundaries}
\chaptermark{Supersymmetric $\sigma$-models with Boundaries}
\label{boundarysusy}

The analysis of chapter \ref{nonabelianBI} showed that derivative terms played a crucial role
in the non-abelian case.  The complexity of the results for the effective action in that chapter
makes one wonder whether a closed expression to all orders in $\alpha'$ might ever be obtainable.
If such a closed expression exists, one should, by taking the abelian limit, obtain a closed expression
for the abelian effective action including all derivative corrections! So before tackling the full
non-abelian problem, perhaps a more modest aim is to find all-order results for the abelian case with
derivatives. Significant progress in this direction has been made in \cite{wyllard} where the
bosonic sector was calculated up to 4 derivatives using the boundary state formalism; see also
section \ref{abeliander}. Moreover, in
\cite{wyllard2} a conjecture has been made about higher-derivative terms.
As we saw in section \ref{roads}, one of the other ways to
obtain the effective action, the Weyl invariance method, involves the calculation of the $\beta$-functions.
While all calculations until now have been done in $X$-space, it would be interesting to try a superspace
calculation\footnote{In \cite{weylbergshoeff} an attempt was made to show through an $N=1$ superspace
$\beta$-function calculation that all derivative corrections would vanish. This is obviously not true.}.
Just as the method based on the boundary state formalism, the Weyl invariance method is rather
powerful in the sense that it works order by order in the number of derivatives while all orders in $\alpha'$
are calculated at once.  A first step is the construction of the $N=2$ superspace with boundaries, which
is the subject of this chapter.

Without boundaries, the supersymmetric extension of the Polyakov action in non-trivial
backgrounds suitable for describing type II
strings --- we will treat left- and right-movers
symmetrically --- is the $N=(1,1)$ non-linear sigma-model. Adding a second supersymmetry for each chirality turns out
to be sufficient to ensure space-time supersymmetry.  Such $N=(2,2)$ supersymmetry
yields restrictions on the geometry of target space.
In the absence of boundaries this geometry has been
intensively studied in the past, e.g. \cite{N2studies}\cite{N2troost}.

Much less is known about the case with boundaries, which is relevant for
type I string theories and type II string theories with D-branes.
Partial results on supersymmetric $\sigma$-models with boundaries
were known already for some time, e.g.\ \cite{boundsusy}\cite{alvarezTdual}. In any case
the boundary breaks a linear combination of the left- and right-moving supersymmetry such that only $N=1$ or
$N=2$ supersymmetry is left. Only more recently a systematic
study was performed, \cite{stock1} and \cite{stock2}, resulting in the most general boundary conditions compatible with
$N=1$ supersymmetry. Subsequently, these results were extended to $N=2$ supersymmetry \cite{zab}.
See also \cite{zab2} for some specific applications to WZW models, \cite{lrn} for a different approach, and
\cite{stock3} for a review of the $N=1$ and $N=2$ case and further applications to WZW models.

However, the results of \cite{stock1} and \cite{stock2} remain somewhat surprising. Not only are the derivations
quite involved, but the presence of a Kalb-Ramond background seems to require a {\em non-local}
superspace description of the model; see for instance eq.~(4.10) of \cite{stock2}, which contains the
awkward $1/D_+$ factor where $D_+$ is a covariant derivative in superspace. A non-trivial Kalb-Ramond background
already occurs in the case of interest for us, namely open strings in a trivial gravitational background, but in a non-trivial
electromagnetic background. But clearly, in order to study the open string effective action through
the calculation of supergraphs, a local superspace description is called for.

In this chapter we reanalyse the models studied in \cite{stock1} and \cite{stock2}, and we resolve many of the
difficulties encountered there. These complications stem from the desire to keep the $N=(1,1)$ bulk supersymmetry
manifest although the presence of the boundary breaks this supersymmetry to an $N=1$ supersymmetry anyway.

After a first section with definitions, we start with a boundary superspace formulation that is manifestly invariant under only one
combination of the two bulk supersymmetries. This approach was inspired by the methods used in \cite{oliverjoanna},
in quite a different setting however, and independently from us suggested in \cite{lrn} although not worked out
therein.
In this way the analysis of boundary conditions compatible with $N=1$ supersymmetry
is greatly facilitated and one finds that, just as for the case without boundaries,
$N=1$ introduces no new conditions on the target space geometry compared to $N=0$. In addition, no
non-local terms are needed and the cases with or without Kalb-Ramond are treated on the same footing.
The price we pay is that we loose manifest bulk $d=2$ super-Lorentz covariance.

In the third section we investigate under which conditions the $N=1$ supersymmetry
can be promoted to an $N=2$ supersymmetry. As for the case without
boundaries, one needs two separately integrable covariantly
constant complex structures. The metric has to be hermitian with
respect to both of them. However, the presence of boundaries
requires that one of them is expressed in terms of the other one
and the remainder of the geometric data.
We study the $N=2$ superspace formulation, which seems to work
in only one case.  Fortunately, it is exactly the case of interest.

\section{Definitions}
\label{superspaceconventions}

We study the general case in curved space with a metric $G_{ij}(X)$
and a Kalb-Ramond tensor $B_{ij}(X)=-B_{ji}(X)$. Only in further work we will specialize
to the case of interest for the effective action calculations i.e.\ flat metric and
closed $F_{ij}= \frac{1}{2 \pi \alpha'} B_{ij}$ on the boundary.

Denoting the world-sheet coordinates by $\tau$ and $\sigma$, we define
light-cone coordinates
\begin{equation}
\sigma^{\neq} =  \tau +\sigma , \; \sigma ^== \tau - \sigma \quad \Rightarrow \quad
\partial_{\neq} = \frac 1 2 ( \partial_ \tau + \partial_ \sigma ), \;
\partial_= =\frac 1 2 ( \partial_ \tau - \partial_ \sigma ).
\end{equation}
The target space coordinates are denoted by $X^i$, $i\in\{1,\cdots d\}$.
The torsion is given by the curl of the Kalb-Ramond field:
\begin{equation}
T_{ijk}=-\frac 3 2 B_{[ij,k]}.
\end{equation}
We introduce two connections
\begin{equation}
\Gamma^{\ i}_{(\pm) jk} = \big\{{}^i{}_{jk}\big\}\pm T^i{}_{jk},
\label{torsieconn}
\end{equation}
where $\big\{{}^i{}_{jk}\big\}$ is the standard Christoffel connection. The
connections are used to define covariant derivatives
\begin{equation}
\begin{split}
\nabla^{(\pm)}_iV^j&= \partial_i V^j+\Gamma^{\ j}_{(\pm) ki}V^k,\\
\nabla^{(\pm)}_iV_j&= \partial_i V_j-\Gamma^{\ k}_{(\pm) ji}V_k\,.
\end{split}
\end{equation}
We define curvature tensors as follows:
\begin{equation}
{[} \nabla_i^{(\pm)},\nabla_j^{(\pm)}{]}V^k=\frac 1 2 V^l R_{(\pm)lij}^{\ k}\pm
T^l{}_{ij}\nabla^{(\pm)}_lV^k.\label{int}
\end{equation}
The curvature tensors $R^{(\pm)}_{ijkl}$ are anti-symmetric in the first and the last two indices, and they also
satisfy
\begin{equation}
R^+_{ijkl}=R^-_{klij}.
\end{equation}

We will start from the $N=(1,1)$ superspace formalism and introduce Ma\-jo\-ra\-na-Weyl fermionic coordinates $\theta^+$ and $\theta^-$,
which satisfy $(\theta^+)^*=\theta^+$ and $(\theta^-)^*=\theta^-$.  The supersymmetry
generators and covariant derivatives are given by:
\begin{equation}
\begin{split}
Q_+ = \frac{\partial}{\partial \theta^+} + i \theta^+ \partial_{\neq} \, , \quad &
Q_- = \frac{\partial}{\partial \theta^-} + i \theta^- \partial_{=}\, , \\
D_+ = \frac{\partial}{\partial \theta^+} - i \theta^+ \partial_{\neq} \, , \quad &
D_- = \frac{\partial}{\partial \theta^-} - i \theta^- \partial_{=} \, ,
\end{split}
\end{equation}
and satisfy
\begin{equation}
Q_+^2 = -D_+^2 = i \partial_{\neq}, \quad Q_-^2 = - D_-^2 = i \partial_{=} \, .
\end{equation}
The real coordinate superfield $\Phi^i$ is defined as
\begin{equation}
\Phi^i = X^i - i \theta^+ \psi_+^i - i \eta \theta^- \psi_-^i + i \theta^- \theta^+ F^i \, ,
\label{superfield}
\end{equation}
where $X^i$ are the target space coordinates, $\psi_+^i$ and $\psi_-^i$ real world-sheet fermions
and $F^i$ an auxiliary field.
The sign $\eta \in \{+1,-1\}$ will not play any essential role but we introduced it to be
able to better compare with \cite{stock1} and \cite{stock2}.
The factors of $i$ appear
to ensure the reality of $X^i$. The $N=(1,1)$ supersymmetry transformation generated
by $\epsilon^+ Q_+ + \epsilon^- Q_-$ is given by
\begin{equation}
\begin{split}
\delta X^i & = -i \epsilon^+ \psi_+^i - i \eta \epsilon^- \psi_-^i, \\
\delta \psi_+^i & = \epsilon^+ \partial_{\neq} X^i + \epsilon^- F^i, \\
\delta \psi_-^i & = \eta \epsilon^- \partial_{=} X^i - \eta \epsilon^+ F^i, \\
\delta F^i & = + i \eta \epsilon^+ \partial_{\neq} \psi_-^i - i \epsilon^- \partial_{=} \psi_+^i \, ,
\end{split}
\end{equation}
which is indeed a symmetry of the {\em bulk} non-linear $\sigma$-model, which reads in $X$-space\footnote{With
respect to action~\eqref{susypolyakov} we absorbed $(\alpha')^{-\frac{1}{2}}$ into the $X^i$ and removed
an overall factor of $\frac{1}{2\pi}$.}
\begin{equation}
\begin{split}
{\cal L}_{\text{bulk}}& = 2(G_{ij}+B_{ij}) \partial_{\neq} \,X^i \partial_= X^j+
2i\, G_{ij}\,\psi^i_+\nabla_=^{(+)} \psi ^j_+
+2i\, G_{ij}\,\psi ^i_-\nabla_{\neq}^{(-)} \psi ^j_- \\
& +R^{(-)}_{ijkl} \psi _-^i \psi _-^j \psi _+^k \psi _+^l
+2 (F^i-i \eta \Gamma ^{\ i}_{(-)kl} \psi ^k_- \psi ^l_+ )G_{ij}
(F^j-i \eta \Gamma ^{\ j}_{(-)mn} \psi ^m_- \psi ^n_+ ).\label{lag11}
\end{split}
\end{equation}

\section{$N=1$ Supersymmetry}

\subsection{Superspace Formulation}

In \cite{susyboundary} the most general boundary superspace action was constructed.
Subsequently, its coefficients were fixed by requiring that it should reproduce the $X$-space
non-linear $\sigma$-model in the bulk.  We will not do so here, but instead derive the action
from the bulk $N=(1,1)$ action. This trick was independently from us suggested in \cite{lrn} and
worked out in the appendix of \cite{susyboundary}.

The well-known superspace $N=(1,1)$ bulk action reads
\begin{equation}
{\cal S}_{\text{bulk}}= 2\, \int d \tau d\sigma \left[ D_+ D_- \left((G_{ij}+B_{ij}) D_+ \Phi^i D_- \Phi^j\right)\right] \, ,
\label{bulksup}
\end{equation}
where the square brackets denote putting $\theta^+=\theta^-=0$.
The demonstration of supersymmetry invariance uses partial integration.  Since for open strings
there is a boundary at $\sigma=0$ and $\sigma=\pi$,
a total derivative with respect to $\sigma$ will introduce boundary terms.
As a consequence, ${\cal S}_{\text{bulk}}$ is {\em not} supersymmetric. In \cite{stock1}\cite{stock2} this was remedied
by applying the boundary conditions to make the boundary variation vanish.

Instead, we are looking for an action that is
\begin{itemize}
\item manifestly supersymmetric without using the boundary conditions;
\item in the bulk equivalent with \eqref{bulksup}.
\end{itemize}

To proceed we introduce new fermionic coordinates, and the corresponding covariant derivatives and supersymmetry
generators:
\begin{equation}
\begin{split}
&\theta  = \theta^+ + \theta^-, \qquad D = \frac{1}{2} (D_{+}+D_{-}), \qquad Q=\frac{1}{2}(Q_+ + Q_-) \, , \\
&\tilde{\theta} = \theta^+ - \theta^-, \qquad \tilde{D} = \frac{1}{2}(D_{+}-D_{-}), \qquad \tilde{Q}=\frac{1}{2}(Q_+ - Q_-) \, ,
\label{qd1}
\end{split}
\end{equation}
such that
\begin{subequations}
\begin{align}
Q^2=\tilde{Q}^2 &= +\frac{i}{4}\frac{\partial}{\partial \tau} \, , \\
\label{qd2-2}
D^2=\tilde{D}^2 &= -\frac{i}{4}\frac{\partial}{\partial \tau} \, , \\
\{Q,\tilde{Q}\} & = + \frac{i}{2}\frac{\partial}{\partial \sigma} \, , \\
\{D,\tilde{D}\} & = - \frac{i}{2}\frac{\partial}{\partial \sigma} \, .
\end{align}
\end{subequations}
Consider the following ``improved'' action:
\begin{equation}
\begin{split}
{\cal S}_{\text{N=1}} & = -4 \int d \tau d \sigma \left[ D \tilde{D} \left((G_{ij}+B_{ij}) D_+ \Phi^i D_- \Phi^j\right)\right] \\
         & = -4 \int d \tau d \sigma d \theta \left[ \tilde{D} \left((G_{ij}+B_{ij}) D_+ \Phi^i D_- \Phi^j\right) \right] \, ,
\label{improvedbulksup}
\end{split}
\end{equation}
where the square brackets in the first line denote $\theta=\tilde{\theta}=0$ and in the
second line $\tilde{\theta}=0$.
This is explicitly supersymmetric on the boundary under the supersymmetry $Q$
and using $D_+D_-= -2 D \tilde{D} - \frac{i}{2} \frac{\partial}{d\sigma}$ modulo
boundary terms equivalent to (\ref{bulksup}).
We define two $N=1$ superfields, a bosonic\footnote{We denote the bosonic $N=1$ superfield and the target space
coordinates by the same symbol $X^i$. Since both are used in different contexts i.e.\ $X$-space versus superspace, there should be
no confusion.} and a fermionic, as follows:
\begin{equation}
\begin{split}
\left[\Phi^i \right]& = X^i \, , \\
i \left[\tilde{D} \Phi^i\right] & = \Psi^i \, ,
\label{bulktoboundary}
\end{split}
\end{equation}
where the square brackets denote again $\tilde{\theta}=0$.
Working out the $\tilde{D}$ derivative we find the full model in $N=1$ superspace:
\begin{equation}
\begin{split}
{\cal S}_{\text{N=1}} =&\int d \tau d \sigma d \theta \left( 2i G_{ij}DX^i \dot X^j
-4i B_{ij} DX^i  X^j{}'
-4 G_{ij} \Psi^i X^j{}'
+8G_{ij} \nabla \Psi^i \Psi^j \right. \\
&\left. -\frac{8i}{3}T_{ijk}\Psi^i\Psi^j\Psi^k
-8iT_{ijk}\Psi^i DX^j DX^k\right),\label{finac}
\end{split}
\end{equation}
where the covariant derivative $\nabla \Psi^i $ is given by
\begin{equation}
\nabla \Psi ^i=D \Psi^i+\Big\{{}^i{}_{jk}\Big\} DX^k \Psi^j .
\end{equation}
The transition to the non-linear $\sigma$-model in $X$-space can be made by
integrating out the $\theta$ coordinate using
\begin{equation}
\begin{split}
\psi^i_+ & = \left[i DX^i+ \Psi ^i\right], \\
\psi^i_- & = \eta \left[i DX^i- \Psi ^i\right] ,\label{id1}
\end{split}
\end{equation}
which we infer from eq.~\eqref{superfield}.  Here the square brackets denote putting $\theta=0$.

Another possibility is the action (\ref{improvedbulksup}) with $D$ and $\tilde{D}$ interchanged.
Using \eqref{bulktoboundary} and \eqref{id1} we find
\begin{equation}
\begin{split}
\psi^i_+ =& i \left[D\Phi^i + \tilde{D}\Phi^i\right], \\
\psi^i_- =& i \eta \left[D\Phi^i -\tilde{D}\Phi^i\right] \, ,
\end{split}
\end{equation}
from which we see that interchanging $D$ and $\tilde{D}$ is in fact equivalent to flipping the sign $\eta \rightarrow -\eta$.

Of course, the argument in the preceding paragraphs does not replace the exhaustive analysis in \cite{susyboundary}
because it is not a priori clear that the most general boundary model could be written in the form (\ref{improvedbulksup}).
In fact, this procedure does not work when one tries to derive $\sigma$-models in $N=2$ boundary
superspace from the $N=(2,2)$ superspace $\sigma$-models.

\subsection{Boundary Conditions}

Varying eq.~(\ref{finac}) yields a boundary term
\begin{equation}
-4\int d \tau d \theta  \left(\Psi^i G_{ij}+iDX^i B_{ij}\right)\delta X^j.
\end{equation}
This vanishes if we take Neumann boundary conditions in all directions,
\begin{equation}
\Psi ^i=i\,B^i{}_j DX^j,
\end{equation}
or Dirichlet boundary conditions in all directions,
\begin{equation}
\delta X^i=0.
\end{equation}
The more general case, which involves both Dirichlet and Neumann boundary conditions requires the introduction
of an almost product structure $ {\cal R}^i{}_j(X)$ satisfying
\begin{equation}
{\cal R}^i{}_k{\cal R}^k{}_j= \delta ^i{}_j.\label{RRisone}
\end{equation}
This allows us to construct the projection operators ${\cal P}_\pm$:
\begin{equation}
{\cal P}_\pm^i{}_j = \frac{1}{2}\left( \delta ^i{}_j\pm {\cal R}^i{}_j\right).\label{bproj}
\end{equation}
Using these projection operators we impose Neumann,
\begin{equation}
{\cal P}_+^i{}_j \left (\Psi ^j-i\,B^j{}_k DX^k \right)=0,\label{bn}
\end{equation}
and Dirichlet boundary conditions,
\begin{equation}
{\cal P}_-^i{}_j \delta X^j=0.\label{bd}
\end{equation}
From eqs.~(\ref{bn}) and (\ref{bd}), we obtain
\begin{equation}
\begin{split}
\Psi^i & = {\cal P}^i_-{}_j \Psi^j
+i {\cal P}^i_+{}_j B^j{}_k {\cal P}_+^k{}_l DX^l, \\
\delta X^i & = {\cal P}^i_+{}_j \delta X^j.
\label{bfin}
\end{split}
\end{equation}
From this one observes that, as was to be expected, $ \delta X$ is completely frozen in the
Dirichlet directions while $ \Psi $ gets a component in the Neumann directions when
there is a non-trivial Kalb-Ramond background.
Eq.~(\ref{bd}) implies
\begin{equation}
{\cal P}^i_-{}_j \dot X^j={\cal P}^i_-{}_j D X^j=0.\label{bdex}
\end{equation}
This equation leads to certain compatibility conditions when requiring
eq.~\eqref{qd2-2}. Indeed, acting with $D$ on the second equation in eq.~(\ref{bdex}),
we get
\begin{equation}
0=-\frac i 4 {\cal P}^i_-{}_j \dot X^j+{\cal P}^i_-{}_{l,m} {\cal P}^l_+{}_j {\cal P}^m_+{}_k D X^k D X^j \, .
\end{equation}
This is indeed consistent with the first expression
in eq.~(\ref{bdex}) provided
\begin{equation}
{\cal P}_+^l{}_{[i} {\cal P}_+^m{}_{j]} {\cal P}^k_-{}_{l,m}=0
\end{equation}
or equivalently
\begin{equation}
{\cal P}_+^l{}_{[i} {\cal P}_+^m{}_{j]} {\cal P}^k_+{}_{l,m}=0
\label{integrability}
\end{equation}
holds. Put in this way it looks as if the above argument requires supersymmetry. However, instead of requiring
that $D^2=-(i/4)\partial/\partial \tau$ still holds on the boundary as we did above, one could demand
the same for $[\delta,\partial/\partial\tau]=0$, which works also without supersymmetry. This would lead to the same integrability
condition~\eqref{integrability}.

Armed with the projection operators we can rewrite the boundary term in the variation as
\begin{equation}
\begin{split}
&\int d \tau d \theta  \left(\Psi^i+iDX^k B_k{}^i\right) G_{ij}\delta X^j\\
&=\int d \tau d \theta \left( {\cal P}_+^i{}_k +{\cal P}_-^i{}_k \right)
\left(\Psi^k+iDX^l B_l{}^k\right)
G_{ij} \delta X^j \\
& =\int d \tau d \theta
\left( {\cal P}_+^i{}_k
\left(\Psi^k+iDX^l B_l{}^k\right)G_{ij} \delta X^j +
\left(\Psi^i+iDX^l B_l{}^i\right)G_{ij}
{\cal P}_-^j{}_k\delta X^k\right),
\end{split}
\end{equation}
where in the last step we had to impose the invariance
of the metric under the $(1,1)$-tensor ${\cal R}$:
\begin{equation}
{\cal R}^k{}_i {\cal R}^l{}_j G_{kl}=G_{ij}.\label{metinv}
\end{equation}
Using eq.~(\ref{RRisone}) this gives
\begin{equation}
{\cal R}_{ij}= {\cal R}_{ji},
\end{equation}
where we defined ${\cal R}_{ij}=G_{ik}{\cal R}^k{}_j$.
Imposing the Neumann, eq.~(\ref{bn}), and the Dirichlet, eq.~(\ref{bd}), boundary conditions the boundary
term in the variation of the action indeed vanishes.

Summarizing, we can have mixed Neumann and Dirichlet boundary conditions
provided there exists a $(1,1)$-tensor $ {\cal R}$ that satisfies
\begin{subequations}
\begin{align}
& {\cal R}^i{}_k {\cal R}^k{}_j= \delta ^i{}_j, \label{c1} \\
& {\cal R}^k{}_i {\cal R}^l{}_j G_{kl}=G_{ij}, \label{c2}\\
& {\cal P}_+^l{}_{[i} {\cal P}_+^m{}_{j]} {\cal P}^k_+{}_{l,m}=0. \label{c3}
\end{align}
\end{subequations}
Eq.~(\ref{c1}) tells us that $ {\cal R}$ is an almost product structure, for which according to eq.~(\ref{c2}) the metric is preserved.
The last condition, eq.~(\ref{c3}), tells us that the projection operator $ {\cal P}_+$ is integrable. Note that this is
weaker than requiring that $ {\cal R}$ is integrable. The latter would require that
\begin{equation}
{\cal R}^i{}_l{\cal R}^l{}_{[j,k]}+{\cal R}^l{}_{[j}{\cal R}^i{}_{k],l}=0
\end{equation}
holds. This is equivalent to the integrability of both $ {\cal P}_+$ and $ {\cal P}_-$.

Performing the same analysis for the $N=0$ model would lead to the same conditions such that
we can conclude that {\em any} $N=0$ non-linear $ \sigma $-model with given boundary conditions
allows for an $N=1$ supersymmetric extension i.e. the one given in
eq.~(\ref{finac}).

We now briefly compare our results to those obtained in \cite{stock1} and \cite{stock2}. In the present derivation,
whether or not a Kalb-Ramond background is present does not play any role.
When the Kalb-Ramond background vanishes, $B_{ij}=0$, eqs.~\eqref{c1}-\eqref{c3} precisely agree with
the conditions derived in \cite{stock1}.
However, as supersymmetry is kept manifest, the derivation of these conditions is tremendously simplified. Contrary to
\cite{stock1} we remained off-shell all the time.
A drawback compared to \cite{stock1} is the loss of manifest $d=2$ bulk super-Lorentz covariance in the present
formulation.
For a non-trivial Kalb-Ramond background the comparison with the results
in \cite{stock2} is a bit more involved.
A first bonus compared to \cite{stock2} is that we have a regular superspace formulation,
i.e.\ no non-local terms are needed here.
Combining eqs.~(\ref{id1}), (\ref{bn}) and (\ref{bdex})
we obtain the following boundary condition for the fermions (in matrix notation):
\begin{equation}
\begin{split}
\psi_- & =\eta \, \frac{ {\cal R}-  B_{++}}{1+ B_{++}}\psi_+ \\
       & =\eta \, ({\cal P}_+ \frac{1 - B_{++}}{1+B_{++}} - {\cal P}_-) \, ,
\end{split}
\label{bcfermions}
\end{equation}
where $\psi_-$ and $\psi_+$ describe the right- and left-moving modes respectively
and $\left(B_{++}\right)^i{}_j$ stands for $ {\cal P}_+^i{}_k B^k{}_l {\cal P}_+^l{}_j$.
Comparing with the boundary condition~(2.1) in \cite{stock2}, we see that
our (1,1)-tensor $(1+  B_{++})^{-1}( {\cal R}- B_{++})$ should be identified
with the (1,1)-tensor $R$ in their paper. Furthermore, our ${\cal R}$, ${\cal P}_+$ and ${\cal P}_-$
are to be identified with $r$, $\pi$ and $Q$ respectively.
It is then straightforward to show that eqs.~\eqref{c1}-\eqref{c3} are equivalent to the conditions in eq.~(3.22) of
\cite{stock2}.

\section{$N=2$ Supersymmetry}
\subsection{Promoting the $N=1$ to an $N=2$ Supersymmetry}

The action~\eqref{finac} is manifestly invariant under the supersymmetry transformation
\begin{equation}
\delta X^i= \varepsilon Q X^i,\qquad \delta \Psi ^i = \varepsilon Q \Psi ^i,
\end{equation}
where the supersymmetry generator $Q$ was defined in eq.~\eqref{qd1}.
Let us now turn our attention to the conditions under which the action~\eqref{finac}
exhibits a {\em second} supersymmetry.
The most general transformation rules consistent with the dimensional analysis\footnote{$X$ has world-sheet mass dimension 0, $\Psi$ and $D$ have
dimension $1/2$ and derivatives to $\tau$ and $\sigma$ have dimension $1$.} and statistics that we can write down are
\begin{equation}
\begin{split}
\delta X^i =&\,\hat\varepsilon {\cal J}^{\ i}_{(1)j}(X)DX^j+ \hat\varepsilon{\cal J}^{\ i}_{(2)j}(X) \Psi^j, \\
\delta \Psi^i =& \,\hat\varepsilon{\cal K}^{\ i}_{(1)j}(X)D \Psi^j+ \hat\varepsilon{\cal K}^{\ i}_{(2)j}(X) \dot X ^j+
\hat\varepsilon{\cal K}^{\ i}_{(3)j}(X)  X'{}^j+ \\
&\, \hat\varepsilon{\cal L}_{(1)jk}^{\ i}(X) \Psi ^j \Psi ^k + \hat\varepsilon{\cal L}_{(2)jk}^{\ i}(X) \Psi ^j DX^k
+ \hat\varepsilon{\cal L}_{(3)jk}^{\ i}(X) DX ^j DX^k .\label{2ndsusy}
\end{split}
\end{equation}
In \cite{susyboundary} we required the variation of the action to vanish
and the supersymmetry algebra
\begin{equation}
{[} \delta (\hat\varepsilon_1), \delta (\hat\varepsilon_2){]}X^i=\frac i 2 \hat\varepsilon_1\hat\varepsilon_2\dot X^i,\qquad
{[} \delta (\hat\varepsilon_1), \delta (\hat\varepsilon_2){]} \Psi ^i=\frac i 2 \hat\varepsilon_1\hat\varepsilon_2\dot \Psi ^i
\label{closure}
\end{equation}
to hold {\em on-shell}.  The calculation is rather lengthy but straightforward.

Here we will make the results plausible by ``deriving'' them from the second supersymmetries in the bulk, which look
in $N=(1,1)$ superspace language like
\begin{equation}
\delta \Phi^i=\hat{\varepsilon}_+ J^i{}_j D_{+} \Phi^j + \hat{\varepsilon}_- \bar{J}^i{}_j D_{-} \Phi^j \, ,
\label{bulk2ndsusy}
\end{equation}
where $J$ and $\bar{J}$ satisfy
\begin{equation}
\begin{split}
G_{i(j} \; J^i{}_{k)} = & G_{i(j} \; \bar{J}^i{}_{k)}=0, \\
\nabla^+_k J^i{}_j  = & \nabla^-_k \bar{J}^i{}_j=0 \, ,\label{cs1}
\end{split}
\end{equation}
and
\begin{equation}
J^i{}_j J^j{}_k = \bar{J}^i{}_j \bar{J}^j{}_k = - \delta^i_j\, , \quad N^i{}_{jk}[J,J]=N^i{}_{jk}[\bar{J},\bar{J}]=0 \, ,\label{cs2}
\end{equation}
with the Nijenhuistensor $N[A,B]$ given by
\begin{equation}
N^i{}_{jk}[A,B]=A^l{}_{[j}B^i{}_{k],l} + A^i{}_{l} B^l{}_{[j,k]} + B^l{}_{[j}A^i{}_{k],l} +B^i{}_{l} A^l{}_{[j,k]} \, .\label{cs2b}
\end{equation}
The latter conditions eq.~\eqref{cs2}-\eqref{cs2b} are needed for the second supersymmetry to satisfy the
algebra {\em on-shell} while the former, eq.~\eqref{cs1}, ensures invariance of the bulk action~\eqref{bulksup}.
Eq.~\eqref{cs2} is the condition for $J$ and $\bar{J}$ to be complex structures, while
according to eq.~\eqref{cs1} the metric is hermitian with respect to both complex structures
and the $J$ and $\bar{J}$ are covariantly constant with respect to the {\em different} connections
including {\em torsion} defined in eq.~\eqref{torsieconn}.

Since only one of the supersymmetries \eqref{bulk2ndsusy} will survive on the boundary we
put $\hat{\varepsilon}_+=\hat{\varepsilon}_-=\frac{1}{2}\hat\varepsilon$.
Going over to the $N=1$ language we find from the bottom and top components of eq.~\eqref{bulk2ndsusy} the tensors
in eq.~\eqref{2ndsusy}.
However, there is more freedom in the boundary $N=1$ superspace than in the bulk $N=(1,1)$.
When requiring only on-shell closure one finds that one can add the following transformation
to the second supersymmetry:
\begin{alignat}{2}
\delta X^i & = 0, && \nonumber \\
\delta \Psi^i & = \hat{\varepsilon} K^i{}_j & \left( \frac{1}{2} D \Psi^j - \frac{1}{8} X^j{}'
+ \frac{1}{2} K^{li} \{ljk\} \Psi^j DX^k + \right. & \nonumber \\
& &  \left. \frac{i}{4} K^{li} T_{ljk} \left(\Psi^j \Psi^k + DX^j DX^k \right) \right) \, ,&
\label{eomterm}
\end{alignat}
where $K$ is an arbitrary antisymmetric tensor, $G_{i(j}K^i{}_{k)}=0$.
Note that $K$ in fact multiplies an equation of motion so that this transformation vanishes on-shell.
Using the antisymmetry of $K$ it is also easy to show that the action eq.~\eqref{finac} is invariant under
this additional transformation .

Putting it all together, the most general second supersymmetry transformation looks like eq.~\eqref{2ndsusy}
with
\begin{equation}
\begin{split} {\cal J}_{(1)} & = \frac{1}{2}(J +\bar{J}), \quad i
{\cal J}_{(2)}=-4 {\cal K}_{(2)}= \frac{1}{2}(J-\bar{J}) \, ,\\
{\cal K}_{(1)}&= \frac{1}{2}(J+\bar{J}+K), \quad {\cal K}_{(3)}= -\frac{1}{8}(2J+2\bar{J}+K) \, , \\
{\cal L}^{i}_{(1)jk}& = - \frac{1}{2} \left(\partial_{[j}
J^i{}_{k]}+\partial_{[j} \bar{J}^i{}_{k]}\right)
+ \frac{i}{4} G^{il}\; K^m{}_l \; T_{jkm} \, ,  \\
{\cal L}^{i}_{(2)jk}&=  \frac{i}{2} \left(\partial_{j} J^i{}_{k}+\partial_{j} \bar{J}^i{}_{k}-i G^{il}\;K^m{}_l \;\big\{mjk\big\} \right) \, ,\\
{\cal L}^{i}_{(3)jk} &=  \frac{i}{4} G^{il}\;K^m{}_l \; T_{jkm} \, .\label{2ndsusyb}
\end{split}
\end{equation}
The algebra closes off-shell iff
\begin{itemize}
\item ($[J,\bar{J}]=0$ and $K=0$) or
\item $K=-2J -2 \bar{J}$.
\end{itemize}
So we see that the algebra can be made to close {\em off-shell} even if $[J,\bar{J}]\neq0$.
This has to be contrasted with the case without boundaries where the $N=(2,2)$ algebra only
closes modulo terms proportional to $[J,\bar{J}]$ times an equation of motion. This suggests an $N=2$
superspace description could be possible without introducing further auxiliary fields even if $[J,\bar{J}]\neq 0$.
Unfortunately, we were only able to construct an $N=2$ superspace if $J=\bar{J}$.

We now turn to the boundary term in the supersymmetry variation of
the action. Using an obvious matrix like notation one shows that
this term vanishes provided\footnote{The integrable projection operator
${\cal P}_+$ defines a {\em foliation}, i.e. a set of D-branes
that together fill the whole target space.  We could restrict to
one (or two) of these D-branes and call its (their total)
world-volume $\gamma$. If we require the endpoints of the open
string to lie on the submanifold $\gamma$, the boundary will
always be part of $\gamma$. Conditions (\ref{ibdy}) and
(\ref{compa}) then only hold on $\gamma$. We will not follow this
approach here and require these conditions on the whole of target
space.}
\begin{equation}
\begin{split}
{\cal P}_-(J-\bar J) {\cal P}_- &=0,\\
{\cal P}_+(J-\bar J) {\cal P}_+ &= {\cal P}_+{[}B, J + \bar J{]} {\cal P}_++ {\cal P}_+B\,(J-\bar J)B\, {\cal P}_+,\label{ibdy}
\end{split}
\end{equation}
holds.
Invariance of the boundary conditions, eqs.~(\ref{bn}) and (\ref{bd}), under the $N=2$ supersymmetry transformations requires
\begin{equation}
\begin{split}
{\cal P}_- (J+\bar J) {\cal P}_+ &= - {\cal P}_-(J-\bar J)B {\cal P}_+, \\
{\cal P}_+ (J+\bar J) {\cal P}_- & = {\cal P}_+B(J-\bar J) {\cal P}_-.\label{compa}
\end{split}
\end{equation}
Using the antisymmetry of $J$ and $B$ and the symmetry of $ {\cal R}$, it is clear that the second equation in
eq.~(\ref{compa}) is the transpose of the first one.
It is surprising that conditions (\ref{ibdy}) and (\ref{compa}) are strictly algebraic.  Indeed, carrying
out the straightforward but tedious calculation, one would find that all derivative terms eventually cancel
using the integrability condition~\eqref{integrability}.
Using the above conditions together with the previously obtained equations, we can express $\bar J$ in terms of $J$:
\begin{equation}
\begin{split}
\bar J  = & \; (1+B_{++})^{-1}(1-B_{++})J_{++}(1+B_{++})(1-B_{++})^{-1}+J_{--} \\
& -(1+B_{++})^{-1}(1-B_{++})J_{+-}-
J_{-+}(1+B_{++})(1-B_{++})^{-1} \\
 = & \; {\cal M}J {\cal M}^{-1},\label{btype}
\end{split}
\end{equation}
with
\begin{equation}
{\cal M}= \frac{ {\cal R}-B_{++}}{1+B_{++}},\qquad
{\cal M}^{-1}= \frac{ {\cal R}+B_{++}}{1-B_{++}}.
\end{equation}
In the above we used the notation
$B_{++} = {\cal P}_+B {\cal P}_+$, $J_{-+} = {\cal P}_-J
{\cal P_+}$, etc. Using the conditions involving $J$, it is quite
trivial to show that $\bar J$ is indeed an almost complex
structure with respect to which the metric $G$ is hermitian. However, the
covariant constancy and integrability of $J$ does not imply that
$\bar J$ as given in eq.~(\ref{btype}) is
covariantly constant or integrable. This imposes further, presumably rather complicated,
conditions on the allowed boundary conditions, geometry and torsion!

In the case there is no torsion, $T=0$, we find from eqs.~(\ref{cs1}) and (\ref{cs2}) that the
geometry is K\"ahler.  Without further conditions on the geometry there can be only one independent
K\"ahler form so that $J=\bar{J}$ or $J=-\bar{J}$.
If $J=\bar{J}$, called {\em B-type boundary conditions}, one
finds the results
\begin{equation}
[J, {\cal R}]=[J, B_{++}]=0.
\label{btypeBC}
\end{equation}
This means that both ${\cal R}$ and $B_{++}$ are holomorphic with respect to $J$.
If $J=-\bar{J}$, called {\em A-type boundary
conditions}, one gets
\begin{equation}
B_{++}J_{++}B_{++}=J_{++},\qquad J_{--}=B_{++}J_{+-}=J_{-+}B_{++}=0.
\end{equation}
This implies the existence of a second almost complex structure $\tilde J$,
\begin{equation}
\tilde J = B_{++}J_{++}+J_{+-}+J_{-+},
\end{equation}
which is integrable in the case of a space filling D-brane.
The following relation exists between the dimension of the D-brane and the rank of $B$:
\begin{equation}
{\rm dim}({\text{D-brane}}) = \frac{1}{2} (d + {\rm rank}(B)) \, .
\end{equation}
In the special case $B=0$ also $J_{++}=0$ and the D-brane
world-volume becomes a Lagrangian submanifold.  For a more
detailed treatment we refer to \cite{zab}.

Concluding, we find that a second supersymmetry is allowed provided two almost complex structures, $J$ and $\bar J$, exist that are
separately integrable and covariantly constant, albeit with two different connections. Up to this point, this is exactly equal to the
situation without boundaries. However,
when boundaries are present it turns out that one of the two complex structures can be expressed in terms of the other one and the remainder
of the geometric data.

\subsection{Generalized Boundary Conditions}

Having at our disposal $J$ and $\bar{J}$
we can generalize eq.~\eqref{id1} to
\begin{equation}
\begin{split}
\psi^i_+&= e^{\alpha J} \left[i DX^i+ \Psi ^i\right], \\
\psi^i_-&= \eta \, e^{\pm \alpha \bar{J}} \left[i DX^i- \Psi ^i\right] ,\label{idnew}
\end{split}
\end{equation}
where $\alpha$ is an arbitrary angle. This amounts to applying an R-rotation to the
original $\psi_+$ and $\psi_-$. Using eqs.~(\ref{cs1}) and (\ref{cs2}) one can show that
both possibilities are symmetries of the bulk action. However, only the one with the plus sign survives on the boundary while
the one with the minus sign leads to a new model with the boundary condition, eq.~(\ref{bcfermions}),
replaced by
\begin{equation}
\begin{split}
\psi_- & = \eta \, e^{ \alpha \bar{J}} \, \frac{ {\cal R}-  B_{++}}{1+ B_{++}} \, e^{ \alpha J} \psi_+ \\
       & = eta \, \frac{ {\cal R}-  B_{++}}{1+ B_{++}} \, e^{ 2 \alpha J} \psi_+,
\end{split}
\end{equation}
where we used eq.~(\ref{btype}).

\subsection{$N=2$ Superspace}
The fact that the supersymmetry algebra, eqs.~\eqref{2ndsusy} and \eqref{2ndsusyb}, closes off-shell
makes one hope for the existence of an $N=2$ superspace formulation without the need of introducing further
auxiliary fields.  However, the structure of eqs.~\eqref{2ndsusy} and \eqref{2ndsusyb} shows that the constraints
on the $N=2$ superfields will generically be non-linear. This looks very problematic, so we will limit ourselves to linear
constraints.

We denote the fermionic coordinates of $N=2$ superspace by $ \theta $ and $\bar \theta $ and introduce the fermionic
derivatives $D$ and $\bar D$, which satisfy
\begin{equation}
\{D,\bar D\}=-i \partial_\tau,\quad D^2=\bar D^2=0.
\end{equation}
We now want to introduce superfields that upon integrating out the extra fermionic coordinate reduce to the fields
of the previous sections.
In order to achieve this we introduce the derivative $\hat D$ --- it corresponds to the $D$ in the previous sections ---
and the ``extra'' derivative $\check D$,
\begin{equation}
\hat D = \frac 1 2 \left(D+\bar D\right),\qquad \check D = \frac i 2 \left(D-\bar D\right),
\end{equation}
which satisfy
\begin{equation}
\hat D^2=\check D^2=-\frac i 4 \partial_\tau,\qquad \{\hat D,\check D\}=0.\label{nis2d}
\end{equation}
Furthermore, we introduce the $N=2$ superfields $X^i$ and $ \Psi^i$.
To avoid introducing extra auxiliary degrees of freedom
when passing from $N=2$ to $N=1$ superspace, the $\check D$-derivatives of the fields
should satisfy constraints. The most general {\em linear} constraints one can write down are
\begin{equation}
\begin{split}
\check D X^i & = {\cal C}_1^i{}_j\hat D X^j+ {\cal C}_2^i{}_j \Psi ^j, \\
\check D \Psi ^i & = {\cal C}_3^i{}_j\hat D \Psi ^j+ {\cal C}_4^i{}_j \dot X ^j+ {\cal C}_5^i{}_j X^j{}',\label{suspacon}
\end{split}
\end{equation}
where the $ {\cal C}_n$, $n\in\{1,\cdots ,5\}$ are constant. Eqs.~(\ref{nis2d}) imply the integrability conditions
\begin{equation}
\begin{split}
& {\cal C}_1^2=-{\bf 1}+4i {\cal C}_2 {\cal C}_4,\quad {\cal C}_3^2=-{\bf 1}+4i {\cal C}_4 {\cal C}_2, \\
& {\cal C}_2 {\cal C}_5= {\cal C}_5 {\cal C}_2=0, \\
& {\cal C}_1 {\cal C}_2= {\cal C}_2 {\cal C}_3,\quad
{\cal C}_3 {\cal C}_5= {\cal C}_5 {\cal C}_1,\quad
{\cal C}_3 {\cal C}_4= {\cal C}_4 {\cal C}_1,\label{cint}
\end{split}
\end{equation}
which allow to solve the constraints, eq.~(\ref{suspacon}), in terms of an
unconstrained, fermionic, dimension -1/2 superfield $\Lambda$ and an unconstrained, bosonic, dimension 0
superfield $Y$:
\begin{equation}
\begin{split}
X & = (\check D- {\cal C}_1\hat D)\Lambda+ {\cal C}_2Y, \\
\Psi & = (\check D- {\cal C}_3\hat D)Y+ {\cal C}_4\dot\Lambda+ {\cal C}_5\Lambda'.
\end{split}
\end{equation}
Motivated by the results for the second supersymmetries in the previous section, we propose the following
parameterizations for the tensors $ {\cal C}_n$, $n\in\{1,\cdots
5\}$
\begin{equation}
\begin{split}
 {\cal C}_1&=\frac 1 2 (J+\bar J),\qquad {\cal C}_2=-\frac i 2 (J-\bar J),\qquad {\cal C}_3=\frac 1 2 (J+\bar J+K), \\
 {\cal C}_4&=-\frac 1 8 (J-\bar J),\qquad {\cal C}_5=-\frac 1 8 (2J+2\bar J+K),
\label{suspacon2}
\end{split}
\end{equation}
where $J^2=\bar J^2=-1$.
In order that eq.~(\ref{cint}) is satisfied, one needs
\begin{equation}
K^2=-\{J+\bar J,K\},\qquad 2[J,\bar J]=K(J-\bar J)=(\bar J-J)K.
\end{equation}
This has two obvious solutions:
\begin{equation}
K= -2(J+\bar J)\label{sol1},
\end{equation}
or
\begin{equation}
K=0 \mbox{ and } [J,\bar J]=0.\label{sol2}
\end{equation}
Obviously, eq.~(\ref{suspacon}) with eq.~(\ref{suspacon2}) is,
modulo non-linear terms, identified with the supersymmetry
transformation rules eq.~\eqref{2ndsusy} with eq.~\eqref{2ndsusyb}.

If we stick to linear constraints we read off from eqs.~(\ref{2ndsusy}) and (\ref{2ndsusyb}) that we need to opt for
eq.~(\ref{sol2})! In that case the two commuting integrable
structures $J$ and $\bar{J}$ are simultaneously integrable and diagonalizable \cite{N2troost}. We
choose complex coordinates so that
\begin{equation}
\begin{split}
J^{\alpha}{}_{\beta} & = \bar{J}^{\alpha}{}_{\beta} = i \delta^{\alpha}{}_{\beta},  \quad
J^{\bar{\alpha}}{}_{\bar{\beta}} = \bar{J}^{\bar{\alpha}}{}_{\bar{\beta}} = - i \delta^{\bar{\alpha}}{}_{\bar{\beta}}, \quad
\quad \alpha,\beta \in\{1,\cdots r\}, \\
J^{\mu}{}_{\nu} & = - \bar{J}^{\mu}{}_{\nu} = i \delta^{\mu}{}_{\nu},  \quad
J^{\bar{\mu}}{}_{\bar{\nu}} = - \bar{J}^{\bar{\mu}}{}_{\bar{\nu}} = - i \delta^{\bar{\mu}}{}_{\bar{\nu}}, \quad
\quad \mu,\nu \in\{1,\cdots n\}.
\end{split}
\end{equation}
In these coordinates, where we denote the bosonic superfield now by $Z$, eq.~(\ref{suspacon})
with eq.~(\ref{suspacon2}) takes the form
\begin{equation}
\begin{split}
&\check D Z^ \alpha =+i\,\hat D Z^ \alpha ,\quad \check DZ^{\bar \alpha }=-i\,\hat D Z^{\bar \alpha }, \\
&\check D \Psi^ \alpha =+i\, \hat D\Psi^ \alpha -\frac i 2 Z^ \alpha {}',\quad
\check D \Psi^{\bar \alpha }=-i\,\hat D\Psi^{\bar \alpha }+\frac i 2 Z^{\bar \alpha }{}',
\quad \alpha \in\{1,\cdots r\},
\end{split}
\label{csf}
\end{equation}
or equivalently
\begin{equation}
\bar D Z^ \alpha =D Z^{\bar \alpha }=0,\quad \bar D \Psi^ \alpha=\frac 1 2 Z^ \alpha {}',\quad
D \Psi^{\bar \alpha }= \frac 1 2 Z^{\bar \alpha }{}', \label{chiralsf}
\end{equation}
and
\begin{equation}
\check D Z^ \mu =+\Psi ^ \mu ,\quad \check DZ^{\bar \mu }=-\Psi^{\bar \mu },\quad
\check D\Psi^\mu=-\frac i 4 \dot Z^ \mu ,\quad \check D\Psi^{\bar\mu}=+\frac i 4 \dot Z^{\bar \mu },\quad
\mu \in\{1,\cdots n\}.
\label{tcsf}
\end{equation}
Eqs.~(\ref{csf}) and (\ref{tcsf}) are the boundary analogs of the two-dimensional chiral and twisted
chiral superfields respectively.

We will consider the case where only one type of superfields is present.  Having only chiral (twisted chiral)
superfields results in a
K\"ahler geometry with B(A)-type supersymmetry. Taking exclusively chiral superfields ($n=0$), we introduce
two potentials $K(Z,\bar Z)$ and $V(Z, \bar Z)$ and the action
\begin{equation}
{\cal S}_{\text{B}} =
\int d^2 \sigma d^2 \theta\, K(Z,\bar Z)_{ ,\alpha \bar \beta }\left(-2i D Z^ \alpha \bar D Z^ {\bar \beta }-8i
\Psi^ \alpha \Psi^{\bar \beta }\right)+
\int d \tau d^2 \theta\, V(Z,\bar Z).
\end{equation}
Passing to $N=1$ superspace one gets the action \eqref{finac} with
\begin{equation}
\begin{split}
&G_{ \alpha \bar \beta }=K_{ ,\alpha \bar \beta },\quad B_{ \alpha \bar \beta }=-\frac 1 2 V_{, \alpha \bar \beta },\\
&G_{ \alpha  \beta }=G_{\bar \alpha \bar \beta }=0,\quad B_{ \alpha \beta }= B_{\bar \alpha \bar \beta }=0.
\end{split}
\end{equation}
Solving the constraints in terms of the unconstrained superfields $ \Lambda $ and $Y$,
\begin{equation}
Z^ \alpha =\bar D \Lambda ^ \alpha ,\quad Z^{\bar \alpha }= D \Lambda ^{\bar \alpha },\quad
\Psi ^ \alpha = \bar D Y^ \alpha + \frac 1 2 \Lambda ^ \alpha {}',\quad
\Psi^{\bar \alpha }=DY^{\bar \alpha }+ \frac 1 2 \Lambda ^{\bar \alpha }{}',
\end{equation}
and varying the action with respect to these unconstrained superfields, we get the boundary term
\begin{equation}
\int d \tau  d^2 \theta \, \left(\delta \Lambda ^ \alpha \left(-4i K_{, \alpha \bar \beta }\Psi^{\bar \beta }+
V_{, \alpha \bar \beta }\bar DZ^{\bar \beta }\right)
+\delta \Lambda ^{\bar \alpha} \left(4i K_{, \bar\alpha  \beta }\Psi^{ \beta }+
V_{, \bar\alpha \beta }D Z^{\beta }\right)
\right).
\end{equation}
To make the boundary term vanish, we introduce an almost product structure $ {\cal R}$ that satisfies
\begin{equation}
\begin{split}
&{\cal R}^ \alpha {}_{ \bar \beta } = {\cal R}^{\bar \alpha }{}_ \beta =0, \\
& {\cal R}_{ \alpha \bar \beta } =
{\cal R}_{\bar \beta \alpha  } .
\end{split}
\end{equation}
Using the almost product structure to construct projection operators $ {\cal P}_+$ and $ {\cal P}_-$, we find that
the boundary term in the variation indeed vanishes if we impose
\begin{equation}
\begin{split}
&{\cal P}_-^{ \alpha }{}_{ \gamma } \delta \Lambda ^{ \gamma }=
{\cal P}_-^{ \bar\alpha }{}_{\bar \gamma } \delta \Lambda ^{\bar \gamma }=0, \\
& {\cal P}_+^{ \alpha }{}_{ \beta }\left( \Psi^{ \beta }-\frac i 4 G^{ \beta \bar \gamma }V_{, \bar \gamma \delta }D Z^{ \delta }\right) =
{\cal P}_+^{\bar \alpha }{}_{ \bar\beta }\left( \Psi^{\bar \beta }+\frac i 4 G^{\bar \beta  \gamma }V_{, \gamma \bar\delta }\bar D Z^{\bar \delta }
\right)=0.
\end{split}
\end{equation}
Demanding compatibility of the first two equations with ${\cal P}_-^ \alpha {}_{ \beta } \delta Z^ \beta  = {\cal P}_-^ {\bar \alpha} {}_{ \bar \beta }\delta Z^{\bar \beta}  =0$ requires
\begin{equation}
{\cal P}^ \alpha _+{}_{ [\delta , \bar \varepsilon] } {\cal P}^ \delta _+{}_ \beta {\cal P}_+^{\bar \varepsilon }{}_{\bar \gamma }={\cal P}^{\bar \alpha} _+{}_{ [\bar \delta , \varepsilon] } {\cal P}^{\bar \delta} _+{}_ {\bar \beta} {\cal P}_+^{ \varepsilon }{}_{ \gamma }=0.
\end{equation}
 Finally from $D Z^ \alpha = {\cal P}^ \alpha _+{}_ \beta D Z^ \beta $ and $D^2=0$ and likewise from $\bar D Z^ {\bar \alpha} = {\cal P}^{\bar \alpha} _+{}_ {\bar \beta} \bar D Z^ {\bar \beta} $ and $\bar D^2=0$, we get
\begin{equation}
{\cal P}^ \alpha _+{}_{ [\delta , \varepsilon ]} {\cal P}^ \delta _+{}_ \beta {\cal P}_+^{ \varepsilon }{}_{ \gamma }={\cal P}^{\bar \alpha} _+{}_{ [\bar \delta , \bar \varepsilon] } {\cal P}^{\bar \delta} _+{}_ {\bar \beta} {\cal P}_+^{ \bar \varepsilon }{}_{\bar \gamma }=0.
\end{equation}
The conditions obtained here are completely equivalent to those in eqs.~(\ref{c1}-\ref{c3}) and (\ref{btypeBC}).

We briefly turn to the case where we take exclusively twisted chiral superfields ($r=0$). The action would be:
\begin{equation}
{\cal S}_{\text{A}}=\int d^2 \sigma d^2 \theta \left(-8K_{,\mu\bar\nu}\Psi^{\mu}\hat D Z^{\bar\nu}+
8K_{,\bar\mu\nu}\Psi^{\bar\mu}\hat D Z^{\nu}
+ 2 K_{,\mu} Z^{\mu'} - 2 K_{,\bar{\mu}} Z^{\bar{\mu}'} \right).
\end{equation}
Although it correctly reproduces the bulk theory, it introduces an extra boundary term when comparing to eq.~\eqref{finac} while
at the same time it seems to be impossible to introduce a non-zero $B$ at the boundary. In other words, the $N=2$ superspace
description of type A boundary conditions remains unknown.

Concluding, the $N=2$ description of type A boundary conditions remains unknown, let alone mixed boundary conditions.
Nevertheless, the action for type B boundary conditions in flat space-time, but with a non-trivial electromagnetic field,
\begin{equation}
{\cal S}_{\text{B}} = \int d^2 \sigma d^2 \theta\, \sum_\alpha\left(-2i D Z^ \alpha \bar D Z^ {\bar \alpha }-8i
\Psi^ \alpha \Psi^{\bar \alpha }\right)+
\int d \tau d^2 \theta\, V(Z,\bar Z),\label{finalaction}
\end{equation}
with the chiral superfields defined in eq.~\eqref{chiralsf} is a good starting point for the
study of the $\beta$-functions directly in superspace.  The easiest would be to take Neumann boundary conditions
in all directions. The simplicity of eq.~\eqref{finalaction} indicates that a systematic study of the derivative
corrections to the effective action might be possible.

\chapter{Concluding Remarks}
\label{conclusions}

The main theme of this work is the construction of the D-brane
effective action.  In the abelian case and in the limit of constant
field strengths this action is already known for a long time to all
orders in $\alpha'$: it is the Born-Infeld action. When several branes
coincide the gauge group becomes non-abelian and much less is know
about the effective action i.e.\ only the first few orders.
The introductory chapter \ref{landscape} gives an overview of past attempts
and successes in constructing more of the effective action.

In this thesis we proposed a new method for constructing the effective
action and applied it to the abelian case with derivative corrections
and to the non-abelian case.  The method is based on an extension of the
self-duality equation in four-dimensional Super-Yang-Mills theory.
When the field strength is self-dual, the equations of motion are
automatically solved. Moreover, these self-dual configurations leave some
supersymmetry unbroken and saturate a Bogomolny bound; we call them
{\em BPS states}. In more than four dimensions the BPS equations turn out to be
the equations found by Corrigan et al.\ \cite{shm}.
We argued that also for the unknown complete D-brane effective action, which is a perturbation
of Yang-Mills theory, there must exist a deformation of the BPS equations so that
the equations of motion are automatically solved. This constraint is actually
enough to construct both the effective action and the deformed BPS equations.
We introduced these BPS equations in chapter \ref{equations} and generalized
them to Born-Infeld theory, demonstrating the saturation of the Bogomolny bound
and the vanishing of the supersymmetry variation for certain spinors.

If a deformation of the Yang-Mills action is supersymmetric there probably exists
a BPS equation as well.  We think that our method is in fact a shortcut to the Noether
method, which tries to construct the effective action by requiring supersymmetry.
Although more involved, the Noether method has one advantage over our algorithm in that
the fermionic terms are also determined.  The role of the supersymmetry transformation
in the Noether method is played by the BPS equation in our method.

In chapter \ref{nonabelianBI} we showed how we constructed the non-abelian D-brane effective action
up to order $\alpha'{}^4$. The result at order $\alpha'{}^3$ has in the
meantime been checked by a number of other groups in a variety of different
ways \cite{groningen}\cite{brasil}\cite{sym1}\cite{sym2}\cite{sym3}.
Furthermore, as we discussed in chapter \ref{checks}, we checked that the spectrum of intersecting branes
calculated with the effective action is exactly what we expected from string theory.
This check has also been extended to our $\alpha'{}^4$ result \cite{testalpha4}\cite{naga1}.

At order $\alpha'{}^3$ there was one coefficient left undetermined: the overall constant at this order. From
the 4-point scattering amplitude in string theory we learned that it contains $\zeta(3)$, which
is shown by Ap\'ery  \cite{zeta3} to be irrational.  Since our method can only produce rational
numbers at this order, we could never have found this coefficient. From the point of view of the
Noether method an undetermined coefficient signifies that there is an independent supersymmetric invariant.
Moreover, from the 4-point amplitude we learn that at every odd order $\alpha'{}^{2r+1}$ a factor of $\zeta(2r+1)$
appears. As we discussed at the end of chapter \ref{nonabelianBI}, this implies that either there will start a new independent
supersymmetric invariant at each odd order or either --- more boldly --- that there should exist relations with
rational coefficients between $\frac{\zeta(2r+1)}{\pi^{2r+1}}$ for different $r$ or, in the most extreme case, that all
$\frac{\zeta(2r+1)}{\pi^{2r+1}}$ are rational.  Although this is rather unlikely, it would be very interesting
to check explicitly! The first occurrence
would be at order $\alpha'{}^5$ where $\zeta(3)$ and $\pi^2 \zeta(5)$ come together.
Unfortunately, we lack the computer power at the moment.

With respect to the undetermined coefficients there is an interesting difference between the abelian
and the non-abelian case.  In the abelian case, there are a lot more of them. For instance, the overall
coefficient of the terms with four derivatives,
determined by Wyllard \cite{wyllard}, is left unfixed. Indeed, in the slowly varying field strength limit these terms
are even scaled away and the abelian Born-Infeld action is supersymmetric by itself.  In the non-abelian case terms
with a different number of derivatives become connected and this coefficient is fixed at precisely the right
value predicted by the string 4-point amplitude. It cannot be scaled away as in the abelian case
which means that, assuming that our method is indeed equivalent to the Noether method,
the symmetrized trace Born-Infeld action is {\em not supersymmetric} on its own!

In the non-abelian case a large part of the problem are the derivative corrections. In fact, the
analysis of chapter \ref{nonabelianBI} shows that the ordering ambiguity of the field
strengths can be done away with at the price of introducing more covariant derivatives.
A playground where we can study derivative terms in a simpler setting is the
abelian case.  In that sense the result in section \ref{abeliander}
is encouraging. There we made use of the result of \cite{wyllard} where
the abelian effective action up to four derivatives was calculated to all orders in
$\alpha'$.  We showed that also in this case BPS equations can be constructed that automatically
solve the equations of motion. Although the calculation is very involved, the end result for the BPS
equations can be written in a compact and suggestive form.  We have good hope that a better
understanding of this result could help to refine and support the conjecture made in \cite{wyllard2}
for corrections with more derivatives. From the form of the BPS equations we could also work back and try to
construct the supersymmetry transformation. In the end we hope to write down an explicitly $N=2$
supersymmetry invariant action and establish $\kappa$-symmetry with derivative corrections.
Furthermore, we think that we can relatively easily extend the calculation at the end of section~\ref{abeliander}
to the non-abelian case. This would lead to a partial all-order result in $\alpha'$, i.e. one with four covariant derivatives.
This would be orthogonal to the result in \cite{brasil2} which studies the superstring 4-point amplitude and thus generates
terms with four field strengths. However, one should be very careful with these results because, as explained
in section~\ref{nacomplexity}, in the non-abelian case it is difficult to find a suitable limit in which the terms one has constructed
stay large while the others do not.

In this thesis we focused on the BPS equations themselves as a way to construct the effective action, for the most
part neglecting solutions.
Only in chapter \ref{solutions} we gave (already known) examples of configurations satisfying these equations.
There we also described work in progress on the construction of a D-brane configuration corresponding to the octonionic
instanton.

In chapter \ref{boundarysusy} we changed tack and introduced an $N=2$ boundary superspace which
gives the initial impetus to use another method for the construction of the abelian effective action,
namely the Weyl invariance method. We focus on the abelian case. The aim is to calculate the $\beta$-functions directly in superspace since
we expect the calculation to be simpler than in $X$-space. The $\beta$-functions can be interpreted as (equivalent to) equations of
motion and we can construct the corresponding effective action.

Summarizing, in this thesis we calculated the non-abelian D-brane effective action up to order $\alpha'{}^4$.
The results at this order are however so involved that an all-order expression
still seems out of reach. On the other hand,
the fact that it could be constructed from such simple principles
as requiring the BPS solutions or requiring supersymmetry, and that it is
expected to have a lot more properties like electromagnetic
duality, Seiberg-Witten duality and a simple spectrum seems to be
tantalizing evidence for a deeper underlying structure.
A large part of the problem are the derivative corrections so that
one could try to find all-order results in the abelian case
with derivative corrections and hope that these could be extended
in some way to the non-abelian case.

\newpage
\thispagestyle{empty}

\appendix

\chapter{Conventions}
\label{conventions}
\def\theequation{\thechapter.\arabic{equation}}

Throughout the text $m$ will denote the order in $\alpha'$ and $q$
the number of derivatives. We will use square brackets, $[$ and $]$,
for antisymmetrization of indices and round brackets, $($ and $)$,
for symmetrization. Our convention is that we sum over all permutations
and divide by the number of permutations so that $[[i_1,\cdots,i_n]]=[i_1,\cdots,i_n]$.
Space-time coordinate indices are labelled by $\mu, \nu, \rho, \ldots \in \{0,\ldots,9\}$.
Our Minkowski metric is ``mostly
plus''. Spatial coordinates are denoted by $i,j,k,\ldots \in \{1,\ldots,9\}$. String
world-sheet coordinates are labelled by $a,b \in \{0,1\}$. $a,b,c \in \{0,\ldots,p\}$ are also
used for D-brane world-volume coordinates if the distinction with target
space coordinates is important.
In fact, this is only the case in section \ref{DBIWZ} while
in the rest of the text the indices $\mu,\nu,\rho,\ldots$ (now running
from $0,\cdots,p$) and the indices $i,j,k$ (running from $1,\cdots p$)
are used for the D-brane world-volume coordinates. $p$ will denote the
(spatial) dimension of the D-branes. The dimensions of the D-branes will
often be even in which case we define $k$ such that $p=2k$ and we
frequently use complex coordinates $z^\alpha $, $\alpha
\in\{1,\ldots k\}$:
\begin{equation}
z^\alpha = \frac{1}{\sqrt 2}\left(x^{2\alpha -1}+ix^{2\alpha
}\right),\quad
\bar z^{\bar\alpha} = \frac{1}{\sqrt 2}
\left(x^{2\alpha -1}-ix^{2\alpha }\right).
\label{cc}
\end{equation}
In flat space the metric is then $g_{\alpha \beta}=g_{\bar\alpha\bar\beta}=0$,
$g_{\alpha \bar\beta}=\delta_{\alpha \bar\beta}$.  Consider a $2k \times 2k$
matrix $A$. If this matrix is holomorphic, $A_{\alpha\beta}=A_{\bar{\alpha}\bar{\beta}}=0$,
it is useful to define the complexification $A^c$ as a $k \times k$ matrix
with components
\begin{equation}
\left(A^c\right)_{\alpha\beta} = A_{\alpha\bar{\beta}} \, .
\label{complexification}
\end{equation}

We introduce $\Gamma$ matrices:
\begin{equation}
\{\Gamma_{\mu},\Gamma_{\nu}\}=2 \eta_{\mu\nu} \, .
\end{equation}
Our convention is such that the $\Gamma$ matrices with spatial indices are hermitian:
$\left(\Gamma^i\right)^{\dagger}=\Gamma^i$.
In complex coordinates they read
\begin{equation}
\Gamma_{\alpha} = \frac{1}{\sqrt{2}} (\Gamma_{2\alpha-1} - i \Gamma_{2\alpha}), \quad
\Gamma_{\bar{\alpha}} = \frac{1}{\sqrt{2}} (\Gamma_{2\alpha-1} + i \Gamma_{2\alpha}),
\end{equation}
and satisfy
\begin{equation}
\{ \Gamma_{\alpha}, \Gamma_{\beta} \} = \{ \Gamma_{\bar{\alpha}}, \Gamma_{\bar{\beta}} \} = 0, \qquad
\{ \Gamma_{\alpha}, \Gamma_{\bar{\beta}} \} = 2 \delta_{\alpha\bar{\beta}} \, .
\end{equation}
With this algebra the $\Gamma_{\alpha}$ can be considered as fermionic annihilation
operators while $\Gamma_{\bar{\alpha}}$ are the creation operators.
We define the vacuum $|0 \rangle$ as the state annihilated by all $\Gamma_{\alpha}$:
\begin{equation}
\Gamma_{\alpha} |0\rangle =0, \qquad \forall \alpha=1,\ldots k \, ,
\end{equation}
and satisfying $\langle 0|0\rangle=1$. The completely filled state is then
defined as
\begin{equation}
|\!\uparrow \rangle =2^{-\frac{k}{2}} \Gamma_{\bar{1}}\cdots\Gamma_{\bar{k}}|0\rangle,
\end{equation}
and is normalized so that $\langle\uparrow\!|\!\uparrow\rangle=1$.

Indices $A,B,\ldots \in \{1,\ldots,N\}$ run over the gauge indices where
the gauge group is $U(N)$. We denote the corresponding algebra with $u(N)$.
We choose anti-hermitian matrices $T_A$ for the generators of the algebra.
They satisfy the orthonormality condition
\begin{equation}
\label{generatortrace}
\Tr T_A T_B = -\delta_{AB},
\end{equation}
and the algebra
\begin{equation}
[T_A,T_B] = f_{ABC} T_C,
\end{equation}
where $f_{ABC}$ are the completely antisymmetric structure constants.
We define $A_{\mu} = A_{\mu}^A T_A$ and $F=F_{\mu\nu}^A T_A$, which are
thus anti-hermitian. However, when working with abelian gauge theories, i.e.\ in section \ref{BPSBI}
and chapter \ref{abelianBI}, it will be more convenient to work with a real gauge
field.
Furthermore the field strength and covariant
derivative are given by
\begin{equation}
\begin{split}
F_{\mu\nu} & = \partial_\mu A_\nu-\partial_\nu A_\mu+{[}A_\mu,A_\nu{]}, \\
D_\mu\cdot & = \partial_\mu\cdot+{[}A_\mu,\cdot{]}.
\end{split}
\end{equation}
Note that we scaled our $A_{\mu}$ field so that the coupling $g_{\text{YM}}$
does not appear in the above formulae, but rather as an overall factor in the
Yang-Mills action as is the custom in the string theory literature.

For the superspace conventions see section \ref{superspaceconventions}.

\chapter{Samenvatting}
\selectlanguage{dutch}
\label{samenvatting}

Snaartheorie is \'e\'en van de weinige kandidaten voor een kwantumtheorie van
de zwaartekracht.  Bovendien heeft ze het voordeel dat ze niet alleen gravitatie
maar ook ijktheorie"en, waarmee in het Standaardmodel de andere drie krachten worden
beschreven, op een natuurlijke manier omvat.  Het spectrum van de gesloten snaar bevat
inderdaad (onder andere) een graviton terwijl het spectrum van de open snaar een massaloos
ijkboson bevat, de drager van ijktheorie"en. Op dit moment is snaartheorie dus {\em de} kandidaat
voor unificatie.

Snaartheorie in de strikte zin is echter geen volledige theorie: in feite
is het slechts een auxiliaire kwantumveldentheorie die toelaat om perturbatieve verstrooiingsamplitudes
van een onderliggende theorie te berekenen. Vanuit onze ervaring met kwantumveldentheorie weten
we echter dat perturbatieve informatie alleen niet het volledige plaatje geeft. Sinds 1995 is men erin
geslaagd beetje bij beetje een tip van de sluier op te lichten en heeft men allerlei manieren gevonden
om niet-perturbatieve informatie te onttrekken onder andere via D-branen en S-dualiteit. Daardoor zijn de meeste
snaartheoreten ervan o\-ver\-tuigd geraakt dat de dieperliggende theorie wel degelijk bestaat en heeft ze zelf een naam
gekregen: M-theorie. Behalve de connectie met snaartheorie en de lage-energie limiet (supergravitatie)
is er echter weinig over bekend. De fundamentele vrij\-heids\-gra\-den zijn ook geen snaren meer.

Of M-theorie inderdaad de wer\-ke\-lijk\-heid beschrijft en geen louter wiskundige constructie is, valt af
te wachten. Vermits de energieschaal van deze theorie zo ontzettend hoog is dat ze moeilijk bereikt kan
worden via deeltjesversneller, is de meeste hoop op experimentele verificatie tegenwoordig gericht op haar
kos\-mo\-lo\-gische en astrofysische voorspellingen. Ook als wiskundige constructie hebben snaartheorie en M-theorie
hun verdienste. Zo hebben ze geleid tot een beter inzicht in ijktheorie"en en zijn er tal van connecties
met gebieden uit de wiskunde.

Deze thesis handelt over het massaloze vector boson in het spectrum van de open snaar.
We trachten de effectieve actie op te stellen voor dit deeltje.  De effectieve actie wordt bekomen
door te integreren over alle deeltjes in lussen, zowel de massieve als massaloze. Het komt erop neer
dat de Feynman diagrammen van de effectieve actie zonder lusdiagrammen de S-matrix van snaartheorie moeten
reproduceren.  Verder zitten de eindpunten van een open snaar vast op een hypervlak, de D-braan.
Vermits alle velden in de open-snaar effectieve actie op deze D-braan leven, spreekt men ook van de D-braan
effectieve actie. De lage-energie limiet van de effectieve actie is de Yang-Mills actie.
Deze effectieve actie moet allerlei interessante eigenschappen hebben zoals onder andere supersymmetrie, bestaan
van supersymmetrische oplossingen, electromagnetische
dualiteit, Seiberg-Witten dualiteit en T-dualiteit, zodat de vraag is in welke mate deze actie door al die
eigenschappen wordt vastgelegd.

Hoofdstuk \ref{introduction} bevat een voorwoord over de status
van snaartheorie en niet-perturbatieve effecten en
vervolgens de Engelse versie van deze samenvatting.

Hoofdstuk \ref{landscape} begint met een brede inleiding tot
snaartheorie. Het bespreekt eerst perturbatieve snaartheorie"en
en behandelt dan niet-perturbatieve effecten met de nadruk op
D-branen. Het laatste deel van dit hoofdstuk leidt het eigenlijke
onderwerp van de thesis in: de lage-energie effectieve actie
van een enkele D-braan (het abelse geval) en van meerdere samenvallende
D-branen (het niet-abelse geval). Het geeft een overzicht van de pogingen
en successen in het verleden tot constructie van de effectieve acties.
De methode die in deze thesis ontwikkeld wordt, is gebaseerd op een uitbreiding
van de instanton vergelijkingen.

Hoofdstuk \ref{equations} bouwt geleidelijk op. Het begint met de vertrouwde instanton
ver\-ge\-lij\-kin\-gen in vier-dimensionale Yang-Mills theorie en breidt die eerst uit tot meer
dan vier dimensies.  Daarbij blijkt dat er twee belangrijke gevallen zijn: het complexe
en het octonionische geval. In het complex geval vinden we als ver\-ge\-lij\-king de conditie
voor het holomorf zijn van de Yang-Mills bundel en een vergelijking die de DUY conditie
genoemd wordt.
Het nieuwe werk in dit hoofdstuk begint wanneer deze vergelijkingen
op hun beurt worden uitgebreid van Yang-Mills tot Born-Infeld theorie.

In hoofdstuk \ref{solutions} bekijken we bestaande oplossingen van deze vergelijkingen en tonen
we aan dat die een interpretatie hebben vanuit D-braan fysica. Daarom is het redelijk om te eisen
dat deze configuraties ook een oplossing moeten vormen van de volledige D-braan effectieve
actie, die we op deze manier hopen op te stellen. De configuratie van snijdende D-branen zal later ook
belangrijk zijn om te dienen als een test op de door onze geconstrueerde actie. We merken op
dat er ook oplossingen bestaan die geen limiet hebben voor kleine velden --- zij zijn dus geen oplossing
van de Yang-Mills theorie --- en we bespreken kort werk in de steigers aan de octonionische
veralgemening van het BIon.

Hoofdstuk \ref{abelianBI} behandelt de aanpak van \cite{uniqueabelian}, waarin de abelse Born-Infeld
actie opnieuw werd afgeleid door te eisen dat oplossingen van vergelijkingen die een deformatie
vormen van de DUY conditie, de bewegingsvergelijkingen moeten oplossen. Vertrekkende
van de actie uit \cite{wyllard}, leiden we ook de termen in de DUY conditie met 4 afgeleiden af tot
op alle ordes in $\alpha'$.

Hoofdstuk \ref{nonabelianBI} vat artikels \cite{alpha3} en \cite{alpha4} samen, waar de niet-abelse
D-braan effectieve actie werd opgesteld tot op orde $\alpha'{}^4$. Het computerprogramma dat geschreven
werd om de zware berekeningen uit te voeren, vindt men op de bijgevoegde CD-ROM, samen met een beknopte
handleiding \cite{manual} en een uitgebreid staal van de output. In het laatste stuk van dit hoofdstuk
weiden we uit over de betekenis van de co"effici"enten die we niet konden vastleggen en speculeren we over
de mo\-ge\-lijk\-heid om relaties met rationale co"effici"enten te vinden tussen de $\zeta(2r+1)/\pi^{2r+1}$
waarbij $r$ integer en $\zeta$ de Riemann zeta functie.

In hoofdstuk \ref{checks} beschrijven we de spectrum test, uitgevoerd in \cite{testalpha3} en \cite{testfermion},
op zowel de bosonische als de fermionische  sector. Die laatste werd gevonden door de Groningen
groep in \cite{groningen}. De test is gebaseerd op het spectrum van snaren die eindigen op snijdende D-branen.

Terwijl al deze hoofdstukken tot hiertoe een geheel vormen, lijkt hoofdstuk \ref{boundarysusy} wat af te wijken van
de rode draad. Toch is het ultieme doel van de inspanningen in dat hoofdstuk weer de constructrie van de D-braan
effectieve actie, maar nu via de oude methode van de Weyl invariantie van het niet-lineaire $\sigma$-model. We willen
in $N=2$ superspace werken omdat we verwachten dat de berekeningen dan gemakkelijker zijn. Aangezien er in het verleden
een aantal problemen rezen om een superspace op te stellen wanneer er randen zijn, moesten we dit eerst uitpluizen, wat
precies het onderwerp vormt van artikel \cite{susyboundary} en dit hoofdstuk.

Alle artikels waarvan ik coauteur ben, worden aangeduid met letters terwijl andere gerefereerde artikels
aangeduid worden met getallen. \cite{corfutalk} is een bijdrage tot de proceedings van een workshop in Korfoe, waar
ik een lezing gaf, en vat artikels \cite{uniqueabelian}, \cite{alpha3} en \cite{testalpha3} samen.
\selectlanguage{english}
\newpage
\thispagestyle{empty}


\begin{thebibliography}{999}
\bibitem[{a}]{uniqueabelian} L.~De Foss\'e, P.~Koerber and A.~Sevrin, {\em The uniqueness of the abelian Born-Infeld action},
\npb{603}{2001}{413}, \hepth{0103015}.
\bibitem[{b}]{alpha3} P.~Koerber and A.~Sevrin, {\em The non-abelian open superstring effective action through order $\alpha'^3$},
\jhep{0110}{2001}{003}, \hepth{0108169}.
\bibitem[{c}]{testalpha3} P.~Koerber and A.~Sevrin, {\em Testing the $\alpha'^3$ term in the non-abelian open superstring effective action},
\jhep{0109}{2001}{009}, \hepth{0109030}.
\bibitem[{d}]{corfutalk} P.~Koerber and A.~Sevrin, {\em Getting the D-brane effective action from BPS configurations},
Fortsch.\ Phys.\ {\bf 50} (2002) 923, \hepth{0112210}.
\bibitem[{e}]{testfermion} M.~de Roo, M.G.C.~Eenink, P.~Koerber and A.~Sevrin,
{\em Testing the fermionic terms in the non-abelian D-brane
effective action through order $\alpha'{}^3$},
\jhep{0208}{2002}{011}, \hepth{0207015}.
\bibitem[{f}]{alpha4} P.~Koerber and A.~Sevrin, {\em The non-abelian D-brane effective action through order $\alpha'{}^4$},
\jhep{0210}{2002}{046}, \hepth{0208044}.
\bibitem[{g}]{susyboundary} P.~Koerber, S.~Nevens and A.~Sevrin, {\em Supersymmetric non-linear $\sigma$-models with boundaries revisited},
\jhep{0311}{2003}{066}, \hepth{0309229}.
\bibitem[{h}]{manual} P.~Koerber, {\em Short manual of {\tt CalculateAction} and related tools}, Manual
of the Java program written to construct the D-brane effective
action. It can be found on {\href{http://tena4.vub.ac.be/ThesisOnEA}{\tt http://tena4.vub.ac.be/ThesisOnEA}} or on the
included CD-ROM.
\bibitem{wyllard} N.~Wyllard, {\em Derivative corrections to D-brane actions with constant
background fields}, \npb{598}{2001}{247}, \hepth{0008125}.
\bibitem{groningen} A.~Collinucci, M.~de Roo and M.G.C.~Eenink,
{\em Supersymmetric Yang-Mills theory at order $\alpha'{}^3$}, \jhep{0206}{2002}{024},
\hepth{0205150}.
\bibitem{bookGSW} M.~Green, J.~Schwarz and E.~Witten, {\em Superstring Theory Vol. 1 and 2}, Cambridge
University Press (Cambridge 1987).
\bibitem{bookpolchinski} J.~Polchinski, {\em String Theory Vol.~I and II}, Cambridge University Press
(Cambridge 1998).
\bibitem{cvj} C.V.~Johnson, {\em D-brane primer}, \hepth{0007170};
C.V.~Johnson, {\em D-branes}, Cambridge University Press (Cambridge 2002).
\bibitem{polchinski} J.~Polchinski, {\em Dirichlet-branes and Ramond-Ramond charges},
\prl{75}{1995}{4724}, \hepth{9510017}.
\bibitem{nambugoto} Y.~Nambu, {\em Duality and hydrodynamics}, lectures at the Copenhagen symposium,
unpublished (1970); T.~Goto, {\em Relativistic quantum mechanics of one-dimensional mechanical
continuum and subsidiary condition of dual resonance model},
\ptp{46}{1971}{1560}.
\bibitem{dz} S.~Deser and B.~Zumino, {\em A complete action for the spinning string},
\plb{65}{1976}{369}.
\bibitem{bdh} L.~Brink, P.~Di Vecchia and P.S.~Howe,
{\em A locally supersymmetric and reparametrization invariant action for the spinning string},
\plb{65}{1976}{471}.
\bibitem{polyakov} A.M.~Polyakov, {\em Quantum geometry of bosonic strings},
\plb{103}{1981}{207}; A.M.~Polyakov, {\em Quantum geometry of fermionic strings},
\plb{103}{1981}{211}.
\bibitem{ginsparg} P.~Ginsparg, {\em Applied conformal field theory},
lectures at Les Houches Summer School (1988), \hepth{9108028}.
\bibitem{fms} D.~Friedan, E.J.~Martinec and S.H.~Shenker,
{\em Conformal invariance, supersymmetry and string theory}, \npb{271}{1986}{93}.
\bibitem{wittenSFT}
E.~Witten, {\em Noncommutative geometry and string field theory},
\npb{268}{1986}{253}.
\bibitem{tachyon} A.~Sen, {\em Universality of the tachyon potential},
\jhep{9912}{1999}{027}, \hepth{9911116};
A.~Sen and B.~Zwiebach, {\em Tachyon condensation in string field theory},
\jhep{0003}{2000}{002}, \hepth{9912249};
For a review see W.~Taylor and B.~Zwiebach, {\em D-branes, tachyons, and string field theory},
lectures at TASI 2001, \hepth{0311017}.
\bibitem{GSO} F.~Gliozzi, J.~Scherk and D.I.~Olive,
{\em Supergravity and the spinor dual model}, \plb{65}{1976}{282};
F.~Gliozzi, J.~Scherk and D.I.~Olive,
{\em Supersymmetry, supergravity theories and the dual spinor model},
\npb{122}{1977}{253}.
\bibitem{CPfactors} J.E.~Paton and H.M.~Chan,
{\em Generalized Veneziano model with isospin},
\npb{10}{1969}{516};
J.H.~Schwarz,
{\em Gauge groups for type I superstrings},
{\it Proc. of the Johns Hopkins Workshop on Current Problems in High-Energy Particle Theory 6}, Florence (1982), p.~233;
N.~Marcus and A.~Sagnotti,
{\em Tree level constraints on gauge groups for type I superstrings},
\plb{119}{1982}{97}.
\bibitem{neveuscherk} A.~Neveu and J.~Scherk,
{\em Connection between Yang-Mills fields and dual models},
\npb{36}{1972}{155}.
\bibitem{RNS} P.~Ramond, {\em Dual theory for free fermions},
\prd{3}{1971}{2415};
A.~Neveu and J.H.~Schwarz, {\em Factorizable dual model of pions},
\npb{31}{1971}{86}.
\bibitem{heterotic}
D.J.~Gross, J.A.~Harvey, E.J.~Martinec and R.~Rohm,
{\em The heterotic string},
\prl{54}{1985}{502};
D.J.~Gross, J.A.~Harvey, E.J.~Martinec and R.~Rohm,
{\em Heterotic string theory (I). The free heterotic string},
\npb{256}{1985}{253};
D.J.~Gross, J.A.~Harvey, E.J.~Martinec and R.~Rohm,
{\em Heterotic string theory  (II). The interacting heterotic string},
\npb{267}{1986}{75}.
\bibitem{hoker} For a review see
E.~D'Hoker and D.H.~Phong,
{\em Lectures on two-loop superstrings},
\hepth{0211111}.
\bibitem{mandelstam}
S.~Mandelstam, {\em The n-loop string amplitude: explicit formulas, finiteness and absence of ambiguities},
\plb{277}{1992}{82}. For an accessible discussion see chapter 9 of \cite{bookpolchinski}.
\bibitem{fluxes} For a review see
A.R.~Frey,
{\em Warped strings: self-dual flux and contemporary compactifications},
\hepth{0308156}. See also references therein.
\bibitem{douglas}
M.R.~Douglas, {\em The statistics of string/M theory vacua},
\jhep{0305}{2003}{046}, \hepth{0303194}.
\bibitem{anthropic}
T.~Banks, M.~Dine and E.~Gorbatov,
{\em Is there a string theory landscape?},
\hepth{0309170}.
\bibitem{greenschwartz}
M.B.~Green and J.H.~Schwarz,
{\em Covariant description of superstrings},
\plb{136}{1984}{367}.
\bibitem{berkovitsaction}
N.~Berkovits, {\em Super-Poincar\'e covariant quantization of the superstring},
\jhep{0004}{2000}{018}, \hepth{0001035}.
For a review see N.~Berkovits,
{\em ICTP lectures on covariant quantization of the superstring},
published in {\em Trieste 2002, Superstrings and related matters} p.~57-107, \hepth{0209059}.
\bibitem{vannieuwenhuizenaction}
For a review see P.A.~Grassi, G.~Policastro and P.~van Nieuwenhuizen,
{\em An introduction to the covariant quantization of superstrings},
\cqg{20}{2003}{S395}, \hepth{0302147}. See also references therein.
\bibitem{grossperiwal}
D.J.~Gross and V.~Periwal,
{\em String perturbation theory diverges},
\prl{60}{1988}{2105}.
S.H.~Shenker, {\em The strength of non-perturbative effects in string theory},
presented at the {\it Carg\`ese Workshop on Random Surfaces, Quantum Gravity and Strings}, Carg\`ese, France (1990).
\bibitem{tdualityreview}
For a review see
A.~Giveon, M.~Porrati and E.~Rabinovici,
{\em Target space duality in string theory},
\prep{244}{1994}{77}, \hepth{9401139}.
\bibitem{firstDbranes}
J.~Dai, R.G.~Leigh and J.~Polchinski,
{\em New connections between string theories},
\mpla{4}{1989}{2073}.
\bibitem{dinstantons}
J.~Polchinski,
{\em Combinatorics of boundaries in string theory},
\prd{50}{1994}{6041}, \hepth{9407031}.
\bibitem{deuclidean}
K.~Becker, M.~Becker and A.~Strominger,
{\em Five-branes, membranes and nonperturbative string theory},
\npb{456}{1995}{130}, \hepth{9507158}.
\bibitem{wittenolive}
E.~Witten and D.I.~Olive,
{\em Supersymmetry algebras that include topological charges},
\plb{78}{1978}{97}.
\bibitem{bogo} E.B.~Bogomolny,
{\em Stability of classical solutions},
{\it Sov.\ J.\ Nucl.\ Phys.\ } {\bf 24} (1976) 449,
[{\it Yad.\ Fiz.\ } {\bf 24} (1976) 861].
Reprinted in Rebbi, C. (ed.), Soliani, G. (ed.): {\em Solitons and Particles}, p.~389-394.
\bibitem{wittenmatrixval}
E.~Witten, {\em Bound states of strings and p-branes},
\npb{460}{1996}{335}, \hepth{9510135}.
\bibitem{myersreview} For a review see
R.C.~Myers, {\em Non-abelian phenomena on D-branes},
\cqg{20}{2003}{S347}, \hepth{0303072}.
\bibitem{dgeometry}
M.R.~Douglas, {\em D-branes in curved space},
\atmp{1}{1998}{198}, \hepth{9703056};
M.R.~Douglas, {\em D-branes and matrix theory in curved space},
\npps{68}{1998}{381}, \hepth{9707228};
M.R.~Douglas, A.~Kato and H.~Ooguri,
{\em D-brane actions on K\"ahler manifolds},
\atmp{1}{1998}{237}, \hepth{9708012}.
\bibitem{raamsdonk}
W.~Taylor and M.~Van Raamsdonk,
{\em Multiple D0-branes in weakly curved backgrounds},
\npb{558}{1999}{63}, \hepth{9904095};
W.~Taylor and M.~Van Raamsdonk,
{\em Multiple Dp-branes in weak background fields},
\npb{573}{2000}{703}, \hepth{9910052}.
\bibitem{deboer} J.~de Boer and K.~Schalm,
{\em General covariance of the non-abelian DBI-action},
\jhep{0302}{2003}{041}, \hepth{0108161};
J.~de Boer, K.~Schalm and J.~Wijnhout,
{\em General covariance of the non-abelian DBI-action: checks and balances},
\hepth{0310150}.
\bibitem{GSanomaly}
M.B.~Green and J.H.~Schwarz,
{\em Infinity cancellations in SO(32) superstring theory}, \plb{151}{1985}{21}.
\bibitem{pbranes}
G.T.~Horowitz and A.~Strominger,
{\em Black strings and $p$-branes},
\npb{360}{1991}{197}.
\bibitem{pbranescat}
S.S.~Gubser, A.~Hashimoto, I.R.~Klebanov and J.M.~Maldacena,
{\em Gravitational lensing by $p$-branes},
\npb{472}{1996}{231}, \hepth{9601057};
M.R.~Garousi and R.C.~Myers,
{\em Superstring scattering from D-branes},
\npb{475}{1996}{193}, \hepth{9603194}.
\bibitem{bion1}
C.G.~Callan and J.M.~Maldacena,
{\em Brane dynamics from the Born-Infeld action},
\npb{513}{1998}{198}, \hepth{9708147}.
\bibitem{bion2}
G.W.~Gibbons,
{\em Born-Infeld particles and Dirichlet $p$-branes},
\npb{514}{1998}{603}, \hepth{9709027}.
\bibitem{diaconescu}
D.E.~Diaconescu,
{\em D-branes, monopoles and Nahm equations},
\npb{503}{1997}{220}, \hepth{9608163}.
\bibitem{maldacenaconj}
J.M.~Maldacena,
{\em The large N limit of superconformal field theories and supergravity},
\atmp{2}{1998}{231}, \hepth{9711200}. For a review see
O.~Aharony, S.S.~Gubser, J.M.~Maldacena, H.~Ooguri and Y.~Oz,
{\em Large N field theories, string theory and gravity},
\prep{323}{2000}{183}, \hepth{9905111}.
\bibitem{thooftlargeN}
G.~'t Hooft,
{\em A planar diagram theory for strong interactions},
\npb{72}{1974}{461}.
\bibitem{dijkgraaf}
R.~Dijkgraaf and C.~Vafa,
{\em A perturbative window into non-perturbative physics},
\hepth{0208048}.
\bibitem{dijkgraafproof}
R.~Dijkgraaf, M.T.~Grisaru, C.S.~Lam, C.~Vafa and D.~Zanon,
{\em Perturbative computation of glueball superpotentials},
\plb{573}{2003}{138}, \hepth{0211017};
F.~Cachazo, M.R.~Douglas, N.~Seiberg and E.~Witten,
{\em Chiral rings and anomalies in supersymmetric gauge theory},
\jhep{0212}{2002}{071}, \hepth{0211170}.
\bibitem{seibergwitten}
N.~Seiberg and E.~Witten,
{\em String theory and noncommutative geometry},
\jhep{9909}{1999}{032}, \hepth{9908142}.
\bibitem{nonanti}
J.~de Boer, P.A.~Grassi and P.~van Nieuwenhuizen,
{\em Non-commutative superspace from string theory},
\plb{574}{2003}{98}, \hepth{0302078}.
\bibitem{wittenduality} E.~Witten,
{\em String theory dynamics in various dimensions},
\npb{443}{1995}{85}, \hepth{9503124}.
\bibitem{hulltownsend} C.M.~Hull and P.K.~Townsend,
{\em Unity of superstring dualities},
\npb{438}{1995}{109}, \hepth{9410167}.
\bibitem{montonenolive}
C.~Montonen and D.I.~Olive, {\em Magnetic monopoles as gauge particles?},
\plb{72}{1977}{117}.
\bibitem{osborn}
H.~Osborn,
{\em Topological charges for N=4 supersymmetric gauge theories and monopoles of spin 1},
\plb{83}{1979}{321}.
\bibitem{emduality}
G.W.~Gibbons and D.A.~Rasheed,
{\em Electric-magnetic duality rotations in nonlinear electrodynamics},
\npb{454}{1995}{185}, \hepth{9506035};
M.B.~Green and M.~Gutperle, {\em Comments on three-branes},
\plb{377}{1996}{28}, \hepth{9602077}.
\bibitem{tseytlinselfdual}
A.A.~Tseytlin, {\em Self-duality of Born-Infeld action and Dirichlet 3-brane of type IIB superstring theory},
\npb{469}{1996}{51}, \hepth{9602064}.
\bibitem{matrixtheory}
T.~Banks, W.~Fischler, S.H.~Shenker and L.~Susskind,
{\em M theory as a matrix model: a conjecture},
\prd{55}{1997}{5112}, \hepth{9610043}.
\bibitem{BornInfeld}
M.~Born and L.~Infeld,
{\em Foundations of the new field theory},
\prsla{144}{1934}{425}.
\bibitem{dirac}
P.A.M.~Dirac, {\em An extensible model of the electron},
\prsla{268}{1962}{57}.
\bibitem{groningenTdual}
E.~Bergshoeff and M.~De Roo,
{\em D-branes and T-duality},
\plb{380}{1996}{265}, \hepth{9603123}.
\bibitem{GHTWZ}
M.B.~Green, C.M.~Hull and P.K.~Townsend,
{\em D-brane Wess-Zumino actions, T-duality and the cosmological constant},
\plb{382}{1996}{65}, \hepth{9604119}.
\bibitem{douglasWZ} M.R.~Douglas, {\em Branes within branes},
\hepth{9512077}.
\bibitem{li}
M.~Li, {\em Boundary states of D-branes and Dy-Strings},
\npb{460}{1996}{351}, \hepth{9510161}.
\bibitem{alvarezTdual}
E.~Alvarez, J.L.F.~Barbon and J.~Borlaf,
{\em T-duality for open strings},
\npb{479}{1996}{218}, \hepth{9603089}.
\bibitem{greenWZ}
M.B.~Green, J.A.~Harvey and G.W.~Moore,
{\em I-brane inflow and anomalous couplings on D-branes},
\cqg{14}{1997}{47}, \hepth{9605033};
Y.K.~Cheung and Z.~Yin, {\em Anomalies, branes, and currents},
\npb{517}{1998}{69}, \hepth{9710206}.
\bibitem{curvaturecorr}
C.P.~Bachas, P.~Bain and M.B.~Green,
{\em Curvature terms in D-brane actions and their M-theory origin},
\jhep{9905}{1999}{011}, \hepth{9903210}.
\bibitem{thorlacius}
L.~Thorlacius,
{\em Born-Infeld string as a boundary conformal field theory},
\prl{80}{1998}{1588}, \hepth{9710181}.
\bibitem{fradkintseytlin1}
E.S.~Fradkin and A.A.~Tseytlin,
{\em Nonlinear electrodynamics from quantized strings},
\plb{163}{1985}{123}.
\bibitem{ACNY}
A.~Abouelsaood, C.G.~Callan, C.R.~Nappi and S.A.~Yost,
{\em Open strings in background gauge fields},
\npb{280}{1987}{599}.
\bibitem{leigh} R.G.~Leigh,
{\em Dirac-Born-Infeld action from Dirichlet $\sigma$-model},
\mpla{4}{1989}{2767}.
\bibitem{susyBIcederwall}
M.~Cederwall, A.~von Gussich, B.E.W.~Nilsson and A.~Westerberg,
{\em The Dirichlet super-three-brane in ten-dimensional type IIB  supergravity},
\npb{490}{1997}{163}, \hepth{9610148};
M.~Cederwall, A.~von Gussich, B.E.W.~Nilsson, P.~Sundell and A.~Westerberg,
{\em The Dirichlet super-p-branes in ten-dimensional type IIA and IIB supergravity},
\npb{490}{1997}{179}, \hepth{9611159}.
\bibitem{susyBIaganagic}
M.~Aganagic, C.~Popescu and J.H.~Schwarz,
{\em D-brane actions with local kappa symmetry},
\plb{393}{1997}{311}, \hepth{9610249};
M.~Aganagic, C.~Popescu and J.H.~Schwarz,
{\em Gauge-invariant and gauge-fixed D-brane actions},
\npb{495}{1997}{99}, \hepth{9612080}.
\bibitem{susyBIbergshoeff}
E.~Bergshoeff and P.K.~Townsend, {\em Super D-branes},
\npb{490}{1997}{145}, \hepth{9611173}.
\bibitem{bilal1}
A.~Bilal, {\em Higher-derivative corrections to the non-abelian Born-Infeld action},
\npb{618}{2001}{21}, \hepth{0106062}.
\bibitem{abelian4derivative} O.D.~Andreev and A.A.~Tseytlin, {\em Partition function representation for the open
superstring effective action: cancellation of M\"obius infinities and derivative corrections to
Born-Infeld Lagrangian}, \npb{311}{1988}{205}.
\bibitem{wyllard2} N.~Wyllard,
{\em Derivative corrections to the D-brane Born-Infeld action: non-geodesic embeddings and the Seiberg-Witten map},
\jhep{0108}{2001}{027}, \hepth{0107185}.
\bibitem{groningen3} M.~de Roo, M.G.C.~Eenink,
{\em The effective action for the 4-point functions in abelian open superstring theory},
\jhep{0308}{2003}{036}, \hepth{0307211}.
\bibitem{brasil2} O.~Chandia and R.~Medina,
{\em 4-point effective actions in open and closed superstring theory},
\jhep{0311}{2003}{003}, \hepth{0310015}.
\bibitem{grosswitten}
D.J.~Gross and E.~Witten,
{\em Superstring modifications of Einstein's equations},
\npb{277}{1986}{1}.
\bibitem{tseytlinvectorfield}
A.A.~Tseytlin, {\em Vector field effective action in the open superstring theory},
\npb{276}{1986}{391}, Erratum \npb{291}{1987}{876}.
\bibitem{tseytlinSTr}
A.A.~Tseytlin, {\em On non-abelian generalisation of the Born-Infeld action in string  theory},
\npb{501}{1997}{41}, \hepth{9701125}.
\bibitem{HTspectrum}
A.~Hashimoto and W.~Taylor, {\em Fluctuation spectra of tilted and intersecting D-branes from the
Born-Infeld action},
\npb{503}{1997}{193}, \hepth{9703217};
\bibitem{DST}
F.~Denef, A.~Sevrin and J.~Troost,
{\em Non-abelian Born-Infeld versus string theory},
\npb{581}{2000}{135}, \hepth{0002180}.
\bibitem{kitazawa}
Y.~Kitazawa, {\em Effective Lagrangian for open superstring from five-point function},
\npb{289}{1987}{599}.
\bibitem{sym1} D.T.~Grasso, {\em Higher order contributions to the effective action
of $N=4$ super Yang-Mills}, \jhep{0211}{2002}{012}, \hepth{0210146}.
\bibitem{sym2} A.~Refolli, A.~Santambrogio, N.~Terzi and D.~Zanon, {\em $F^5$
contributions to the non-abelian Born-Infeld action from a supersymmetric
Yang-Mills five-point function}, \npb{613}{2001}{64}; Erratum
\npb{648}{2003}{453}, \hepth{0105277}.
\bibitem{brasil} R.~Medina, F.T.~Brandt, F.R.~Machado,
{\em The open superstring 5-point amplitude revisted},
\jhep{0207}{2002}{071}, \hepth{0208121}.
\bibitem{testalpha4}
A.~Sevrin and A.~Wijns,
{\em Higher order terms in the non-abelian D-brane effective action and  magnetic background fields},
\jhep{0308}{2003}{059}, \hepth{0306260}.
\bibitem{Smatrix}
J.~Scherk and J.~H.~Schwarz, {\em Dual models for nonhadrons},
\npb{81}{1974}{118};
T.~Yoneya, {\em Connection of dual models to electrodynamics and gravidynamics},
\ptp{51}{1974}{1907}.
\bibitem{bergshoeffnoether}
E.~Bergshoeff, M.~Rakowski and E.~Sezgin,
{\em Higher derivative super Yang-Mills theories},
\plb{185}{1987}{371}.
\bibitem{metsaevsusy}
R.R.~Metsaev and M.A.~Rakhmanov,
{\em Fermionic terms in the open superstring effective action},
\plb{193}{1987}{202}.
\bibitem{groningensevrin}
E.A.~Bergshoeff, A.~Bilal, M.~de Roo and A.~Sevrin,
{\em Supersymmetric non-abelian Born-Infeld revisited},
\jhep{0107}{2001}{029}, \hepth{0105274}.
\bibitem{fradkintseytlin0}
E.S.~Fradkin and A.A.~Tseytlin,
{\em Effective field theory from quantized strings},
\plb{158}{1985}{316};
E.S.~Fradkin and A.A.~Tseytlin,
{\em Quantum string theory effective action},
\npb{261}{1985}{1};
E.S.~Fradkin and A.A.~Tseytlin,
{\em Effective action approach to superstring theory},
\plb{160}{1985}{69}.
\bibitem{tseytlinsubtle}
A.A.~Tseytlin,
{\em Ambiguity in the effective action in string theories},
\plb{176}{1986}{92};
A.A.~Tseytlin,
{\em Renormalization of M\"obius infinities and partition function representation
for the string theory effective action},
\plb{202}{1988}{81}.
\bibitem{boundarystate} For a review see
P.~Di Vecchia and A.~Liccardo,
{\em D branes in string theory, I},
lectures at NATO-ASI on {\em Quantum Geometry} in Akureyri, Iceland (1999),
\hepth{9912161};
P.~Di Vecchia and A.~Liccardo,
{\em D-branes in string theory, II},
lectures at the YITP Workshop on {\em Developments in Superstring and M-theory}, Kyoto, Japan (1999),
\hepth{9912275}.
\bibitem{tseytlin2loop}
R.R.~Metsaev and A.A.~Tseytlin,
{\em Two-loop $\beta$-function for the generalized bosonic sigma model},
\plb{191}{1987}{354}.
\bibitem{weylloop}
C.G.~Callan, C.~Lovelace, C.R.~Nappi and S.A.~Yost,
{\em String loop corrections to $\beta$-functions},
\npb{288}{1987}{525}.
\bibitem{FS}
W.~Fischler and L.~Susskind,
{\em Dilaton tadpoles, string condensates and scale invariance II},
\plb{173}{1986}{262}.
\bibitem{groningen2} A.~Collinucci, M.~de Roo, M.G.C.~Eenink,
{\em Derivative corrections in 10-dimensional super-Maxwell theory},
\jhep{0301}{2003}{039}, \hepth{0212012}.
\bibitem{goteborg}
M.~Cederwall, B.E.W.~Nilsson and D.~Tsimpis,
{\em The structure of maximally supersymmetric Yang-Mills theory:  constraining higher-order corrections},
\jhep{0106}{2001}{034}, \hepth{0102009};
M.~Cederwall, B.E.W.~Nilsson and D.~Tsimpis,
{\em D = 10 super-Yang-Mills at ${\cal O}(\alpha'^2)$},
\jhep{0107}{2001}{042}, \hepth{0104236}.
\bibitem{sym3} J.M.~Drummond, P.J.~Heslop, P.S.~Howe,
{\em Integral invariants in $N=4$ SYM and the effective action for coincident D-branes},
\jhep{0308}{2003}{016}, \hepth{0305202}.
\bibitem{nonabeliankappa}
E.A.~Bergshoeff, M.~de Roo and A.~Sevrin,
{\em Non-abelian Born-Infeld and kappa-symmetry}, \jmp{42}{2001}{2872}, \hepth{0011018}.
\bibitem{cornalba}
L.~Cornalba, {\em The general structure of the non-abelian Born-Infeld action},
\atmp{4}{2002}{1259}, \hepth{0006018}.
\bibitem{stieberger}
S.~Stieberger and T.R.~Taylor,
{\em Non-abelian Born-Infeld action and type I - heterotic duality (I): heterotic
$F^6$ terms at two loops},
\npb{647}{2002}{49}, \hepth{0207026};
S.~Stieberger and T.R.~Taylor,
{Non-abelian Born-Infeld action and type I - heterotic duality (II):
nonrenormalization theorems},
\npb{648}{2003}{3}, \hepth{0209064}.
\bibitem{BPSTmonopole}
A.A.~Belavin, A.M.~Polyakov, A.S.~Shvarts and Y.S.~Tyupkin,
{\em Pseudoparticle solutions of the Yang-Mills equations},
\plb{59}{1975}{85}.
\bibitem{prasad}
M.K.~Prasad and C.M.~Sommerfield,
{\em An exact classical solution for the 't Hooft monopole and the Julia-Zee Dyon},
\prl{35}{1975}{760}.
\bibitem{hooftmono}
G.~'t Hooft,
{\em Magnetic monopoles in unified gauge theories},
\npb{79}{1974}{276}.
\bibitem{polmono}
A.~M.~Polyakov,
{\em Particle spectrum in quantum field theory},
JETP Lett.\  {\bf 20} (1974) 194.
\bibitem{olivealgebra}
D.I.~Olive,
{\em The electric and magnetic charges as extra components of four-momentum},
\npb{153}{1979}{1}.
\bibitem{shm} E.~Corrigan, C.~Devchand, D.B.~Fairlie and J.~Nuyts, {\em First order equations for
gauge fields in spaces of dimension greater than four}, \npb{214}{1983}{452}.
\bibitem{bak}
D.~Bak, K.~Lee and J.-H.~Park,
{\em BPS equations in six and eight dimensions},
\prd{66}{2002}{025021}, \hepth{0204221}.
\bibitem{acharyaBPS}
B.S.~Acharya and M.~O'Loughlin,
{\em Self-duality in D $\le$ 8-dimensional Euclidean gravity},
\prd{55}{1997}{4521},
\hepth{9612182}.
\bibitem{egh}
T.~Eguchi, P.B.~Gilkey and A.J.~Hanson,
{\em Gravitation, gauge theories and differential geometry},
\prep{66}{1980}{213}.
\bibitem{duy} K.~Uhlenbeck and S.T.~Yau,
{\em On the existence of hermitian Yang-Mills connections on stable vectorbundles},
{\it Comm. Pure Appl. Math.} {\bf 39} (1986) 257 and
{\em A note on our previous paper: on the existence of hermitian Yang-Mills connections on
stable vectorbundles}, {\it Comm. Pure Appl. Math.} {\bf 42} (1989) 703;
S.K.~Donaldson, {\em Infinite determinants, stable bundles and curvature},
{\it Duke Math. J.} {\bf 54} (1987) 231; see also chapter 15 in the second volume of \cite{bookGSW}.
\bibitem{octonions} For a review see J.C.~Baez, {\em The octonions}, \Math{RA}{0105155}.
\bibitem{angles2} E.~Bergshoeff, R.~Kallosh, T.~Ortin and G.~Papadopoulos,
{\em $\kappa$-symmetry, supersymmetry and intersecting branes},
\npb{502}{1997}{149}, \hepth{9705040}.
\bibitem{angles1}  M.~Berkooz, M.R.~Douglas, R.~Leigh,
{\em Branes intersecting at angles}, \npb{480}{1996}{265}, \hepth{9606139}.
\bibitem{jabbari}
H.~Arfaei and M.M.~Sheikh Jabbari, {\em Different D-brane
interactions}, \plb{394}{1997}{288}, \hepth{9608167}; M.M.~Sheikh
Jabbari, {\em Classification of different branes at angles},
\plb{420}{1998}{279}, \hepth{9710121}.
\bibitem{moduliangles} V.~Balasubramanian and R.G.~Leigh,
{\em D-branes, moduli and supersymmetry},
\prd{55}{1997}{6415}, \hepth{9611165}.
\bibitem{Mbrane}
J.P.~Gauntlett, G.W.~Gibbons, G.~Papadopoulos and P.K.~Townsend,
{\em Hyper-K\"ahler manifolds and multiply intersecting branes},
\npb{500}{1997}{133}, \hepth{9702202};
P.K.~Townsend, {\em M-branes at angles},
\npps{67}{1998}{88}, \hepth{9708074};
N.~Ohta and P.K.~Townsend,
{\em Supersymmetry of M-branes at angles},
\plb{418}{1998}{77}, \hepth{9710129}.
\bibitem{wittenBPS} 
B.~Chen, H.~Itoyama, T.~Matsuo and K.~Murakami,
{\em $p-p'$ system with $B$ field, branes at angles and noncommutative geometry},
\npb{576}{2000}{177}, \hepth{9910263};
M.~Mihailescu, I.Y.~Park and T.A.~Tran,
{\em D-branes as solitons of an $N=1$, $D=10$ non-commutative gauge theory},
\prd{64}{2001}{046006}, \hepth{0011079};
E.~Witten, {\em BPS bound states of D0-D6 and D0-D8
systems in a $B$-field}, \jhep{0204}{2002}{012}, \hepth{0012054}.
\bibitem{boundaryfield} C.G.~Callan, C.~Lovelace, C.R.~Nappi and S.A.~Yost,
{\em Loop corrections to superstring equations of motion},
\npb{308}{1988}{221}.
\bibitem{derrick}
G.H.~Derrick, {\em Comments on nonlinear wave equations as models for
elementary particles},
\jmp{5}{1964}{1252}.
\bibitem{zeta3} R.~Ap\'ery, {\em Irrationalit\'e de $\zeta(2)$ et $\zeta(3)$},
{\it Ast\'erisque} {\bf 61} (1979) 11-13.
\bibitem{hashtaylor}
K.~Hashimoto and W.~Taylor,
{\em Strings between branes},
\jhep{0310}{2003}{040}, \hepth{0307297}.
\bibitem{spectrumtroost}
J.~Troost, {\em Constant field strengths on $T^{2n}$},
\npb{568}{2000}{180} \hepth{9909187}.
\bibitem{spectrumbaal} P.~van Baal,
{\em $SU(N)$ Yang-Mills solutions with constant field strength on $T^4$},
\cmp{94}{1984}{397}
and {\em Some results for $SU(N)$ gauge fields on the hypertorus},
\cmp{85}{1982}{529}.
\bibitem{STT}
A.~Sevrin, J.~Troost and W.~Troost,
{\em The non-abelian Born-Infeld action at order $F^6$},
\npb{603}{2001}{389}, \hepth{0101192}.
\bibitem{uranga} A.M.~Uranga, {\em Chiral four-dimensional string compactifications with
intersecting D-branes}, \cqg{20}{2003}{S373}, \hepth{0301032}.
\bibitem{blumenhagen}
R.~Blumenhagen, {\em String unification of gauge couplings with intersecting D-branes},
contributed to {\em Proc. of 2nd International Conference on String Phenomenology}, Durham (2003),
\hepth{0309146}.
\bibitem{epplelust}
F.~Epple and D.~Lust, {\em Tachyon condensation for intersecting branes at small and large angles},
\hepth{0311182}.
\bibitem{hana1}
K.~Hashimoto and S.~Nagaoka,
{\em Recombination of intersecting D-branes by local tachyon condensation},
\jhep{0306}{2003}{034}, \hepth{0303204}.
\bibitem{naga2}
S.~Nagaoka,
{\em Higher dimensional recombination of intersecting D-branes},
\jhep{0402}{2004}{063}, \hepth{0312010}.
\bibitem{naga1}
S.~Nagaoka,
{\em Fluctuation analysis of non-abelian Born-Infeld action in the background of
intersecting D-branes}, \ptp{110}{2004}{1219}, \hepth{0307232}.
\bibitem{weylbergshoeff}
E.~Bergshoeff, E.~Sezgin, C.N.~Pope and P.K.~Townsend,
{\em The Born-Infeld action from conformal invariance of the open superstring},
\plb{188}{1987}{70}.
\bibitem{N2studies}
B.~Zumino,
{\em Supersymmetry and K\"ahler manifolds},
\plb{87}{1979}{203};
L.~Alvarez-Gaume and D.Z.~Freedman,
{\em Geometrical structure and ultraviolet finiteness in the supersymmetric $\sigma$-model},
\cmp{80}{1981}{443};
S.J.~Gates, C.M.~Hull and M.~Ro\u{c}ek,
{\em Twisted multiplets and new supersymmetric nonlinear $\sigma$-models},
\npb{248}{1984}{157};
P.S.~Howe and G.~Sierra,
{\em Two-dimensional supersymmetric nonlinear $\sigma$-models with torsion},
\plb{148}{1984}{451};
C.M.~Hull, {\em Lectures on nonlinear $\sigma$-models and strings},
lectures given at {\it Super Field Theories Workshop}, Vancouver, Canada (1986);
M.~Ro\u{c}ek, K.~Schoutens and A.~Sevrin,
{\em Off-shell WZW models in extended superspace},
\plb{265}{1991}{303};
I.T.~Ivanov, B.~Kim and M.~Ro\u{c}ek,
{\em Complex structures, duality and WZW models in extended superspace},
\plb{343}{1995}{133}, \hepth{9406063};
J.~Bogaerts, A.~Sevrin, S.~van der Loo and S.~Van Gils,
{\em Properties of semi-chiral superfields},
\npb{562}{1999}{277}, \hepth{9905141}.
\bibitem{N2troost}
A.~Sevrin and J.~Troost,
{\em Off-shell formulation of N = 2 non-linear $\sigma$-models},
\npb{492}{1997}{623}, \hepth{9610102}.
\bibitem{boundsusy}
H.~Ooguri, Y.~Oz and Z.~Yin,
{\em D-branes on Calabi-Yau spaces and their mirrors},
\npb{477}{1996}{407}, \hepth{9606112};
K.~Hori, A.~Iqbal and C.~Vafa,
{\em D-branes and mirror symmetry},
\hepth{0005247};
J.~Borlaf and Y.~Lozano,
{\em Aspects of T-duality in open strings},
\npb{480}{1996}{239}, \hepth{9607051};
P.~Haggi-Mani, U.~Lindstr\"om and M.~Zabzine,
{\em Boundary conditions, supersymmetry and $A$-field coupling for
 an open string in a $B$-field background},
\plb{483}{2000}{443}, \hepth{0004061}.
\bibitem{stock1} C.~Albertsson, U.~Lindstr\"om, M.~Zabzine, {\em N = 1 supersymmetric $\sigma$-model with boundaries, I},
\cmp{233}{2003}{403}, \hepth{0111161}.
\bibitem{stock2} C.~Albertsson, U.~Lindstr\"om, M.~Zabzine, {\em N = 1 supersymmetric $\sigma$-model with boundaries, II},
\npb{678}{2004}{295}, \hepth{0202069}.
\bibitem{zab} U.~Lindstr\"om, M.~Zabzine, \emph{N = 2 boundary conditions for nonlinear sigma models and
Landau-Ginzburg models}, \jhep{0302}{2003}{006}, \hepth{0209098}.
\bibitem{zab2}
U.~Lindstr\"om and M.~Zabzine,
{\em D-branes in N = 2 WZW models},
\plb{560}{2003}{108}, \hepth{0212042}.
\bibitem{lrn}
U.~Lindstr\"om, M.~Ro\u{c}ek and P.~van Nieuwenhuizen,
{\em Consistent boundary conditions for open strings},
\npb{662}{2003}{147}, \hepth{0211266}.
\bibitem{stock3}
C.~Albertsson, U.~Lindstr\"om and M.~Zabzine,
{\em Superconformal boundary conditions for the WZW model},
\jhep{0305}{2003}{050}, \hepth{0304013}.
\bibitem{oliverjoanna}
O.~DeWolfe, D.Z.~Freedman and H.~Ooguri,
{\em Holography and defect conformal field theories},
\prd{66}{2002}{025009}, \hepth{0111135};
J.~Erdmenger, Z.~Guralnik and I.~Kirsch,
{\em Four-dimensional superconformal theories with interacting boundaries or defects},
\prd{66}{2002}{025020}, \hepth{0203020}.
\end{thebibliography}
\end{document}